%% file: thesis.tex
\documentclass[bibliography=totoc, a4paper,11pt,twoside,numbers=noenddot]{scrbook}

\usepackage{a4wide}
\usepackage{fancyhdr}
\usepackage{graphicx}
\usepackage{amsmath}
\usepackage{amssymb}
\usepackage{accents}
\usepackage{hyperref}
\usepackage{subcaption}
\usepackage{exscale}
\usepackage{latexsym}
\usepackage{amsfonts}
\usepackage{amsbsy}
\usepackage{listings}
\usepackage{epstopdf}
\graphicspath{{./figures/},{./}}

\usepackage[american]{babel}
\usepackage{lscape}
\usepackage{wasysym} 
\newcommand{\mathsym}[1]{{}}

\usepackage{sidecap}
\usepackage[utf8]{inputenc}
\usepackage{pbox}
\usepackage{xcolor}
\definecolor{light-gray}{gray}{0.7}

\newcommand{\beq}{\begin{equation}}
\newcommand{\eeq}{\end{equation}}
\newcommand{\beqn}{\begin{eqnarray}}
\newcommand{\eeqn}{\end{eqnarray}}
\def\bea#1\eea{\begin{align}#1\end{align}}
\newcommand{\nn}{\nonumber}

\newcommand{\rmd}{\mathrm{d}}
\newcommand{\rmt}{\mathrm{t}}
\newcommand{\tu}{\tilde{u}}
\newcommand{\tih}{\tilde{h}}
\newcommand{\tD}{\tilde{\Delta}}
\newcommand{\tZ}{\tilde{Z}}
\newcommand{\tX}{\tilde{X}}
\newcommand{\tY}{\tilde{Y}}
\newcommand{\tPsi}{\tilde{\Psi}}
\newcommand{\tF}{\tilde{F}}
\newcommand{\hu}{\hat{u}}
\newcommand{\du}{\dot{u}}
\newcommand{\sgn}{\text{sgn}}
\newcommand{\dw}{\dot{w}}
\newcommand{\df}{\dot{f}}
\newcommand{\ti}{{t_{\rm i}}}
\newcommand{\tf}{{t_{\rm f}}}
\newcommand{\tm}{{t_{\rm m}}}
\newcommand{\tmp}{{t'_{\rm m}}}
\newcommand{\xf}{{x_{\rm f}}}
\newcommand{\xxi}{{x_{\rm i}}}
\newcommand{\xm}{{x_{\rm m}}}
\newcommand{\Rdr}{\mathbb{R}}
\newcommand{\hc}{{\hat{c}}}
\newcommand{\hP}{{\hat{P}}}
\newcommand{\hJ}{{\hat{J}}}
\newcommand{\hF}{{\hat{F}}}
\newcommand{\hD}{{\hat{D}}}
\newcommand{\mO}{\mathcal{O}}
\newcommand{\mD}{\mathcal{D}}
\newcommand{\hsh}{\hat{\mathfrak{s}}}
\newcommand{\mfs}{{\mathfrak{s}}}
\newcommand{\uD}{\underline{\Delta}}
\newcommand{\uu}{\underline{u}}
\newcommand{\ulx}{\underline{x}}
\newcommand{\ult}{\underline{t}}
\newcommand{\ueta}{\underline{\eta}}
\newcommand{\usigma}{\underline{\sigma}}
\newcommand{\utu}{\underline{\tilde{u}}}
\newcommand{\udu}{\underline{\dot{u}}}
\newcommand{\uhu}{\underline{\hat{u}}}
\newcommand{\tree}{\mathrm{tree}}
\newcommand{\onel}{\mathrm{1-loop}}
\newcommand{\dtree}{\mathrm{tree,dim}}

\newcommand{\dsf}{\mathsf{d}}

\renewcommand{\chaptermark}[1]%
         {\markboth{\thechapter.\ #1}{}}

\renewcommand{\sectionmark}[1]%
         {\markright{\thesection\ #1}}

\newcommand{\LMUTitle}[9]{

  \thispagestyle{empty}

  \vspace*{\stretch{1}}

  {\parindent0cm

   \rule{\linewidth}{.7ex}}

  \begin{flushright}

    \vspace*{\stretch{1}}

    \sffamily\bfseries\LARGE #1\\

    \vspace*{\stretch{0.5}}

    \large #9\\

    \vspace*{\stretch{1}}

    \sffamily\bfseries\large

    #2

    \vspace*{\stretch{1}}

  \end{flushright}

  \rule{\linewidth}{.7ex}

  \vspace*{\stretch{5}}

  \begin{center}

\includegraphics[width=1.5in]{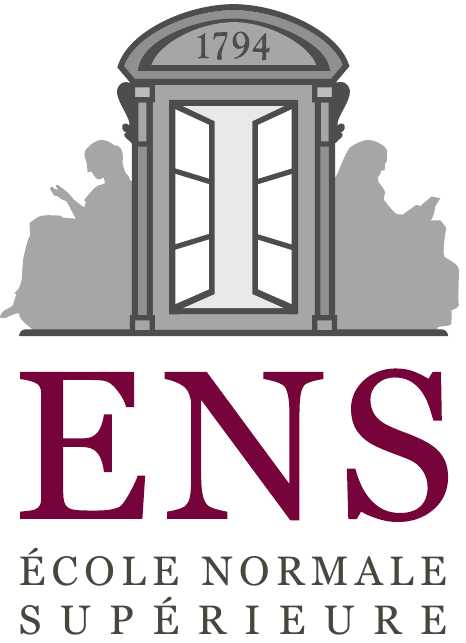}

  \end{center}

  \vspace*{\stretch{1}}

  \begin{center}\sffamily\LARGE{#5}\end{center}

  \newpage

  \thispagestyle{empty}

  \cleardoublepage

  \thispagestyle{empty}

\vspace*{\stretch{3}}

  {\parindent0cm

  \rule{\linewidth}{.7ex}}

  \begin{flushright}

    \vspace*{\stretch{1}}

    \sffamily\bfseries\LARGE #1\\

    \vspace*{\stretch{0.5}}

    \large #9\\

    \vspace*{\stretch{1}}

    \sffamily\bfseries\large

    #2

    \vspace*{\stretch{1}}

  \end{flushright}

  \rule{\linewidth}{.7ex}

  \vspace*{\stretch{3}}

  \begin{center}

  \vspace*{\stretch{1}}

    \Large Thèse de Doctorat\\

    \Large au Département de Physique de l'Ecole Normale Supérieure, Paris\\

    \Large Ecole Doctorale 107 -- Physique de la Région Parisienne\\

    \vspace*{\stretch{3}}

    \Large soutenue le 30. septembre 2013 devant le jury composé de \\

    \vspace*{\stretch{1}}

    \Large \begin{tabular}{ l l }
G. Biroli & Examinateur \\
G. Blatter & Rapporteur \\
P. Le Doussal & Invité (co-directeur de thèse) \\
T. Giamarchi & Examinateur \\
M. Müller & Examinateur \\
A. Rosso & Examinateur \\
K. Wiese & Directeur de thèse \\
S. Zapperi & Rapporteur
\end{tabular}
		\\

    \vspace*{\stretch{2}}

  \end{center}

  \newpage

  \thispagestyle{empty}

  \vspace*{\stretch{1}}

  \begin{flushleft}

    \large Cette thèse a été préparée au \\
		\large Laboratoire de Physique Théorique de l'Ecole Normale Supérieure -- CNRS UMR 8549 \\
		\large 24 rue Lhomond, 75005 Paris, France \\
		\large et financée par une bourse doctorale du CNRS

  \end{flushleft}

  \cleardoublepage

}

\begin{document}

  \frontmatter

  \LMUTitle
      {Systèmes désordonnés et théorie des champs \\ Field theory of disordered systems}
      {Alexander Dobrinevski}                      
      {Minsk}                             
      {Département de Physique}              
      {Paris 2013}                        
      {30 septembre 2013}                               
      {Kay Jörg Wiese}                        
      {Pierre Le Doussal}                       
      {Avalanches d'une interface élastique en milieu aléatoire \\ Avalanches of an elastic interface in a random medium}  

  
  \include{abstract}

  \include{acknowledgements}

  \selectlanguage{american}
      
  \tableofcontents


  \cleardoublepage

  \mainmatter\setcounter{page}{1}

\include{Introduction}
\include{MeanField}
\include{OneLoop}
\include{KPZ}

\include{Summary}

  \appendix

    \include{AppendixNotations}

  \markboth{}{}

  \backmatter

 \bibliographystyle{unsrt}
 \bibliography{thesis} 




  \newpage

  \thispagestyle{empty}

  \vspace*{\stretch{1}}

  \newpage

  \thispagestyle{empty}

  \vspace*{\stretch{1}}

  \begin{flushleft}
  \end{flushleft}

\end{document}

%% file: abstract.tex
\chapter*{Abstract}
In this thesis I discuss analytical approaches to disordered systems using field theory. Disordered systems are characterized by a random energy landscape due to heterogeneities, which remains fixed on the time scales of the phenomena considered. I focus specifically on elastic interfaces in random media, such as pinned domain walls in ferromagnets containing defects, fluid contact lines on rough surfaces, etc. The goal is to understand static properties (e.g. the roughness) of such systems, and their dynamics in response to an external force. The interface moves in avalanches, triggered at random points in time, separated by long periods of quiescence. For magnetic domain walls, this phenomenon is known as Barkhausen noise. I first study the model of a particle in a Brownian random force landscape, applicable to high-dimensional interfaces. Its exact solution for a monotonous, but time-varying driving force is obtained. This allows computing the joint distribution of avalanche sizes and durations, and their spatial and temporal shape. I then generalize these results to a short-range correlated disorder. Using the functional renormalization group, I compute the universal distributions of avalanche sizes, durations, and their average shape as a function of time. The corrections with respect to the Brownian model become important in sufficiently low dimension (e.g. for a crack front in the fracture of a solid). I then discuss connections to the rough phase of Kardar-Parisi-Zhang nonlinear surface growth, and an application to the variable-range hopping model of electric conductance in a disordered insulator.

\vspace{1cm}

Keywords: avalanches, disorder, renormalization, field theory, pinning, Barkhausen noise

\vspace*{\stretch{1}}

%% file: acknowledgements.tex
\chapter*{Acknowledgements}

In conclusion of this thesis, I first wish to express my gratitude to my advisors Kay Wiese and Pierre Le Doussal for an incredible amount of support during all stages of this work. I learned a lot in the process of doing research and writing publications with them, I enjoyed very much the open and constructive atmosphere of our discussions and the rich supply of challenging questions and projects which they led to. More than three years after being introduced by them to the theory of disordered systems and to avalanches, I find these topics more fascinating than ever and I am very happy to have profited from their experience and encouragement while doing research in this field. 
I am also very thankful for the opportunities they gave me to visit conferences, schools and workshops all over the world (particularly memorable being Santa Barbara, Cargèse, Obergurgl, Trieste and Les Houches). I profited enormously, both for my scientific work and beyond it, from these possibilities to meet like-minded scientists, broaden the horizon of my knowledge, and present research results. Lastly, I appreciate very much their time and efforts spent on the less scientific and less fascinating aspects of the thesis supervision, such as writing reference letters, funding applications, and other formalities.

I would like to thank Markus Müller for inviting me to two extremely pleasant stays at the ICTP Trieste, and especially for the stimulating discussions on directed polymers and quantum disordered systems during these visits. The entire chapter \ref{sec:KPZ} profited tremendously from his input. 
I am also very thankful to Alberto Rosso for intriguing talks on a vast range of subjects, ranging from earthquakes to numerical simulations to the life of a scientist.
More broadly, I would like to thank all members of my thesis commitee (Giulio Biroli, Gianni Blatter, Pierre Le Doussal, Thierry Giamarchi, Markus Müller, Alberto Rosso, Kay Wiese and Stefano Zapperi) for taking the time to read this manuscript, provide feedback on it, and participate in the thesis defense itself.

My three years at the ENS have been greatly enriched by the contact with many fellow students, from the LPT as well as from other labs. Not being able to list them all individually, I would like to thank particularly Thomas Gueudré for the productive and fun collaboration on the ``Explore-or-Exploit'' project, many discussions on physics and other topics, enjoyable common activities, and finally his patience and help with my French. I am likewise very obliged to Mathieu Delorme and Thimothée Thiery for many interesting discussions on avalanches and renormalization, and for valuable feedback on parts of this manuscript.

My thanks also goes to all of my family and friends, in Paris, Munich, and other places, without whom life would be much more bland. At the same time, I would like to express my gratitude to my parents for the countless invitations to Munich and to Aix-en-Provence, their support in many practical matters, and for being there whenever needed. Finally, the role of my love Eva in my life, during the time of my thesis and beyond it, needs no explanation.

%% file: Introduction.tex
\chapter{Introduction to elastic interfaces in disordered media\label{sec:Introduction}}
Equilibrium statistical physics analyzes the effects of thermal fluctuations, which arise from environmental noise. There, the system is observed on time scales much larger than those of such fluctuations. For example, when measuring the magnetization of a typical ferromagnet in a typical external field, at two times separated by an ``everyday'' interval (say a second), one expects the thermal fluctuations to be uncorrelated.
Disordered systems, on the other hand, are characterized by the presence of \textit{quenched randomness}. 
In contrast to thermal or quantum fluctuations, this is disorder which is ``frozen'', and does not change significantly on the typical time scales we are interested in. Examples are defects in ferromagnetic materials, which do not move on the time scale of the motion of magnetic domain walls, or rough surfaces, which do not deform on the time scale on which they are wetted by an advancing fluid. On the time scales of these phenomena, one observes a single realization of the disordered environment; performing the average over quenched disorder requires averaging over many samples. In many cases of practical relevance, such quenched disorder gives rise to new phenomena. Examples are
\begin{itemize}
	\item Localization of electrons, and transition between metals (weak disorder) and insulators (strong disorder).
	\item A wide variety of mechanical properties of heterogeneous materials, ranging from ductile materials (for weak disorder) to brittle materials (for strong disorder).
	\item Glasses and jamming in granular media.
\end{itemize}

In this thesis, I focus on the special case of \textit{elastic interfaces in a random environment}. A few important examples are magnetic domain walls, crack fronts, and fluid contact lines. These are simple disordered systems, since they do not exhibit geometrical frustration\footnote{I.e.~constraints which cannot be satisfied simultaneously, such as in antiferromagnets on a triangular lattice or in hard sphere packings.} (in contrast to e.g.~glasses and spin glasses). Furthermore, in many of these cases (such as fluid contact lines) a single interface can be isolated, so that there are no many-body interactions (in contrast e.g.~to electrons in disorder). 
 I begin by explaining experimental observations of such systems in section \ref{sec:ExpAvalanches}. I focus in particular on systems where \textit{avalanches} are observed. These manifest themselves by a non-smooth, strongly fluctuating response to the application of a small external force, also known as \textit{crackling noise} \cite{SethnaDahmenMyers2001}. I will discuss in particular detail Barkhausen noise (section \ref{sec:Barkhausen}) and fracture of brittle materials (section \ref{sec:Fracture}).

In chapter \ref{sec:BFM}, I discuss the theory of the mean-field universality class of elastic interfaces in disordered media. This corresponds to the Brownian Force Landscape, a generalization of the \textit{Alessandro-Beatrice-Bertotti-Montorsi (ABBM)} model, and has been the subject of many theoretical studies in  recent years (reviewed in section \ref{sec:ReviewABBM}). I then discuss new results on the dynamics of interfaces in the mean-field universality class, which have been obtained during this thesis and published in two articles \cite{DobrinevskiLeDoussalWiese2012,DobrinevskiLeDoussalWiese2013}. In section \ref{sec:BFMABBM} I explain the exact solution of the Brownian Force Model, for monotonous but not necessarily stationary driving, using statistical field theory. This reduces the computation of arbitrary disorder-averaged observables in the BFM and in the ABBM model to the solution of a nonlinear ordinary differential equation. This approach is also applicable to some generalizations of the ABBM model, such as the ABBM model with retardation discussed in section \ref{sec:BFMRetardation}. It allows a complete understanding of the avalanche statistics.

In chapter \ref{sec:OneLoop}, I explain how corrections to the mean-field behaviour, which become relevant in sufficiently low dimensions, can be obtained using the functional renormalization group. In particular, I compute the leading-order corrections to critical exponents characterizing the avalanche statistics. I also obtain analytical results for universal scaling functions, such as the distribution of avalanche durations (section \ref{sec:OneLoopDurations}), and the average avalanche shape (sections \ref{sec:OneLoopShapeTime}, \ref{sec:OneLoopShapeSize}). These functions exhibit characteristic features (asymmetry in the avalanche shape, bump in the avalanche duration distribution), which are not present in the mean-field case. These results are useful for testing the applicability the elastic-interface model to experimentally accessible systems.

In chapter \ref{sec:KPZ}, I discuss the relationship between elastic interfaces and directed polymers, as well as the Kardar-Parisi-Zhang equation for surface growth. I review the state of the art and fundamental questions regarding the nature of the low-temperature phase of this model in sections \ref{sec:KPZOverview} - \ref{sec:KPZHD}. In section \ref{sec:KPZHD} I also discuss some new results on the universal behaviour of moments of the directed polymer partition sum. In section \ref{sec:KPZComplex}, I then present some new results obtained during this thesis and published in \cite{DobrinevskiLeDoussalWiese2011} on the directed polymer with random complex weights, which has an interesting relationship to conductance fluctuations of a Mott insulator in the variable-range-hopping regime.

Finally, I summarize the results of this thesis and discuss possible extensions in chapter \ref{sec:Summary}.

\section{Experimental observations\label{sec:ExpAvalanches}}

\subsection{Magnetic Domain Walls and Barkhausen Noise\label{sec:Barkhausen}}
\begin{figure}%
\centering
         \centering
         \begin{subfigure}[t]{0.45\textwidth}
                 \centering
                 \includegraphics[width=\textwidth]{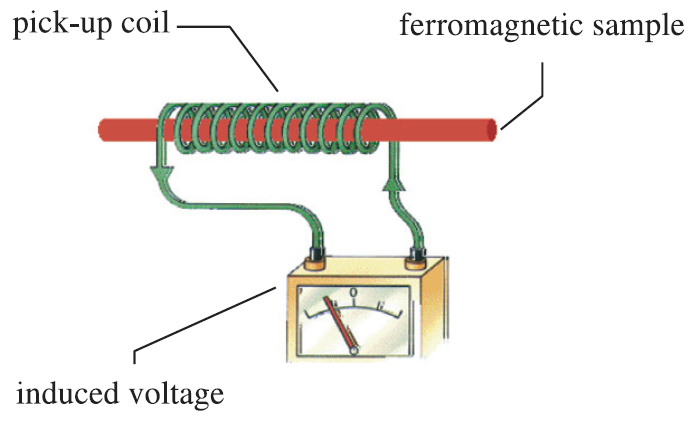}
                 \caption{Sketch of an experimental setup for measuring Barkhausen noise. Figure reprinted with permission from \cite{Colaiori2008}. Copyright \copyright \, 2008 Taylor \& Francis.}
                 \label{fig:BarkhausenExpSetup}
         \end{subfigure}\quad
         \begin{subfigure}[t]{0.45\textwidth}
                 \centering
                 \includegraphics[width=\textwidth]{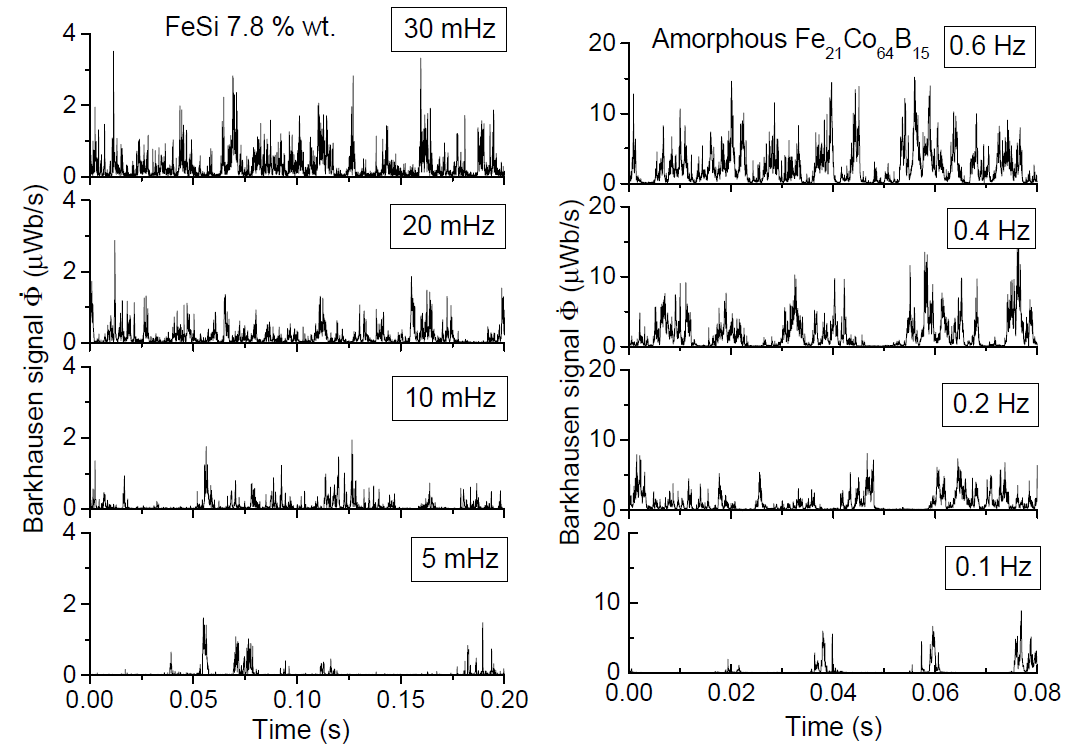}
                 \caption{Example of Barkhausen noise measurements (Figure reprinted with permission from \cite{DurinZapperi2006b}. Copyright \copyright \, 2006 Elsevier). The magnetic flux change $\dot{\Phi}$ on the vertical axis is proportional to the induced voltage in the pickup coil in figure \ref{fig:BarkhausenExpSetup}. From top to bottom, the oscillation frequency of the external magnetic field (shown in the box in the top-right corner of each curve) is lowered. Barkhausen jumps overlap and form a crackling pulse for high driving frequencies, but become well-separated in time for low driving frequencies. Their sizes and durations are defined in figure \ref{fig:SizeDurDef}.}%
                 \label{fig:BarkhausenMeas}
         \end{subfigure}
\caption{Barkhausen noise experiments}
\end{figure}

In 1919, H. Barkhausen first recorded the crackling electromagnetic signal emitted when slowly magnetizing a ferromagnet (\cite{Barkhausen1919}, for a translation to English see \cite{DurinZapperi2006b}), which is today known as \textit{Barkhausen noise}.\footnote{For a recent review on this phenomenon, and theoretical approaches to it, see \cite{Colaiori2008}. In this section, I extend this review with a more detailed discussion of amorphous vs. polycristalline ferromagnets and recent results on the Barkhausen pulse shape. Progress on the theoretical side beyond the results reviewed in \cite{Colaiori2008}, mostly due to a new field-theoretical approach, is summarized in section \ref{sec:ReviewABBM}.} A typical experimental setup (see \cite{Bertotti1998,DurinZapperi2006b,Colaiori2008} and figure \ref{fig:BarkhausenExpSetup}) consists in two induction coils wrapped around a ferromagnetic sample. One coil applies a time-varying external field to the sample. The resulting change in magnetization leads to a change in the magnetic flux permeating the second coil (``pick-up'' coil). The induced voltage in the pick-up coil is then measured, and provides a signal proportional to the magnetization rate of the sample.\footnote{For more details on experimental setups, see \cite{DurinZapperi2006b}.}

\begin{figure}%
				         \begin{subfigure}[t]{0.45\textwidth}
                 \centering
                 \includegraphics[width=\textwidth]{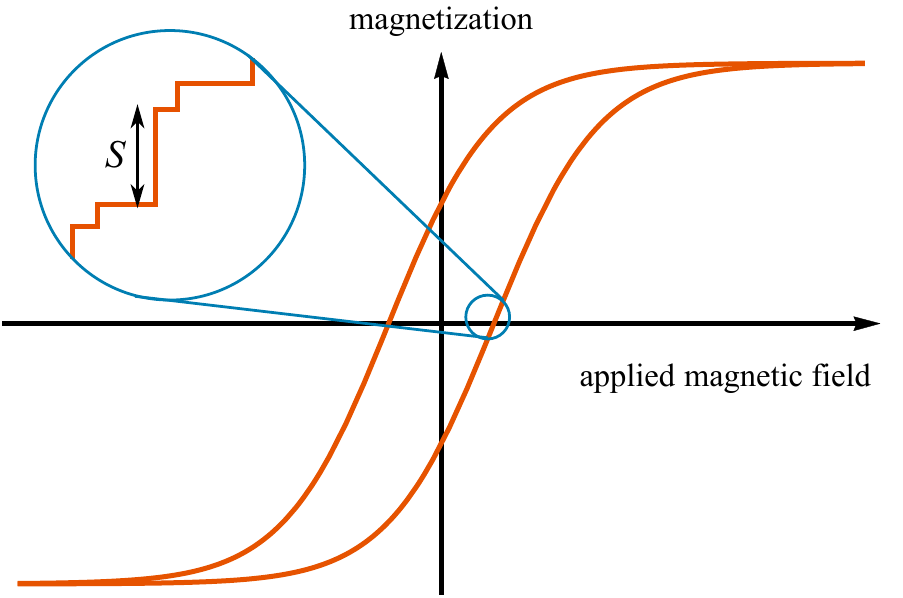}
                 \caption{Sketch of the hysteresis loop of a ferromagnet, and of the Barkhausen jumps observed when the applied magnetic field is varied slowly.}
                 \label{fig:IntroHystLoopSizes}
         \end{subfigure}\quad
         \begin{subfigure}[t]{0.45\textwidth}
                 \centering
                 \includegraphics[width=\textwidth]{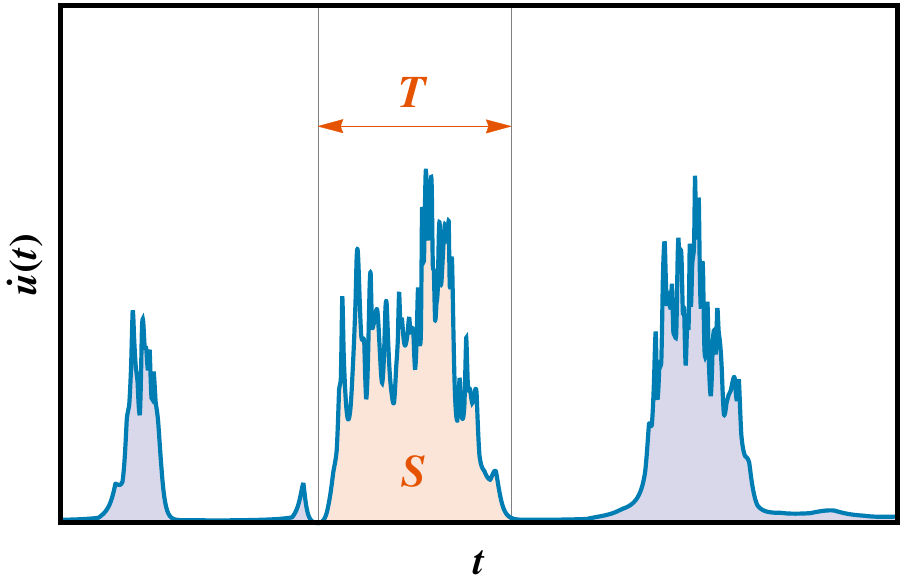}
                 \caption{Definition of the size $S$ and the duration $T$ of a Barkhausen jump, from the measured magnetization rate $\dot{u}(t)$. The magnetization rate is proportional to the flux change rate $\dot{\Phi}(t)$ shown in figure \ref{fig:BarkhausenMeas}.
								}%
                 \label{fig:SizeDurDef}
         \end{subfigure}
				\caption{Barkhausen jumps\label{fig:BarkhausenJumps}}
\end{figure}

The observed Barkhausen signal (see figure \ref{fig:BarkhausenMeas} for an example), even for slowly varying external fields, is jagged and non-smooth. It consists in a series of \textit{Barkhausen jumps} of varying sizes and durations, followed by long periods of quiescence (see figure \ref{fig:BarkhausenJumps}). This behaviour is similar across a wide range of ferromagnetic materials, and sample geometries \cite{DurinZapperi2006b,Colaiori2008}. Let us focus on so-called \textit{soft magnets}, where the Barkhausen effect is a result of non-smooth motion of domain walls pinned by defects.\footnote{The other class are \textit{hard magnets}, consisting in well-separated grains. There magnetization proceeds through spin reversal of entire grains, and not through motion of domain walls \cite{DurinZapperi2006b}. Although this also leads to a crackling Barkhausen signal and avalanches, their physics is quite different. In particular, nucleation plays an important role \cite{BassoEtAl2004}, and hence the elastic interface models discussed in this thesis are not applicable. Possibly, disordered spin models such as the random-field Ising model would provide a better description \cite{SethnaEtAl1993}, but a good theoretical understanding is still lacking \cite{SethnaEtAl1993,DurinZapperi2006b,Colaiori2008}.}
These domain walls can be described as two-dimensional elastic interfaces in a three-dimensional random medium.

\begin{figure}
         \centering
         \begin{subfigure}[t]{0.45\textwidth}
                 \centering
                 \includegraphics[width=\textwidth]{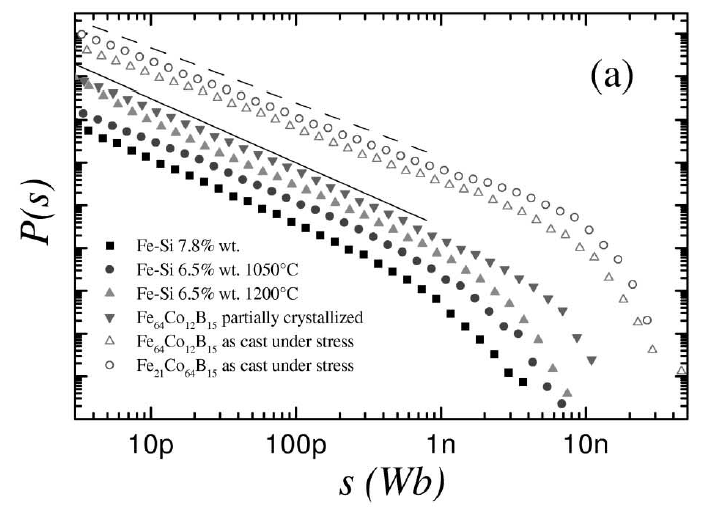}
                 \caption{$P(S)$}
                 \label{fig:BarkhausenMeasPofS}
         \end{subfigure}%
         ~ 
         \begin{subfigure}[t]{0.45\textwidth}
                 \centering
                 \includegraphics[width=\textwidth]{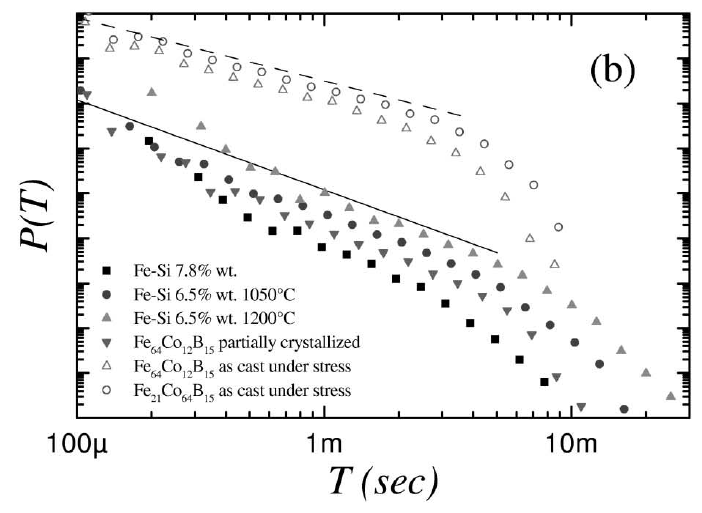}
                 \caption{$P(T)$}
                 \label{fig:BarkhausenMeasPofT}
         \end{subfigure}
         \caption{Experimentally measured distributions of Barkhausen jump sizes $S$ and durations $T$ (Figure reprinted with permission from \cite{DurinZapperi2000}. \copyright \, 2000 by the American Physical Society). The solid lines are exponents $\tau=3/2, \alpha=2$ characterizing polycristalline ferromagnets, and the mean-field universality class. The dashed lines are exponents $\tau \approx 1.27, \alpha \approx 1.5$ characterizing amorphous ferromagnets, and a non-mean-field universality class. Observe that all measurements can be attributed clearly to one of these two universality classes. \label{fig:BarkhausenMeasPofST}}
\end{figure}

Understanding how the Barkhausen signal is related to material properties is important for applications such as non-destructive testing: it allows to deduce, for example, residual stresses \cite{GauthierKrauseAtherton1998,StewartStevensKaiser2004} or grain sizes \cite{RanjanJilesBuckThompson1987,YamauraFuruyaWatanabe2001} in metallic materials. 
This requires a quantitative characterization of the irregular Barkhausen signal, such as shown in figure \ref{fig:BarkhausenMeas}. 
For that, natural observables to consider  are the probability distributions of instantaneous signal amplitudes $\du$, and of sizes $S$ and durations $T$ of Barkhausen jumps, or \textit{avalanches} (see figure \ref{fig:SizeDurDef}). These avalanches are delimited by regions of quiescence, where the Barkhausen signal vanishes. A precise analysis requires a careful introduction of a small-amplitude cutoff (\cite{DurinZapperi2006b} and references therein) and a fixed choice of the location on the hysteresis loop, at which the signal is measured \cite{WiegmanStege1978,BertottiFiorilloSassi1981,DurinZapperi2006}. Usually this is the region shown in figure \ref{fig:IntroHystLoopSizes}, where the magnetization passes through zero and the hysteresis loop is linear. One then finds that the experimentally determined distributions of $\du$, $S$ and $T$, for slow driving, take the following forms \cite{DurinZapperi2006b}, which are largely sample-independent:
\bea
\label{eq:IntroBNScalingForms}
P(\du) \sim \du^{-a} f_{\du} (\du/v_0),\quad\quad\quad P(S)\sim S^{-\tau} f_S ( S/S_0 ), \quad\quad\quad P(T) \sim T^{-\alpha} f_T ( T/T_0 ).
\eea
In other words, they exhibit characteristic power-law divergences for small avalanches with exponents $a, \tau$ and $\alpha$. For large avalanches, the power-law distributions are cut off on some scale $v_0, S_0, T_0$, which varies depending on the sample geometry, material properties, and dimensions (see section \ref{sec:ReviewABBM} and \cite{DurinZapperi2006b} for more details). Some examples of experimental measurements are shown in figure \ref{fig:BarkhausenMeasPofST}. By comparing the exponents $a, \tau, \alpha$ and the scaling of $v_0, S_0, T_0$ one observes \cite{DurinZapperi2000, DurinZapperi2006b} that different materials and experimental setups can be described by a small set of \textit{universality classes}\footnote{Here, I define two systems to be in the same universality class if all of their critical exponents (such as $a, \tau, \alpha$ in \eqref{eq:IntroBNScalingForms}, and the roughness and dynamical exponents $\zeta$ and $z$ in \eqref{eq:IntroDefZetaZ}) coincide.}. For example, two such universality classes, corresponding to amorphous alloys, and polycristalline materials, have been identified (\cite{DurinZapperi2000} and figure \ref{fig:BarkhausenMeasPofST}). 
\begin{itemize}
	\item For polycristalline materials, the exponents for slow driving are given by $a=1$, $\tau=3/2$, $\alpha=2$, and decrease as the driving rate $c$ of the external field is increased. They correspond to the mean-field universality class of elastic interface models, as discussed in chapter \ref{sec:BFM}, and in particular to the Alessandro-Beatrice-Bertotti-Montorsi model \cite{AlessandroBeatriceBertottiMontorsi1990,AlessandroBeatriceBertottiMontorsi1990b} (see section \ref{sec:ReviewABBM}). Both can be solved exactly, and yield predictions in quantitative agreement with experiments. The mean-field nature can be understood from the presence of long-range dipolar interactions, which correspond to an interface model with a slowly decaying elastic kernel (see \cite{CizeauZapperiDurinStanley1997} and section \ref{sec:InterfacePhenomenology} for details).
\item  For amorphous materials, the exponents are given by $\tau \approx 1.27, \alpha \approx 1.5$, and no rate dependence \cite{DurinZapperi2000}. This can be explained by the absence of long-range dipolar interactions \cite{DurinZapperi2000, DurinZapperi2006b}. This makes an interface model with short-ranged elasticity, and non-mean-field behaviour, relevant (see section \ref{sec:InterfacePhenomenology}). 
\end{itemize}

\subsubsection{Average shapes of Barkhausen pulses\label{sec:BarkhausenShape}}
\begin{figure}%
         \centering
         \begin{subfigure}[t]{0.25\textwidth}
                 \centering
                 \includegraphics[width=\textwidth]{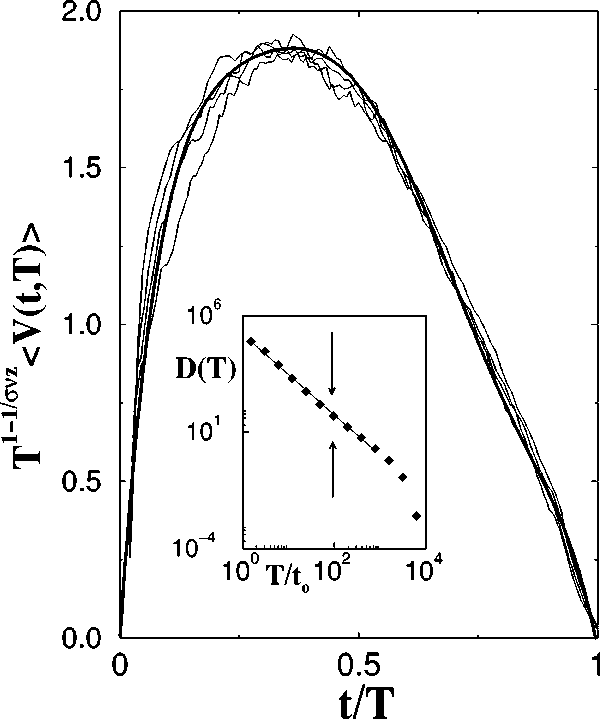}
                 \caption{Figure reprinted with permission from \cite{MehtaMillsDahmenSethna2002}. Copyright \copyright \, 2002 by the American Physical Society.}
                 \label{fig:BarkhausenMeasShape1}
         \end{subfigure}%
         ~ 
         \begin{subfigure}[t]{0.35\textwidth}
                 \centering
                 \includegraphics[width=\textwidth]{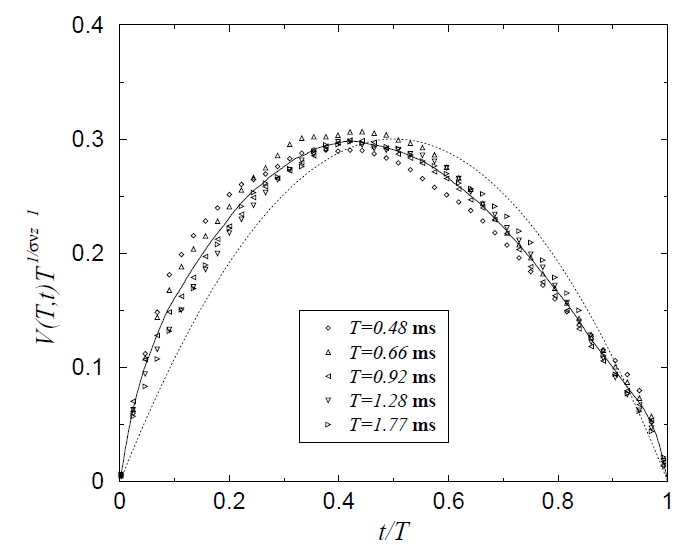}
                 \caption{Figure reprinted with permission from \cite{DurinZapperi2002}. Copyright \copyright \, 2002 by Elsevier.}
                 \label{fig:BarkhausenMeasShape2}
         \end{subfigure}
         ~ 
         \begin{subfigure}[t]{0.35\textwidth}
                 \centering
                 \includegraphics[width=\textwidth]{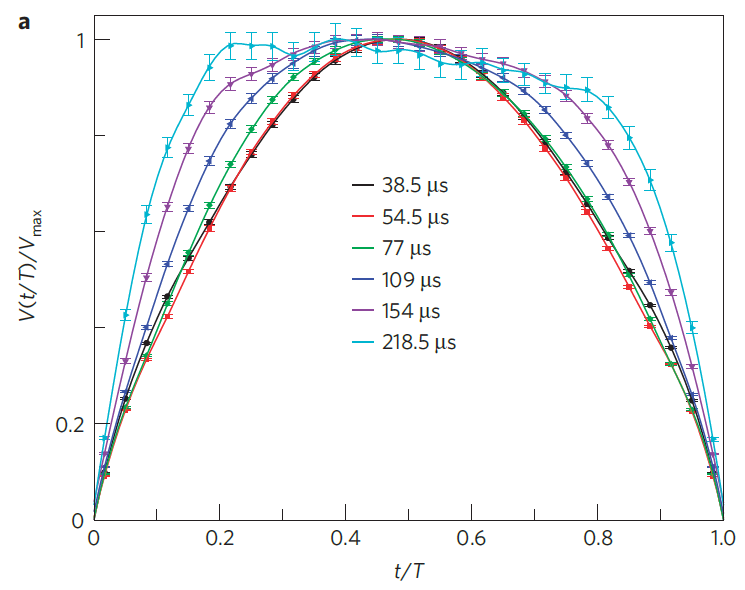}
                 \caption{Figure reprinted with permission from \cite{PapanikolaouBohnSommerDurinZapperiSethna2011}. Copyright \copyright \, 2011 by Macmillan Publishers Ltd.}
                 \label{fig:BarkhausenMeasShape3}
         \end{subfigure}%
         \caption{Experimentally measured average Barkhausen pulse shapes, for a fixed pulse duration. In each case, the vertical axis shows the voltage measured on the pickup coil (cf. figure \ref{fig:BarkhausenExpSetup}), proportional to the rate of change of the magnetization of the sample.}
				\label{fig:BarkhausenMeasShapes}
\end{figure}

The distributions of Barkhausen jump sizes and durations discussed above and shown in figure \ref{fig:BarkhausenMeasPofST} have some characteristic power-law divergences for small avalanches, and an exponential cutoff for large avalanches, but are otherwise rather featureless. It is thus interesting to consider more refined observables, which provide a stronger reflection of the underlying domain wall dynamics. 
One such example, proposed among others in \cite{SpasojevicBukvicMilosevicStanley1996,SethnaDahmenMyers2001,ZapperiCastellanoColaioriDurin2005}, is the average shape of a Barkhausen noise pulse. It is obtained by considering the voltage signal as a function of time, $\du(t)$, for pulses whose total duration is in a narrow window $[T;T+dT]$.\footnote{In some cases, where one expects pulse shapes to be scale-invariant, one may choose a larger window and rescale the time axis for each pulse by its duration, in order to improve the statistics.}
It has been measured in various experiments \cite{SpasojevicBukvicMilosevicStanley1996,DurinZapperi2002,MehtaMillsDahmenSethna2002,ZapperiCastellanoColaioriDurin2005,PapanikolaouBohnSommerDurinZapperiSethna2011}, some of which are shown in figure \ref{fig:BarkhausenMeasShapes}.

One observes that for small avalanches\footnote{I.e.~for avalanches whose size is much smaller than the cutoff in \eqref{eq:IntroBNScalingForms}, and which are in the ``critical'' power-law scaling regime of the size and duration distributions.}, the average pulse shape shows a scaling collapse and only depends on the universality class (i.e. is the same for systems where the critical power-law exponents $\alpha, \tau, ...$ coincide). In that limit, it is well described by a parabolic shape, as predicted by the mean-field ABBM model (see section \ref{sec:ReviewABBM} and \ref{sec:BFMShape}).\footnote{Although we will see in section \ref{sec:OneLoopShapeTime} that in the non-mean-field universality class (amorphous ferromagnets) the theory predicts some corrections to the parabolic shape of the ABBM model, they are small and difficult to measure.} However, for longer avalanches, the average shape is sensitive to the details of the model. For typical Barkhausen-noise experiments, it shows a clear leftward asymmetry. This can be explained by the presence of eddy currents, which provide a negative effective mass to the domain wall \cite{ZapperiCastellanoColaioriDurin2005}. These can be modelled by introducing an additional retardation term into the ABBM model (see \cite{ZapperiCastellanoColaioriDurin2005} and section \ref{sec:BFMRetardation}). 
While this does not modify its universality class (i.e. the scaling exponents, and the average shape, for short avalanches), it reproduces the experimentally observed asymmetry of the average shape for long avalanches. 
Thus, the average avalanche shape for long avalanches is an observable that is sensitive to non-universal features. It allows to learn details specific to the motion of magnetic domain-walls (like the presence of eddy currents), and to distinguish between different models in the same universality class.

\subsection{Fracture of heterogeneous media\label{sec:Fracture}}
Another physical process where crackling noise is observed is fracture of brittle materials \cite{HerrmanRoux1990,ChakrabartiBenguigui1997,Bouchaud1997,BonamySantucciPonson2008,Bonamy2009}. In such materials, due to intrinsic heterogeneities, the response to an external loading is an irregular, non-smooth crack propagation on a wide range of scales.
This can be observed post-mortem on the roughness of fracture surfaces \cite{MaloyHansenHinrichsenRoux1992,Bouchaud1997,SantucciEtAl2007,GuerraScheibertBonamyDalmas2012}, through crackling acoustic emissions (see \cite{MinozziCaldarelliPietroneroZapperi2003,DavidsenStanchitsDresen2007,RostiIllaKoivistoAlava2009} and references therein) or by direct observation of the dynamics of the crack front.

\begin{figure}%
         \centering
         \begin{subfigure}[t]{0.4\textwidth}
                 \centering
                 \includegraphics[width=0.6\textwidth]{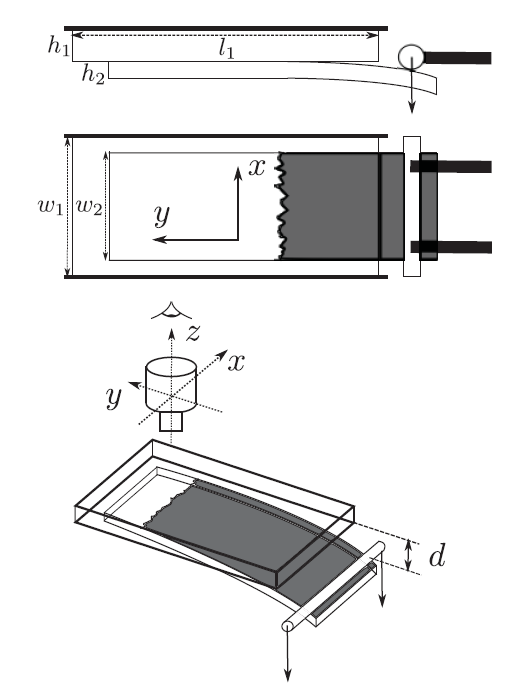}
                 \caption{Experimental setup for observing crack fronts in mode I fracture of a sintered PMMA block. Figure reprinted with permission from \cite{TallakstadEtAl2011}. Copyright \copyright \, 2011 by the American Physical Society.}
                 \label{fig:FractureExp1}
         \end{subfigure}
				\quad
				         \begin{subfigure}[t]{0.5\textwidth}
                 \centering
                 \includegraphics[width=\textwidth]{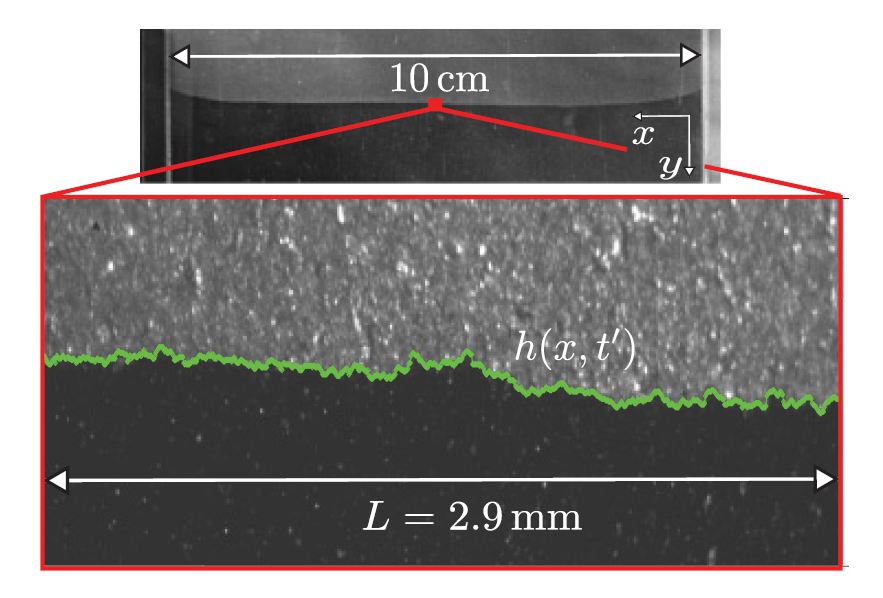}
                 \caption{Experimentally measured crack front profile. $h(x,t)$ is the displacement of the crack front along the $y$-axis for given $x,t$. Figure reprinted with permission from \cite{TallakstadEtAl2011}. Copyright \copyright \, 2011 by the American Physical Society.}
                 \label{fig:FractureExpMeas}
         \end{subfigure}%
\caption{Fracture experiments.\label{fig:FractureExp}}%
\end{figure}

The latter is possible, for example, for mode I fracture\footnote{I.e. prying open a block by pulling two sides in opposite directions, see figure \ref{fig:FractureExp1}.} of two sandblasted, sintered PMMA blocks \cite{SchmittbuhlMaloy1997,DelaplaceSchmittbuhlMaloy1999}. Due to the transparency of the material, the crack front between the fractured and the intact parts of the surface can be observed in real time using optical methods (see figure \ref{fig:FractureExp} and \cite{SchmittbuhlMaloy1997,DelaplaceSchmittbuhlMaloy1999} for experimental details). The measured crack fronts are self-affine: The height profile $h(x,t)$ (see figure \ref{fig:FractureExpMeas} for a definition) has typical fluctuations scaling like
\bea
\label{eq:FractureRoughness}
\overline{\left[h(x)-h(x')\right]^2} \sim |x-x'|^{2\zeta},
\eea
where the overline denotes an average over many samples. The experimentally obtained roughness exponent $\zeta$ is $\zeta=\zeta^{-} \sim 0.63$ on small scales \cite{SchmittbuhlMaloy1997,DelaplaceSchmittbuhlMaloy1999,MaloySantucciSchmittbuhlToussaint2006,SantucciEtAl2010} and a different value $\zeta = \zeta^{+} \sim 0.35$ on large scales \cite{Bonamy2009,SantucciEtAl2010}. The small-scale roughness $\zeta^{-}$ can be attributed to the microscopic preparation of the disorder \cite{SantucciEtAl2010}. On the other hand, the large-scale roughness $\zeta^{+}$ is universal and can be explained by the model of a one-dimensional interface, with long-range elastic interactions, in a two-dimensional random medium \cite{SchmittbuhlRouxVilotteMaloy1995,BonamySantucciPonson2008,SantucciEtAl2010}. This non-zero roughness exponent indicates a non-mean-field universality class, such as discussed in chapter \ref{sec:OneLoop}.

Apart from the crack-front roughness, optical measurements on mode I PMMA fracture (such as in figure \ref{fig:FractureExpMeas}) allow one to observe its local and global dynamics in real time \cite{MaloySantucciSchmittbuhlToussaint2006,BonamySantucciPonson2008,Bonamy2009,TallakstadEtAl2011}. Just as in the case of Barkhausen noise (section \ref{sec:Barkhausen}), for slow loading the crack front propagates in well-separated jumps or \textit{avalanches}. Due to the long-range elastic interaction, these decompose into multiple disconnected local \textit{clusters}. The statistics of local and global velocities, as well as avalanche and cluster sizes and durations, exhibit characteristic power-law distributions similar to the Barkhausen noise statistics discussed above. In particular, one observes that local velocities $v_{\text{loc}}$, and cluster sizes $S_{\text{cl}}$ are distributed as \cite{MaloySantucciSchmittbuhlToussaint2006,BonamySantucciPonson2008,Bonamy2009,TallakstadEtAl2011,TallakstadToussaintSantucciMaloy2013}
\bea
\label{eq:FractureExpExponents}
P(v_{\text{loc}})\sim v_{\text{loc}}^{-a_{\text{loc}}},\quad\quad a_{\text{loc}} \approx 2.55,\quad\quad P(S_{\text{cl}}) \sim S_{\text{cl}}^{-\tau_{\text{cl}}},\quad\quad \tau_{\text{cl}} \approx 1.56
\eea
(Note that in the notations of \cite{MaloySantucciSchmittbuhlToussaint2006,BonamySantucciPonson2008,Bonamy2009,TallakstadEtAl2011}, $v=v_{\text{loc}}$, $\eta = a_{\text{loc}}$, $S=S_{\text{cl}}$, $\gamma = \tau_{\text{cl}}$). These observations are in good quantitative agreement with numerical simulations of a one-dimensional interface, with long-ranged elastic interactions, in a two-dimensional random medium (see \cite{BonamySantucciPonson2008,Bonamy2009,LaursonSantucciZapperi2010} and section \ref{sec:NumericsLR} below). Furthermore, detailed information on the geometry and anisotropy of the avalanche clusters can be obtained experimentally \cite{TallakstadEtAl2011} and numerically \cite{Bonamy2009,LaursonSantucciZapperi2010}.

Another important aspect of fracture experiments is the \textit{out-of-plane} roughness: In a real experiment, the crack front is not strictly confined to a two-dimensional fracture plane, but can fluctuate around it. This is not described by a standard elastic interface model, but requires generalizations, such as the path equation discussed in \cite{Bonamy2009}. These will not be discussed here, but investigating the applicability of the analytical methods discussed in this thesis to a more complete discussion of the roughness of fracture surfaces (cf. the thesis \cite{Ponson2006}) is an interesting avenue for future work.

\subsection{Earthquakes\label{sec:Earthquakes}}
\begin{figure}%
\centering
\includegraphics[width=0.8\columnwidth]{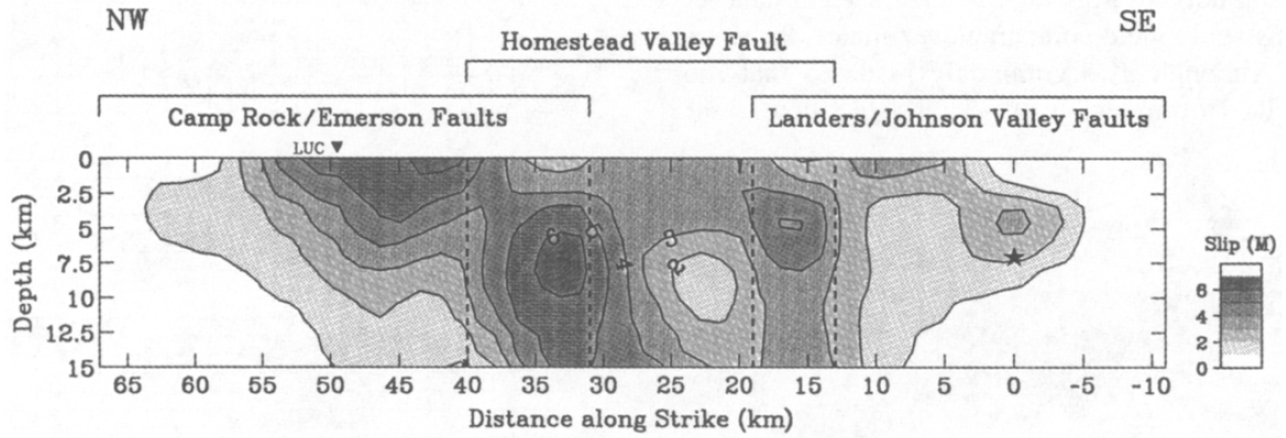}%
\caption{Lateral displacement profile along the fault during the $M_w=7.3$ 1992 Landers strike-slip earthquake. Figure reprinted with permission from \cite{WaldHeaton1994}. Copyright \copyright \, 1994 by the Seismological Society of America.}%
\label{fig:RuptureFault}%
\end{figure}

Qualitatively, earthquakes are a striking example of avalanche phenomena. While the bulks of tectonic plates drift slowly (with velocities on the scale of $mm/y$), the faults between them are locked by friction. The accumulated stress is partially released during rare, sudden earthquakes. During these, the two plate boundaries touching at the fault are displaced relatively to each other in some region of the fault (\textit{rupture area}). The displacement varies in space, and can be sideways (\textit{strike-slip} earthquakes, see e.g. figure \ref{fig:RuptureFault}) or vertical (\textit{normal} or \textit{thrust} earthquakes).
The discussion of self-organized criticality in earthquakes has a long history, starting with the discovery by Gutenberg and Richter \cite{GutenbergRichter1944,GutenbergRichter1956} of the power-law distribution of earthquake sizes. The size of an earthquake is measured by its \textit{seismic moment} $M_0$, which is the total energy released by an earthquake. By linear elasticity theory, it can be written as $M_0 = \mu D A$, where $\mu$ is the shear modulus of the rock, $D$ is the total slip (lateral displacement) along the fault, and $A$ is the area of the rupture (see \cite{Scholz2002}, section 4.3.1). 
The \textit{magnitude} $M_w$ of an earthquake (which is the number typically reported in newspapers) is defined via the moment magnitude scale \cite{HanksKanamori1979,Kagan2002,Scholz2002} as $M_w = \frac{2}{3}\log_{10}( M_0) - 6.0 (1)$, where $M_0$ is in units of $N m$. For a typical large earthquake (e.g. the $M_w=7.3$ 1992 Landers earthquake, see \cite{WaldHeaton1994} and figure \ref{fig:RuptureFault}), the lateral displacement of the fault is on the scale of meters, the duration is on the scale of tens of seconds, and recurrence times are on the scale of hundreds of years.
Gutenberg and Richter found that the probability distribution of $M_w$ decays exponentially \cite{GutenbergRichter1944,GutenbergRichter1956,Kagan2002,Scholz2002},
\bea
P( M_w) \propto 10^{-b M_w}.
\eea
This corresponds to a power-law distribution of seismic moments (earthquake sizes),
\bea
P(M_0 ) \propto M_0^{-\frac{2}{3}b - 1}.
\eea
 This \textit{Gutenberg-Richter (G-R) distribution} and the value of the exponent $b$ are reproduced qualitatively by the mean-field elastic interface model \cite{FisherDahmenRamanathanBenZion1997,DahmenErtasBenZion1998}. Identifying $M_0$ with the avalanche size $S$ in an elastic interface model, the mean-field value of the avalanche size exponent $\tau=3/2$ gives $b=\frac{3}{2}(\tau-1)=\frac{3}{4}$. This is not far from the value $b\approx 1$ reported for some earthquake catalogs \cite{Kagan2002,Scholz2002}. However, a closer inspection reveals important discrepancies.
The Gutenberg-Richter distribution describes reasonably well the overall seismic activity in a large region with many faults, but for smaller regions there is a strong variability in the observed distributions. In particular, the G-R $b$-value varies between $0.4$ and $2.0$ \cite{WiemerWyss2002,SchorlemmerWiemerWyss2005}. On a single fault, seismicity is usually better described by a characteristic distribution of earthquakes \cite{WesnouskyEtAl1983,Wesnousky1994,Scholz1998,Scholz2002}, which are all similar in size and recur periodically, up to small fluctuations. This is seen clearly e.g. on the San Andreas fault in southern California, both for large \cite{SchwartzCoppersmith1984} and for small \cite{NadeauFoxallMcEvilly1995} earthquakes. For other faults, such as the San Jacinto fault, the picture is less clear. There are claims that when isolating a single fault segment, characteristic earthquakes are observed \cite{Wesnousky1994}, but that complicated fault structures consisting of multiple segments lead to a distribution resembling the G-R law \cite{Wesnousky1994,Scholz1998,Scholz2002}. Though some controversies regarding this subject persist \cite{SteinNewman2004}, it is clear that the Gutenberg-Richter distribution is only an effective description for regional seismicity which arises from many interacting faults or sub-faults \cite{Scholz2002}. The situation is similar for other observables: It has been claimed that the distribution of waiting times between earthquakes is universal \cite{Corral2004}, but a more detailed analysis shows this is due to averaging of different, nonuniversal distributions from different seismogenic regions \cite{TouatiNaylorMain2009,TouatiNaylorMainChristie2011}. 
Thus, it is not clear whether it is reasonable to expect universality in the sequence of earthquake events on an idealized, isolated, flat two-dimensional fault plane (which one may be tempted to interpret as an ``elastic interface'' with the relative displacement as the transversal variable, as in \cite{FisherDahmenRamanathanBenZion1997,DahmenErtasBenZion1998,MehtaDahmenBenZion2006}). 

		The dynamics of a single fault is phenomenologically well-described by complicated, nonlinear friction laws, the so-called \textit{rate-and-state friction} \cite{Ruina1983,Dieterich1992,CochardMadariaga1996,Scholz1998,Marone1998,Scholz2002}. These depend strongly on material parameters and the history of the fault, and vary between faults and along a fault. There are regions with velocity-weakening friction (which accelerate the slip once it starts) and regions with velocity-strengthening friction (which slow down the slip, and inhibit the propagation of earthquakes). In particular, velocity-weakening friction is only found in the brittle layer of the earth crust (\textit{schizosphere}, depths below $\sim 20km$). In depths $> 20km$, due to increased pressure and temperature, the earth crust becomes ductile (\textit{plastosphere}), and propagation of earthquakes is inhibited. 
Note that this is not simply a random pinning force as in an elastic interface model; this is a variation in the constitutive law for the dependence of the average friction force on the slip velocity. These nonlinear friction laws lead -- when material parameters are varied -- to a rich variety of observed earthquake behaviour, including postseismic afterslip \cite{MaroneScholtzBilham1991}, frictional healing \cite{Marone1998b,Marone1998} and varying earthquake size distributions \cite{CochardMadariaga1996}.

\begin{figure}%
\centering
\begin{minipage}[c]{0.35\textwidth}
\includegraphics[width=\columnwidth]{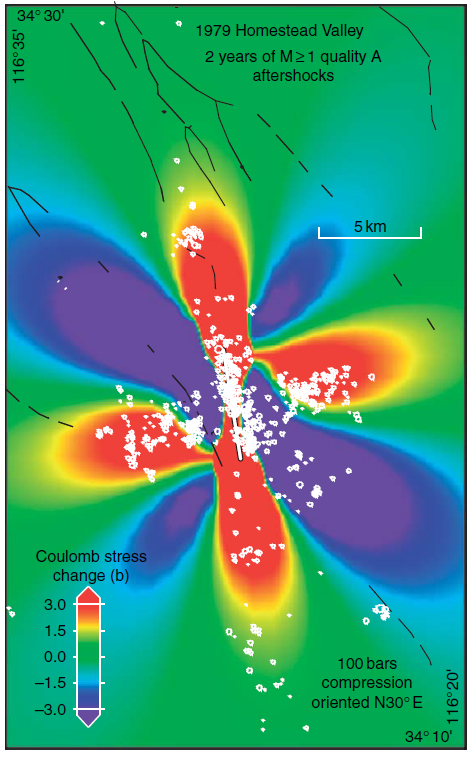}%
\end{minipage}\quad\quad
\begin{minipage}[b]{0.4\textwidth}
\caption{Correlation between static stress change and aftershocks for the 1979 Homestead Valley earthquake. Figure reprinted with permission from \cite{King2007}. Copyright \copyright \, 2007 by Elsevier.}%
\label{fig:DeltaCFS}%
\end{minipage}
\end{figure}

Just as earthquakes themselves, at first sight \textit{aftershocks} satisfy a simple universal law: The rate of aftershock events after a main shock decays as a power-law with time, with an exponent close to one (the so-called \textit{Omori-Utsu law}) \cite{Omori1894,UtsuEtAl1995,Scholz1998,Scholz2002}.
However, a detailed analysis shows that aftershocks are actually very diverse and caused by different physical phenomena in different cases \cite{Freed2005}. Most of these triggering mechanisms rely on interactions between different faults, and depend on their geometry. For example, many aftershocks are caused by the modification of the static stress on one fault due to elastic deformations of the earth crust after an earthquake on another fault \cite{Stein1999}. This change in static stress can be positive or negative (depending on the geometry and relative orientation of the faults), and can be estimated using linear elasticity theory. One observes a clear correlation: In regions where stress is increased, seismic activity rises and one observes many aftershocks. In regions where stress is decreased, one has ``negative aftershocks'', and the seismic activity actually decreases below the background level (see figure \ref{fig:DeltaCFS} and \cite{Stein1999,KingCocco2000,King2007}). Likewise, aftershock triggering via dynamical stress transfer -- essentially the propagation of seismic waves triggered by the rapid deformation of the earth crust during an earthquake -- involves multiple interacting faults and depends on their orientation and geometry \cite{KilbGombergBodin2000,Gomberg2001}. Finally, there are other mechanisms (such as stress changes due to the displacement of fluids and the resulting change in pore pressure), which also trigger aftershocks, and are even more complicated \cite{Freed2005}.

To summarize, the large-scale statistics of seismic activity certainly shows power-law distributions resembling self-organized criticality and mean-field avalanche models.  While such mean-field models show reasonable results, to become more precise, they need to incorporate the statistics of fault systems and interactions between different faults, which are not easy to describe. For example, distributions of fault sizes are, themselves, ``critical'' power-laws \cite{Scholz2002}. Thus, it is unclear whether the random pinning thresholds on each fault, the complex interactions between the faults, or the random fault geometries (as suggested in \cite{Scholz1998b}), lead to the observed universal distributions (GR, Omori-Utsu, etc.) for large-scale seismicity.
On the other hand, the behaviour on smaller scales, and especially for a single fault, is much more complex, less universal, and strongly influenced by the nonlinear dynamics of friction, which depends on the local geological properties. It is not clear which, if any, of the nontrivial (and highly variable) friction properties of a single fault emerge as a collective, and possibly universal, effect of microscopical disorder, and which are simply consequences of a complex microscopic dynamics. In particular, models of single faults as elastic interfaces in disorder \cite{FisherDahmenRamanathanBenZion1997,DahmenErtasBenZion1998,MehtaDahmenBenZion2006} typically observe the same overall fault motion, as the microscopic friction law which is put into the model.
For example, in order to produce the Omori law for aftershocks, \cite{MehtaDahmenBenZion2006} supposes a logarithmical aging law for the friction force, i.e. for the pinning thresholds, on the microscopic level. As another example, the ``characteristic'' earthquake distribution observed with the velocity-weakening friction law in \cite{DahmenErtasBenZion1998} is the same stick-slip motion which would be observed with this velocity-weakening friction law even in the absence of disorder. So, while the fluctuations introduced by the disorder in such models are certainly interesting, they do not seem to explain the nontrivial constitutive laws for the \textit{total} friction force, which lead to the rich earthquake phenomenology. 
		Comparison of such models to observations is also difficult due to practical limitations: recurrence times of earthquakes on a typical fault are on the order of hundreds of years \cite{SchwartzCoppersmith1984,Scholz2002}, making it difficult to obtain statistically relevant data sets on earthquakes from a single fault.

\subsection{Other experimental systems\label{sec:OtherExp}}
It would be beyond the scope of this introduction to discuss in detail all physical systems which are described by elastic interfaces in disorder. Let me  give a few more examples beyond those mentioned above:
\begin{enumerate}
	\item \textbf{Fluid contact lines.} It has been observed that the contact line of a fluid on a rough substrate, which separates the dry from the wet region, shows a self-affine roughness, similar to the crack front roughness in fracture, eq.~\eqref{eq:FractureRoughness}; see e.g. \cite{SchafferWong2000,MoulinetGuthmannRolley2002,LeDoussalWieseMoulinetRolley2009} and the references therein. The contact-line dynamics as the fluid front advances or recedes also exhibits avalanches with power-law distributions, which agree well with the prediction from elastic interface models \cite{LeDoussalWieseMoulinetRolley2009}. 
	\item \textbf{Imbibition.} Another related system where avalanche dynamics is observed is forced-flow imbibition of a disordered medium by a viscous fluid \cite{PlanetSantucciOrtin2009,SantucciPlanetMaloyOrtin2011}. A typical experimental setup is a \textit{Hele-Shaw cell} consisting of two parallel transparent plates with a narrow gap between them. The plates can be etched, giving a well-controlled disordered medium. As fluid is forced into the gap, the interface can be observed with optical methods in real-time \cite{PlanetSantucciOrtin2009}. One finds avalanches with power-law distributions of sizes and durations, as well as non-gaussian stationary velocity distributions \cite{PlanetSantucciOrtin2009,SantucciPlanetMaloyOrtin2011}. These observations resemble a lot the avalanche phenomenology of Barkhausen noise and fracture discussed above.
	There is nevertheless a significant difference to the elastic interfaces discussed in this thesis. The total volume of the fluid in the Hele-Shaw cell is conserved, and does not fluctuate with the disorder. Hence,	the average position of the fluid-air interface is fixed \cite{DubeEtAl1999,RostLaursonDubeAlava2007,PradasEtAl2009}, in contrast to elastic interface models. This constraint is taken into account in a so-called \textit{phase-field model} \cite{DubeEtAl1999,DubeEtAl2000}, which is different from a standard elastic interface model, and outside the scope of this thesis. Nevertheless, since the avalanche phenomenology is similar (see also the numerical results in \cite{RostLaursonDubeAlava2007,PradasEtAl2009}), it would be interesting to see if some of the theoretical methods discussed here can also be applied in that case.
	\item \textbf{Vortex lattices in superconductors.} When the external magnetic field applied to a type-II superconductor exceeds a certain threshold, the superconductor is penetrated by flux lines, or \textit{vortices}, of the magnetic field. Vortices repel each other; hence, in a pure system, they arrange themselves in a regular Abrikosov lattice \cite{Abrikosov1957}. However, in the presence of disorder, vortices can be pinned to defects. The resulting lattice displacement is an elastic ``interface'' with two transversal components and two internal dimensions. Since in the following chapters, I will focus on elastic interfaces with $d$ internal dimensions but one transversal component only, the results will not be directly applicable to vortex lattices. Nevertheless, the phenomenology of vortex lattices shows significant similarities, at least qualitatively. In particular, the depinning transition of vortex lattices \cite{BlatterEtAl1994,DiScalaEtAl2012} resembles closely that of an elastic interface model (cf.~section \ref{sec:IntroInterfaceModel}). The applicability of elastic manifold models, and their predictions (in particular, a \textit{moving glass phase} \cite{GiamarchiLeDoussal1996,LeDoussalGiamarchi1998}), were discussed in \cite{GiamarchiLeDoussal1994,GiamarchiLeDoussal1995,GiamarchiLeDoussal1997,OliveEtAl2003,OliveFilySoret2009,DiScalaEtAl2012}. For an extended review of vortex lattice pinning see also the reviews \cite{BlatterEtAl1994,NattermannScheidl2000}, the thesis \cite{Chauve2000} and the references therein. Let us also note that pinned vortex lattices move intermittently in avalanches \cite{Altshuler2004,OlsonReichhardtNori1997,BasslerPaczuski1998,AransonGurevichVinokur2001}, exhibiting critical behaviour like in Barkhausen noise or in fracture discussed above. Avalanche phenomena were also observed at the onset of the dendritic flux instability \cite{DenisovEtAl2006,DenisovEtAl2006a,VestgardenEtAl2011}.
\end{enumerate}

\section{Elastic Interface Model\label{sec:IntroInterfaceModel}}
As we saw in the previous section \ref{sec:ExpAvalanches}, a wide range of interesting and experimentally accessible avalanche phenomena can be described using models of an elastic interface in a disordered medium. 
Let us represent an interface with $d$ internal dimensions, in a $d+1$ embedding space, via a \textit{height function} $u(x,t)$. $x\in \mathbb{R}^d$ is the internal coordinate along the interface, and $u(x,t)$ is the transversal displacement. The assumption that such a single-valued $u(x,t)$ describes the physical interface means that overhangs can be neglected. Historically the first and the simplest model for the dynamics of $u$ is the quenched Edwards-Wilkinson equation \cite{Feigelman1983,BruinsmaAeppli1984}
\bea
\label{eq:InterfaceQEW}
\eta\partial_t u_{xt} = f_{xt} + \nabla^2_x u_{xt} + F(u_{xt},x),
\eea
where $F(u,x)$ is a short-range correlated random force (modelling the random pinning in the disordered medium), and $f$ is the applied driving force. From here on I frequently denote space and time arguments by subscripts, i.e. $u(x,t) =: u_{xt}$, for compactness (see appendix \ref{sec:AppendixNotations} for a summary of notations). \eqref{eq:InterfaceQEW} describes an interface with \textit{short-ranged} elasticity: Its elastic energy $\int \rmd^d x (\nabla u_{xt})^2$ is a local functional of the height profile $u_{xt}$. 
For applications to magnetic domain walls (section \ref{sec:Barkhausen}) and to fracture (section \ref{sec:Fracture}), it is interesting to consider a generalization of \eqref{eq:InterfaceQEW} with a \textit{long-ranged}, power-law elastic kernel with exponent\footnote{The exponent $\mu$ defined here is the same $\mu$ as in \cite{DurinZapperi2000}. In the notations of \cite{LeDoussalWieseChauve2002}, $\mu=\alpha$ (but here we use $\alpha$ for the exponent of the avalanche duration distribution), in the notations of \cite{LeDoussalWiese2013}, $\mu=\gamma$.} $\mu+d$
\bea
\label{eq:InterfaceQEWLR}
\eta\partial_t u_{xt} = f_{xt} + \int_y \frac{u_{yt}-u_{xt}}{|x-y|^{\mu+d}} + F(u_{xt},x).
\eea
From now on we will frequently use the shorthand $\int_x := \int \rmd^d x$; see appendix \ref{sec:AppendixNotations} for a summary of notations. For sufficiently large $\mu$, the divergence at $x=y$ in \eqref{eq:InterfaceQEWLR} needs to be regularized by some small-scale (UV) cutoff, e.g. by a discrete lattice spacing.
A slightly different model is obtained if the driving force $f_{xt}$ in \eqref{eq:InterfaceQEW}, \eqref{eq:InterfaceQEWLR} is replaced by a harmonic force (a spring) with spring constant $m^2$, pulled along a trajectory $w(x,t)$
\bea
\label{eq:InterfaceEOMi}
\eta\partial_t u_{xt} = - m^2(u_{xt}-w_{xt}) + \nabla^2_x u_{xt} + F(u_{xt},x),\\
\label{eq:InterfaceEOMLRi}
\eta\partial_t u_{xt} = - m^2(u_{xt}-w_{xt}) + \int_y \frac{u_{yt}-u_{xt}}{|x-y|^{\mu+d}} + F(u_{xt},x).
\eea
\eqref{eq:InterfaceEOMi}, \eqref{eq:InterfaceEOMLRi} correspond to an ensemble where instead of fixing the driving force $f$, we fix the position of the driving spring $w$. They reduce to \eqref{eq:InterfaceQEW}, \eqref{eq:InterfaceQEWLR} in the limit $m \to 0, m^2 w_{xt} \to f_{xt}$, so we expect their universal properties (which are obtained in the limit of large systems and $m\to 0$) to be related. 

Magnetic domain walls, discussed in section \ref{sec:Barkhausen}, have an internal dimension $d=2$. In polycristalline ferromagnets, dipolar interactions lead to long-ranged elastic interactions between different parts of the domain wall. This class of domain walls is described by \eqref{eq:InterfaceEOMLRi} with $\mu = 1$ \cite{CizeauZapperiDurinStanley1997,ZapperiCizeauDurinStanley1998}. On the other hand, in amorphous ferromagnets, dipolar interactions are absent and they are described better by $\mu=2$ or \eqref{eq:InterfaceEOMi} \cite{DurinZapperi2000}. In both cases, the domain-wall is affected by a demagnetizing field, which acts similarly to the harmonic confinement in \eqref{eq:InterfaceEOMi}, \eqref{eq:InterfaceEOMLRi}.\footnote{However, it applies to the global displacement $\int_x u_{xt}$ and not to the local displacement $u_{xt}$ \cite{CizeauZapperiDurinStanley1997,ZapperiCizeauDurinStanley1998,DurinZapperi2000}. The precise form of the infrared cutoff in \eqref{eq:InterfaceEOMi}, \eqref{eq:InterfaceEOMLRi} is thus not necessarily accurate.}

Crack fronts in fracture (see section \ref{sec:Fracture}) are described by \eqref{eq:InterfaceEOMLRi} with $d=1$ and $\mu=1$ \cite{BonamySantucciPonson2008}. There, the role of the harmonic confinement is played by the elastic deformation of the bulk of the material.\footnote{However, as in the case of magnetic domain walls, it is argued to acts on the total displacement of the crack front only \cite{BonamySantucciPonson2008}.} Fluid contact lines (see section \ref{sec:OtherExp}) are also described by \eqref{eq:InterfaceEOMLRi} with $d=1$ and $\mu=1$ \cite{MoulinetGuthmannRolley2002,LeDoussalWiese2010}. The case $\mu=1$ will be called ``long-ranged elasticity'' from now on.

Let us defer a detailed discussion of the analytical approaches to \eqref{eq:InterfaceQEW} and \eqref{eq:InterfaceQEWLR} to the following chapters. For now, I will only give a brief overview of its phenomenology (section \ref{sec:InterfacePhenomenology}), numerical results (section \ref{sec:Numerics}), and some general monotonicity properties of these elastic interface models (section \ref{sec:InterfaceMonot}).

\subsection{Phenomenology\label{sec:InterfacePhenomenology}}
\begin{figure}%
\centering
\includegraphics[width=0.4\columnwidth]{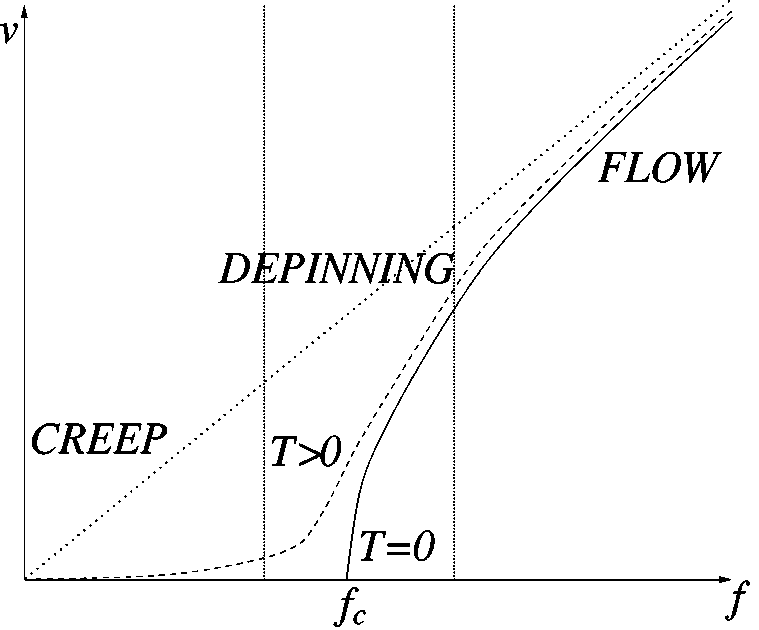}%
\caption{Typical force-velocity characteristics of an elastic interface model. Figure reprinted with permission from \cite{ChauveGiamarchiLeDoussal2000}. Copyright \copyright \, 2000 by the American Physical Society.}%
\label{fig:ForceVelocity}%
\end{figure}
The equation \eqref{eq:InterfaceQEW} is straightforward to simulate numerically, by discretizing the interface along its $d$ internal directions (i.e. by discretizing the $x$ variable in \eqref{eq:InterfaceQEW}). Applying a constant force $f_{xt}=: f$, one observes that after an initial transient (which exhibits interesting aging phenomena, see \cite{SchehrLeDoussal2005,KoltonSchehrLeDoussal2009}, but shall not be discussed here in detail), the interface motion becomes stationary. The interface then moves with a constant average velocity $v$, which depends on the applied force $f$ via the velocity-force characteristics indicated in figure \ref{fig:ForceVelocity} \cite{NattermannStepanowTangLeschhorn1992,LeschhornNattermannStepanow1997,Chauve2000,ChauveGiamarchiLeDoussal2000}. In particular, there is a finite threshold $f_c$ (the \textit{critical force}), below which $v=0$ and the interface remains stuck. At $f=f_c$, one observes a \textit{depinning transition}. Similar to a normal phase transition, it is characterized by critical power laws in the moving phase $f> f_c$: for example, the average velocity grows as $v \sim |f-f_c|^{\beta}$, and the interface shows a self-affine roughness,
\bea
\label{eq:IntroDefZetaZ}
\overline{\left[u(x,t)-u(x',t)\right]^2}^c \sim |x-x'|^{2\zeta},\quad\quad\quad \overline{\left[u(x,t)-u(x,t')\right]^2}^c \sim |t-t'|^{2\zeta/z}
\eea
where $\overline{\cdot}^c$ means ``cumulant'' or ``connected average'' over all realizations of the random force $F$\footnote{The connected average is required to remove the average $\overline{u(x,t)-u(x,t')} = v (t-t')$ for the temporal correlations.}. $\zeta$ is the \textit{roughness exponent}, and $z$ is the \textit{dynamical exponent} of the interface model. Furthermore, the interface motion proceeds in \textit{avalanches}, similar to the experimental observations discussed above \cite{DahmenSethna1996,KuntzSethna2000,SethnaDahmenMyers2001,RossoLeDoussalWiese2009}. The response to a small step in the driving force $f_{xt}$ is random, and fluctuates over a wide range of scales. Most of the time the interface remains pinned in a metastable state and hardly moves, but sometimes it performs a large jump. The distributions of their sizes and durations have the critical forms \eqref{eq:IntroBNScalingForms}, and computing them analytically will be the main subject of the following chapters \ref{sec:BFM} and \ref{sec:OneLoop}. These critical properties are visible for $|x-x'|$ smaller than a certain cutoff scale (the \textit{correlation length}) $\xi$, which becomes large as $f \searrow f_c$, $\xi \sim |f-f_c|^{-\nu}$. In this sense, the point $f=f_c, v=0^+$ is a dynamical critical point.

If one considers \eqref{eq:InterfaceEOMi}, i.e. an interface driven at a constant driving velocity $\dw_{xt}=v$ by the harmonic confinement, the behavior is similar. The  force applied by the harmonic well $f_t = \frac{1}{L^d}m^2\int_x (w_{xt}-u_{xt})$ now fluctuates, but its average $\overline{f}$ is still given by the force-velocity curve in figure \ref{fig:ForceVelocity}. As one lowers $v \searrow 0$, the mean pinning force approaches the critical force $\overline{f} \searrow f_c$. Avalanches arise when the harmonic well, i.e. $w$, is moved, which changes the applied force $f_{xt} =m^2 w_{xt}$. As we shall see in section \ref{sec:InterfaceMonot}, monotonous motion of the harmonic well, $\dw_{xt} \geq 0$, makes the interface traverse monotonously an ordered set of metastable states (\textit{Middleton states}). A related, but different, problem is the study of the static ground state of the interface, and its jumps (shocks, or static avalanches) as $w$ is varied. The differences and similarities between these two cases will be discussed further in section \ref{sec:FRGReviewOneLoop}. For now, let us focus on the overdamped dynamics \eqref{eq:InterfaceEOMi}/quasi-static depinning.

The sharp depinning transition is smeared out in a finite-size system, or by thermal fluctuations (see the $T>0$ curve in figure \ref{fig:ForceVelocity}). These also lead to interesting (and experimentally relevant) \textit{creep} phenomena for $f < f_c$ \cite{FeigelmanGeschkenbeinLarkinVinokur1989,BlatterGeshkenbeinVinokur1991,BlatterEtAl1994,LeDoussalVinokur1995,LemerleEtAl1998,ChauveGiamarchiLeDoussal2000,MuellerGorokhovBlatter2001,KoltonRossoGiamarchi2005,KoltonRossoGiamarchiKrauth2006,KoltonRossoGiamarchiKrauth2009} and universal features of thermal rounding \cite{Middleton1992a,BustingorryKoltonGiamarchi2008,BustingorryKoltonGiamarchi2012,BustingorryKoltonGiamarchi2012b}. Quantum fluctuations, which are important e.g.~in vortex lattices in superconductors (cf.~section \ref{sec:OtherExp}), lead to the similar but distinct \textit{quantum creep} for $f< f_c$ \cite{BlatterGeschkenbein1993}, and quantum vortex lattice melting \cite{BlatterIvlev1993}. These are, however, outside the scope of this thesis. Instead, we shall focus on the zero-temperature classical depinning transition and analyze in detail the avalanche phenomena arising near it. 

The interface roughness exponent $\zeta$, and the typical avalanche scales tend to zero as the internal dimension of the interface $d$ approaches a certain critical dimension $d_c$. For $d \geq d_c$, the interface is macroscopically flat ($\zeta=0$), and mean-field theory (discussed in chapter \ref{sec:BFM}) is valid.\footnote{At $d=d_c$, there are logarithmic corrections. For a more detailed discussion see \cite{FedorenkoStepanow2003,LeDoussalWiese2013} and section \ref{sec:AvalancheAction}.} For $d < d_c$, $\zeta > 0$ and the interface is rough. This case will be discussed in chapter \ref{sec:OneLoop}.
For short-ranged elasticity \eqref{eq:InterfaceEOMi}, we will see in section \ref{sec:FRGReview} that $d_c=4$, whereas for a long-ranged elastic kernel \eqref{eq:InterfaceEOMLRi}, $d_c = 2 \mu$ (see section \ref{sec:OneLoopLongRange}). We see that for domain walls in polycristalline ferromagnets, where $\mu=1$, $d=d_c=2$, and hence mean-field theory applies \cite{CizeauZapperiDurinStanley1997,ZapperiCizeauDurinStanley1998}. On the other hand, for amorphous ferromagnets, a model with short-ranged elasticity is adequate and hence $d=2<4=d_c$. For these materials, one would expect the domain walls to have a nonzero self-affine roughness exponent $\zeta$, in contrast to the polycristalline case where $\zeta=0$ and the domain walls are macroscopically flat. It would be interesting to see if this hallmark of non-mean-field behaviour can be observed experimentally.
Fracture, where $\mu=1$, has $d=1<2=d_c$ and is also in a non-mean-field universality class.

Before we proceed with analytical considerations, let us briefly review numerical work on the interface models \eqref{eq:InterfaceEOMi}, \eqref{eq:InterfaceEOMLRi}.

\subsection{Numerical results\label{sec:Numerics}}

\subsubsection{Short-ranged elasticity}
Early numerical work \cite{Leschhorn1993,LeschhornNattermannStepanow1997} has focused on the quenched Edwards-Wilkinson interface with short-ranged elasticity \eqref{eq:InterfaceQEW}, and on the roughness exponent $\zeta$ and the dynamical exponent $z$. In \cite{RossoKrauth2001,RossoHartmannKrauth2003} a new and very efficient algorithm for computing $\zeta$ was developed. It is based on jumping directly to the next quasi-static state, without simulating the dynamics in between. In $d=1$ a recent high-precision study using GPUs \cite{FerreroBustingorryKolton2013} gives the (currently most precise) values $\zeta=1.250\pm 0.005$ and $z = 1.433 \pm 0.007$. Other reported values for $\zeta$ and $z$ are shown in table \ref{tbl:ExpSRSims}.

Avalanche observables are less well known. The avalanche size exponent in $d=1$ was determined in \cite{RossoLeDoussalWiese2009} as $\tau = 1.08 \pm 0.02$.  In this work \cite{RossoLeDoussalWiese2009}, the entire avalanche size distribution $P(S)$ was also computed, and compared to the analytical predictions (see section \ref{sec:OneLoopSize}).\footnote{Note also the related study \cite{LeDoussalMiddletonWiese2009}, which obtains similar (but slightly different) values for $\tau$ for static avalanches in a random-field landscape. Since it is known that statics and depinning differ (see section \ref{sec:FRGReview}, this is not surprising.}
Additional numerical work is in progress \cite{KoltonEtAl2013inpr} towards measuring the avalanche exponents (in particular, $\alpha$ and $a$) and scaling functions directly.

A particularly interesting case is $d=2$, which corresponds to magnetic domain walls with short-ranged elastic interactions (i.e.~amorphous soft magnets as discussed in section \ref{sec:Barkhausen}). Numerical simulations performed in \cite{DurinZapperi2000} give $\tau \approx 1.27$ and $\alpha \approx 1.5$, in agreement with experiments as discussed in \cite{DurinZapperi2000} and with scaling relations between $\zeta, z$ and $\alpha, \tau$ (see section \ref{sec:FRGScalingRelations}, and table \ref{tbl:ExpSRSims}). 
For the related problem of the static zero-temperature ground state of the interface, the avalanche size exponent was determined in \cite{LeDoussalMiddletonWiese2009} to be $\tau \approx 1.25 \pm 0.02$ in $d=1$ and $\tau \approx 1.37 \pm 0.03$ in $d=2$.

\begin{center}
\begin{table}
 \footnotesize
\begin{center}
   \begin{tabular}{| c || c | c | c | c | c |}
	\hline
	$d$ & $\zeta$ & $z$ & $\tau$ & $\alpha$ & $a$ \\
	\hline
      $1$ & \pbox{20cm}{
			$1.25 \pm 0.01$ \cite{Leschhorn1993} \\
			$1.25 \pm 0.01$ \cite{LeschhornNattermannStepanow1997} \\
			$1.26 \pm 0.01$ \cite{RossoHartmannKrauth2003}  \\
			\colorbox{light-gray}{$1.250\pm 0.005$} \cite{FerreroBustingorryKolton2013}
			} & \pbox{20cm}{
			$1.42 \pm 0.03$ \cite{Leschhorn1993} \\
			$1.42 \pm 0.04$ \cite{LeschhornNattermannStepanow1997} \\
			\colorbox{light-gray}{$1.433 \pm 0.007$} \cite{FerreroBustingorryKolton2013}
			} & \pbox{20cm}{
			$1.08 \pm 0.02$ \cite{RossoLeDoussalWiese2009} \\
			$1.111\pm 0.002 \,^*$
			} &
			$1.174 \pm 0.004 \,^*$ &
			$-0.45 \pm 0.03 \,^*$
			\\
	\hline
			$2$ & 
			\pbox{20cm}{
			$0.75\pm 0.02$ \cite{Leschhorn1993} \\
			$0.75\pm 0.02$ \cite{LeschhornNattermannStepanow1997} \\
			\colorbox{light-gray}{$0.753 \pm 0.002$} \cite{RossoHartmannKrauth2003} 
			} & \pbox{20cm}{
			$1.58 \pm 0.04$ \cite{Leschhorn1993} \\
			\colorbox{light-gray}{$1.56\pm 0.06$} \cite{LeschhornNattermannStepanow1997}
			}
			& \pbox{20cm}{
			$1.27$ \cite{DurinZapperi2000} \\
			$1.274 \pm 0.001 \,^*$
			} & \pbox{20cm}{
			$1.5$ \cite{DurinZapperi2000} \\
			$1.48 \pm 0.02 \,^*$
			} & 
			$0.32 \pm 0.08 \,^*$
			\\
	\hline
			$3$ & 
			\pbox{20cm}{
			$0.35 \pm 0.01$ \cite{LeschhornNattermannStepanow1997} \\
			\colorbox{light-gray}{$0.355 \pm 0.01$} \cite{RossoHartmannKrauth2003} 
			} &
			\colorbox{light-gray}{$1.75\pm 0.15$} \cite{LeschhornNattermannStepanow1997}
			&
			$1.404 \pm 0.002 \,^* $
			&
			$1.77 \pm 0.07 \,^* $
			&
			$0.8 \pm 0.1 \,^*$ \\
			\hline
   \end{tabular}
	\end{center}
   \caption{Exponent values for short-ranged elasticity. $\zeta$ and $z$ are the roughness and dynamical exponents defined in \eqref{eq:IntroDefZetaZ}, $a, \tau$ and $\alpha$ are the total velocity, avalanche size and avalanche duration exponents defined in \eqref{eq:IntroBNScalingForms}. Values with references are taken from the corresponding numerical simulations, values with $*$ are obtained using the scaling relations in section \ref{sec:FRGScalingRelations} from the ``best'' values for $\zeta$ and $z$ highlighted with a gray background.\label{tbl:ExpSRSims}}
\end{table}
\end{center}

\subsubsection{Long-ranged elasticity\label{sec:NumericsLR}}
Now let us turn to long-ranged elasticity, $\mu = 1$ in \eqref{eq:InterfaceQEWLR}. Here the only dimension of interest is $d=1$, since for $d \geq 2$ we expect mean-field behaviour. Table \ref{tbl:ExpLRSims} summarizes the known results on the scaling exponents from numerics, scaling relations, and compares them to experiments.

Avalanche sizes in $d=1$ for long-ranged elasticity were first simulated in \cite{BonamySantucciPonson2008}, giving the exponent $\tau \approx 1.25$ for total avalanche sizes. As mentioned in section \ref{sec:Fracture}, due to the long-ranged elastic interactions in this case avalanches may decompose into disconnected clusters. These have different scaling exponents, in particular the distribution of cluster sizes $S_{\text{cl}}$ has an exponent $\tau_{\text{cl}}$ given in numerical simulations by $\tau_{\text{cl}} \approx 1.65 \pm 0.05$ \cite{BonamySantucciPonson2008} or $1.52 \pm 0.05$ \cite{LaursonSantucciZapperi2010}. \cite{LaursonSantucciZapperi2010} also gives an heuristic argument for $\tau_{\text{cl}} = 2\tau-1$ based on the assumption that during an avalanche, the number of active cluster evolves as a random walk. However, it is not clear if this argument is exact and can be made more precise. $\tau_{\text{cl}}$ agrees well with fracture experiments (see eq.~\eqref{eq:FractureExpExponents}), but no published values for the global $\tau$ in fracture experiments are available. The exponent $a_{\text{loc}}$ for the distribution of local velocities has not been measured in numerical simulations, but can be obtained from $\zeta$ and $z$ through the scaling relation \eqref{eq:OneLoopScalingExpLR} as $a_{\text{loc}} \approx 4.6$. In fracture experiments, $a_{\text{loc}}$ is observed to be $2.55 \pm 0.15$ \cite{MaloySantucciSchmittbuhlToussaint2006,TallakstadEtAl2011,TallakstadToussaintSantucciMaloy2013}. Although both values are rather quickly decaying power laws, the absolute values differ significantly. This discrepancy remains to be understood; in particular, it would be interesting to have a direct numerical measurement of $a_{\text{loc}}$ in a simulation.

\begin{center}
\begin{table}%
\begin{tabular}{| c || c | c | c | }
\hline
Exponent	& Numerics & Scaling Relations & Experiments \\
\hline
$\zeta$ & 
\pbox{20cm}{
			\colorbox{light-gray}{$0.388 \pm 0.002$} \cite{RossoKrauth2002} \\
			$0.385 \pm 0.005$ \cite{DuemmerKrauth2007} 
			}
			&	n/a &
			\pbox{20cm}{
			$0.35 \pm 0.05$ (fracture, \cite{SantucciEtAl2010}) \\
			$0.51 \pm 0.03$ (fluid contact lines \cite{MoulinetGuthmannRolley2002})
			}
\\
\hline
$z$ & \colorbox{light-gray}{$0.770 \pm 0.005$} \cite{DuemmerKrauth2007}& n/a &   \\
\hline
$\tau$ & 
\pbox{20cm}{ 
$1.25$ \cite{BonamySantucciPonson2008} \\
$1.25 \pm 0.05$ \cite{LaursonSantucciZapperi2010}
} & $1.280 \pm 0.001$ &
 \\
\hline
$\alpha$ & 1.43 \cite{BonamySantucciPonson2008} & $1.504 \pm 0.004$ & 
\\
\hline
$\tau_{\text{cl}}$ & 
\pbox{20cm}{ 
$1.65 \pm 0.05$ \cite{BonamySantucciPonson2008} \\
$1.52 \pm 0.05$ \cite{LaursonSantucciZapperi2010}
} & & $1.56 \pm 0.04$ (fracture, \cite{TallakstadEtAl2011}) \\
\hline
$a$ & & $0.38 \pm 0.01$ & \\
\hline
$a_{\text{loc}}$ & & $4.62 \pm 0.04$ & $2.55 \pm 0.15$ (fracture, \cite{MaloySantucciSchmittbuhlToussaint2006}) \\
\hline
\end{tabular}
   \caption{Exponent values for long-ranged elasticity, $\mu=1$ and $d=1$. The values in the ``scaling relation'' column are obtained using the scaling relations in section \ref{sec:OneLoopLRScaling} from the ``best'' values for $\zeta$ and $z$ highlighted with a gray background.\label{tbl:ExpLRSims}}
\end{table}
\end{center}


Now let us turn towards an analytical discussion of the model \eqref{eq:InterfaceQEW}, \eqref{eq:InterfaceQEWLR}.

\subsection{Monotonicity and Middleton states \label{sec:InterfaceMonot}}
In the following sections we will heavily use several monotonicity properties of the elastic interface models defined above. These are discussed at length e.g. in \cite{Rosso2002}, but let us review them here for completeness. We generalize \eqref{eq:InterfaceEOMLRi} with an arbitrary linear elastic kernel, and also take a derivative in order to obtain an equation-of-motion for the interface velocity $\du_{xt}$
\bea
\label{eq:InterfaceEOMGenEl}
\eta\partial_t u_{xt} =& - m^2(u_{xt}-w_{xt}) + \int_y c_{xy}\left(u_{yt}-u_{xt}\right) + F(u_{xt},x), \\
\label{eq:InterfaceEOMGenElVel}
\eta\partial_t \du_{xt} =& - m^2(\du_{xt}-\dw_{xt}) + \int_y c_{xy}\left(\du_{yt}-\du_{xt}\right) + \partial_t F(u_{xt},x). 
\eea
We assume the elastic kernel to be convex, $c_{xy} \geq 0$.

\begin{figure}%
\includegraphics[width=\columnwidth]{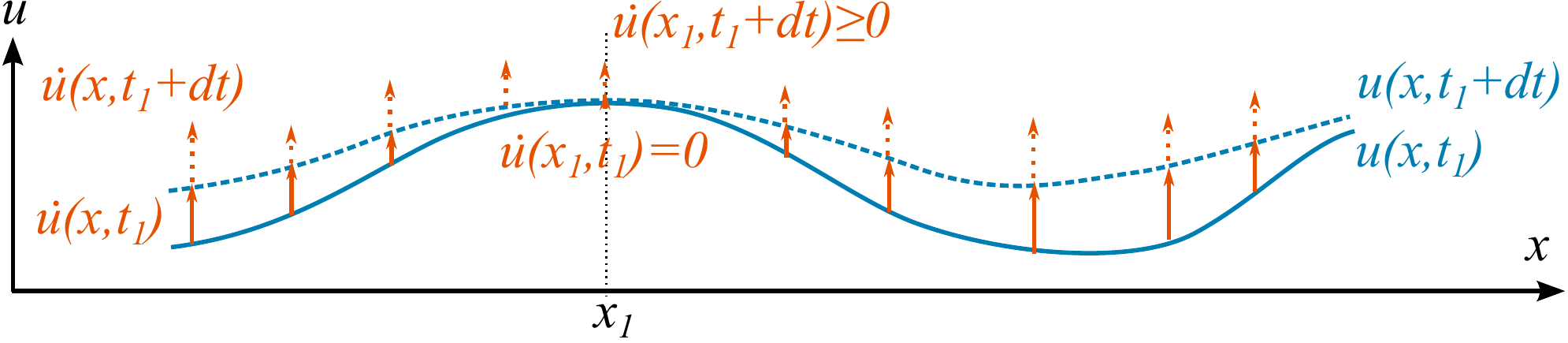}%
\caption{Monotonicity of interface motion. The point $(x_1,t_1)$ is the first one where the velocity $\du$ becomes zero. Hence, the only change in $\du$ at this point comes from the elastic and driving forces. Since the rest of the interface is still moving forward and the driving is monotonous, the new velocity $\du(x_1,t_1+dt)$ will be non-negative.\label{fig:MonotInterface1}}
\end{figure}

The first important result is that \textit{solutions of \eqref{eq:InterfaceEOMGenEl} are monotonous}, $\du_{xt} \geq 0$ for all $x,t \geq \ti$, under two assumptions:
\begin{itemize}
	\item The initial condition at some initial time $\ti$ satisfies $\du_{x,\ti} \geq 0$ for all $x$, and
	\item the driving is monotonous everywhere, $\dw_{xt} \geq 0$ for all $x$, $t$.
\end{itemize}
This result is a variant of \textit{Middleton's theorem} or ``no-passing rule'' \cite{Middleton1992,BaesensMacKay1998}.\footnote{As a mathematical statement, it is known since \cite{Hirsch1985}.} 
To see that the claim holds, consider eq.~\eqref{eq:InterfaceEOMGenElVel} for the local interface velocities. Before the local velocity becomes negative, it must (by continuity) pass through zero. Let us take the point $x_1$ and the time $t_1$ where this happens for the first time (see also figure \ref{fig:MonotInterface1}). Since all other points are still moving forward, the elastic term $\int_y c_{x_1 y}\left(\du_{yt}-\du_{x_1 t}\right) \geq 0$. The same holds for a local elastic term, such as $\nabla^2 \du_{x_1,t_1}$ in \eqref{eq:InterfaceEOMi}, and in fact for any, not necessarily linear, elastic kernel, as long as it is convex, i.e.~moving one point forward never decreases the force on another point \cite{Hirsch1985}. By the choice of $x_1, t_1$ we have $\du_{x_1,t_1}=0$ and hence also $\partial_t\big|_{x=x_1,t=t_1} F(u_{xt},x) = \du_{x_1,t_1}\partial_u\big|_{u=u_{x_1,t_1}}F(u,x_1) = 0$.\footnote{As such, this holds only for smooth (differentiable) disorder. However, even for non-differentiable Brownian disorder which will be discussed in chapter \ref{sec:BFM}, we have $\partial_t F(u_{xt},x) = \sqrt{\du_{xt}}\xi_{xt}$ (see section \ref{sec:BFMVelocity}) vanishing when $\du_{xt}=0$, so the argument stays true as long as the disorder is continuous.} Inserting this into eq.~\eqref{eq:InterfaceEOMGenElVel} shows that the increment of the velocity $\partial_t\big|_{t=t_1,x=x_1}\du_{xt} \geq m^2 \dw_{x_1,t_1} \geq 0$, so the local velocity $\du_{xt}$ has no way of becoming negative. 

Let us now consider \textit{metastable states}, i.e. configurations $u_x$ for which the right-hand side of \eqref{eq:InterfaceEOMGenEl} vanishes for a fixed position $w_{x}$ of the harmonic confinement.\footnote{In this definition, we do not require metastable states to be local minima of the energy landscape; they can, in principle, also be saddle points or maxima.} The second important property of \eqref{eq:InterfaceEOMGenEl} at monotonous driving, which will be shown in the following, is:

For any configuration $w_{x}^{\mathrm{f}}$ of the harmonic confinement, there is a \textit{unique leftmost metastable state (Middleton state)} \cite{Middleton1992} $u_{x}^{\mathrm{f}}[w^{\mathrm{f}}]$. This is the state reached for $t\to\infty$ after preparing the interface at some initial $u_{x,\ti}^{\mathrm{i}} \ll w_{x}^{\mathrm{f}}$, and driving it monotonously until the harmonic confinement rests in the final configuration $w_{x,t} = w_x^{\mathrm{f}}$, for all $t\geq \tf$. This asymptotic state $\lim_{t\to\infty} u_{x,t}$ at fixed $w_x^{\mathrm{f}}$ is independent of the initial condition (as long as the assumption $u_{x,\ti}^{\mathrm{i}} \ll w_{x}^{\mathrm{f}}$ holds), and of the form of the driving for $t \leq \tf$ (as long as it is monotonous).

To see this, take any metastable state $u_x^{0}$ for $w = w_x^{\mathrm{f}}$ and consider the trajectory $u_{x,t}$ for the driving $w_{x,t}$ which terminates at $w_{x,t}=w_x^{\mathrm{f}}$ for $t\geq \tf$. We claim that $u_x^{0} \geq u_{x,t}$ for all $x,t$, i.e. that the trajectory can not cross \textit{any metastable state} (defined as above) corresponding to the final location of the harmonic well. Assume this were not the case. Clearly, the inequality $u_x^{0} \geq u_{x_i,t_i}$ holds for the initial condition at $u_{x,\ti}^{\mathrm{i}} \ll w_{x}^{\mathrm{f}}$.
This means, that at some time and at some point $x_1,t_1$ the trajectory $u_{x,t}$ must have crossed the metastable state $u_x^{0}$ for the first time. At this point, by monotonicity of the driving, $w_{x_1,t_1} \leq w^{\mathrm{f}}_{x_1}$. Then, since the right-hand side of \eqref{eq:InterfaceEOMGenEl} vanishes at $x=x_1$ for the metastable state $u_x^{0}$ and for $w=w^{\mathrm{f}}_{x_1}$, and since for all other $x$ $u_{x,t_1} \leq u_x^{0}$, by convexity as above we have $\partial_t \big|_{t_1}\du_{x_1,t} \leq 0$, i.e. the interface is pulled backwards and cannot cross the final state $u_x^{0}$. Thus, the asymptotic state for $t\to\infty$ with driving $w_{x,t}$ will be $u_x^{\mathrm{f}} \leq u_x^{0}$. Since this holds for any metastable state $u_x^{0}$ (in particular for the asymptotic state of $u_{x,t}$ as $t\to\infty$ for any other choice of the driving $w$ which terminates at $w_x^{\mathrm{f}}$), this shows that the final state is unique, independent of the driving, and of the initial condition (as long as the assumption $u_{x,\ti}^{\mathrm{i}} \ll w_{x}^{\mathrm{f}}$ holds).
This argument explains why it also makes sense to describe the Middleton state as the \textit{leftmost metastable state} for the given position $w^{\mathrm{f}}$ of the harmonic well, although in general metastable states of \eqref{eq:InterfaceEOMGenEl} are not ordered and may cross.
In practice, the Middleton state can be obtained by fixing the harmonic well at $w_{x}^{\mathrm{f}}$ for all times and starting at some initial $u_{x,\ti}^{\mathrm{i}} \ll w_{x}^{\mathrm{f}}$. It is also the state reached at $t=0$ when driving the interface quasi-statically starting from $t=-\infty$, according to $w_{xt} = vt + w_{x}^{\mathrm{f}}$ and $v\to 0^+$.

Recall that the Middleton state is defined as the limiting state reached for long times for a fixed $w_x^\mathrm{f}$ from a \textit{generic} initial condition (we only require the initial condition to be far on the left of $w^{\mathrm{f}}_x$). Thus, we expect the Middleton state to be not only the leftmost state with vanishing right-hand-side of \eqref{eq:InterfaceEOMGenEl} (as shown above), but actually a local minimum of the energy landscape (since else its attractor would only consist of isolated points). However, showing this rigorously is more difficult and will not be attempted here.

%% file: MeanField.tex
\chapter{Elastic Interfaces in a Brownian Force Landscape\label{sec:BFM}}
Let us consider the quenched Edwards-Wilkinson equation \eqref{eq:InterfaceEOMi}, as motivated in the preceding introduction
\bea
\label{eq:InterfaceEOM}
\eta\partial_t u_{xt} = - m^2(u_{xt}-w_{xt}) + \nabla^2_x u_{xt} + F(u_{xt},x).
\eea

In this chapter I discuss the special case where the increments of the pinning force $F(u,x)$ in the $u$ direction are white noises, independent at each point $x$ in space
\bea
\label{eq:BFM}
\overline{\partial_u F(u,x) \partial_{u'}F(u',x')} = \delta^d(x-x')\partial_u \partial_{u'} \Delta(u,u') = 2\sigma \delta^d(x-x')\delta(u-u').
\eea
Since this leads to non-differentiable disorder, the choice of discretization in \eqref{eq:InterfaceEOM} is important. We will henceforth always interpret \eqref{eq:InterfaceEOM} in the sense of an It\^{o} stochastic differential equation.
The uncorrelated force increments \eqref{eq:BFM} lead to force correlations growing linearly in $u$,
\bea
\label{eq:BFM2}
\overline{\left[F(u,x)-F(u',x)\right]^2} \sim \sigma |u-u'|.
\eea
This linear growth of correlations is the hallmark of Brownian motion, and the model \eqref{eq:InterfaceEOM} with \eqref{eq:BFM} is hence called the \textit{Brownian Force Model} (BFM) \cite{LeDoussalWiese2011,LeDoussalWiese2011b,LeDoussalWiese2013,DobrinevskiLeDoussalWiese2012,DobrinevskiLeDoussalWiese2013inpr}. \eqref{eq:BFM2} shows that for this model, the pinning force has infinite-ranged correlations in the $u$ direction. This is a priori unrealistic; one would expect the correlations of the physical pinning force be short-ranged, i.e. to decay to zero as $|u-u'| \to \infty$. 

Nevertheless, the BFM is physically interesting. Through phenomenological arguments for the center-of-mass of an interface (see \cite{CizeauZapperiDurinStanley1997,ZapperiCizeauDurinStanley1998,Colaiori2008} and section \ref{sec:ReviewABBM}) or a field-theoretic treatment valid for more general, space-dependent observables (\cite{LeDoussalWiese2008c,LeDoussalMiddletonWiese2009,LeDoussalWiese2011b,LeDoussalWiese2011,LeDoussalWiese2013} and section \ref{sec:AvalancheAction}), the force field \eqref{eq:BFM} of the BFM can be shown to be the mean-field limit of a disorder with short-ranged correlations, as long as one considers only small displacements from a Middleton state. Thus, studying the BFM allows understanding avalanches of elastic interfaces in physical, short-ranged, disorder at or above a certain critical dimension $d_c$.
In particular, it is applicable to magnetic domain walls in polycristalline ferromagnets, where $d_c=2$ due to the presence of long-range dipolar elastic forces, as discussed in sections \ref{sec:Barkhausen} and \ref{sec:InterfacePhenomenology}.
For short-ranged disorder below the critical dimension, the BFM can be used as the basis for a perturbative expansion and renormalization-group treatment \cite{LeDoussalWiese2008c,LeDoussalWiese2011,LeDoussalWiese2011b,LeDoussalWiese2013,DobrinevskiLeDoussalWiese2013inpr}. This will be discussed in detail in the following chapter \ref{sec:OneLoop}. Before that, let me review existing literature and discuss the new results obtained on the BFM during my thesis. On this simple and solvable case, I will define the concepts and observables used to study avalanches analytically in section \ref{sec:BFMAvalanches}. These definitions will then be used throughout the rest of the thesis.

\section{Introduction and review\label{sec:ReviewABBM}}
The Brownian Force landscape first arose in a phenomenological model for Barkhausen noise (\textit{ABBM model}) proposed by Alessandro, Beatrice, Bertotti, Montorsi \cite{AlessandroBeatriceBertottiMontorsi1990,AlessandroBeatriceBertottiMontorsi1990b}.\footnote{For historical details and an extensive review of the ABBM model, see \cite{Colaiori2008}.} They postulated an effective equation of motion for the total magnetization $u(t)$ of a ferromagnetic sample 
\bea
\label{eq:ABBM}
\Gamma \du(t) = 2  I_s\big[H(t) - k u(t) + F(u(t))\big].
\eea
I follow here the conventions of \cite{ZapperiCastellanoColaioriDurin2005,ColaioriDurinZapperi2007} and \cite{DobrinevskiLeDoussalWiese2013}.
$I_s$ is the saturation magnetization, and $H(t)$ the external field which drives the domain-wall motion. A typical choice is a constant ramp rate $ c$, $ H(t)=c\, t = k v\, t$, which leads to a constant average magnetization rate $\overline{\du} = v  = c/k$ \cite{AlessandroBeatriceBertottiMontorsi1990}. $k$ is the demagnetizing factor characterizing the strength of the demagnetizing field $-k u$ generated by effective free magnetic charges on the sample boundary \cite{AlessandroBeatriceBertottiMontorsi1990,Colaiori2008}. The domain-wall motion induces a voltage proportional to its velocity $\du(t)$, which is the measured Barkhausen-noise signal. Finally, $F(u(t))$ is a random local pinning force. 
ABBM made the guess that the effective disorder $F(u)$ acting on the total magnetization $u$ is a Brownian motion. In other words, it is a Gaussian with mean 0 and correlations
\bea
\label{eq:DisorderABBM}
\overline{\left[F(u)-F(u')\right]^2} = 2\sigma |u-u'|.
\eea
This guess can be understood by a simple phenomenological argument \cite{CizeauZapperiDurinStanley1997,ZapperiCizeauDurinStanley1998,Colaiori2008}: In the mean-field limit, elasticity is strong and the interface is essentially flat. There are no correlations between distant points of the interface; it decomposes into small ``correlation volumes'' which are independent. Thus, after some coarse-graining, the interface can be described by a discrete model with $N$ lattice sites (corresponding to correlation volumes of the original interface), and local disorder $\eta(u_i)$ (with mean zero, independent for each $i=1,...,N$, and uncorrelated in the $u$ direction, as expected physically). To move from total magnetization $u$ to total magnetization $u'$, a certain number $n < N$ of sites unpin and move. For small $|u-u'|$, since the sites are independent, we can assume $n \propto |u-u'|$ (for large $|u-u'|$ this will no longer hold, since $n$ must saturate at $N$). Due to the absence of correlations, after the jump the pinning forces of these $n$ sites are replaced by new independent forces $\eta(u_j')$. The change in the effective pinning force is then given by $\Delta F = F(u')-F(u)= \sum_{j=1}^n \left[\eta(u_j')-\eta(u_j)\right]$. Since the increments $\eta(u_j')-\eta(u_j)$ are independent and have mean zero, $\Delta F$ has mean zero and variance $\propto n $. Recall that $n \propto |u-u'|$, under the assumption $|u-u'|$ is small. Thus, $\Delta F \propto |u-u'|$, and the effective pinning field is a Brownian (at least on small scales). Another, more formal, way to recover the Brownian correlations in the mean-field limit is by considering the tree limit of the corresponding field theory \cite{LeDoussalWiese2011b,LeDoussalWiese2011,LeDoussalWiese2013,DobrinevskiLeDoussalWiese2013inpr}, and will be discussed further in section \ref{sec:AvalancheAction}.

\begin{figure}%
         \centering
         \begin{subfigure}[t]{0.4\textwidth}
                 \centering
                 \includegraphics[width=\textwidth]{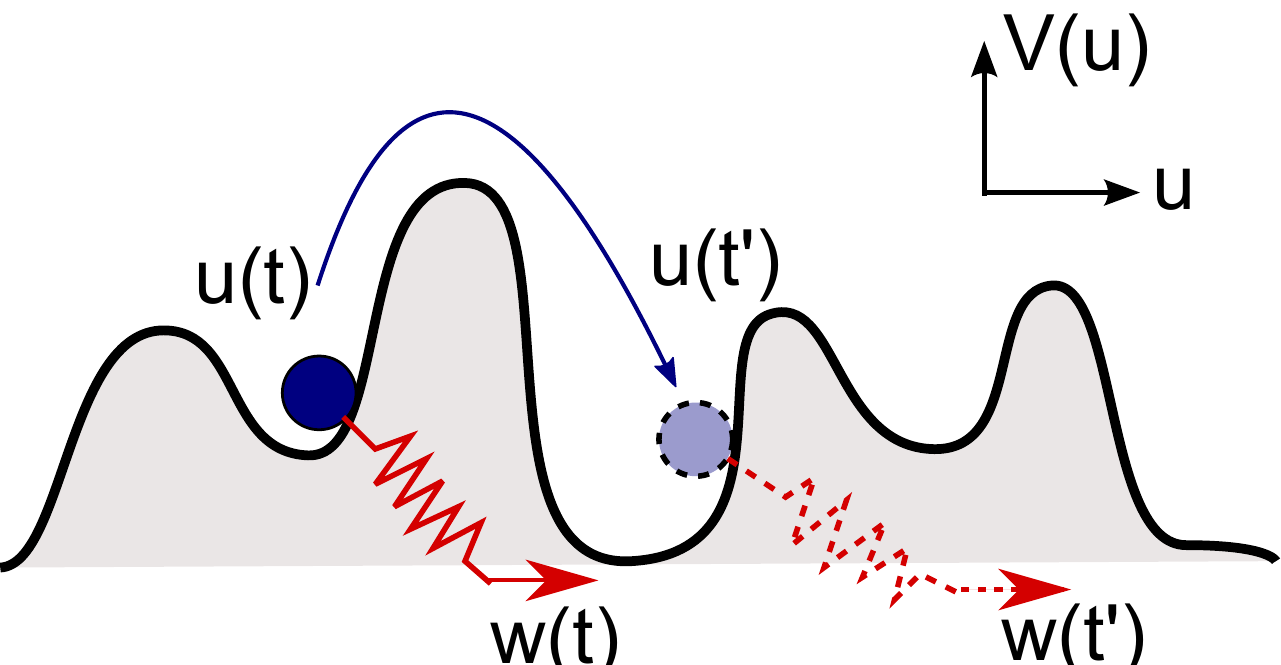}
                 \caption{Particle dynamics and avalanches in the ABBM model \eqref{eq:ABBM}.}
                 \label{fig:PlotABBMLandscape}
         \end{subfigure}
				\quad
			 \begin{subfigure}[t]{0.5\textwidth}
                 \centering
                 \includegraphics[width=\textwidth]{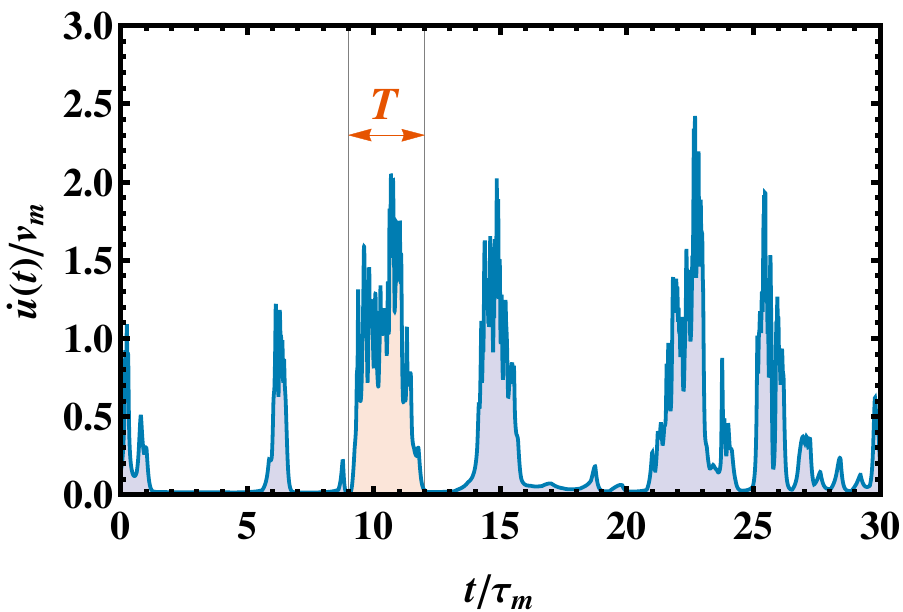}
                 \caption{Numerical simulation results of the ABBM model \eqref{eq:ABBM}. $v_m$ and $\tau_m$ are typical avalanche velocity and time scales, which will be discussed in section \ref{sec:BFMUnits}. The driving was at a constant velocity $\dw_t = 0.3v_m$.}
                 \label{fig:PlotABBMTrajSim}
         \end{subfigure}
				\caption{ABBM model}
\end{figure}

The solution of \eqref{eq:ABBM} can be visualized as the motion of an overdamped particle (at position $u(t)$, proportional to the average displacement of the domain wall), driven by a spring, and subject to the position-dependent force $F(u)$. This is shown graphically in figure \ref{fig:PlotABBMLandscape}, where the force $F(u) = V'(u)$ is the derivative of the rough potential $V(u)$ drawn. \eqref{eq:ABBM} is straightforward to simulate numerically, and the resulting trajectories (figure \ref{fig:PlotABBMTrajSim}) look qualitatively similar to experimental signals of Barkhausen noise (figure \ref{fig:BarkhausenMeas}). This led Alessandro et al. to conjecture that the ABBM model \eqref{eq:ABBM} is a good description for Barkhausen noise \cite{AlessandroBeatriceBertottiMontorsi1990}.
For a quantitative comparison ABBM \cite{AlessandroBeatriceBertottiMontorsi1990} analyzed the SDE \eqref{eq:ABBM} by mapping it to a Fokker-Planck equation \cite{Gardiner}. They determined the stationary distribution $P(\du)$ of the magnetization rate $\du(t)$ (which is proportional to the induced voltage measured in a Barkhausen experiment), for a constant ramp rate $c = \dot{H}$. The result (which will be re-derived in section \ref{sec:BFMABBMStat}) has the form \eqref{eq:IntroBNScalingForms},
\bea
P(\du)\rmd \du = \frac{1}{\Gamma(v/v_m)}(\du/v_m)^{-1+v/v_m}e^{-\du/v_m} \frac{\rmd \du}{v_m},
\eea
where $v:= c/k$ is the average magnetization rate, and $v_m$ is a scale for it defined in section \ref{sec:BFMUnits}. Both the rate-dependent power-law exponent near $\du=0$, and the exponential cutoff, agree well with experimental measurements \cite{AlessandroBeatriceBertottiMontorsi1990,AlessandroBeatriceBertottiMontorsi1990b} (at least for polycristalline magnets in the mean-field universality class, see section \ref{sec:Barkhausen}).
Alessandro et al. also computed the two-time correlation function $\overline{\du(t) \du(t')}$ (which is an exponential), and the power spectrum (which is Lorentzian). This was confirmed in some experiments \cite{AlessandroBeatriceBertottiMontorsi1990,AlessandroBeatriceBertottiMontorsi1990b}.

\begin{figure}%
         \centering
         \begin{subfigure}[t]{0.45\textwidth}
                 \centering
                 \includegraphics[width=\textwidth]{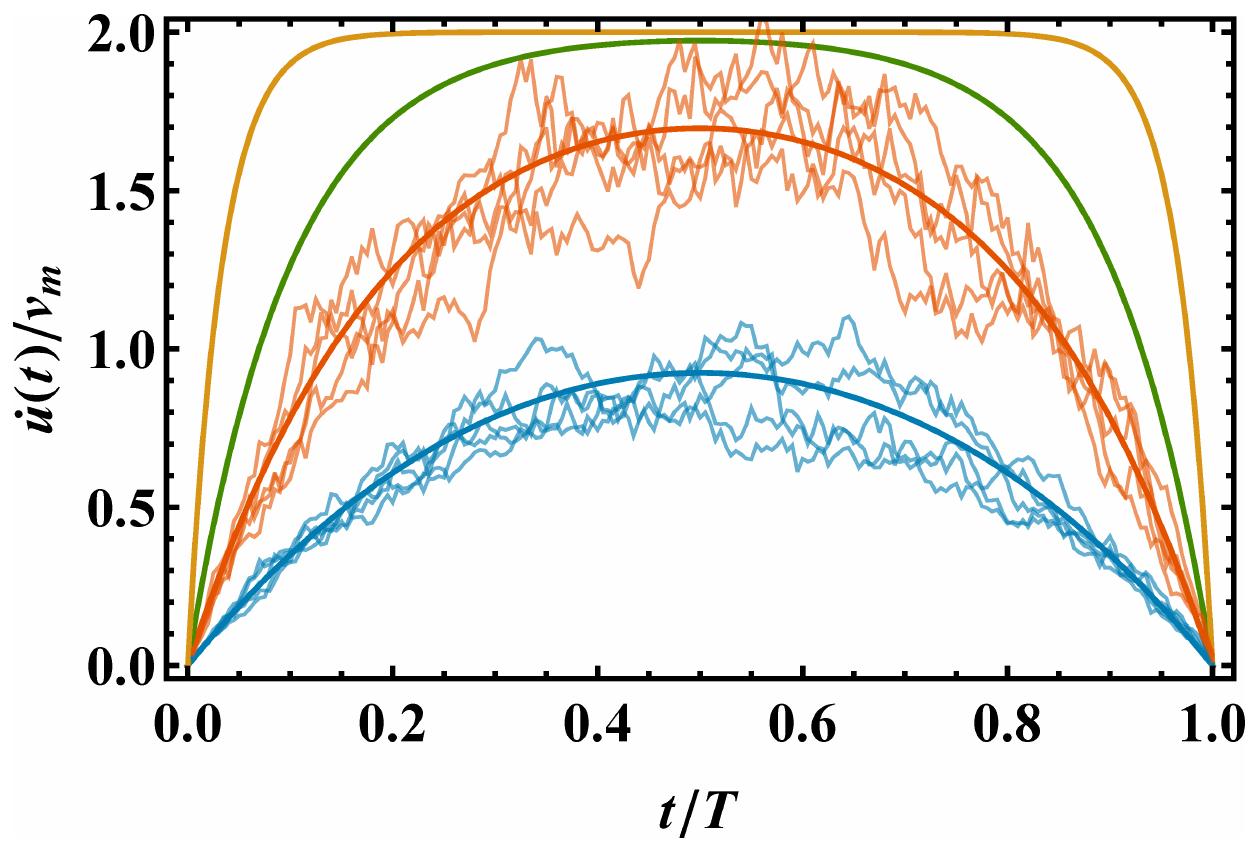}
                 \caption{Average avalanche shape in the ABBM model. Thick solid lines show \eqref{eq:BFMShapeFixedT} for (from bottom to top) $T=2 \tau_m, 5 \tau_m, 10 \tau_m, 30 \tau_m$. Thin solid lines show results of numerical simulations of \eqref{eq:ABBM}, averaged over $20$ realizations. }
                 \label{fig:ABBMShapeAnal}
         \end{subfigure}%
         ~ 
					         \begin{subfigure}[t]{0.45\textwidth}
                 \centering
                 \includegraphics[width=\textwidth]{figures/BarkhausenMeasShape3}
                 \caption{Experimental measurements of average Barkhausen pulse shapes. Figure reprinted with permission from \cite{PapanikolaouBohnSommerDurinZapperiSethna2011}. Copyright \copyright \, 2011 by Macmillan Publishers Ltd.}
                 \label{fig:ABBMShapeMeas}
         \end{subfigure}%
				                \caption{ \label{fig:ABBMShape}}
\end{figure}

Pulling the particle with a very soft spring, i.e. for $k$ small, the particle tends to stick in deep valleys of the potential landscape for a long time. When sufficient energy is accumulated in the extension of the spring, it can make a jump from one valley to the next one. These Barkhausen jumps, or \textit{avalanches} 
 in the ABBM model have then been analyzed by Bertotti, Durin and Magni \cite{BertottiDurinMagni1994}. Their sizes $S$ and durations $T$ (see figure \ref{fig:PlotABBMTrajSim}) are random. For small avalanches, their densities exhibit power-law ``critical'' behaviour \eqref{eq:IntroBNScalingForms}
\bea
\label{eq:BFMSizeDurExp}
P(S) \sim S^{-\tau},\quad \tau=3/2 \quad\quad\quad P(T)\sim T^{-\alpha},\quad \alpha = 2.
\eea
This agrees well with experimental measurements of Barkhausen jumps on polycristalline alloys \cite{DurinZapperi2000} (see also section \ref{sec:Barkhausen}). These exponents are robust and characterize the mean-field universality class of avalanches; their values $\tau = 3/2$, $\alpha=2$ for the ABBM model were first identified in \cite{BertottiDurinMagni1994} through scaling arguments.
In fact, the avalanche size and duration densities \eqref{eq:IntroBNScalingForms} for the ABBM model have closed expressions for arbitrary $S, T$:
\bea
\label{eq:BFMSizeDurIntr}
P(S)\rmd S = \frac{1}{\sqrt{4\pi}(S/S_m)^{3/2}}e^{-(S/S_m)/4} \frac{\rmd S}{S_m},\quad\quad\quad P(T) = \frac{e^{T/\tau_m}}{(e^{T/\tau_m}-1)^2} \frac{\rmd T}{\tau_m}.
\eea
Here $S_m$, $\tau_m$ are avalanche size and duration scales which will be made explicit in section \ref{sec:BFMUnits}.
The expressions \eqref{eq:BFMSizeDurIntr} were derived e.g.~in \cite{LeDoussalWiese2009,LeDoussalWiese2011,DobrinevskiLeDoussalWiese2012,LeblancAnghelutaDahmenGoldenfeld2013} and will be discussed further in sections \ref{sec:BFMSize}, \ref{sec:BFMDuration}.
As we saw in section \ref{sec:BarkhausenShape}, another interesting and experimentally accessible observable is the average avalanche shape. It is obtained by averaging the magnetization rate $\du$ at an intermediate coordinate (i.e. at a fixed intermediate time, or a fixed intermediate magnetization), over all avalanches of a given total size, or a given total duration. 
For the ABBM model, it was first considered in \cite{BaldassarriColaioriCastellano2003,ColaioriZapperiDurin2004}. For the average shape as a function of magnetization $u$, at fixed total size $S$, they obtained the exact result \cite{BaldassarriColaioriCastellano2003}
\bea
\label{eq:BFMUShapeFixedS}
\mathfrak{s}(u,S) = E\left[\du(u) \Bigg| \int_0^\infty  \du(t) \rmd t = S\right] = \frac{4}{\sqrt{\pi}} v_m \sqrt{\frac{u}{S}\left(1-\frac{u}{S}\right)}.
\eea
An exact analytical calculation for the average avalanche shape as a function of time $t$, at fixed duration $T$, was first performed for the ABBM model in \cite{PapanikolaouBohnSommerDurinZapperiSethna2011} and independently in \cite{LeDoussalWiese2011}. It gives the result 
\bea
\label{eq:BFMShapeFixedT}
\mathfrak{s}(t,T) := E[\du(t)|\min\{t'|\du(t')=0\}=T] =  v_m\frac{4 \sinh \frac{t}{2\tau_m} \sinh \frac{T-t}{2\tau_m}}{\sinh \frac{T}{2\tau_m}}.
\eea
Here $v_m$, $\tau_m$ are avalanche velocity and duration scales which will be made explicit in section \ref{sec:BFMUnits}. 
This compares reasonably well to some Barkhausen noise experiments (see \cite{PapanikolaouBohnSommerDurinZapperiSethna2011} and figure \ref{fig:ABBMShape}). The derivation of the results \eqref{eq:BFMShapeFixedT} and \eqref{eq:BFMUShapeFixedS} will be revisited in section \ref{sec:BFMShape}.

However, most experimental Barkhausen measurements exhibit avalanche shapes which are noticeably different, and skewed to the left (see \cite{SpasojevicBukvicMilosevicStanley1996,DurinZapperi2002,ZapperiCastellanoColaioriDurin2005,Colaiori2008} and figure \ref{fig:BarkhausenMeasShapes}). This can be explained by the buildup and relaxation of eddy currents \cite{ZapperiCastellanoColaioriDurin2005}. In the ABBM model, this effect can be included via a retarded memory kernel. This so-called ABBM model with retardation will be discussed at length in section \ref{sec:BFMRetardation}.

Finally, \cite{LeblancAnghelutaDahmenGoldenfeld2013} provides an extensive study of avalanche durations, sizes and shapes at arbitrary constant driving velocity $\dot{H}>0$. The authors also obtained interesting results on the distribution of the maximum velocity during an avalanche \cite{LeblancAnghelutaDahmenGoldenfeld2012,LeblancAnghelutaDahmenGoldenfeld2013}.

All these results are based on the mapping of \eqref{eq:ABBM} to a Fokker-Planck equation \cite{Gardiner}. A complementary approach is to use statistical field theory \cite{LeDoussalWiese2011,LeDoussalWiese2011b,DobrinevskiLeDoussalWiese2012,LeDoussalWiese2013}. It generalizes the ABBM model to the BFM, an elastic interface with $d$ internal dimensions, moving according to \eqref{eq:InterfaceEOM} in the Brownian disorder \eqref{eq:BFM} generalizing the ABBM guess \eqref{eq:DisorderABBM}.\footnote{It was first introduced in \cite{LeDoussalWiese2008c,LeDoussalWiese2011b} in the context of the study of the equilibrium ground state of an elastic interface.} As we will see in section \ref{sec:BFMSolABBM}, the theory for the total velocity of an extended interface in the BFM is identical to the ABBM model. 
This field-theoretical approach leads to a number of new results, including the complete distribution (and not just the power-law scaling) of avalanche sizes \cite{LeDoussalWiese2009,LeDoussalMiddletonWiese2009,LeDoussalWiese2011b}, and durations \cite{LeDoussalWiese2011}, as well as their joint distribution \cite{DobrinevskiLeDoussalWiese2012}. Likewise, it allows calculating the mean-field predictions (expected to be valid also for short-range-correlated disorder in $d\geq d_c$) for space-dependent observables \cite{LeDoussalWiese2013}, such as local avalanche sizes, which are not described by the ABBM model.
It also provides a basis for a perturbative expansion around the ABBM model, including nontrivial correlations of the disorder which become important below a critical dimension. This will be discussed in detail in chapter \ref{sec:OneLoop}; let us stick with the BFM for now.

In the course of this thesis, I advanced the theoretical understanding of the BFM and the ABBM model using statistical field theory in two published articles \cite{DobrinevskiLeDoussalWiese2012,DobrinevskiLeDoussalWiese2013}, as well as some additional results in preparation. In \cite{DobrinevskiLeDoussalWiese2012} (reviewed in section \eqref{sec:BFMABBM}), I show that the closed-form expressions \eqref{eq:BFMSizeDurIntr}, \eqref{eq:BFMShapeFixedT} 
for avalanche sizes, durations and shapes in the ABBM model are not accidental. All these exact results (and other observables) can be deduced from a simple formula for the generating functional of domain-wall velocities, eq.~\eqref{eq:BFMSolGenFct}. It gives the exact solution of the ABBM model in the sense that the computation of disorder-averaged observables is reduced to the solution of a simple nonlinear differential equation \eqref{eq:BFMInstanton}, first introduced in \cite{LeDoussalWiese2011,LeDoussalWiese2011b}.

This result is a significant generalization of previous results on the ABBM model in several respects:
\begin{itemize}
	\item It provides an exact solution not only for the model of one particle in a Brownian potential (ABBM eq. \eqref{eq:BFMABBM}), but for an arbitrary number of particles in independent Brownian potentials, coupled by linear elasticity.\footnote{I.e. by an elastic force of the form $\int_y c_{xy} \left(u_{xt}-u_{yt}\right)$ as in \eqref{eq:InterfaceEOMGenEl}. The kernel $g_{xy}$ can be an arbitrary nonnegative function in space, but the force must be linear in the displacement $u_{xt}-u_{yt}$.}
	In the continuum limit (on which we will focus below), it gives the exact solution of an elastic interface in the Brownian Force Landscape \eqref{eq:InterfaceEOM}, \eqref{eq:BFM}. This allows computing space-dependent observables in the BFM (e.g. local avalanche sizes), which are not described by the ABBM model. 
	\item This exact solution of the BFM also shows (see section \ref{sec:BFMSolABBM}) that the center-of-mass of an extended elastic interface with $d$ internal dimensions, in the BFM disorder \eqref{eq:BFM}, moves exactly as a single particle in the ABBM disorder \eqref{eq:DisorderABBM}.
	\item It provides an exact solution for \textit{any time- and space-dependent driving velocity} $\dw(x,t)$ in \eqref{eq:InterfaceEOM}, as long as the local interface velocity is non-negative at all times. This provides a natural way of delimiting avalanches, by applying a step in the driving field and then stopping the driving. This solution for arbitrary monotonous $\dw_{xt}$ also generalizes existing results on the BFM from \cite{LeDoussalWiese2011b,LeDoussalWiese2011,LeDoussalWiese2013}, which focused on the case $\dw=const.=0^+$.
\end{itemize}
Apart from these fundamental aspects, in this thesis several previously unknown results on the standard ABBM model were obtained using \eqref{eq:BFMABBM}, e.g. a formula for the joint distribution of avalanche sizes and durations $P(S,T)$ and the average avalanche shape at fixed avalanche size (see section \ref{sec:BFMSize} and \cite{DobrinevskiLeDoussalWiese2012,DobrinevskiLeDoussalWiese2013}).

Another advantage of the field-theory approach is that it can be generalized to more general types of dynamics. In particular, in section \ref{sec:BFMRetardation} I will show how it leads to an analytical treatment \cite{DobrinevskiLeDoussalWiese2013} for the ABBM model with retardation. This generalization of the ABBM model has been introduced to model eddy current relaxation in magnetic domain walls and to explain the observed asymmetry of average avalanche shapes (see \cite{ZapperiCastellanoColaioriDurin2005} and section \ref{sec:BarkhausenShape}). Before the present work it had only been studied numerically \cite{ZapperiCastellanoColaioriDurin2005,Colaiori2008}.

All these results are restricted to monotonous motion, where the interface moves forward at all times in each realization of the disorder. This holds for the standard equation of motion \eqref{eq:InterfaceEOM}, and the retarded dynamics discussed in section \ref{sec:BFMRetardation}, as a consequence of the Middleton theorem (see section \ref{sec:InterfaceMonot}), as long as the driving is monotonous, $\dw_{xt} \geq 0$.
Generalizing the present results to the case of non-monotonous motion is very important for the study of hysteresis \cite{DurinZapperi2006b}, but is also difficult. A first step to this end is the calculation of the quasi-static avalanche size distribution for all points on the hysteresis loop of the ABBM model, performed in section \ref{sec:BFMNonMon} and unpublished so far.

\section{First-order dynamics: ABBM model and its extension to interfaces\label{sec:BFMABBM}}
In this section, I review the results in \cite{DobrinevskiLeDoussalWiese2012} leading to an exact solution of the BFM \eqref{eq:InterfaceEOM}, \eqref{eq:BFM} for monotonous driving; see also \cite{LeDoussalWiese2011,LeDoussalWiese2013}. I then use this solution to recover known results on the ABBM model with stationary driving (section \ref{sec:BFMABBMStat}), and some known and some novel results for avalanches (section \ref{sec:BFMAvalanches}).

\subsection{The BFM velocity theory for monotonous motion\label{sec:BFMVelocity}}
Let us start by taking a derivative of eq.~\eqref{eq:InterfaceEOM},
\bea
\label{eq:IntVelEOM}
\eta\partial_t \du_{xt} = - m^2(\du_{xt}-\dw_{xt}) + \nabla^2_x \du_{xt} + \partial_t F(u_{xt},x).
\eea
For general disorder\footnote{I.e. any disorder where the increments are not independent as in the Brownian case \eqref{eq:BFM}, for example in physical disorder with short-ranged correlations.}, the quenched pinning force $F(u_{xt},x)$ depends on $u_{xt}$ and hence on the history of the local velocity in the past.
However, for the special case of the BFM correlations \eqref{eq:BFM}, and for monotonous motion, \eqref{eq:IntVelEOM} can be rewritten as a closed equation for the instantaneous local velocities $\du_{xt}$ \cite{DobrinevskiLeDoussalWiese2012}:
\bea
\label{eq:IntVelEOMBFM}
\eta\partial_t \du_{xt} = - m^2(\du_{xt}-\dw_{xt}) + \nabla^2_x \du_{xt} + \sqrt{\du_{xt}} \xi_{xt},
\eea
where $\xi_{xt}$ is a white noise in space and time,
\bea
\label{eq:BFMCorrWhiteNoise}
\overline{\xi(x,t)\xi(x',t')} = 2\sigma \delta(t-t')\delta^{d}(x-x').
\eea
In other words, the BFM disorder corresponds to a noise in time, whose variance is linear in the local interface velocity. Physically, this mapping from a quenched to an annealed noise is possible, since the increments of the Brownian disorder \eqref{eq:BFM} are independent. For the ABBM model, this fact was already noted in \cite{AlessandroBeatriceBertottiMontorsi1990,Bertotti1998,PapanikolaouBohnSommerDurinZapperiSethna2011}.
In order to obtain \eqref{eq:IntVelEOMBFM} from \eqref{eq:IntVelEOM}, consider the correlations of $\partial_t F(u_{xt},x)$ (following \cite{LeDoussalWiese2011,DobrinevskiLeDoussalWiese2012}):
\bea
\overline{\partial_t F(u_{xt},x) \partial_{t'} F(u_{x't'},x')} = \partial_t \partial_{t'} \Delta(u_{xt},u_{x't'})\delta^d(x-x') = -\sigma\du_{xt}\partial_{t'}\sgn(u_{xt}-u_{xt'})\delta^d(x-x').
\eea
In the last equality we used the BFM correlations \eqref{eq:BFM}. Now, let us assume monotonous motion, so that $u_{xt} \leq u_{xt'} \Leftrightarrow t \leq t'$. Then, $\sgn(u_{xt}-u_{xt'}) = \sgn(t-t')$ and we obtain
\bea
\label{eq:BFMForceVel}
\overline{\partial_t F(u_{xt},x) \partial_{t'} F(u_{x't'},x')} = 2\sigma\du_{xt}\delta(t-t')\delta^d(x-x') = \overline{\sqrt{\du_{xt}}\xi(x,t)\sqrt{\du_{x't'}}\xi(x',t')}.
\eea
This equality proves \eqref{eq:IntVelEOMBFM}, under the crucial assumption of the monotonicity of each trajectory, i.e. $\du_{xt} \geq 0$ for all $x, t$ and for each realization of the disorder. As discussed in section \ref{sec:InterfaceMonot}, as a consequence of the Middleton theorem, this holds for the solutions of the equation of motion \eqref{eq:InterfaceEOM}, under two assumptions:
\begin{itemize}
	\item The initial condition at some time $\ti$ satisfies $\du_{x,\ti} \geq 0$ for all $x$.
	\item The driving is monotonous everywhere, $\dw_{xt} \geq 0$ for all $x$, $t$.
\end{itemize}

\subsection{Solution for general monotonous driving\label{sec:BFMSol}}
The complete distribution of local interface velocities is encoded in the generating functional
\bea
\label{eq:GenFct}
G[\lambda_{xt},\dw_{xt}] &:= \overline{\exp\left(\int_{xt}\lambda_{xt}\du_{xt}\right)}^{\dw_{xt}}.
\eea
Here $\overline{O[\du]}^{\dw_{xt}}$ denotes the average of the observable $O$, a general functional of the solutions $\du_{xt}$ of \eqref{eq:InterfaceEOM}, over all realizations of the disorder $F$ in \eqref{eq:InterfaceEOM}, for a fixed choice of the driving velocity $\dw_{xt}$.
The exact solution of the Brownian-Force Model is an explicit formula for the generating functional $G$, which will be shown in section \ref{sec:BFMSolDeriv}:
\bea
\label{eq:BFMSolGenFct}
G[\lambda_{xt},\dw_{xt}] = & \exp\left( m^2 \int_{x,t} \dw_{xt} \tu_{xt}[\lambda] \right) = \exp\left( \int_{x,t} \df_{xt} \tu_{xt}[\lambda] \right),
\eea
where $\df_{xt} := m^2 \dw_{xt} \geq 0$ is the ramp rate of the driving force, non-negative due to the assumption of monotonous driving, and $\tu[\lambda]$ is the solution of the following non-linear ordinary differential equation \cite{LeDoussalWiese2011}
\bea
\label{eq:BFMInstanton}
\eta \partial_t \tu_{xt} + \nabla_x^2 \tu_{xt} - m^2 \tu_{xt} + \sigma \tu_{xt}^2 = -\lambda_{xt}.
\eea
Formula \eqref{eq:BFMSolGenFct} supposes an initial condition at $\ti=-\infty$, prepared in the ``Middleton state'' (see section \ref{sec:BFMSolIC} and \ref{sec:InterfaceMonot}). Then, the monotonicity property discussed in section \ref{sec:InterfaceMonot} ensures that the assumptions made in section \ref{sec:BFMVelocity} hold, and that the velocities $\du_{xt}$ satisfy the \textit{closed} equation \eqref{eq:IntVelEOMBFM}. In this case, the initial positions $u_{x,\ti}$ play no role. For a more detailed discussion and a variant of \eqref{eq:BFMSolGenFct} for more general initial conditions, see section \ref{sec:BFMSolIC}.

Causality implies that for a source $\lambda_{xt}$ localized in time (i.e. vanishing for $t > \tf$), the resulting $G$ cannot depend on $\dw_{xt'}$ for $t' > \tf$. Hence, the correct boundary condition to be used for solving \eqref{eq:BFMInstanton} in this case is $\tu_{xt'}=0$, for $t' > \tf$. Situations where the source $\lambda_{xt}$ is not localized in time require a case-by-case discussion.

\eqref{eq:BFMSolGenFct} remains valid if the short-range elastic kernel $\nabla^2$ in \eqref{eq:InterfaceEOM} is replaced by any linear elastic kernel (such as \eqref{eq:InterfaceEOMGenEl}), as long as it is positive definite (i.e. moving any part of the interface forward never pulls another part of the interface backward). In this case, as discussed in section \ref{sec:InterfaceMonot}, the Middleton property still holds and interface motion is monotonous. The only modification is to the instanton equation \eqref{eq:BFMInstanton}, where $\nabla^2$ needs to be replaced by the general elastic kernel as in \eqref{eq:InterfaceEOMGenEl}. For details, see \cite{DobrinevskiLeDoussalWiese2012} section IV B.

Generalizing \eqref{eq:BFMSolGenFct} to different types of dynamics\footnote{I.e. different dynamical operators on the left-hand-side of \eqref{eq:InterfaceEOM}} is more difficult. Adding inertia -- i.e. a $\partial_t^2 u_{xt}$ term -- breaks the Middleton property, and sometimes the interface moves backwards. The resulting model can be treated with some approximations \cite{LeDoussalPetkovicWiese2012}, but it is much harder to get precise results. Only one very particular generalization of the dynamics \eqref{eq:BFMSolGenFct} preserving the Middleton property is known: A retarded memory kernel, linear in the interface velocity. This BFM with retardation will be discussed in detail in section \ref{sec:BFMRetardation}.

\subsubsection{Special case: ABBM model\label{sec:BFMSolABBM}}
Let us now make the connection to the ABBM model discussed in section \ref{sec:ReviewABBM}. The ABBM equation \eqref{eq:ABBM} can be rewritten in the conventions of elastic interfaces, as in \eqref{eq:InterfaceEOM}, as
\bea
\label{eq:BFMABBM}
\eta \partial_t u_t = -m^2(u_t-w_t) + F(u(t)).
\eea
Here we set the demagnetizing factor $k =: m^2$, and $\Gamma/(2I_s) =: \eta$. With these identifications, and keeping in mind the Brownian disorder \eqref{eq:DisorderABBM}, the ABBM equation is the same as a ``particle'' version of the BFM, an interface with internal dimension $d=0$. Going to the velocity theory as in \eqref{eq:IntVelEOMBFM}, we obtain
\bea
\label{eq:IntVelEOMBFMABBM}
\eta\partial_t \du_{t} = - m^2(\du_{t}-\dw_{t}) + \sqrt{\du_{t}} \xi_{t},
\eea
Already before the ABBM model, this stochastic process was also considered by Feller \cite{Feller1951} as a population model and later for other applications\footnote{Among others, the Feller model for firing of single neurons \cite{LanskyLanska1987}, and the Cox-Ingersoll-Ross model for the term structure of interest rates \cite{CoxIngersollRoss1985}. See also \cite{MasoliverPerello2012} and references therein.}. Using the field-theory formulation above and in \cite{DobrinevskiLeDoussalWiese2012}, \eqref{eq:IntVelEOMBFMABBM} is solved by the zero-dimensional version of \eqref{eq:BFMSolGenFct}: The generating functional is
\bea
\label{eq:ABBMSolGenFct}
G^{\mathrm{ABBM}}[\lambda_{t},\dw_{t}] := \overline{\exp \left( \int_t \du_t \right) }^{\dw_t}= \exp\left( m^2 \int_{t} \tu_{t}[\lambda] \dw_t  \right) .
\eea 
Here $\tu[\lambda]$ is the space-independent version of \eqref{eq:BFMInstanton}, an ordinary differential equation of Riccati form,
\bea
\label{eq:ABBMInstanton}
\partial_t \tu_t - m^2 \tu_t + \sigma \tu_t^2 = -\lambda_t.
\eea

Now let us return to an interface with $d$ internal dimensions in the BFM, and consider an observable that only depends on its total velocity 
\bea
\du^{\rmt}_t := \int_x \rmd^d x \; \du_{x,t}.
\eea
The superscript $\cdot^{\rmt}$ signifies ``total'', i.e. integrated over space.
Then, $\lambda_{x,t}$ in the generating functional \eqref{eq:GenFct} is independent of $x$, $\lambda_{xt} = \lambda_t$. The solution $\tu[\lambda]$ of the BFM instanton equation \eqref{eq:BFMInstanton} is also $x$-independent, and reduces to the zero-dimensional ABBM instanton \eqref{eq:ABBMInstanton}. The BFM generating functional $G$ \eqref{eq:BFMSolGenFct} can then be expressed in terms of the ABBM generating functional $G^{\mathrm{ABBM}}$ \eqref{eq:ABBMSolGenFct}  as
\bea
\overline{\exp \left(\int_t \lambda_t \du^{\rmt}_t \right)}^{\dw_{xt}} = G[\lambda_{xt} = \lambda_t,\dw_{xt}] = \exp \left(\int_{t}\tu_t[\lambda]\int_x \dw_{x,t}\right) = G^{\mathrm{ABBM}}[\lambda_t,\dw^{\rmt}_t].
\eea
Here we defined the total driving $\dw^\rmt$ acting on the domain wall 
\bea
\label{eq:BFMTotalDriving}
\dw^{\rmt}(t) := \int_x \rmd^d x\, \dw(x,t).
\eea

We observe that in this case the BFM -- independently of the elastic kernel -- reduces to the ABBM model for magnetic domain walls:
\textit{The total magnetization $\du^{\rmt}(t)$ moves as a single particle in a Brownian Force landscape $F(u^{\rmt}_t:=\int_x u_{x,t})$, driven by the total field acting on the domain-wall $\dw^{\rmt}(t)$}. In particular, the distribution of the total interface velocity $\du^\rmt_t$ in the BFM is the same for any spatial distribution of driving forces $\dw_{x,t}$, as long as the total driving $\dw^\rmt$ is the same. Altogether, this shows that the effective potential felt by the center-of-mass of any number of linearly coupled particles in independent Brownian landscapes is again a Brownian. This amazing property is easy to check explicitly e.g.~for 2 particles, see \cite{DobrinevskiLeDoussalWiese2012} section IV A.

This statement is complementary to the applicability of the ABBM model in the mean-field limit, discussed in \cite{CizeauZapperiDurinStanley1997,ZapperiCizeauDurinStanley1998,Colaiori2008} and section \ref{sec:AvalancheAction}: There, it was shown that the ABBM model describes the total velocity of an interface in short-range correlated disorder, for sufficiently strong (or infinite-range) elasticity and sufficiently high dimension. On the other hand, here we see that the ABBM model also describes the total velocity of an interface in the BFM disorder \eqref{eq:BFM}, for any kind of elastic kernel, and in any dimension.

In the following sections, I will first briefly repeat the proof of \eqref{eq:BFMSolGenFct} and its ABBM variant \eqref{eq:ABBMSolGenFct} (originally given in \cite{DobrinevskiLeDoussalWiese2012}). Then I will discuss how these formulae allow one to compute avalanche observables (in particular, the avalanche size and duration distributions).

\subsubsection{Computing the generating functional using the Martin-Siggia-Rose formalism\label{sec:BFMSolDeriv}}
A simple way to derive \eqref{eq:BFMSolGenFct} is to use the Martin-Siggia-Rose (MSR) statistical-field-theory formalism \cite{MartinSiggiaRose1973}. Since this will be a useful starting point for the perturbative RG treatment of short-ranged disorder in chapter \ref{sec:OneLoop}, let us sketch the derivation here. For a more detailed discussion, including a first-principle derivation in discrete time, see \cite{DobrinevskiLeDoussalWiese2012}.

The generating functional \eqref{eq:GenFct} can be expressed as an average over solutions of \eqref{eq:IntVelEOMBFM} following the approach of MSR
\bea
\nn
&G[\lambda_{xt},\dw_{xt}] = \int \mD [\xi]\int \mD [\du] \,e^{\int_{xt} \lambda_{xt}\du_{xt}} \prod_{x,t}\delta\left[-\eta\partial_t \du_{xt} - m^2(\du_{xt}-\dw_{xt}) + \nabla^2_x \du_{xt} + \sqrt{\du_{xt}} \xi_{xt}\right] \\
\nn
&= \int \mD [\xi]\int \mD [\du,\tu] \,e^{\int_{xt} \lambda_{xt}\du_{xt}} \exp \int_{xt}\tu_{xt}\left[-\eta\partial_t \du_{xt} - m^2(\du_{xt}-\dw_{xt}) + \nabla^2_x \du_{xt} + \sqrt{\du_{xt}} \xi_{xt}\right] 
\eea
The MSR auxiliary field $\tu_{xt}$ is integrated over the imaginary axis; it yields the $\delta$-functional enforcing the equation of motion \eqref{eq:IntVelEOMBFM} at all times. We can now average over the noise using the fact that $\xi$ is Gaussian with correlations given by \eqref{eq:BFMCorrWhiteNoise}, and hence $\int \mD[\xi] \exp\left( \int_{xt}Q_{xt}\xi_{xt}\right) = \exp \left(\sigma \int_{xt} Q_{xt}^2\right)$.
\bea
\label{eq:BFMGenFctTemp1}
G[\lambda_{xt},\dw_{xt}]= & \int \mD [\du,\tu] \,\exp \int_{xt}\left\{ \tu_{xt}\left[-\eta\partial_t \du_{xt} - m^2(\du_{xt}-\dw_{xt}) + \nabla^2_x \du_{xt}\right] + \lambda_{xt}\du_{xt} + \sigma\tu_{xt}^2\du_{xt} \right\} .
\eea
The fact that there is no Jacobian factor, i.e. that equality holds in the last line follows from a more careful discretization procedure (for a general discussion, see \cite{AltlandSimons2010}, for this specific case \cite{Chauve2000}). This also uses the fact that we are working in the It\^{o} representation. 
We see that the generating functional \eqref{eq:GenFct}, \eqref{eq:BFMGenFctTemp1} can be written as a path integral 
\bea
\label{eq:DefGSources}
G[\lambda_{xt},\dw_{xt}] = \int \mD[\du,\tu] \,\exp\left(-S[\du,\tu] + \int_{xt} \lambda_{xt} \du_{xt} + \int_{xt} \df_{xt} \tu_{xt}\right),
\eea
where the dynamical action of the BFM is given by \cite{LeDoussalWiese2011,DobrinevskiLeDoussalWiese2012,LeDoussalWiese2013}
\bea
\label{eq:BFMAction}
S[\du,\tu] = \int_{xt}\left\{\tu_{xt}\left[\eta\partial_t \du_{xt} + m^2 \du_{xt} - \nabla^2_x \du_{xt} \right] - \sigma \tu_{xt}^2\du_{xt}\right\}.
\eea
The path integral formula \eqref{eq:DefGSources} for $G$ is valid, beyond the BFM, for any elastic interface model in Gaussian disorder, if the action \eqref{eq:BFMAction} is correspondingly modified. We will use this in  chapter \ref{sec:OneLoop} (in particular, see eq.~\eqref{eq:InterfaceAction}) for analyzing short-range-correlated disorder.
Recall from \eqref{eq:BFMSolGenFct} that the driving force ramp rate $\df$ is related to the driving velocity as $\df_{xt} := m^2 \dw_{xt}$. Here we see that $\df$ is the source for the MSR response field $\tu$, just as $\lambda$ is the source for $\du$. 
In order to evaluate $G$ for the BFM, we continue from \eqref{eq:BFMGenFctTemp1} and note that the exponential is linear in $\du$. Thus, the path integral over $\du$ can be performed, giving again a $\delta$-functional,
\bea
\nn
G[\lambda_{xt},\dw_{xt}] = & \int \mD [\tu] \,e^{ m^2\int_{xt}\tu_{xt}\dw_{xt}}\prod_{xt}\delta\left(\partial_t \tu_{xt} - m^2 \tu_{xt} + \nabla^2 \tu_{xt} + \sigma \tu_{xt}^2 + \lambda_{xt}\right) \\
\label{eq:BFMMSRSolTemp1}
= & \exp\left( m^2\int_{xt}\dw_{xt} \tu_{xt}[\lambda]\right) = \exp\left( \int_{xt}\df_{xt} \tu_{xt}[\lambda]\right).
\eea
Here $\tu_{xt}[\lambda]$ is the solution of the instanton equation \eqref{eq:BFMInstanton}. To fix the normalization factor in the last line, note that the Jacobian introduced by evaluating the $\delta$-functional is independent of $\dw$. Since for $\dw=0$, there is no motion and hence $G[\lambda,\dw=0]=1$, we find that the Jacobian is actually one. This concludes the proof of formula \eqref{eq:BFMSolGenFct}.
\footnote{Note that this depends on the preparation of the initial condition discussed in section \ref{sec:BFMSolIC}. Also, this does not necessarily hold for the position theory \cite{DobrinevskiLeDoussalWiese2012}.}

\begin{figure}%
         \centering
         \begin{subfigure}[t]{0.95\textwidth}
                 \centering
                 \includegraphics[width=0.5\textwidth]{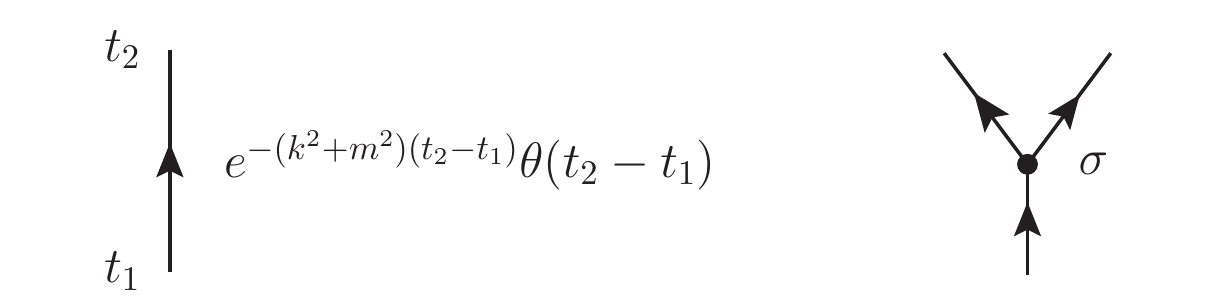}
                 \caption{Propagator and interaction vertex of the MSR action \eqref{eq:BFMAction} for the Brownian Force Model.}
                 \label{fig:BFMVertices}
         \end{subfigure}
				
         \begin{subfigure}[t]{0.95\textwidth}
                 \centering
                 \includegraphics[width=\textwidth]{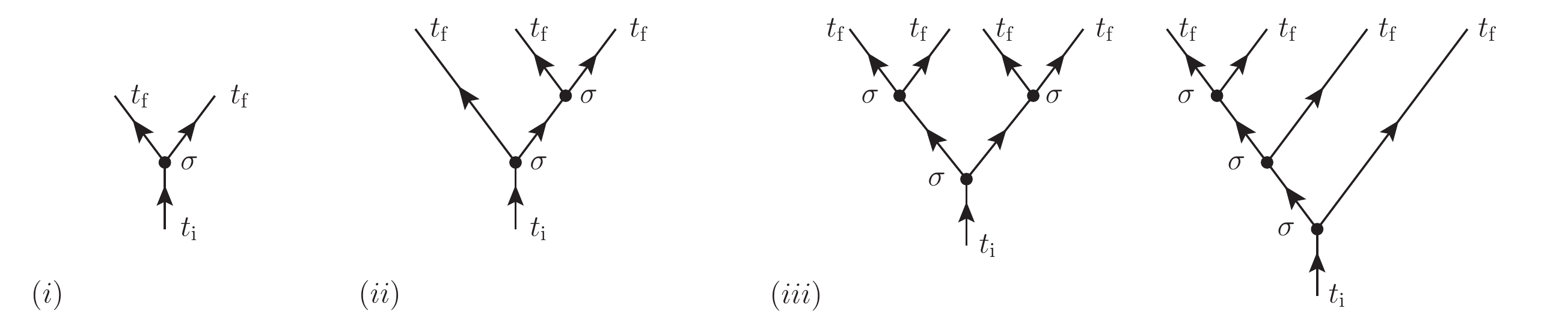}
                 \caption{Feynman diagrams contributing to the observables $\overline{\tu_{\ti}\du_{\tf}^n}$, for (i) $n=2$, (ii) $n=3$, (iii) $n=4$. Up to permutations of vertices, these are \textit{all} the diagrams arising for these observables; in particular, there are no loops due to the causal structure of the interaction. These observables are the $n$-th moments of velocity $\du_\tf$, following an infinitesimal step in the driving force at $\ti$: $\overline{\tu_{\ti}\du_{\tf}^n} = \partial_w\big|_{w=0}\overline{\du_{\tf}}^{\dw_t = w\delta(t-\ti)}$ (this follows from \eqref{eq:GenFct} and \eqref{eq:DefGSources}).}
                 \label{fig:BFMCorrel}
         \end{subfigure}
\caption{Diagrammatics of the Brownian Force model.\label{fig:BFMDiagrammatics}}
\end{figure}

\eqref{eq:BFMSolGenFct} can also be obtained by diagrammatic methods \cite{LeDoussalWiese2008c,LeDoussalWiese2011b,LeDoussalWiese2013,LeDoussalWiese2011}. The action \eqref{eq:BFMAction} contains a propagator going from a $\df$ source to a $\lambda$ source, and a cubic interaction vertex $\sigma \du\tu^2$ which has one incoming and two outgoing lines (see figure \ref{fig:BFMVertices}). Expanding an observable perturbatively in $\sigma$ yields a series of diagrams, all of which are trees (e.g.~the diagrams for the first four moments of $\du$ are shown in figure \ref{fig:BFMCorrel}). These can be enumerated and resummed for specific cases (for example the avalanche size distribution $P(S)$ \cite{LeDoussalWiese2008c,LeDoussalWiese2011b}), confirming the results of \eqref{eq:BFMSolGenFct} discussed in section \ref{sec:BFMAvalanches} below.

\subsubsection{Initial condition\label{sec:BFMSolIC}}
The solution \eqref{eq:BFMSolGenFct} supposes that one is starting from a Middleton state (as defined in section \ref{sec:InterfaceMonot}) at $t=-\infty$. This was used in order to ensure that the Jacobian factor introduced by evaluating the delta-functional in \eqref{eq:BFMMSRSolTemp1} is equal to 1. This assumes that for $\dw=0$, $\du_{xt}=0$ for all $x$ and $t$ and hence $G[\lambda,\dw=0]=1$, which is only true when starting from a metastable state (cf. section \ref{sec:InterfaceMonot}).
However, starting from any metastable state is not sufficient. For obtaining the closed equation-of-motion for the interface velocity in section \ref{sec:BFMVelocity}, we also crucially used the independence of increments of the pinning force for all times. This only holds if the interface never revisits a point where it has been in the past. This would fail if one starts with some fixed value $u_{x,\ti}$, lets the interface relax to some arbitrary metastable state (which involves some backward and some forward motion in different regions of the interface), and then takes that metastable state as the initial condition at $t=-\infty$. This property would also fail, if one moves backward, waits for the interface to relax into some metastable state, and then uses that as the initial condition for the following forward motion. The case pertinent for us, where the independence of the pinning force increments from the initial condition does hold, is starting from a Middleton state (for forward motion) at some $\ti$ or at $\ti=-\infty$, and moving monotonously forward. Another way to realize a Middleton starting state for the BFM is to take $F(u,x)$ a non-stationary Brownian (see appendix \ref{sec:BFMPositionTheory}), starting from $F(u=0,x)=0$, an initial state $u(x,\ti)=0$ (which is then automatically a metastable state), and considering motion for positive $u$ only.

The first condition $G[\lambda,\dw=0]=1$, can be relaxed. One can consider monotonous motion starting from a Middleton state at $t=-\infty$, but fix the velocity at some initial time $t=\ti$ to a given value $\du_{x}^\mathrm{i}$ (or an arbitrary distribution $P[\du^{\mathrm{i}}_x]$). This can be done by adding a $\delta$-peak to the driving force, $\df_{xt} \to \df_{xt} + \delta(t-\ti) \du_{x}^\mathrm{i}$ (assuming that $\df_{xt}=0$ for $t<\ti$). This imposes $\du_{x,\ti^+}=\du^{\mathrm{i}}_x$ on the solutions of \eqref{eq:InterfaceEOM}.
The generalization of \eqref{eq:BFMSolGenFct} to an initial condition with fixed velocity (see also \cite{DobrinevskiLeDoussalWiese2012}) then reads
\bea
\label{eq:BFMSolGenFctGenIC}
G[\lambda_{xt},\dw_{xt}]  = \int \mD[\du^{\mathrm{i}}] P[\du^{\mathrm{i}}] \exp\left( \int_x \int_{\ti}^\infty \rmd t\, \df_{xt} \tu_{xt}[\lambda] + \int_{x} \tu_{x,\ti}[\lambda] \du_{x}^\mathrm{i} \right),
\eea
where $\tu[\lambda]$ is still the solution of \eqref{eq:BFMInstanton}.

\subsubsection{Units\label{sec:BFMUnits}}
The equation of motion \eqref{eq:InterfaceEOM}, its solution (the generating functional formula \eqref{eq:BFMSolGenFct}) and the instanton equation \eqref{eq:BFMInstanton} contain explicitly the dimensionful parameters $m^2, \eta, \sigma$. In fact, these parameters only determine the characteristic length and time scales in the solution; the physics is given by a single paramater-free ``dimensionless'' generating functional $G'[\lambda',w']$, where $m=\eta=\sigma=1$. To see this explicitly, let us set
\bea
\label{eq:BFMRescaling}
x = m^{-1} \cdot x', \quad\quad t = \eta m^{-2} \cdot t' \quad\quad \du_{xt} = \frac{\sigma m^{d-2}}{\eta}\cdot \du_{x't'}' \quad\quad \tu_{xt} = \frac{m^2}{\sigma} \cdot \tu_{x't'}'.
\eea
Inserting eq.~\eqref{eq:BFMRescaling} into the action \eqref{eq:BFMAction}, one obtaines the action in the rescaled units
\bea
S[\du,\tu] = S'[\du',\tu'] = \int_{x't'}\left\{{\tu'}_{x't'}\left[\partial_{t'} {\du'}_{x't'} + {\du'}_{x't'} - \nabla^2_{x'} {\du'}_{x't'} \right] - {\tu'}_{x't'}^2 {\du'}_{x't'}\right\}.
\eea
Thus, in rescaled (primed) units we effectively have $m=\eta=\sigma=1$. The corresponding rescaling for the sources $\lambda$, $\dw$ (or equivalently $\df = m^2 \dw$) in \eqref{eq:DefGSources} is
\bea
\lambda_{xt} = m^4/\sigma \lambda'_{x't'}, \quad\quad\quad \dw_{xt} = \frac{\sigma m^{d-2}}{\eta}\cdot \dw_{x't'}', \quad\quad\quad \df_{xt} = \frac{\sigma m^{d}}{\eta}\cdot \df_{x't'}'.
\eea
Using this rescaling, we can write the dimensionful generating functional in terms of the dimensionless one,
\bea
G[\lambda_{xt},\dw_{xt}] = G'\left[\lambda'_{x',t'} = \frac{\sigma}{m^4}\lambda_{mx,(m^2/\eta)t}, \dw'_{x't'} = \frac{\eta}{\sigma m^{d-2}} \dw_{mx,(m^2/\eta)t}\right],
\eea
where $G'$ is obtained using the action $S'$ with $m=\eta=\sigma=1$.
Another way to understand \eqref{eq:BFMRescaling} is that there is a time scale $\tau_m = \eta/m^2$, a transversal length scale $x_m = m^{-1}$, and an avalanche size scale $S_m = \sigma/m^4$. The natural, ``primed'' or \textit{dimensionless}, units express times in units of $\tau_m$, transversal lengths in terms of $x_m$, the total displacement (or avalanche size)  $S = \int_{xt}\du_{xt}$ in terms of $S_m$. This also means that the total velocity $\du^{\rmt} = \int_x \du_{xt}$ is expressed in terms of the total velocity scale $v_m := S_m / \tau_m = \sigma / (\eta m^2)$.

From now on let us use dimensionless units (and drop the primes for simplicity of notation), except where indicated.

\subsection{Stationary velocity distribution, and propagator\label{sec:BFMABBMStat}}
Let us briefly review how \eqref{eq:BFMSolGenFct} allows one to recover known results on the ABBM model for stationary driving, at a constant velocity (more details can be found in \cite{LeDoussalWiese2013,DobrinevskiLeDoussalWiese2012}). Following the discussion in section \ref{sec:BFMSolABBM}, this also generalizes these results to the total velocity of an extended interface in the BFM. Let us assume a constant local driving velocity $\dw_{xt}=v=const.$. The total driving velocity $\dw^{\rmt}(t)$, defined in \eqref{eq:BFMTotalDriving}, is then $\dw^{\rmt} = v^{\rmt} = L^d v$ where $L^d$ is the total internal volume of the $d$-dimensional interface in the BFM\footnote{In the ABBM model of a single particle, $d=0$, and there is no difference between local and global velocities. There, one can drop the superscript $\cdot^{\rmt}$ and just write $v$, $\du$.}. Let us denote the distribution of the instantaneous total velocity $\du^{\rmt}$ of the interface as $P_{v^{\rmt}}(\du^{\rmt})$. Without loss of generality let us consider the velocity at $t=\tf$, i.e. $\du^{\rmt}:=\du^{\rmt}_\tf = \int_x \rmd x\, \du_{x,\tf}$.

The Laplace transform of $P_{v^\rmt}$ is given through the generating functional $G$ defined in \eqref{eq:GenFct} as
\bea
\nn
\int_0^\infty \rmd \du^{\rmt}\, e^{\lambda \du^{\rmt}}P_{v^{\rmt}}(\du^{\rmt}) =& \overline{\exp\left(\lambda\int_x \rmd x\,\du_{x,\tf}\right)} = G[\lambda_{xt} = \lambda \delta(t-\tf), \dw_{xt} = v] \\
\label{eq:DefPvofUdot}
=& \int \mD[\du,\tu]\exp\left(-S[\du,\tu] + \frac{m^2}{L^d}\int_{x,t}v^\rmt \tu_{xt} + \lambda\int_{x}\du_{x,\tf} \right)
\eea
In the last line we used the path integral formulation \eqref{eq:DefGSources} of the generating functional $G$. \eqref{eq:DefPvofUdot} is given, exceptionally, in dimensionful units since it will be reused later in chapter \ref{sec:OneLoop} where the scaling with $m$ will be important.
\eqref{eq:DefPvofUdot} is a general definition, valid for an elastic interface with the equation of motion \eqref{eq:InterfaceEOM} in any kind of disorder (assuming the action $S$ in \eqref{eq:DefGSources} is correspondingly modified). Here we will discuss the mean-field case of the BFM, where we can compute $G$ using \eqref{eq:BFMSolGenFct}, \eqref{eq:BFMInstanton}. The extension to short-range correlated disorder using RG methods was discussed in \cite{LeDoussalWiese2011,LeDoussalWiese2013} and will be revisited in section \ref{sec:FRGAvalanchesObsOneLoop}.  The solution of \eqref{eq:BFMInstanton} with $\lambda_{xt} = \lambda \delta(t-\tf)$ is constant in space, and hence reduces to a solution of \eqref{eq:ABBMInstanton} \cite{LeDoussalWiese2011}
\bea
\label{eq:ABBMOneTimeInstanton}
\tu_{xt} = \frac{\lambda}{\lambda + (1-\lambda)e^{\tf-t}}\theta(\tf-t).
\eea
Inserting this into \eqref{eq:ABBMSolGenFct} we obtain the generating function of instantaneous velocities:
\bea
\overline{\exp\left(\lambda \du^\rmt\right)} = (1-\lambda)^{-v^\rmt}.
\eea
Inverting the Laplace transform, one recovers the stationary velocity distribution of the ABBM model \cite{AlessandroBeatriceBertottiMontorsi1990,Bertotti1998,Colaiori2008}:
\bea
\label{eq:BFMPvofUdot}
P_{v^\rmt}(\du^{\rmt}) = \frac{1}{\Gamma(v^\rmt)}(\du^{\rmt})^{-1+v^\rmt}\exp\left(-\du^{\rmt}\right).
\eea
The behaviour of this distribution near $\du^{\rmt}=0$ shows a transition at $v^\rmt=1$ where individual avalanches merge to form a single big pulse \cite{AlessandroBeatriceBertottiMontorsi1990,Bertotti1998,WhiteDahmen2003,Colaiori2008}. For $v^\rmt<1$, the instantaneous total interface velocity $\du^{\rmt}$ becomes zero infinitely often, and one observes intermittent avalanches. On the other hand, for $v^\rmt>1$, $\du^{\rmt}$ never becomes 0 and one has continuous motion. This transition can also be understood from the recurrence properties of Brownian motion \cite{Colaiori2008}. Observe that for the BFM, the critical value of the microscopical driving is $v=L^{-d}$; thus this transition occurs at a driving velocity which depends on the system size, as noted in \cite{WhiteDahmen2003}.

It is interesting to consider the case of quasi-static driving, $v=0^+$. Recall from \eqref{eq:DefGSources} that the driving $\dw_{xt} = \frac{1}{L^d} v^\rmt$ corresponds to a constant ramp rate of the total driving force $\df^{\rmt}_t := \int_x \df_{xt} = m^2 \int_x \dw_{xt} = m^2 v^\rmt = const.$.
We define the quasi-static velocity density $P(\du^\rmt)$ as the linear response to the force ramp rate $\df^{\rmt}$.\footnote{This definition is given, exceptionally, in dimensionful units since it will be reused later in chapter \ref{sec:OneLoop} where the scaling with $m$ will be important.}
\bea
\nn
P(\du^{\rmt}) := & \partial_{\df^{\rmt}}\big|_{\df^{\rmt}=0} P_{v^{\rmt}}(\du^{\rmt}) = m^{-2} \partial_{v^{\rmt}}\big|_{v^{\rmt}=0} P_{v^{\rmt}}(\du^{\rmt})  \\
\nn
\Rightarrow \int_0^\infty \rmd \du^{\rmt}\,\left(e^{\lambda \du^{\rmt}}-1\right)P(\du^{\rmt}) = & \partial_{\df^{\rmt}}\big|_{\df^{\rmt}=0} G\left[\lambda_{xt} = \lambda \delta(t-\tf), \dw_{xt} = \frac{1}{m^2 L^d}\df^{\rmt}\right] \\
\label{eq:DefPofUdot}
=& \int \mD[\du,\tu]\frac{1}{L^d}\int_{x_i,\ti}\tu_{x_i,\ti}\, \exp\left(-S[\du,\tu] + \lambda\int_{x}\du_{x,\tf} \right)
\eea
In the limit $v\to 0^+$, avalanches are well-separated, and the leading contribution to $P(\du^{\rmt})$ comes from a single avalanche. $P(\du^{\rmt})$ is a density, not a true probability distribution, and not normalized. From \eqref{eq:DefPofUdot}, we see that the avalanche is triggered by an infinitesimal step in the driving force ($\tu_{xt}$ in the field theory formulation), which may have happened at any time $\ti<\tf$ (times $\ti > \tf$ do not contribute to the path integral by causality). Again, \eqref{eq:DefPofUdot} is a general definition, valid for an elastic interface with the equation of motion \eqref{eq:InterfaceEOM} in any kind of disorder. In a diagrammatic language, \eqref{eq:DefPofUdot} shows that the Laplace transform of $P(\du^{\rmt})$ consists in all causal diagrams starting from a single vertex at $\ti$ and terminating at any number of vertices at $\tf$ (since the exponential in $\lambda$ generates all powers of $\du^\rmt_\tf$). For the BFM, these diagrams are trees as discussed in section \ref{sec:BFMSolDeriv}, and can be resummed explicitly \cite{LeDoussalWiese2011,LeDoussalWiese2013}. Here we deduce the result from \eqref{eq:BFMPvofUdot} (note that in dimensionless units, $v^\rmt = \df^\rmt$):
\bea
\label{eq:BFMPofUdot}
P(\du^{\rmt}) = \partial_{v^\rmt}\big|_{v^{\rmt}=0} P_{v^\rmt}(\du^{\rmt}) = (\du^{\rmt})^{-1}\exp\left(-\du^{\rmt}\right).
\eea

Similarly, the joint distribution of velocities at two times $P(\du_1,\du_2)$, for stationary driving, can be obtained by considering the source $\lambda(t) = \lambda_1 \delta(t-t_1) + \lambda_2 \delta(t-t_2)$. This allows one to recover the two-point function $\overline{\du_{t_1} \du_{t_2}}$, and the ABBM propagator at $v=0^+$ \cite{LeDoussalWiese2011} and at finite $v$ \cite{DobrinevskiLeDoussalWiese2012,LeblancAnghelutaDahmenGoldenfeld2013}, which was first computed in \cite{Bertotti1998}. One can also obtain an expression for the $n$-times generating function $\overline{\exp \sum_{i=1}^n \lambda_i \du(t_i)}$ \cite{LeDoussalWiese2011}, and the $n$-times joint distribution $P\left(\du(t_1),...,\du(t_n)\right)$. The latter decomposes into a product of two-time joint distributions $P(\du_1,\du_2)$ due to the Markovian property of the SDE \eqref{eq:BFMABBM}.

Let us now proceed to the case of non-stationary driving, where \eqref{eq:ABBMSolGenFct} allows us to compute new observables.

\subsection{Avalanche sizes, durations and shapes\label{sec:BFMAvalanches}}
A simple way to trigger an avalanche is to apply a step in the driving force at some time $\ti$. Let us apply a step of size $w$ in the position of the harmonic well, $\dw_{x,t} = w\delta(t-\ti)$, $w = \frac{1}{L^d} w^\rmt$, so that the total driving $\dw^\rmt_t$ defined in \eqref{eq:BFMTotalDriving} is $\dw^\rmt_t = w^\rmt \delta(t-\ti)$. We will consider $w^\rmt$ of order $1$.
\footnote{As in the previous section, for the ABBM model of a single particle, $d=0$ and there is no difference between the local and global step size. There we can drop the superscript $\cdot^\rmt$. If we were to take $\dw(x,t) = w\delta(t-\ti)$ for an extended interface in the BFM, with $w \sim 1$ (i.e. without the additional factor $1/L^d$), we would trigger avalanches everywhere in the system. Then, in the thermodynamic limit $L\to\infty$ the mean interface velocity averaged over the system would not fluctuate, and we would not be able to observe avalanche statistics.}
This is equivalent to imposing the initial condition $\du(x,\ti) = m^2 w$.

In the following we will use \eqref{eq:BFMSolGenFct} to compute the exact distributions $P_{w^\rmt}(S), P_{w^\rmt}(T)$ of the size $S$ and duration $T$ of the avalanche following such a finite step in the driving force. We will also compute the joint distribution $P_{w^\rmt}(S,T)$, and the mean shape of the avalanche given its duration or its size. The response to a finite step $w^\rmt$ consists, in principle, of multiple avalanches. To isolate a single avalanche, we will consider the limiting forms of these observables, as $w^\rmt \to 0$. In this limit avalanches triggered by a step in the driving force are equivalent to avalanches in the steady state with quasi-static driving, $\dw = const. = 0^+$. Some of the observables we will compute, in particular $P_{w^\rmt}(T)$ and $P_{w^\rmt}(S,T)$ were previously unknown and published first in \cite{DobrinevskiLeDoussalWiese2012}.

Analyzing avalanches at a nonzero driving velocity is more complicated, since there the motion never stops entirely: Even when the domain wall momentarily stops, it immediately restarts due to the external driving. This requires introducing an artificial absorbing boundary in order to obtain results on avalanche statistics. I will discuss some partial results in that direction in section \ref{sec:ABBMFinVel}. For now let us stick with the case where the driving stops after the avalanche was triggered.

\subsubsection{Avalanche duration distribution\label{sec:BFMDuration}}
The microscopic roughness (non-differentiability) of the BFM force landscape ensures that at some time $\ti + T$, the motion of the interface following a step in the driving force at $t=\ti$ truly stops, so that $\du_{xt} = 0$ for all $x$ and all $t \geq \ti+T$.\footnote{In  any microscopically smooth force landscape, the velocity $\du$ is a differentiable function of time and one needs to introduce a cutoff in order to define the avalanche duration.}
This stopping property allows to express the cumulative distribution $F$ of the avalanche duration $T$, for a step of total size $w^\rmt$ (and local size $w = w^\rmt / L^d$), through the generating functional $G$ as follows:
\bea
\nn
F_{w^\rmt}(\tf-\ti) := P_{w^\rmt}(T \leq \tf-\ti) :=& P\left(\int_{\xf} \du_{\xf,\tf} = 0\right) = \lim_{\lambda \to -\infty} \overline{\exp\left(\lambda \int_{\xf} \du_{\xf,\tf}\right)}^{\dw_{xt} = w \delta(t-\ti)} \\
\label{eq:DefPwofT}
=& \lim_{\lambda \to -\infty} G[\lambda_{xt}=\lambda \delta(t-\tf),\dw_{xt} = w\delta(t-\ti)].
\eea
Just like \eqref{eq:DefPvofUdot}, \eqref{eq:DefPwofT} is a general definition in dimensionful units, valid for any elastic interface model (assuming the action $S$ in the definition \eqref{eq:DefGSources} of $G$ is correspondingly modified). For the BFM considered here, we can compute $G$ using \eqref{eq:BFMSolGenFct} and \eqref{eq:ABBMInstanton}.

The source $\lambda_{x,t} = \lambda \delta(t-\tf)$ considered here is exactly the same as in the previous section \ref{sec:BFMABBMStat}. Thus, the instanton solution $\tu[\lambda \delta(t-\tf)]$ of \eqref{eq:BFMInstanton} is constant in space and given by \eqref{eq:ABBMOneTimeInstanton}:
\bea
\label{eq:ABBMOneTimeInstanton2}
\tu_{xt} = \frac{\lambda}{\lambda + (1-\lambda)e^{\tf-t}}\theta(\tf-t).
\eea
Inserting this into \eqref{eq:BFMSolGenFct}, \eqref{eq:DefPwofT}, and taking the limit $\lambda \to -\infty$ gives:
\bea
\label{eq:BFMFwofT}
P_{w^\rmt}(T \leq \tf-\ti) = & \lim_{\lambda \to -\infty }\exp\left[ w^\rmt \frac{\lambda}{\lambda + (1-\lambda)e^{\tf-\ti}}\right] = \exp\left(\frac{w^\rmt}{1-e^{\tf-\ti}}\right).
\eea
As in the previous section \ref{sec:BFMABBMStat}, the contribution from a single avalanche can be isolated by considering an infinitesimal step in the force. 
Recall from \eqref{eq:DefGSources} that the driving $\dw_{xt} = \frac{1}{L^d} w^\rmt \delta(t-\ti)$ corresponds to a step in the total driving force $\df^{\rmt}_t := \int_x \df_{xt} = m^2 \int_x \dw_{xt} = m^2 w^\rmt \delta(t-\ti) =: f^\rmt \delta(t-\ti)$. We define the cumulative density of avalanche durations $F(\tf-\ti)$ as the leading order of $P_{w^\rmt}(T \leq \tf-\ti)$ in the driving force $f^\rmt$, i.e. via the linear response to small steps in $f^\rmt$:\footnote{This definition is given, exceptionally, in dimensionful units since it will be reused later in chapter \ref{sec:OneLoop} where the scaling with $m$ will be important.}
\bea
\nn
F(\tf-\ti) := & \partial_{f^{\rmt}}\big|_{f^{\rmt}=0} F_{w^\rmt}(\tf-\ti) = \partial_{f^{\rmt}}\big|_{f^{\rmt}=0} \lim_{\lambda \to -\infty} G[\lambda_{xt}=\lambda \delta(t-\tf),\dw_{xt} = \frac{f^\rmt}{m^2 L^d}\delta(t-\ti)] \\
\label{eq:DefPofT}
=& \lim_{\lambda \to -\infty} \int \mD[\du,\tu]\frac{1}{L^d}\int_{\xxi}\tu_{\xxi, \ti}\, \exp\left[-S[\du,\tu] + \lambda\int_{\xf}\du_{\xf,\tf} \right].
\eea
In the last line we used the path integral expression \eqref{eq:DefGSources} for $G$.
As in \eqref{eq:DefPofT}, the avalanche is triggered by an infinitesimal step in the driving force ($\tu_{xt}$ in the field theory formulation). However, it is now triggered at a fixed $\ti$ and not integrated over all $\ti$ as in \eqref{eq:DefPofT}. Again, \eqref{eq:DefPofT} is a general definition, in dimensionful units, valid for an elastic interface in any kind of disorder. The diagrammatic interpretation of \eqref{eq:DefPofT} in the MSR field theory is analogous to \eqref{eq:DefPofUdot}: All diagrams starting with a single vertex at time $\ti$ (which is now fixed), and terminating at $\tf$. 
For the BFM, we obtain the cumulative density for avalanche durations from the small-$w^\rmt$ behavior of \eqref{eq:BFMFwofT} (note that in dimensionless units, $w^\rmt = f^\rmt$):
\bea
\label{eq:BFMFofT}
F(\tf-\ti) =\partial_{f^{\rmt}}\big|_{f^{\rmt}=0} P_{w^\rmt}(T \leq \tf-\ti) = \frac{1}{1-e^{\tf-\ti}}.
\eea
As for \eqref{eq:BFMPofUdot}, this is a cumulative density, not a true cumulative probability distribution, and not normalized.

\begin{figure}%
         \centering
         \begin{subfigure}[t]{0.45\textwidth}
                 \centering
                 \includegraphics[width=\textwidth]{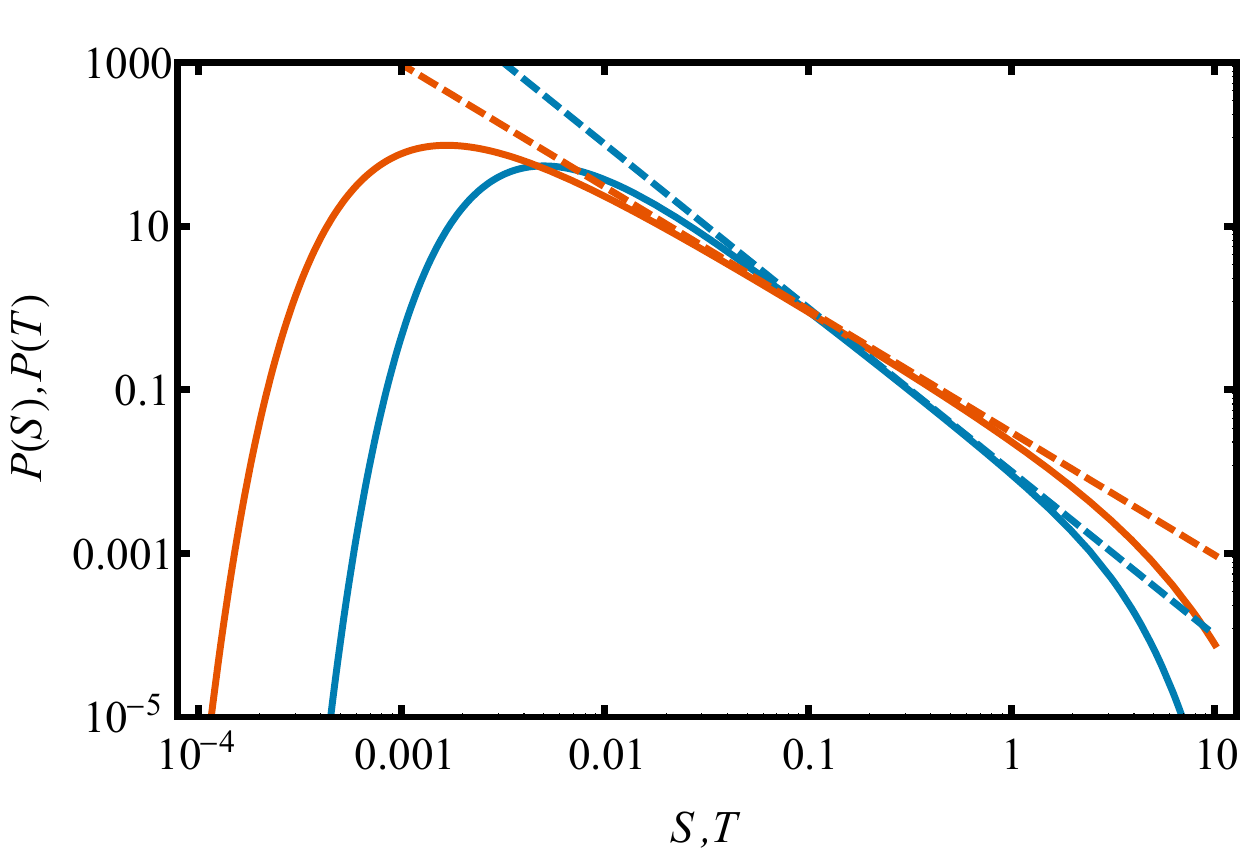}
                 \caption{Distributions $P(S)$ \eqref{eq:BFMPwofS} and $P(T)$ \eqref{eq:BFMPwofT} for avalanche sizes $S$ and durations $T$ in the ABBM model (dimensionless units), following a finite step of size $w^\rmt = 0.01$ for $P(T)$ and $w^\rmt=0.1$ for $P(S)$. Observe that $w^\rmt$ and $\left(w^\rmt\right)^2$, give a lower cutoff for $T$ and $S$, respectively. The upper cutoff $S,T \approx 1$ is given by the harmonic confinement in \eqref{eq:BFMABBM}.}
                 \label{fig:ABBMSizesDurationsSep}
         \end{subfigure}%
         ~ 
         \begin{subfigure}[t]{0.45\textwidth}
                 \centering
                 \includegraphics[width=\textwidth]{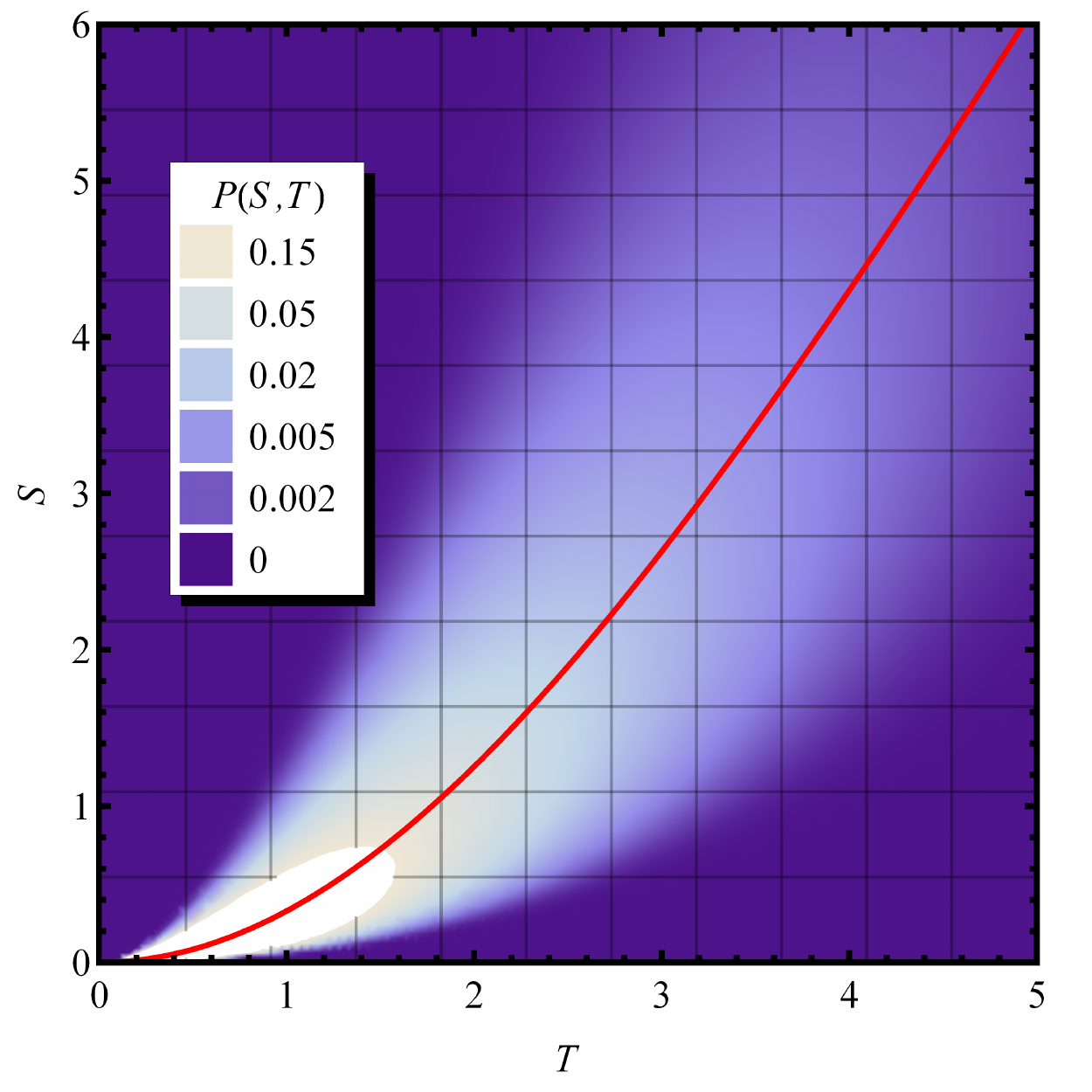}
                 \caption{Joint distribution of quasi-static avalanche sizes and durations in the ABBM model (dimensionless units), \eqref{eq:BFMPwofSTLT} for $w^\rmt \to 0$. Red line: Average avalanche size at fixed duration $\overline{S}(T)$, \eqref{eq:MeanSOfT}.}
                 \label{fig:ABBMSizesDurationsJoint}
         \end{subfigure}
				                \caption{Distributions of avalanche sizes and durations in the ABBM model. \label{fig:ABBMSizeDurDist}}
\end{figure}

Taking a derivative with respect to $\tf$, one obtains the distribution of avalanche durations for finite $w^\rmt$ \cite{DobrinevskiLeDoussalWiese2012}, and the density of avalanche durations for infinitesimal $w^\rmt$ or $f^\rmt$.\footnote{This definition is given, exceptionally, in dimensionful units since it will be reused later in chapter \ref{sec:OneLoop} where the scaling with $m$ will be important.}
\bea
\label{eq:BFMPwofT}
P_{w^\rmt}(T) := & \partial_{\tf}\big|_{\tf = \ti + T} P_{w^\rmt}(T \leq \tf-\ti) = \frac{w^{\rmt}}{\left(2\sinh T/2\right)^2} \exp\left(\frac{w^\rmt}{1-e^{T}}\right) \\
\label{eq:BFMPofT}
P(T) :=  & \partial_{\tf}\big|_{\tf = \ti + T} F(\tf-\ti) = \frac{e^T}{(e^T-1)^2} = \frac{1}{\left(2\sinh T/2\right)^2}.
\eea
This expression for $P(T)$ has also been obtained in \cite{LeDoussalWiese2011} by considering stationary driving $\dw_t = v = 0^+$ and conditioning to the avalanche beginning and ending in certain intervals. On the other hand, \eqref{eq:BFMPwofT} is a new result from \cite{DobrinevskiLeDoussalWiese2012}. We observe that in the BFM, for short avalanches $T\to 0$, $P(T)$ has a power-law divergence
\bea
P(T) \sim T^{-\alpha}, \quad\quad \alpha=2.
\eea
We thus recover the mean-field value of the power-law exponent for avalanche durations at quasi-static driving, $\alpha=2$. It has been derived before for the ABBM model using scaling arguments \cite{NarayanFisher1993,BertottiDurinMagni1994,CizeauZapperiDurinStanley1997,ZapperiCizeauDurinStanley1998,Colaiori2008}. The present results for the BFM \eqref{eq:BFMPwofT}, \eqref{eq:BFMPofT} are much more precise: They also give the exact shape of the small-$T$ cutoff by a finite $w$, and the large-$T$ exponential cutoff (cf.~figure \ref{fig:ABBMSizesDurationsSep}). I will discuss the generalization of these results to short-range correlated disorder using RG methods in section \ref{sec:OneLoopDurations}. 

\subsubsection{Avalanche size, and joint size-duration distribution\label{sec:BFMSize}}
Since we imposed the boundary condition $\du(x,t=-\infty) = 0$ (cf. the discussion in section \ref{sec:BFMSolIC}), we have $\du_{xt}=0$ for $t<\ti$, so that the total avalanche size is $S = \int_{x,t}\du_{x,t}$. The probability distribution of $S$ following a kick of total size $w^\rmt$ (as in the previous section) is $P_{w^\rmt}(S)$. 
The generating function for the total size  of the avalanche triggered by the step at $\ti$ then is
\bea
\nn
\int_0^\infty \rmd S\,e^{\lambda S}P_{w^\rmt}(S) =& \overline{\exp\left(\lambda S\right)}^{\dw_{xt}=w\delta(t-\ti)} = \overline{\exp\left(\lambda\int_{\xf,\tf}\, \du_{\xf,\tf}\right)}^{\dw_{xt}=w\delta(t-\ti)} \\
\label{eq:DefPwofS}
=& G[\lambda_{xt}=\lambda,\dw_{xt}=w\delta(t-\ti)].
\eea
This is a general definition for any elastic interface model. In order to apply \eqref{eq:BFMSolGenFct} for the BFM, we need to solve \eqref{eq:BFMInstanton} with a constant $\lambda_{xt} = \lambda$.
The corresponding instanton solution $\tu[\lambda_{xt}=\lambda]$ is constant in space and time, and reads in dimensionless units
\bea
\label{eq:ABBMSizeInstanton}
\tu_{xt} = \frac{1}{2}\left(1-\sqrt{1-4\lambda}\right).
\eea
Using \eqref{eq:BFMSolGenFct}, we obtain the distribution of avalanche sizes for total step size $w^\rmt$:\footnote{This definition is given, exceptionally, in dimensionful units since it will be reused later in chapter \ref{sec:OneLoop} where the scaling with $m$ will be important.}
\bea
\label{eq:BFMPwofS}
\overline{e^{\lambda S}} = \exp\left[\frac{w^\rmt}{2}\left(1-\sqrt{1-4\lambda}\right)\right] \Leftrightarrow P_{w^\rmt}(S) = \frac{w^\rmt}{\sqrt{4\pi} S^{3/2}}\exp\left[-\frac{(S-w^\rmt)^2}{4 S}\right].
\eea
For a plot, see figure \ref{fig:ABBMSizesDurationsSep}. As for the avalanche durations distribution, \eqref{eq:BFMPwofT}, there is a small-size cutoff $S\sim \left(w^\rmt\right)^2$ and a large-size cutoff $S \sim 1$, in between one observes power-law behavior. 
As in the previous section \ref{sec:BFMABBMStat}, the contribution from a single avalanche can be isolated by considering an infinitesimal step in the force. Analogous to \eqref{eq:DefPofT}, we define the density of avalanche sizes $P(S)$ as the leading order of $P_{w^\rmt}(S)$ in the driving force $f^\rmt = m^2 w^\rmt$:\footnote{This definition is given, exceptionally, in dimensionful units since it will be reused later in chapter \ref{sec:OneLoop} where the scaling with $m$ will be important.}
\bea
\nn
P(S) := & \partial_{f^{\rmt}}\big|_{f^{\rmt}=0} P_{w^\rmt}(S)  \\
\nn
\hat{P}_S(\lambda):=&\int_0^\infty \rmd S\,\left(e^{\lambda S}-1\right) P(S) = \partial_{f^{\rmt}}\big|_{f^{\rmt}=0} G\left[\lambda_{xt} = \lambda, \dw_{xt} = \frac{1}{m^2 L^d}f^{\rmt}\delta(t-\ti)\right] \\
\label{eq:DefPofS}
=& \int \mD[\du,\tu]\frac{1}{L^d}\int_{\xxi}\tu_{\xxi, \ti}\, \exp\left(-S[\du,\tu] + \lambda\int_{\xf,\tf}\du_{\xf,\tf} \right).
\eea
Again, \eqref{eq:DefPofS} is a general definition, in dimensionful units, valid for an elastic interface in any kind of disorder. As in \eqref{eq:DefPofT}, the avalanche is triggered at time $\ti$ by the field $\tu_{x,\ti}$. The diagrammatic interpretation of \eqref{eq:DefPofS} is analogous to \eqref{eq:DefPofUdot} and \eqref{eq:DefPofT}: All diagrams starting with a single vertex at time $\ti$, and terminating at an arbitrary time $\tf$ (which is now, however, integrated over, instead of being fixed). 
For the BFM, we obtain the density for avalanche sizes from the small-$w^\rmt$ behavior of \eqref{eq:BFMPwofS} (note that in dimensionless units, $w^\rmt = f^\rmt$):
\bea
\label{eq:BFMPofS}
P(S) :=  & \partial_{f^{\rmt}}\big|_{f^{\rmt}=0} P_{w^\rmt}(S) = \frac{1}{\sqrt{4\pi} S^{3/2}}\exp\left(-\frac{S}{4}\right).
\eea
We see that in the BFM, for small avalanches $S\to 0$, $P(S)$ diverges as
\bea
P(S) \sim S^{-\tau}, \quad\quad \tau=3/2.
\eea
This mean-field value $\tau=3/2$ of the avalanche size exponent is well-known \cite{BertottiDurinMagni1994,CizeauZapperiDurinStanley1997,ZapperiCizeauDurinStanley1998,LeDoussalWiese2009,Colaiori2008}.
The precise shape for a finite step \eqref{eq:BFMPwofS} has likewise already been known for the particle model (ABBM) \cite{LeDoussalWiese2009}. Here we have shown that it generalizes also to a spatially extended interface in the Brownian Force landscape.

We can also go further and consider the joint distribution $P_{w^\rmt}(S,T)$ of avalanche sizes and durations \cite{DobrinevskiLeDoussalWiese2012}. Combining the ideas from this and the previous section, its Laplace transform from $S$ to $\lambda_1$ is given by
\bea
\nn
\int_0^\infty \rmd S \, e^{\lambda_1 S} P_{w^\rmt}(S,T \leq \tf-\ti) =& \lim_{\lambda_2\to-\infty} \overline{\exp \left(\lambda_1 \int_{-\infty}^\infty \rmd t\, \du^\rmt_t + \lambda_2 \du^\rmt_\tf \right) }^{\dw_{xt} = w \delta(t-\ti)} \\
\nn
=& \lim_{\lambda_2\to-\infty} G[\lambda_{xt} = \lambda_1 + \lambda_2 \delta(t-\tf),\dw_{xt} = w\delta(t-\ti)].
\eea
The instanton solution of \eqref{eq:ABBMInstanton} with the source $\lambda_{xt} = \lambda_1 + \lambda_2 \delta(t-\tf)$ is computed in \cite{DobrinevskiLeDoussalWiese2012}. Taking a derivative one obtains the final result
\bea
\label{eq:BFMPwofSTLT}
\int_0^\infty \rmd S\, e^{\lambda S} P_{w^\rmt}(S,T) =\frac{w^\rmt(1-4\lambda)e^{\frac{w^\rmt}{2}\left(1-\sqrt{1-4\lambda}\coth\frac{T}{2}\sqrt{1-4\lambda}  \right)}}{\left(2\sinh \frac{T}{2}\sqrt{1-4\lambda}\right)^2}.
\eea
This Laplace transform can be inverted analytically in some limits \cite{DobrinevskiLeDoussalWiese2012}, or numerically (see figure \ref{fig:ABBMSizesDurationsJoint} for a plot in the case $w^\rmt \to 0$). One can also can use \eqref{eq:BFMPwofSTLT} directly to calculate the moments of avalanche size at fixed avalanche duration. In particular, for the mean avalanche size as a function of its duration one obtains
\bea
\nonumber
\overline{S}(T) = \frac{\int_0^\infty \rmd S\, S\,P(S,T)}{\int_0^\infty \rmd S\,P(S,T)} = \frac{4-w^\rmt T - 4\cosh T +(2T+w^\rmt)\sinh T}{\cosh T - 1}.
\eea
For a single quasi-static avalanche, $w^\rmt\rightarrow 0$, this reduces to
\beq
\label{eq:MeanSOfT}
 \overline{S}(T) = 2T \coth{\frac{T}{2}} - 4.
\eeq
For small avalanches, $S \ll S_m = 1$, Eq.~\eqref{eq:MeanSOfT} reproduces the expected scaling behaviour \cite{ZapperiCizeauDurinStanley1998,Colaiori2008}, $\overline{S}(T) \sim T^2$. Eq.~\eqref{eq:MeanSOfT} also predicts the deviations of large avalanches from this scaling, and shows that they obey $S\sim T$ instead. This  is in qualitative agreement with experimental observations on Barkhausen noise in polycristalline
FeSi materials \cite{DurinZapperi2006b,ColaioriZapperiDurin2004,Colaiori2008}.

\subsubsection{Average avalanche shape\label{sec:BFMShape}}
The distributions of avalanche sizes and durations discussed in sections \ref{sec:BFMDuration} and \ref{sec:BFMSize} have some characteristic power-law divergences for small avalanches and an exponential cutoff for large avalanches, but are otherwise rather featureless. It is thus interesting to consider more refined observables, which provide a stronger reflection of the underlying model's physics. As we saw in section \ref{sec:BarkhausenShape}, one such example (proposed among others in \cite{SethnaDahmenMyers2001,ZapperiCastellanoColaioriDurin2005}) is the average shape of an avalanche. 

As in the previous sections, let us trigger an avalanche by a finite step of total size $w^\rmt$ in the driving, without loss of generality at $\ti=0$. The average avalanche shape is a scalar function of an intermediate coordinate (time or distance) and a final coordinate (total avalanche duration or size). It is the total interface velocity at the intermediate point, averaged over all realizations of avalanches terminating at the final point.
More precisely, three variants of the avalanche shape can be distinguished.

\paragraph{Average avalanche shape, as a function of time, at fixed avalanche duration}
This is obtained by averaging the instantaneous total interface velocity $\du^\rmt_{\tm}$, at an intermediate time $0 < \tm < T$ during an avalanche of total duration $T$.\footnote{From now on, we set without loss of generality $\ti=0$ except where indicated, in order to simplify notation.}
As in section \ref{sec:BFMDuration}, the restriction to durations $0< T < \tf$ can be imposed by $\du^\rmt_\tf = 0$. So, we can express it as follows:
\bea
\nn
& \mathfrak{s}_{w^\rmt}(\tm,T) := E[\du^\rmt_{\tm}|T] = \frac{\partial_\tf \big|_{\tf = T} E[\du^\rmt_{\tm}, \du^\rmt_\tf = 0]}{P_{w^\rmt}(T)} = \frac{1}{P_{w^\rmt}(T)}\partial_\tf \big|_{\tf = T} \lim_{\lambda \to -\infty} \overline{\du^\rmt_{\tm} e^{\lambda \du^\rmt_\tf}}^{\dw_{xt} = w\delta(t)} =\\
\label{eq:DefShapewFixedT}
& = \frac{1}{P_{w^\rmt}(T)}\partial_\tf \big|_{\tf = T} \lim_{\lambda \to -\infty} \partial_{\mu}\big|_{\mu = 0} G[\lambda_{xt} = \mu \delta(t-{\tm}) + \lambda\delta(t-\tf), \dw_{xt} = w \delta(t)].
\eea
For quasi-static avalanches, as for avalanche durations \eqref{eq:DefPofT} and sizes \eqref{eq:DefPofS} we define the shape by taking an infinitesimal step in the driving force $f^\rmt = m^2 w^\rmt$:\footnote{This definition is given, exceptionally, in dimensionful units since it will be reused later in chapter \ref{sec:OneLoop} where the scaling with $m$ will be important.}
\bea
\nn
&\mathfrak{s}(\tm,T) := \lim_{w^\rmt \to 0} \mathfrak{s}_{w^\rmt}(\tm,T) =  \\
\nn
& = \frac{1}{P(T)}\partial_{f^\rmt}\big|_{f^\rmt = 0}\partial_\tf \big|_{\tf = T} \lim_{\lambda \to -\infty} \partial_{\mu}\big|_{\mu = 0} G\left[\lambda_{xt} = \mu \delta(t-{\tm}) + \lambda\delta(t-\tf), \dw_{xt} = \frac{1}{m^2 L^d}f^\rmt \delta(t)\right] \\
\label{eq:DefShapeFixedT}
& =  \frac{1}{P(T)} \lim_{\lambda\to -\infty}\int \mD[\du,\tu]\frac{1}{L^d}\int_{\xxi}\tu_{\xxi, \ti=0}\int_{\xm}\du_{\xm,\tm}\, \exp\left(-S[\du,\tu] + \lambda\int_{\xf}\du_{\xf,\tf} \right).
\eea
This is a general definition valid for any interface model. For the BFM/ABBM, we can use \eqref{eq:BFMSolGenFct} and \eqref{eq:ABBMInstanton} to compute $G$. The full solution of \eqref{eq:ABBMInstanton} with a two-time source $\lambda(t) = \mu \delta(t-\tm) + \lambda\delta(t-\tf)$ is given in \cite{DobrinevskiLeDoussalWiese2012}. This also allows one to obtain the full distribution of intermediate velocities $P(\du_{t_m}|T)$ \cite{DobrinevskiLeDoussalWiese2012}.\footnote{This was later generalized to finite driving velocity in \cite{LeblancAnghelutaDahmenGoldenfeld2013}, see also section \ref{sec:ABBMFinVel}.} For simplicity, let us restrict ourselves to the average shape, which only requires the solution of \eqref{eq:ABBMInstanton} to linear order in $\mu$. It is obtained by perturbing \eqref{eq:ABBMInstanton} around the known solution \eqref{eq:ABBMOneTimeInstanton} for the case $\mu=0$
\bea
\nn
\lim_{\lambda \to -\infty}\tu_{t}[\mu \delta(t-{\tm}) + \lambda\delta(t-\tf)] = & \tu_t^{(0)} + \mu \tu_t^{(1)} + \mO(\mu)^2 \\
\label{eq:BFMShpTInstAnsatz}
\tu_t^{(0)}= & \lim_{\lambda \to -\infty} \tu_t[\lambda\delta(t-\tf)] = \frac{1}{1 -e^{\tf-t}}.
\eea
The linear order $\tu_t^{(1)}$ satisfies a linear equation obtained by inserting \eqref{eq:BFMShpTInstAnsatz} into \eqref{eq:ABBMInstanton}, and collecting the terms of order $\mu$:
\bea
\label{eq:ABBMShapeLinPert}
\partial_t \tu_t^{(1)} - \tu_t^{(1)} + 2 \tu_t^{(0)}\tu_t^{(1)} = -\delta(t-\tm).
\eea
Solving this we obtain 
\bea
\label{eq:ABBMShapeLinPertSol}
\tu_t^{(1)} = \frac{e^{\tm-t}(e^{\tf-\tm}-1)^2}{(e^{\tf-t}-1)^2}\theta(\tm-t).
\eea
Finally, this gives the average avalanche shape \cite{DobrinevskiLeDoussalWiese2012}
\bea
\nn
\mathfrak{s}(\tm,T) = &\frac{\partial_\tf \big|_{\tf = T} w\tu_{t=0}^{(1)} e^{w^\rmt \tu_{t=0}^{(0)}}}{\partial_\tf \big|_{\tf = T} e^{w^\rmt \tu_{t=0}^{(0)}}} = w^\rmt\tu_{t=0}^{(1)}\big|_{\tf=T} + \frac{\partial_\tf \big|_{\tf = T} \tu_{t=0}^{(1)}}{\partial_\tf \big|_{\tf = T} \tu_{t=0}^{(0)}} = \\
\label{eq:ABBMShapewFixedT}
= & w^\rmt\left[\frac{\sinh\left(\frac{T-\tm}{2}\right)}{\sinh\left(\frac{T}{2}\right)}\right]^2 + \frac{4\sinh \left(\frac{\tm}{2}\right)\sinh \left(\frac{T-\tm}{2}\right)}{\sinh \left(\frac{T}{2}\right)}.
\eea
For $w^\rmt=0$ (quasi-static avalanches), the first term vanishes and \eqref{eq:ABBMShapewFixedT} reduces to \eqref{eq:BFMShapeFixedT} as previously obtained in \cite{PapanikolaouBohnSommerDurinZapperiSethna2011} using Fokker-Planck methods and independently in \cite{LeDoussalWiese2011}. In the limit of small avalanches, $T \to 0$, the shape becomes universal. It is then a simple parabola as a function of $x:= \tm/T$; the only explicit $T$ dependence is a global prefactor:
\bea
\label{eq:ABBMShapeShort}
\mathfrak{s}(\tm = xT,T) = 2T x(1-x) + \mO(T)^2.
\eea
A previous approximate solution \cite{ColaioriZapperiDurin2004} (subsequently also reported in \cite{DurinZapperi2006b,Colaiori2008}) is
\bea
\nn
\mathfrak{s}(\tm,T) \approx \sin \pi \tm/T.
\eea
However, its derivation in \cite{ColaioriZapperiDurin2004} contains some uncontrolled approximations and does not become exact neither for small nor for large avalanches. 

\paragraph{Average avalanche shape, as a function of time, at fixed avalanche size}
In some cases, it may be more adequate to fix the total avalanche size instead of the total avalanche duration. This is the case e.g. for earthquake statistics, where the avalanche size can be obtained from the readily available earthquake magnitude. Such mean moment rate profiles have been studied e.g. in \cite{MehtaDahmenBenZion2006}.
The shape at fixed size is obtained by averaging the instantaneous total interface velocity $\du_{\tm}$, $\tm > 0$, during an avalanche of total size $S = \int_{-\infty}^\infty \du^\rmt_{t}\rmd t$:
\bea
\label{eq:ABBMShapeSizeDef}
\mathfrak{s}_{w^\rmt}(\tm,S) := & E[\du_{\tm}|S] = \frac{E[\du_{\tm}, S]}{P_{w^\rmt}(S)}.
\eea
Since $P_{w^\rmt}(S)$ is known from \eqref{eq:BFMPwofS}, $\mathfrak{s}_{w^\rmt}({\tm},S)$ can be obtained by inverting the following Laplace transform
\bea
\nn
\hsh_{w^\rmt}(\tm,\lambda) := &\overline{e^{\lambda S} P_{w^\rmt}(S) \mathfrak{s}_{w^\rmt}(\tm,S)}^{\dw_{xt} = w\delta(t)} = \overline{\du^\rmt_{\tm}\, \exp\left(\lambda\int_{\tf} \du^\rmt_{\tf} \right)} \\
\label{eq:ABBMShapeSizeLT}
=& \partial_\mu \big|_{\mu=0} G[\lambda_{xt} = \mu \delta(t-{\tm}) + \lambda,\dw_{xt} = w \delta(t)].
\eea
For quasi-static avalanches, in analogy to \eqref{eq:DefShapeFixedT}, we consider an infinitesimal force step $f^\rmt = m^2 w^\rmt$ and obtain\footnote{This definition is given, exceptionally, in dimensionful units since it will be reused later in chapter \ref{sec:OneLoop} where the scaling with $m$ will be important.}
\bea
\nn
\mathfrak{s}(\tm,S) :=& \lim_{f^\rmt \to 0} \mathfrak{s}_{w^\rmt}(\tm,S) = \frac{1}{P(S)}\partial_{f^\rmt}\big|_{f^\rmt = 0} E[\du_{\tm}, S], \\
\nn
\hsh(\tm,\lambda) := & \overline{e^{\lambda S} P(S) \mathfrak{s}(\tm,S)} = \partial_{f^\rmt}\big|_{f^\rmt = 0}\hsh_{w^\rmt}(\tm,\lambda) \\
\label{eq:DefShapeFixedS}
 =&  \int \mD[\du,\tu]\frac{1}{L^d}\int_{\xxi}\tu_{\xxi, \ti=0}\int_{\xm}\du_{\xm,\tm}\, \exp\left(-S[\du,\tu] + \lambda\int_{\xf,\tf}\du_{\xf,\tf} \right).
\eea
Again, \eqref{eq:ABBMShapeSizeLT} and \eqref{eq:DefShapeFixedS} are general definitions valid for any interface model. For the BFM/ABBM, we can use \eqref{eq:BFMSolGenFct} to compute $G$, and $\hsh$. As in the previous section, we perturb \eqref{eq:ABBMInstanton} around the solution \eqref{eq:ABBMSizeInstanton} for $\mu=0$
\bea
\nn
\tu_{t}[\mu \delta(t-{\tm}) + \lambda] = & \tu_t^{(0)} + \mu \tu_t^{(1)} + \mO(\mu)^2 \\
\label{eq:ABBMShapeSize0}
\tu_t^{(0)}= & \tu_t[\lambda_{xt}=\lambda] = \frac{1}{2}\left(1-\sqrt{1-4\lambda}\right).
\eea
$\tu_t^{(1)}$ again satisfies the linear equation \eqref{eq:ABBMShapeLinPert}, but with $\tu_t^{(0)}$ constant and given by \eqref{eq:ABBMShapeSize0}. Its solution is
\bea
\tu_t^{(1)} = e^{-\sqrt{1-4\lambda}(\tm-t)} \theta(\tm-t).
\eea
Finally, we obtain for the Laplace-transformed shape \eqref{eq:ABBMShapeSizeLT}
\bea
\label{eq:ABBMShapeSizeLTRes}
\hsh_{w^\rmt}(\tm,\lambda) = w^\rmt\tu_{t=0}^{(1)}e^{w^\rmt\tu_{t=0}^{(0)}} = w^\rmt \exp\left[ \frac{w^\rmt}{2}\left(1-\sqrt{1-4\lambda}\right) - \sqrt{1-4\lambda} \tm\right].
\eea
Inverting the Laplace transform and dividing by the size distribution $P_{w^\rmt}(S)$ in \eqref{eq:BFMPwofS}, we obtain the shape at fixed size (first given in \cite{DobrinevskiLeDoussalWiese2013})
\bea
\label{eq:ABBMShapeSizeRes}
\mathfrak{s}_{w^\rmt}(\tm,S) = (2\tm+w^\rmt)\exp\left[-\frac{\tm(\tm+w^\rmt)}{S}\right],
\eea
which reduces for $w^\rmt \to 0$ to
\bea
\label{eq:ABBMShapeSizeResSmallW}
\mathfrak{s}(\tm,S)= \lim_{w^\rmt \to 0} \mathfrak{s}_{w^\rmt}(\tm,S)  = 2 \tm \exp\left(-\tm^2/S\right).
\eea

\paragraph{Average avalanche shape, as a function of distance, at fixed avalanche size}
Let us define the time $t_u$ at which the total displacement of the interface is $u$ implicitly via $\int_0^{t_u} \rmd t \, \du^{\rmt}(t) = u$.
Then, the average avalanche shape, as a function of distance $u$ (``$u$-shape''), after a finite step of size $w^{\rmt}=L^d w$ as above, can be expressed as
\bea
\nn
\mathfrak{s}_{w^\rmt}(u,S) := & E[\du^{\rmt}(u)|S]^{\dw_{xt} = w\delta(t)} = E\left[\int_t \du^{\rmt}(t) \delta\left(t-t_u\right)\big|\int_t\du^{\rmt}(t) = S\right]^{\dw_{xt} = w\delta(t)} \\
= & E\left[\int_t \du^{\rmt}(t)^2 \delta\left(u-\int_0^t\rmd t'\,\du^{\rmt}(t')\right)\big|\int_t\du^{\rmt}(t) = S\right]^{\dw_{xt} = w\delta(t)}.
\eea
As for the average shape as a function of time, at fixed size, it is useful to consider the Laplace transform
\bea
\nn
\hsh_{w^\rmt}(\lambda_1,\lambda_2) :=& \overline{\mathfrak{s}_{w^\rmt}(u,S)P_{w^\rmt}(S)\,e^{\lambda_1 u + \lambda_2 S}}^{\dw_{xt} = w\delta(t)} \\
= & \overline{\int_0^\infty \rmd t\, \du^{\rmt}(t)^2\,\exp\left[\lambda_1 \int_0^t \rmd t'\,\du^{\rmt}(t') + \lambda_2 \int_0^\infty \rmd t'\,\du^{\rmt}(t')\right]}^{\dw_{xt} = w\delta(t)}.
\eea
For the BFM, $\hsh_{w^\rmt}$ can be computed by solving the instanton equation \eqref{eq:ABBMInstanton} with the source
\bea
\lambda_t = \lambda_1 \theta(t < t_1) + \mu \delta(t-t_1) + \lambda_2.
\eea
Inverting the Laplace transform, one finds the shape following a step of size $w^{\rmt}$:
\bea
\label{eq:ABBMUShapeSizeRes}
\mathfrak{s}_{w^\rmt}(u=sS,S) = e^{-w^\rmt} \left\{\frac{\left[2 s S+(1-s) (w^\rmt)^2\right] \text{erf}\left(\frac{1}{2} w^\rmt \sqrt{\frac{1-s}{s
   S}}\right)}{w^\rmt}+\frac{2 \sqrt{(1-s) s S} e^{-\frac{(1-s) (w^\rmt)^2}{4 s S}}}{\sqrt{\pi }}\right\}.
\eea
For an infinitesimal step $w^{\rmt} \to 0$, this reduces to \eqref{eq:BFMUShapeFixedS} as known from \cite{BaldassarriColaioriCastellano2003}.

Alternatively, since $w$ is kept fixed after the initial step, \eqref{eq:BFMABBM} shows that $\du(u)$ is a Brownian motion with drift, whose propagator in the presence of an absorbing boundary is known exactly.  This can likewise be used to recover \eqref{eq:ABBMUShapeSizeRes}.

\subsubsection{Local avalanche sizes\label{sec:BFMLocalAv}}
For the discussion of avalanche sizes, durations, and shapes above we only considered the total interface velocity, integrated over space. In these cases, the source $\lambda(x,t)=\lambda$, and the solution $\tu[\lambda_{xt}=\lambda]$ of the instanton equation \eqref{eq:BFMInstanton}, were actually independent of $x$. The observables thus reduced to those of the standard ABBM model, as discussed in section \ref{sec:BFMSolABBM}, up to some additional factors $L^d$.

On the other hand, the solution formula \eqref{eq:BFMSolGenFct} also allows one to compute more general observables depending on local velocities.
One interesting case are local avalanche sizes $S_\phi$ on a hyperplane $\phi$ of dimension $d_\phi<d$ of the interface of total dimension $d$. Denoting $\mathbb{R}^d \owns x =(x_1...x_d)$, the local avalanche size $S_\phi$ is given as
\bea
\label{eq:DefLocSD}
S_\phi = \int_{x_1}\cdots \int_{x_{d_\phi}} \int_t \du_{(x_1,...x_{d_\phi},0...0),t}.
\eea
Clearly, the distribution of $S_\phi$ is independent of which sub-hyperplane of dimension $d_\phi<d$ is chosen. Let us now trigger, as in the previous sections, an avalanche through a step of size $w$ in the local driving, i.e. $\dw_{xt} = w \delta(t-\ti)$.
From \eqref{eq:DefLocSD}, the average $S_\phi$ following this step will be $w L^{d_\phi}$ (for the global avalanche size $d_\phi=d$, we find $w^\rmt = w L^d$ as in section \ref{sec:BFMSize}). Let us thus fix $w^{(d_\phi)}:=w L^{d_\phi}$ and define the distribution of local avalanche sizes $P_{w^{(d_\phi)}}(S_\phi)$ through its Laplace transform
\bea
\label{eq:DefPwofLocS}
\int \rmd S_\phi\, e^{\lambda S_\phi} P_{w^{(d_\phi)}}(S_\phi) := \overline{e^{\lambda S_\phi}}^{\dw_{xt} = w \delta(t-\ti)} = 
G\left[\lambda_{xt} = \lambda \prod_{j=d_\phi+1}^d \delta(x_j),\dw_{xt} = w \delta(t-\ti)\right].
\eea
For the density of local avalanche sizes in a single avalanche, as in section \ref{sec:BFMSize} we take the leading order for small $f^{(d_\phi)} = m^2 w^{(d_\phi)}$. We define\footnote{This definition is given, exceptionally, in dimensionful units since it will be reused later in chapter \ref{sec:OneLoop} where the scaling with $m$ will be important.}
\bea
\nn
P(S_\phi) =& \partial_{f^{(d_\phi)}}\big|_{f^{(d_\phi)}=0} P_{w^{(d_\phi)}}(S_\phi), \\
\label{eq:DefPofLocS}
\int \rmd S_\phi\, (e^{\lambda S_\phi}-1) P(S_\phi) = & \int \mD[\du,\tu] \frac{1}{L^{d_\phi}}\int_{\xxi}\tu_{\xxi,\ti}\exp\left[ \lambda \int_{x_1}\cdots \int_{x_{d_\phi}} \int_{\tf} \du_{(x_1,...x_{d_\phi},0...0),\tf}\right].
\eea
\eqref{eq:DefPwofLocS} and \eqref{eq:DefPofLocS} are general definitions for the local avalanche size distribution. The main difference to the global avalanche size distribution defined in \eqref{eq:DefPwofS}, \eqref{eq:DefPofS} is the source $\lambda$ for $\du$ (which is now localized on a subspace of dimension $d_\phi$), and the scaling of the local step size (which now scales $\sim L^{-d_\phi}$, not  $\sim L^{-d}$).

In order to obtain the distribution of $S_\phi$ in the BFM, one needs to solve the instanton equation \eqref{eq:BFMInstanton} with the source 
\bea
\label{eq:BFMLocalLambda}
\lambda_{xt} = \lambda \delta(x_{d_\phi+1})\cdots\delta(x_d). 
\eea
One case which is solvable analytically are local avalanche sizes on a codimension one hyperplane, $d_\phi = d-1$ (see \cite{LeDoussalWiese2008c}, section IX). In this case, the instanton solution is constant in the directions $x_1\cdots x_{d-1}$. In the direction $y := x_{d}$ we obtain a second-order ODE:
\bea
\label{eq:BFMLocalODE}
\partial_y^2 \tu_y - \tu_y + \tu_y^2 = -\lambda \delta(y)
\eea
For $y>0$, \eqref{eq:BFMLocalODE} can be integrated after multiplying it with $\partial_y \tu_y$. The result is \cite{LeDoussalWiese2008c,Delorme2013inpr}
\bea
\label{eq:BFMLocalInstSol}
\tu_y = \begin{cases}
\frac{3}{1+\cosh\left[y+y_0(\lambda)\right]}=\frac{3}{2 \cosh^2\left[\frac{y+y_0(\lambda)}{2}\right]},&\quad\quad \text{for } \lambda \geq 0, \\
\frac{3}{1-\cosh\left[y+y_0(\lambda)\right]}=-\frac{3}{2 \sinh^2\left[\frac{y+y_0(\lambda)}{2}\right]},&\quad\quad \text{for } \lambda < 0,
\end{cases}
\eea
where $y_0(\lambda)$ is fixed by symmetry and the jump condition on $\partial_y \tu_y$ at $y=0$ due to the source on the right-hand side of \eqref{eq:BFMLocalODE}. Finally this gives an implicit equation for $\tilde{Z}[\lambda \delta(x)] := \frac{1}{L^{d-1}}\int_x \tu_x$ (eq. 217 in \cite{LeDoussalWiese2008c}; note that the prefactor $\frac{1}{L^{d-1}}$ cancels against the $d-1$ integrals over the directions $x_1$...$x_{d-1}$ in which the instanton $\tu$ is constant):
\bea
72 \lambda = \tZ(\tZ-6)(\tZ-12).
\eea

This implicit equation suffices in order to invert the Laplace transform, and obtain the density of the local avalanche size $S_{d-1}$. By \eqref{eq:DefPwofLocS} and \eqref{eq:BFMSolGenFct} we have
\bea
\nn
\int \rmd S_{d-1}\, e^{\lambda S_{d-1}} P_{w^{(d-1)}}(S_{d-1}) =& e^{w^{(d-1)}\tZ(\lambda)}\\
\nn
\Rightarrow P_{w^{(d-1)}}(S_{d-1}) =& \int \frac{\rmd \lambda}{2\pi i} e^{ w^{(d-1)}\tZ(\lambda) - \lambda S_{d-1}} = \int \frac{\rmd \tZ}{2\pi i} \frac{\rmd \lambda(\tZ)}{\rmd \tZ} e^{w^{(d-1)}\tZ-\lambda(\tZ) S_{d-1}} \\
\nn
=&\int \frac{\rmd \tZ}{2\pi i} \left(1-\frac{\tZ}{2}+\frac{\tZ^2}{24}\right) e^{w^{(d-1)}\tZ-\frac{1}{72} S_{d-1}\tZ(\tZ-6)(\tZ-12)} \\
\nn
= & \frac{2 \sqrt[3]{3} e^{6 w^{(d-1)}} w^{(d-1)} \text{Ai}\left(\frac{\sqrt[3]{3} (S_{d-1}+2
   w^{(d-1)})}{\sqrt[3]{S_{d-1}}}\right)}{S_{d-1}^{4/3}}
\eea
This formula was first derived in \cite{LeDoussalWiese2013} (eq. 225). 
The contour for the complex integrals used for the Laplace inversion has to be chosen so that the result is convergent; the correctness of the final result can be verified by comparing its (real-valued and convergent) integral $\int \rmd S_{d-1}\, e^{\lambda S_{d-1}} P_{w^{(d-1)}}(S_{d-1})$.
For a single quasi-static avalanche \eqref{eq:DefPofLocS} yields the density first obtained in \cite{LeDoussalWiese2008c} (as usual, note that in dimensionless units $f^{(d_\phi)}=w^{(d_\phi)}$)
\bea
P(S_{d-1}) =  \partial_{f^{(d-1)}}\big|_{f^{(d-1)}=0} P_{w^{(d-1)}}(S_{d-1}) = \frac{2}{\pi S_{d-1}}K_{1/3}\left(\frac{2 S_{d-1}}{\sqrt{3}}\right).
\eea

This can be generalized to the joint distribution of local avalanches at several points in space \cite{Delorme2013inpr}.
It would be very interesting to obtain similar analytical results or approximations for local velocities. However, this would require solving the time-dependent PDE \eqref{eq:BFMInstanton} with a source local in space and in time, which is difficult.

\subsection{Avalanches at finite driving velocity\label{sec:ABBMFinVel}}
The non-stationary avalanches discussed above are clearly delimited: At some time following the step in the driving force, the motion stops completely. The situation is different for a nonzero driving velocity $0<v<1$: There, the velocity $\du_t$ becomes zero at some times, but then immediately restarts due to the driving. Thus, to delimit avalanches at a finite driving velocity, one needs to manually introduce an absorbing boundary at $\du=0$ in the equation of motion \eqref{eq:IntVelEOMBFMABBM}.\footnote{The natural boundary implicitly assumed in the solution formula \eqref{eq:BFMSolGenFct} is a reflecting boundary. For a more thorough discussion see \cite{DobrinevskiLeDoussalWiese2013}.} 
For the ABBM model, this can be performed using a Fokker-Planck formalism \cite{LeblancAnghelutaDahmenGoldenfeld2013} or by a modification of the MSR/instanton method above \cite{DobrinevskiLeDoussalWiese2013}. One obtains the following result for the distribution of avalanche durations \cite{LeblancAnghelutaDahmenGoldenfeld2013,DobrinevskiLeDoussalWiese2013}, starting from $\du_{\ti} = w$:
\bea
\label{eq:ABBMPofTFiniteV}
P_w(T) = \frac{w^{1-v}}{\Gamma(1-v) (e^T-1)^{2-v}}\exp\left[-\frac{w}{e^T-1}+T\right].
\eea
One can also obtain $P_w(S)$ \cite{DobrinevskiLeDoussalWiese2013}, and the average avalanche shape \cite{LeblancAnghelutaDahmenGoldenfeld2013} at finite driving velocity $v$.

It would be interesting to see if this result applies to the center-of-mass of an extended interface in a Brownian landscape in the same way as the results for avalanches following a step in the driving force, at $v=0$. Possibly, this can be elucidated by generalizing the backward instanton method used to obtain \eqref{eq:ABBMPofTFiniteV} in \cite{DobrinevskiLeDoussalWiese2013} to the case of the BFM.

\section{Dynamics with retardation\label{sec:BFMRetardation}}
The predictions of the BFM/ABBM model discussed above can be compared to Barkhausen-noise experiments. 
As discussed in section \ref{sec:Barkhausen}, for polycristalline ferromagnetic materials, the power-law exponents of avalanche sizes and durations, as well as instantaneous velocities, show very good agreement with the predictions of the BFM mean-field universality class \cite{DurinZapperi2000}. However, other observables, like average avalanche shapes, have more structure and are more sensitive to the details of the model. In particular, average avalanche shapes at a fixed avalanche duration show a marked asymmetry to the left in experiments (see \cite{ZapperiCastellanoColaioriDurin2005} and section \ref{sec:Barkhausen}), in contrast to the symmetric prediction of the BFM \eqref{eq:BFMShapeFixedT}.

Qualitatively, this asymmetry can be ascribed to the presence of eddy currents \cite{ZapperiCastellanoColaioriDurin2005,ColaioriDurinZapperi2007,Colaiori2008}: They build up during the initial phase of the avalanche, and then keep pushing the domain wall forward when it starts slowing down. This way, the avalanche is stretched towards a longer total duration, with a prolonged low-velocity tail at the end. This decreases the average velocity near the end of the avalanche, and yields the observed asymmetry towards the left. 

A generalization of the ABBM model of magnetic domain wall dynamics, including the dissipation of eddy currents, was first proposed and studied numerically in \cite{ZapperiCastellanoColaioriDurin2005}. In this section, I review this generalization, and the analytical treatment for it developed in \cite{DobrinevskiLeDoussalWiese2013}. Quantitatively, \cite{ZapperiCastellanoColaioriDurin2005} proposes to describe the dissipation of eddy currents using a generalization of the ABBM equation \eqref{eq:ABBM}
\beq
\label{eq:EOMZapperi}
 \frac{1}{\sqrt{2\pi}} \int_{-\infty}^t \rmd s\, \mathfrak{f}(t-s) \, \du(s) =  2 I_s\big[H(t) - k u(t) + F(u(t))\big].
\eeq
The response function $\mathfrak{f}$, derived by solving the Maxwell equations in a rectangular sample \cite{Bishop1980,ZapperiCastellanoColaioriDurin2005,ColaioriDurinZapperi2007,Colaiori2008}, is
\beq
\label{eq:EOMZapperi1}
\mathfrak{f}(t) = \sqrt{2\pi}\frac{64 I_s^2}{a b^2 \sigma \mu^2} \sum_{n,m=0}^\infty \frac{e^{-t/\tau_{m,n}}}{(2n+1)^2\omega_b}.
\eeq
$\tau_{m,n}$ are relaxation times for the individual eddy-current modes,
\begin{align*}
& \tau_{m,n}^{-1} = (2m+1)^2 \omega_a + (2n+1)^2 \omega_b \\
& \omega_a = \frac{\pi^2}{\sigma \mu a^2},\quad\quad \omega_b = \frac{\pi^2}{\sigma \mu b^2}.
\end{align*}
They depend on the sample width $a$, thickness $b$, permeability $\mu$ and conductivity $\sigma$.  
\eqref{eq:EOMZapperi} and \eqref{eq:EOMZapperi1} correspond to Eqs.~(13), (17) and (21) in \cite{ColaioriDurinZapperi2007}; we refer the reader there for  details of the derivation. In the following section \ref{sec:BFMRetSol}, we shall see that the exact formula for the generating functional of velocities \eqref{eq:ABBMSolGenFct} at arbitary monotonous driving can be generalized to \eqref{eq:EOMZapperi} for an arbitrary memory kernel $\mathfrak{f}$.

A simplified model is obtained by considering only the leading contributions for small and large relaxation times. The contribution from small relaxation times becomes a term of the form $\partial_t u(t)$, as in the standard ABBM model \eqref{eq:BFMABBM}. The contribution from the longest relaxation time $\tau:=\tau_{0,0}=\frac{\mu\sigma}{\pi^2}\left(\frac{1}{a^2}+\frac{1}{b^2}\right)^{-1}$ gives rise to an exponential memory kernel, and thus to the equation-of-motion
\bea
\label{eq:EOMZapperi2}
\Gamma \du(t) + \frac{\Gamma_0}{\tau}\int_{-\infty}^t \rmd s\, e^{-(t-s)/\tau}\du(s) 
 =  2 I_s\big[H(t) - k u(t) + F(u(t))\big].
\eea
For more details and the expressions for the damping coefficients $\Gamma$ and $\Gamma_0$ see \cite{ZapperiCastellanoColaioriDurin2005,DobrinevskiLeDoussalWiese2013}. In this ABBM model with exponential retardation, analytical results for the instantaneous velocity distribution, and for the avalanche statistics, can be obtained in several limits (see sections \ref{sec:BFMRetStat}, \ref{sec:BFMRetSize}, and \ref{sec:SlowRel} below). We will observe that retardation does not modify the universality class of the model: The power-law exponents of the size and duration distributions for small quasi-static avalanches, as well as the average shape of small avalanches, are identical to those of the standard ABBM model. On the other hand, the behaviour at finite driving velocity, and for long avalanches, \textit{is} modified. This suggests that these non-universal observables are more useful for identifying the role of retardation effects e.g.~in experiments.

\subsection{Monotonicity and MSR solution\label{sec:BFMRetSol}}
For the remainder of this section, let us adopt the units and conventions used in \eqref{eq:BFMABBM} for an elastic interface, following \cite{DobrinevskiLeDoussalWiese2013}. The generalization of \eqref{eq:IntVelEOMBFMABBM} with a retarded memory kernel $f(t)$ reads
\beq
\label{eq:BFMRetEOMU}
\eta \du(t) + a \int_{-\infty}^t \rmd s\,f(t-s) \du(s) = F\big(u(t)\big) + m^2 \big[w(t) - u(t)\big].
\eeq
It describes a particle driven in a force landscape $F(u)$, {\it with retardation}. $f(t)$ is a general memory kernel with the following properties:
\begin{enumerate}
        \item $f(0) = 1$ (without loss of generality, since a constant may be absorbed into the parameter $a$).
        \item $f(x) \to 0$ as $x \to \infty$.
        \item $f'(x) \leq 0$ for all $x$, i.e.~memory of the past trajectory always decays with time. This will be necessary to ensure monotonicity of the domain-wall motion for monotonous driving.
\end{enumerate}

To understand how the retardation term in eq.~\eqref{eq:BFMRetEOMU} modifies the avalanche statistics, we would like to construct a closed theory for the \textit{velocity} of the particle $\du$. As we saw in section \ref{sec:BFMVelocity}, this requires the particle motion to be monotonous at all times, in any realization of the disorder. Amazingly, for monotonous driving this is ensured for any force landscape $F$ by a generalization of the Middleton property shown in section \ref{sec:InterfaceMonot}: Assuming $\dw_t \geq 0$, and $\du_{t\leq \ti} \geq 0$, we also have $\du_{t \geq \ti}\geq 0$ for the solutions of \eqref{eq:BFMRetEOMU}. To see this, let us first take a derivative of Eq.~\eqref{eq:BFMRetEOMU}. This gives an equation of motion for $\du(t)$, instead of $u(t)$:
\bea
\label{eq:BFMRetEOMUdTemp}
\eta \partial_t \du(t) + a \du(t) + a \int_{-\infty}^t \rmd s \,\partial_t f(t-s) \du(s) = \partial_t F(u(t)) + m^2 \left[\dw(t) - \du(t)\right].
\eea
Now, consider as in section \ref{sec:InterfaceMonot} the first time $t_1$ at which $\du(t_1)=0$. Then also $\partial_t\big|_{t=t_1} F(u(t))=0$, and we have from \eqref{eq:BFMRetEOMUd}
\bea
\partial_t\big|_{t=t_1}\du(t) = m^2 \dw(t) - a \int_{-\infty}^t \rmd s \,\partial_t f(t-s) \du(s).
\eea
The first term is non-negative due to the assumption of monotonous driving. The second term is also non-negative, since $\du(s) \geq 0$ for $s \leq t_1$, and since we assumed the memory kernel to be decreasing in time, $\partial_x f(x) \leq 0$ for all $x$. This shows the importance of the assumption 3. above. Thus, after touching $\du=0$, the velocity will not become negative but instead restart to positive values. In that respect, the decay of the memory term acts as an additional positive driving and preserves monotonicity. 

This property is quite non-trivial: Other ways of generalizing the interface model \eqref{eq:InterfaceEOM} with additional degrees of freedom, such as the second-order dynamics (damped Newton equation) discussed in \cite{LeDoussalPetkovicWiese2012}, thermal noise \cite{ChauveGiamarchiLeDoussal2000} or including the domain-wall phase \cite{BarnesEckmannGiamarchiLecomte2012,LecomteBarnesEckmannGiamarchi2009} typically do not preserve this monotonicity property. 

Monotonicity also ensures the existence of a unique ``Middleton'' state $u(w)$ for quasi-static driving, which is the ``leftmost metastable state'' as discussed in section \ref{sec:InterfaceMonot}. In fact, since this is the state reached when $w$ is fixed for a long time, the trajectory $u(t)$ approaches it monotonously and $\du(t) \to 0$ for long times. This means that the memory term in \eqref{eq:BFMRetEOMU} goes to zero, due to the decay of $f$ for long times. Hence, the Middleton state $u(w)$ is the leftmost solution of $F(u)+m^2(w-u)=0$, just as for the equation-of-motion without retardation \eqref{eq:BFMABBM}. The metastable states through which the particle passes under quasi-static driving, in any disorder $F(u)$, are not modified by retardation. In particular, quasi-static avalanche sizes remain the same; only the dynamics inside an avalanche changes.

Let us now specialize to the ABBM disorder (Brownian random forces), i.e. $F(u)$ given by \eqref{eq:DisorderABBM}. Following the discussion in section \ref{sec:BFMVelocity}, we can rewrite \eqref{eq:BFMRetEOMUdTemp} to obtain a closed equation of motion for $\du(t)$,
\bea
\label{eq:BFMRetEOMUd}
\eta \partial_t \du(t) + a \du(t) + a \int_{-\infty}^t \rmd s \partial_t f(t-s) \du(s) = \sqrt{\du(t)}\xi(t) + m^2 \left[\dw(t) - \du(t)\right].
\eea
$\xi(t)$ is a Gaussian white noise, with $\overline{\xi(t_1) \xi({t_2})} = 2\sigma \delta(t_1-t_2)$.  The term $\sqrt{\du(t)}$ comes from rewriting the {\em position-dependent} white noise in terms of a {\em time-dependent} white noise, as in section \ref{sec:BFMVelocity}, using the monotonicity property.

Since the memory term in \eqref{eq:BFMRetEOMU} is linear in $\du$, we can already see that the corresponding Martin-Siggia-Rose field theory will be solvable using the instanton approach in a way similar to section \ref{sec:BFMSol}. The natural generalization of these results is that formula \eqref{eq:ABBMSolGenFct} for the generating functional still holds for arbitrary monotonous driving $\dw_t \geq 0$
\bea
\label{eq:RetSolution}
G[\lambda_t,\dw_t] := \overline{\exp\left( \int_t \lambda_t \du_t \right)}^{\dw_t} = \exp\left( m^2 \int_{t} \dw_t \tu_{t}[\lambda]  \right).
\eea
However, the instanton equation \eqref{eq:ABBMInstanton} is replaced by 
\bea
\label{eq:BFMRetInstanton}
\eta \partial_t \tu(t) - (m^2 + a) \tu(t) + \sigma \tu(t)^2 - a \int^{\infty}_t \rmd s\, \tu(s) \partial_s f(s-t)  = - \lambda(t),
\eea
As discussed in sections \ref{sec:BFMSol} and \ref{sec:BFMSolIC}, eq.~\eqref{eq:RetSolution} is valid for monotonous driving starting from a ``Middleton'' initial condition $u_{\ti}(w_{\ti})$ at $\ti = -\infty$. 

To prove \eqref{eq:RetSolution} we apply the same method as in the absence of retardation (see section \ref{sec:BFMSol} and \cite{LeDoussalWiese2011,DobrinevskiLeDoussalWiese2012,LeDoussalWiese2011,DobrinevskiLeDoussalWiese2013}). Using the Martin-Siggia-Rose formalism, we can write the generating functional for solutions of \eqref{eq:BFMRetEOMUd} as a path integral with the action $S[\du,\tu]$,
\bea
\nn
G[\lambda_t,\dw_t] :=& \overline{\exp\left( \int_t \lambda_t \du_t \right)}^{\dw_t} = \int \mD[\du,\tu]\,\exp\left(-S[\du,\tu] + \int_t \lambda_t \du_t + \int_t \df_t \tu_t \right), \\
S[\du,\tu] = & \int_t \tu_t  \Big[\eta \partial_t \du_{t} + a \du_{t} + a \int_{-\infty}^t \rmd s\, \partial_t f(t-s) \dot u_{s} 
 + m^2 \du_{t}  \Big] - \sigma \int_t \tu_t^2 \du_t.
\eea
Here we defined the generating functional and the source $\df = m^2 \dw$ for the MSR response field $\tu$ as in \eqref{eq:DefGSources}.
The action $S$ is linear in $\du$; integrating over this field as in section \ref{sec:BFMSol} yields the solution \eqref{eq:RetSolution}, \eqref{eq:BFMRetInstanton}.

The generalization to extended elastic interfaces in a Brownian force landscape, as in section \ref{sec:BFMABBM}, is straightforward.

\subsubsection{Exponential relaxation\label{sec:BFMRetExp}}
Solving the instanton equation \eqref{eq:BFMRetInstanton} for a general memory kernel $f$ is complicated.
As we saw in eq.~\eqref{eq:EOMZapperi2}, a special case which captures the phenomenology of eddy currents in magnetic domain-wall motion is an exponential memory kernel 
\bea
f(t) = e^{-t/\tau}.
\eea
In this case, \eqref{eq:BFMRetEOMU} can be re-written as two coupled, \textit{local} equations for the domain-wall velocity $\du(t)$, and the retarded force $h(t)$ (corresponding to the eddy-current pressure for magnetic domain walls),
\bea
\nn
h(t) &=  \frac{1}{\tau}\int_{-\infty}^t \rmd s\,e^{-(t-s)/\tau} \du(s) \\
\nn
\eta \du(t)+ a \tau h(t) &= F\big(u(t)\big) + m^2 \big[w(t) - u(t)\big]\\
 \tau \partial_t h(t) &= \du(t) - h(t) .
\label{eq:BFMRetEOMExp}
\eea
Using the same method as above, one obtains an exact expression for the generating functional of the domain-wall velocity $\du(t)$ and the retarded force $h(t)$,
\bea
\label{eq:BFMRetSolExp}
G[\lambda_t,\mu_t,\dw_t] := \overline{\exp\left(\int_t \lambda_t \du_t + \mu_t h_t \right)}^{\dw_t} = \exp\left(m^2 \int_t \dw_t \tu_t[\lambda,\mu] \rmd t\right).
\eea
The MSR response fields $\tu_t[\lambda,\mu]$, $\tih_t[\lambda,\mu]$ satisfy the 
set of two local instanton equations
\begin{align}
\label{eq:RetInstantonExpTu}
\eta \partial_t \tu_{t} - \left(m^2 + a\right) \tu_{t} + \sigma \tu_{t}^2 + \tih_{t} &=
- \lambda_{t}, \\
\label{eq:RetInstantonExpTh}
\tau\partial_t \tih_{t} -\tih_{t} + a \tu_{t} &=  - \mu_{t}.
\end{align}
For details, see \cite{DobrinevskiLeDoussalWiese2013}.

\subsection{Parameters and units\label{sec:BFMRetUnits}}
In section \ref{sec:BFMUnits} we introduced the natural time scale $\tau_m = \eta/m^2$ and the natural velocity scale $v_m = S_m/\tau_m = \sigma/(\eta m^2)$ for the ABBM model \footnote{Here we consider only a particle model, $d=0$, and do not distinguish between local and global velocities. The generalization to $d>0$ is straightforward.}. Rescaling times and velocities in the ABBM model with retardation \eqref{eq:BFMRetEOMUd} by these scales as in \eqref{eq:BFMRescaling}, we obtain 
\bea
\partial_{t'} \du'(t') + a' \du'(t') + a' \int_{-\infty}^{t'} \rmd {s'} \partial_{t'}f\left[(t'-s')\tau_m\right] \du(s') = \sqrt{\du'(t')}\xi'(t') + \dw'(t') - \du'(t'),
\eea
where $a' := a / m^2$. Recall from section \ref{sec:BFMUnits} that the primed quantities are the dimensionless ones. 
 Assuming that $f$ decays on a time scale $\sim \tau$, i.e. $f(x) = f'(x/\tau)$ where $f'$ is dimensionless, we can also write $f\left[(t'-s')\tau_m\right] = f'\left[(t'-s')/\tau'\right]$ with $\tau' := \tau / \tau_m$.
In the ``primed'' variables (i.e. when times and velocities are expressed in terms of their natural scales), as for the ABBM model, the parameters $\eta$, $\sigma$ and $m$ are all equal to 1. In the model with retardation, we now remain with \textit{two additional dimensionless parameters} \cite{DobrinevskiLeDoussalWiese2013}
\begin{itemize}
	\item The retardation strength $a' := a / m^2$. This determines the strength of retardation effects, compared to the confining harmonic well (for magnetic domain walls, the demagnetizing factor).
	\item The retardation time scale $\tau' := \tau / \tau_m$, compared to the time scale of avalanche motion $\tau_m = \eta/m^2$.
\end{itemize}
For the following discussion, we will drop all primes and use dimensionless units exclusively.

\subsection{Measuring the memory kernel using response and correlation functions\label{sec:BFMRetRespCorr}}
Let us define the response function $r(\tf-\ti,w)$ as the average velocity at time $\tf$ following a step of size $w$ in the driving force at time $\ti$.  For the ABBM model this is a simple exponential, as known already from \cite{AlessandroBeatriceBertottiMontorsi1990,Bertotti1998}. Using the approach of section \ref{sec:BFMABBM}, we can express it as
\bea
r(\tf-\ti,w) = \partial_\lambda\big|_{\lambda=0}G[\lambda_t = \lambda \delta(t-\tf),\dw_t = w\delta(t-\ti)] = w\, e^{-(\tf-\ti)}.
\eea 
This arises from the fact that the disorder term $\sigma \tu^2$ in the instanton equation \eqref{eq:ABBMInstanton} is of order $\lambda^2$ for small $\lambda$, and hence does not contribute to the average of the velocity. The same holds for the ABBM model with retardation and its instanton equation \eqref{eq:BFMRetInstanton}. The linear part of the instanton equation \eqref{eq:BFMRetInstanton} simplifies in Fourier space: The convolution in the memory term becomes a simple product. A short calculation (see \cite{DobrinevskiLeDoussalWiese2013}, section IV) gives the following expression for the Fourier-transformed response function:
\bea
\label{eq:BFMRetResponse}
r(\omega,w) := \int_\ti^\tf \rmd t \,e^{-i \omega (t-\ti)} r(\tf-\ti,w) = \frac{w}{1+i\omega\left[1 + a f(\omega)\right]},
\eea
where $f(\omega):=\int_0^\infty \rmd t \,e^{-i \omega t}f(t)$ is the Fourier transform of the memory kernel.
\eqref{eq:BFMRetResponse} provides a direct way for obtaining the memory kernel $f$ from a measurement of the response function $r$.

\begin{figure}%
         \centering
         \begin{subfigure}[t]{0.45\textwidth}
                 \centering
                 \includegraphics[width=\textwidth]{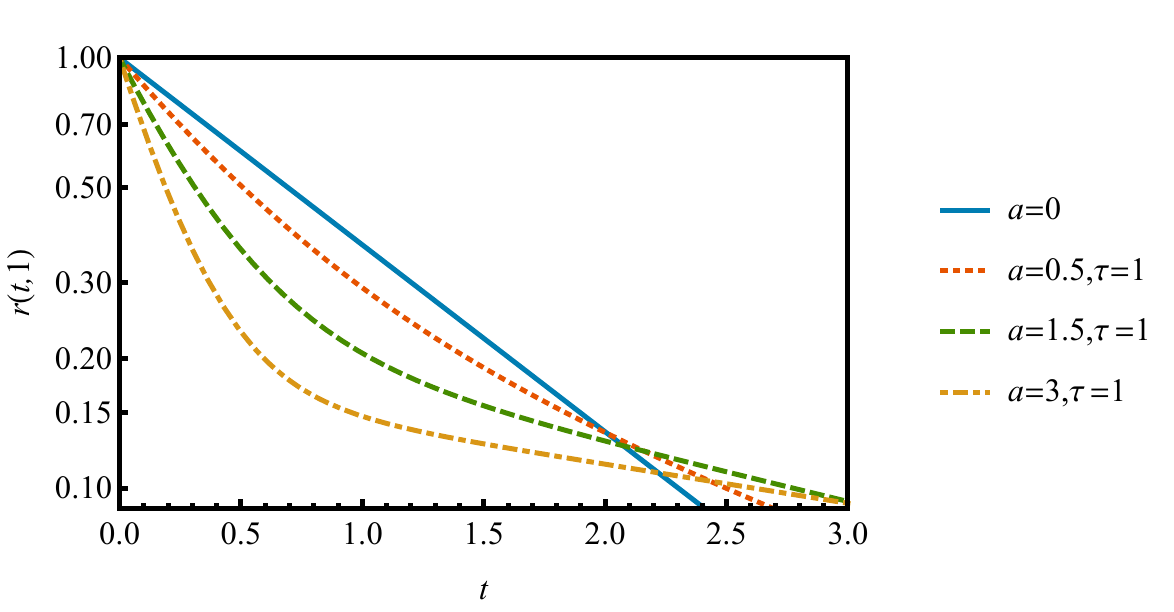}
                 \caption{Response function \eqref{eq:BFMRetResponseFinal} in the ABBM model with exponential retardation, at fixed retardation time scale $\tau=0.5$ for various values of retardation strength $a$.}
                 \label{fig:RetResponseVarA}
         \end{subfigure}%
         ~ 
         \begin{subfigure}[t]{0.45\textwidth}
                 \centering
                 \includegraphics[width=\textwidth]{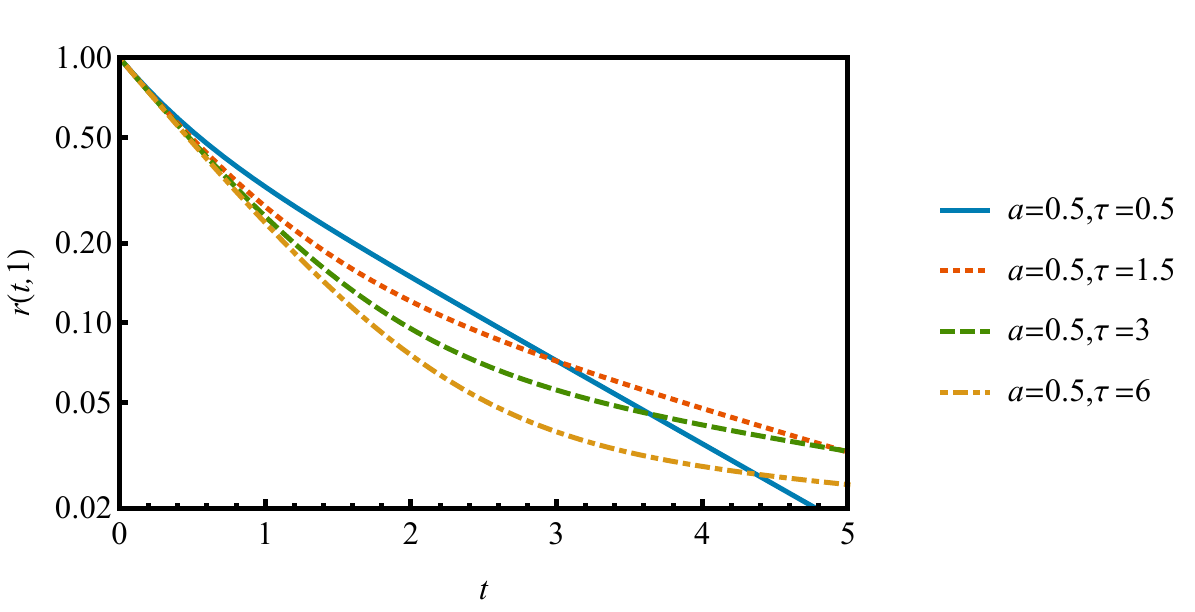}
                 \caption{Response function \eqref{eq:BFMRetResponseFinal} in the ABBM model with exponential retardation, at fixed retardation strength $a=0.5$ and varying retardation time scale $\tau$.}
                 \label{fig:RetResponseVarTau}
         \end{subfigure}\\
				         \begin{subfigure}[t]{0.45\textwidth}
                 \centering
                 \includegraphics[width=\textwidth]{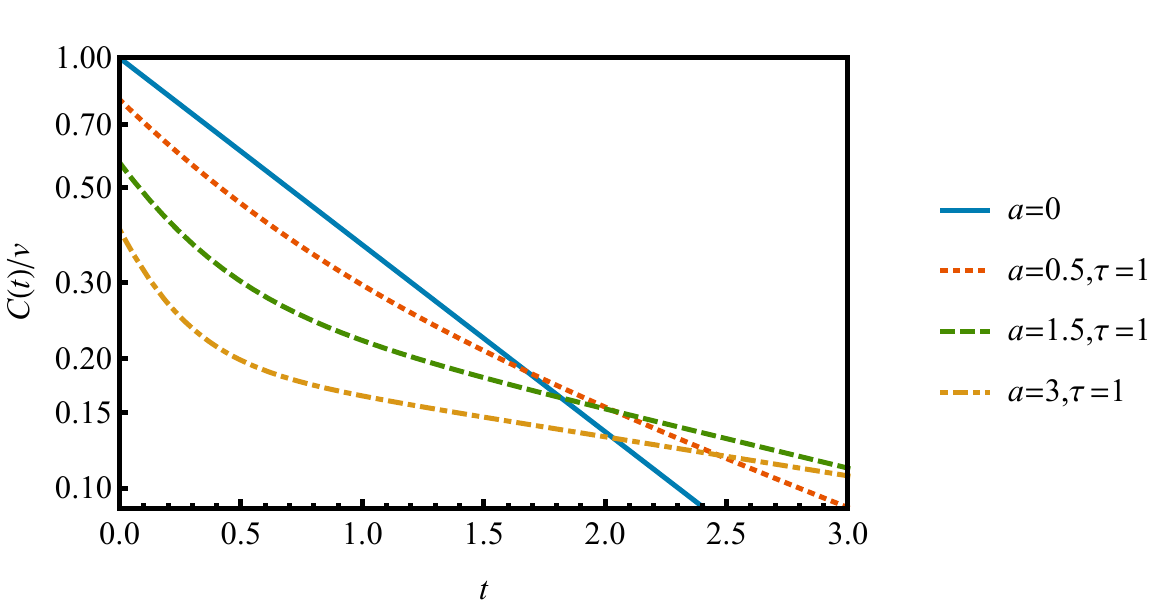}
                 \caption{Correlation function \eqref{eq:BFMRetCorrFinal} in the ABBM model with exponential retardation, at fixed retardation time scale $\tau=0.5$ for various values of retardation strength $a$.}
                 \label{fig:RetCorrVarA}
         \end{subfigure}%
         ~ 
         \begin{subfigure}[t]{0.45\textwidth}
                 \centering
                 \includegraphics[width=\textwidth]{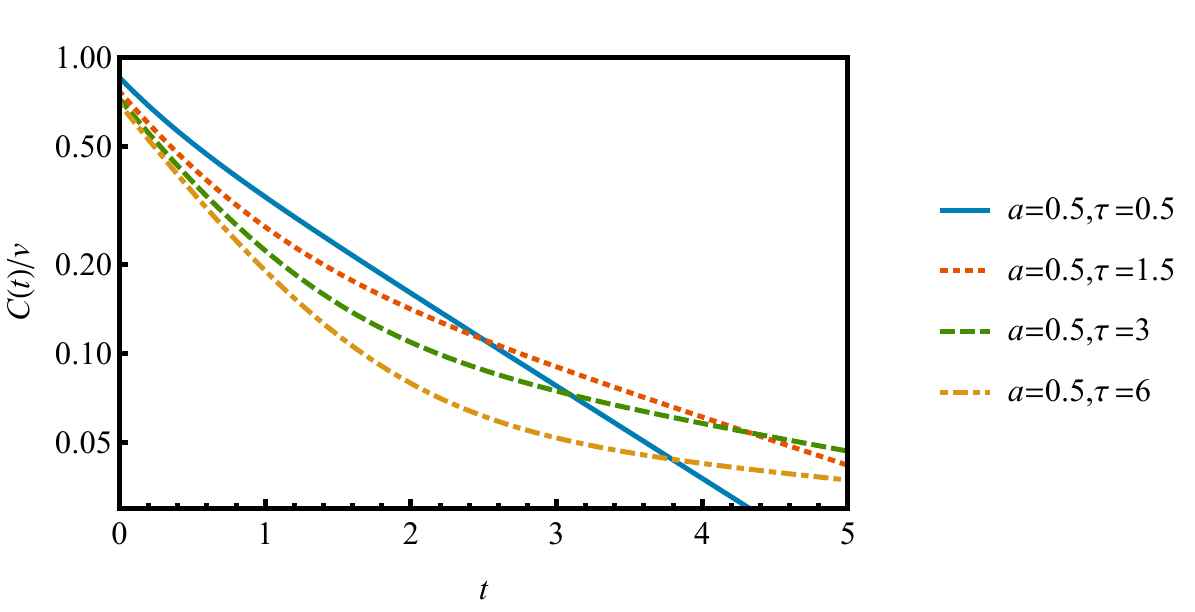}
                 \caption{Correlation function \eqref{eq:BFMRetCorrFinal} in the ABBM model with exponential retardation, at fixed retardation strength $a=0.5$ and varying retardation time scale $\tau$.}
                 \label{fig:RetCorrVarTau}
         \end{subfigure}
				                \caption{Response and correlation functions in the ABBM model with exponential retardation. \label{fig:RetResponseCorr}}
\end{figure}

Another simple observable, that can be computed exactly, is the two-point connected correlation function, $C(t_1-t_2) := \overline{\du_{t_1}\du_{t_2}}^c$, for stationary driving at a constant velocity $v$. One finds
\bea
\label{eq:BFMRetCorr}
C(t_1-t_2) = 2v\int_0^\infty \rmd s \, r(s,w=1) r(s + |t_2-t_1|,w=1).
\eea
For the particularly interesting case of exponential relaxation, 
\bea
f(t) = e^{-t/\tau} \quad\quad\Leftrightarrow \quad\quad f(\omega) = \frac{\tau}{1+i \omega \tau}.
\eea
The Fourier transform \eqref{eq:BFMRetResponse} can be inverted, and the response function is
\bea
\label{eq:BFMRetResponseFinal}
r(t,w) = \frac{w}{2} \left[\left(1-\frac{2-q}{\sqrt{q^2-4 \tau }}\right)
   e^{-\frac{t}{2 \tau
   } \left(q+\sqrt{q^2-4 \tau }\right)}+\left(1+\frac{2-q}{\sqrt{q^2-4 \tau }}\right) e^{-\frac{t}{2 \tau }
   \left(q-\sqrt{q^2-4 \tau }\right)}\right],
\eea
where $q := 1+(1+a)\tau$.
For the two-point function $C(\Delta t := t_2-t_1)$ in \eqref{eq:BFMRetCorr} one finds in the case of exponential relaxation
\bea
\nn
C(\Delta t) =& \frac{v}{2 q \sqrt{q^2-4 \tau }} \left\{\left[(\tau +1) \sqrt{q^2-4 \tau }-q (1-\tau )\right]
   e^{-\frac{|\Delta t|}{2 \tau
   }\left(q+\sqrt{q^2-4 \tau }\right)} \right. \\
	\label{eq:BFMRetCorrFinal}
	& \left. +\left[(\tau +1) \sqrt{q^2-4 \tau }+q (1-\tau )\right]
   e^{-\frac{|\Delta t|}{2 \tau
   } \left(q-\sqrt{q^2-4 \tau }\right)}\right\},
\eea
where as above $q := 1+(1+a)\tau$. The modified response and correlation functions \eqref{eq:BFMRetResponseFinal}, \eqref{eq:BFMRetCorrFinal} are plotted in figure \ref{fig:RetResponseCorr}. One can clearly observe two different slopes on the semi-logarithmic plot, which correspond to the two different exponentials in \eqref{eq:BFMRetResponseFinal}, \eqref{eq:BFMRetCorrFinal}. Comparing these predictions to experiments should provide a simple way to identify the typical retardation time scale, as the time at which the response and correlation functions show a crossover between two different slopes.

\subsection{Stationary velocity distribution\label{sec:BFMRetStat}}
The stationary velocity distribution for the ABBM model with retardation can be obtained from \eqref{eq:BFMRetSolExp} with the sources $\lambda_t = \lambda \delta(t), \mu_t = 0$ (cf. section \ref{sec:BFMABBMStat}). In the two limits $\tau/\tau_m \ll 1$ or $\tau/\tau_m \gg 1$, a boundary-layer analysis of the instanton equations \eqref{eq:RetInstantonExpTu}, \eqref{eq:RetInstantonExpTh} can be performed analytically \cite{DobrinevskiLeDoussalWiese2013}. In both cases, one gets the following result for the stationary distribution of $\du$:
\bea
\label{eq:BFMRetStatVel}
P(\du) = \frac{1}{\Gamma(v/v_m^a)}\frac{1}{\du} \left(\du/v_m^a\right)^{v/v_m^a}e^{-\du/v_m^a},
\eea
i.e. to leading order the shape of the instantaneous velocity distribution is unchanged, up to a modification of the effective velocity scale which is now $v_m^a$ instead of $v_m = \sigma/(\eta m^2)$ in the standard ABBM model. When comparing to experimental data, the microscopic model parameters are not known and typically the velocity scale is obtained by fitting to the measured data. \eqref{eq:BFMRetStatVel} shows that in this case, no difference between the standard ABBM model and the ABBM model with retardation will be observed to leading order. Thus, the stationary velocity distribution is not a good observable for identifying the effects of retardation in experiments. By taking $v\to 0$ in \eqref{eq:BFMRetStatVel}, we also observe that the quasi-static power-law exponent $P(\du)\sim \du^{-1}$ is not modified, showing that the model is still in the same universality class as the standard ABBM model. 
 $v_m^a$ in \eqref{eq:BFMRetStatVel} is given by \cite{DobrinevskiLeDoussalWiese2013}
\bea
v_m^a = 
\begin{cases} 
\frac{\sigma}{\eta(m^2+a)} + \mO(\tau_m/\tau) &\mbox{for} \quad \tau \gg \tau_m \\
\frac{\sigma}{m^2(\eta + a\tau)} + \mO(\tau/\tau_m)^2 &\mbox{for} \quad \tau \ll \tau_m
\end{cases}
\eea
In both cases, $v_m^a = \frac{\sigma}{m^2(\eta+a\tau_f)}$, where $\tau_f$ is the fastest time scale in the system ($\tau_f = \tau$ for $\tau \ll \tau_m$, or $\tau_f = \tau_m = \eta/m^2$ for $\tau_m \ll \tau$).
In the case $\tau \ll \tau_m$ relevant for Barkhausen experiments, the correction from $v_m$ to $v_m^a$ is perturbative, and vanishes as $\tau \to 0$. This leading-order correction corresponds to a rescaling $\eta \to \eta + a \tau$, which is exactly what is obtained if one naively expands the first equation of \eqref{eq:BFMRetEOMExp} to order $\tau$. However, the instanton approach is still useful, providing a basis for a higher-order expansion and allowing to compute more complicated observables \cite{DobrinevskiLeDoussalWiese2013}.

In the case $\tau_m \ll \tau$, the correction from $v_m$ to $v_m^a$ corresponds to  a rescaling of the ``effective'' mass $m^2 \to m^2 + a$. This limit is less relevant for Barkhausen experiments, but shows interesting new physics: the splitting of an avalanche in well-separated subavalanches. This will be discussed further in section \ref{sec:SlowRel}.

\subsection{Avalanche sizes and average shape at fixed size\label{sec:BFMRetSize}}
\begin{figure}%
\centering
         \begin{subfigure}[t]{0.5\textwidth}
                 \centering
                 \includegraphics[width=\textwidth]{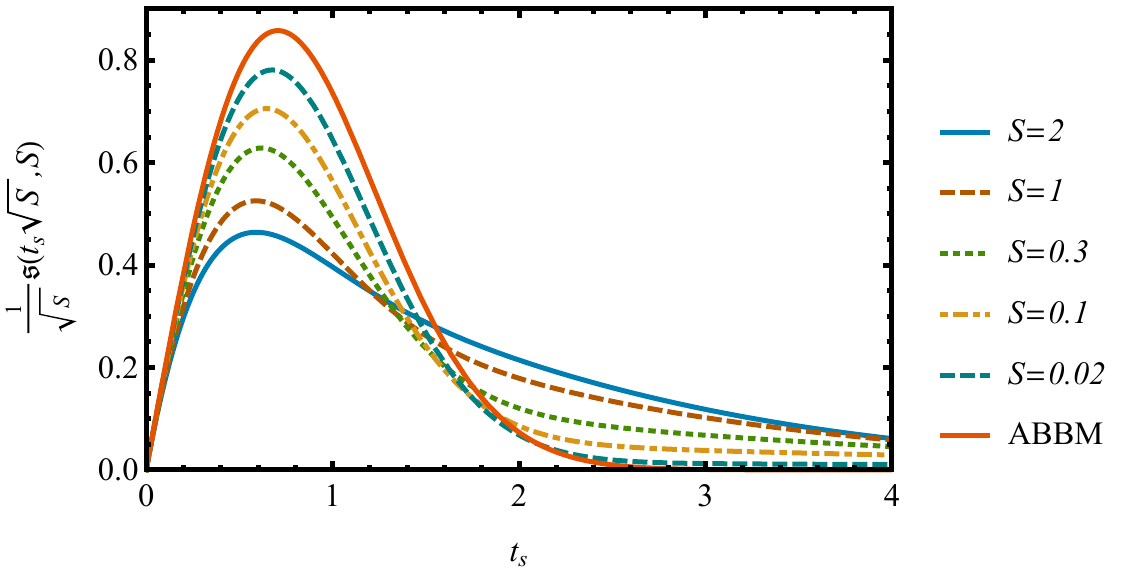}
                 \caption{Rescaled shape for quasi-static avalanches \eqref{eq:BFMRetShpFixedSize2}, for $a=\tau=1$.
. Observe that for small avalanches, $S \to 0$, the rescaled shape tends to the shape of the standard ABBM model, \eqref{eq:BFMRetShapeSmallABBM}. For large $S$, one sees the slower exponential decay for large $t$, as compared to the Gaussian decay of the ABBM model, cf.~figure \ref{fig:rABBMShapeTail}. }
                 \label{fig:RetShapeLargeSmall}
         \end{subfigure}
         \quad 
         \begin{subfigure}[t]{0.43\textwidth}
                 \centering
                 \includegraphics[width=\textwidth]{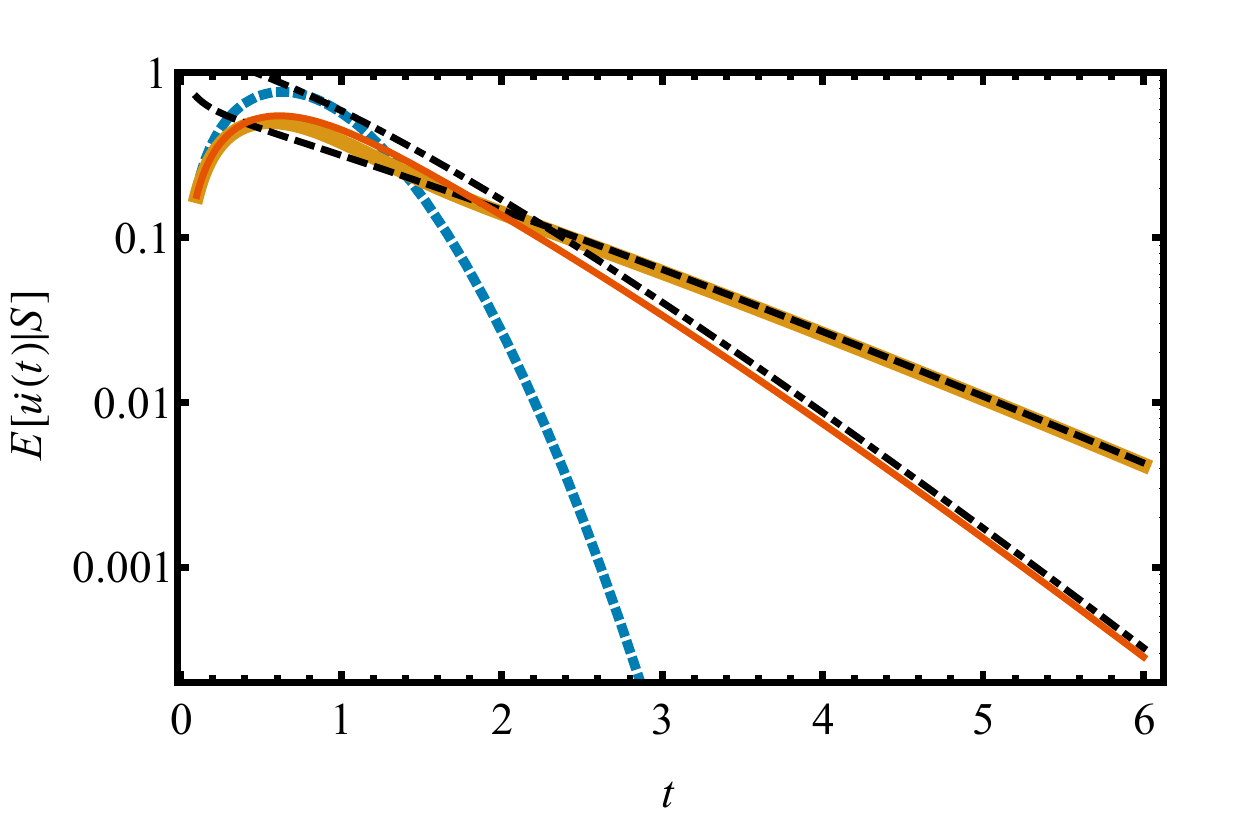}
\caption{Tail of the average avalanche shape, at fixed avalanche size $S=0.8$. Blue dotted line: ABBM model, $a=0$. Solid lines: ABBM model with retardation, \eqref{eq:BFMRetShpFixedSize}, for $\tau=0.4, a=1.$ (thin orange line) and for $\tau=0.7, a=1.$ (thick yellow line). Black dot-dashed and dashed lines: Asymptotics obtained in \cite{DobrinevskiLeDoussalWiese2013}, $\sim e^{-t/\tau}$ for large $t$.}%
\label{fig:rABBMShapeTail}%
         \end{subfigure}%
				\caption{Average avalanche shape, as a function of time, at fixed avalanche size, for the ABBM model with retardation.}
\end{figure}

Let us now consider avalanches following a step of size $w$ in the location of the harmonic well, at $\ti=0$, as in section \ref{sec:BFMAvalanches}. As discussed above in section \ref{sec:BFMRetSol}, the total size of the avalanche $S=\int_0^\infty\rmd t\,\du_t$ is given by the distance of the Middleton states $u(w_i+w)-u(w_i)$, and unchanged by retardation. To see this explicitly, recall from section \ref{sec:BFMSize} that the Laplace transform of the avalanche size distribution is obtained by computing $G[\lambda_t = \lambda,w_t = w\delta(t)]$. For a source $\lambda_t=\lambda$ constant in time, the solution $\tu_t$ of \eqref{eq:BFMRetInstanton} is also constant in time. Using the fact that $f(0)=1$ and $f(\infty)=0$, in this case \eqref{eq:BFMRetInstanton} reduces to \eqref{eq:ABBMInstanton}. Its solution is then still given by \eqref{eq:ABBMSizeInstanton},
\bea
\label{eq:ABBMSizeInstanton2}
\tu_t = \frac{1}{2}\left(1-\sqrt{1-4\lambda}\right).
\eea
This holds for any memory kernel satisfying the properties above.
For the exponential memory kernel discussed in section \ref{sec:BFMRetExp}, we can go further and obtain the average avalanche shape at fixed avalanche size, following the procedure in section \ref{sec:BFMShape}.
The Laplace-transformed shape at fixed size, defined in \eqref{eq:ABBMShapeSizeLT}, is given by the perturbation of the solution \eqref{eq:ABBMSizeInstanton2} of \eqref{eq:BFMRetInstanton} to linear order with an additional source local in time, $\mu\delta(t-\tm)$ ($\tm$ is the time at which the average velocity is observed). This is performed explicitly in \cite{DobrinevskiLeDoussalWiese2013}, section VII B. The result for $\hsh_w$ as defined in \eqref{eq:ABBMShapeSizeLT} is
\bea
\label{eq:BFMRetShpFixedSizeLT}
\hsh_w(\tm,\lambda) = e^{\frac{w}{2}(1-\sqrt{1-4\lambda})} w
\frac{4 a  e^{\frac{2 a \tm}{r}}+r^2 e^{-\frac{r}{2}\tm/\tau}}{4 a \tau+r^2} e^{-\tm/\tau},
\eea
where $r = r(\lambda)$ is any of the two solutions of
\bea
\frac{(2+r)(r-2a\tau)}{2r\tau}=\sqrt{1-4\lambda}.
\eea
For $a=0$, \eqref{eq:BFMRetShpFixedSizeLT} reduces to the standard ABBM result \eqref{eq:ABBMShapeSizeLTRes}.
The shape at fixed size as defined in \eqref{eq:ABBMShapeSizeDef} can be obtained by inverting the Laplace-transformed result \eqref{eq:BFMRetShpFixedSizeLT}, and dividing by the avalanche size distribution $P_w(S)$ in \eqref{eq:BFMPwofS} (note again that it is unchanged by retardation!). The Laplace inversion can be performed using a contour integral  \cite{DobrinevskiLeDoussalWiese2013}:
\bea
\nn
\mathfrak{s}_w(t,S)  =&  \frac{1}{P_w(S)} \int \frac{\rmd \lambda}{2\pi i} \hsh(t,\lambda) e^{-\lambda S} = - \frac{1}{P_w(S)} \int_{r_0 -i\infty}^{r_0 + i\infty} \frac{\rmd r}{2\pi i} \hsh \big(t,\lambda(r)\big) e^{-\lambda(r)S} \frac{\rmd \lambda}{\rmd r} =\\
\nn
= &  \int_{r_0 -i\infty}^{r_0 + i\infty} \frac{\rmd r}{i \sqrt{\pi}}\frac{(r-2 a\tau) (r+2) \left(4 a  \tau e^{\frac{2 a t}{r}} +r^2 e^{-\frac{t r}{2 \tau
   }}\right)}{8 \tau^2 
   r^3 S^{-3/2}} \times \\
\label{eq:BFMRetShpFixedSize}
	& \times \exp \left[\frac{S (r+2)^2(r-2a\tau)^2}{16 r^2\tau^2}-\frac{t}{\tau} + \frac{w^2}{4S}- \frac{w}{2} \frac{(2+r)(r-2a\tau)}{2r\tau} \right].
\eea
$r_0$ fixes the location of the integration contour; it can be chosen arbitrarily, as long as $r_0 \neq 0$ \footnote{This is in order to avoid the singularity of the integrand at $r=0$. We checked that the result of the (numerical) integration of \eqref{eq:BFMRetShpFixedSize} is independent of the value and the sign of $r_0$.}. It was checked in \cite{DobrinevskiLeDoussalWiese2013} that \eqref{eq:BFMRetShpFixedSize} is properly normalized, and reduces to the known result \eqref{eq:ABBMShapeSizeRes} for $a=0$. The integral \eqref{eq:BFMRetShpFixedSize} can be analyzed analytically in various limits \cite{DobrinevskiLeDoussalWiese2013}. In particular, one observes that retardation modifies the Gaussian tail $\mathfrak{s}(t) \sim e^{-t^2/S}$ of the standard ABBM model to an exponential tail, $\mathfrak{s}(t,S) \sim e^{-t/\tau}$ (see \cite{DobrinevskiLeDoussalWiese2013} and figure \ref{fig:rABBMShapeTail}). 

As in section \ref{sec:BFMShape}, a single quasi-static avalanche can be isolated by taking the limit $w \to 0$ in \eqref{eq:BFMRetShpFixedSize},
\bea
\nn
\mathfrak{s}(t,S) = \lim_{w\to 0}\mathfrak{s}_w(t,S) = &\int_{r_0 -i\infty}^{r_0 + i\infty} \frac{\rmd r}{i \sqrt{\pi}}\frac{(r-2 a\tau) (r+2) \left(4 a  \tau e^{\frac{2 a t}{r}} +r^2 e^{-\frac{t r}{2 \tau
   }}\right)}{8 \tau^2 
   r^3 S^{-3/2}} \times \\
\label{eq:BFMRetShpFixedSize2}
	& \quad\quad \times \exp \left[\frac{S (r+2)^2(r-2a\tau)^2}{16 r^2\tau^2}-\frac{t}{\tau}\right].
\eea
It is now interesting to consider the limit of small avalanches, $S \to 0$. To leading order, the shape $\mfs(t,S)$ becomes a function of the scaling variable $t_s := t / \sqrt{S}$. Inserting this and $r =: p \sqrt{S}$ in \eqref{eq:BFMRetShpFixedSize2}, we obtain
\bea
\nn
\mathfrak{s}(t,S)  =&  \sqrt{S}\, h_S(t_s=t/\sqrt{S}) + \mO(S),\\
h_S(t_s) := & 
  \int_{p_0 -i\infty}^{p_0 + i\infty} \frac{\rmd p}{i \sqrt{\pi}}\frac{(-2 a\tau) 2 \left(4 a  \tau e^{\frac{2 a t_s}{p}} \right)}{8 \tau^2 
   p^3 }  \exp \left[\frac{2^2(-2a\tau)^2}{16 p^2\tau^2}\right],
\eea
where we dropped all terms subleading in $S$. Simplifying the expression and performing the integral over $p$ via the substitution $p=1/q$, one finds
\bea
\label{eq:BFMRetShapeSmallABBM}
h_S(t_s) = 2 t_s\,e^{-t_s^2},
\eea
which is the same result as found for the standard ABBM model \eqref{eq:ABBMShapeSizeResSmallW}. This convergence to the ABBM result is confirmed by inverting \eqref{eq:BFMRetShpFixedSize2} numerically for small $S$, cf.~figure \ref{fig:RetShapeLargeSmall}. This fact, and the results from the preceding section \ref{sec:BFMRetStat}, show that universal features of the ABBM model with retardation and those the standard ABBM model coincide: They have the same critical exponents for the distributions of velocities, sizes, etc. of quasi-static avalanches, and the same average shape of short avalanches. On the other hand, the scaling exponents at finite driving velocity (as discussed for $P(\du)$ in the previous section \ref{sec:BFMRetStat}, and for $P(S)$, $P(T)$ in \cite{DobrinevskiLeDoussalWiese2013}), and the shape for long avalanches, differ. These observables are less universal and more sensitive to the details of the model; they allow to obtain more information on the microscopical physics of domain-wall motion.

The predictions of \eqref{eq:BFMRetShpFixedSize} can be compared to experimental results. Preliminary fits of \eqref{eq:BFMRetShpFixedSize} to Barkhausen noise measurements from \cite{PapanikolaouBohnSommerDurinZapperiSethna2011} indicate that including exponential retardation with the additional parameters $a$ and $\tau$ leads to better agreement of average avalanche shapes with measurements, than in the pure ABBM model \cite{Cheninpr}. In the future, it would be interesting to continue this comparison and see if the fit can be made sufficiently precise to distinguish the ABBM model with retardation from other models with additional degrees of freedom (such as the inertial model discussed in \cite{LeDoussalPetkovicWiese2012}). It would also be interesting to see if the form of the retarded memory kernel $f$ in \eqref{eq:BFMRetEOMU} can be determined from experiments, for example by comparing the two-times velocity correlation function to the calculation in \eqref{sec:BFMRetRespCorr}. 

\subsection{Slow-Relaxation limit: Aftershocks and subavalanches\label{sec:SlowRel}}

In this limit, the memory term decreases much slower than the domain wall moves, $\tau \gg \tau_m$. As we already saw in section \ref{sec:BFMRetStat}, stationary velocities show a rescaling of the ``effective'' mass $m^2 \to m^2 + a$. This can be understood qualitatively from the original equation of motion \eqref{eq:BFMRetEOMExp}, for any kind of disorder (not just the Brownian disorder of the ABBM model with retardation) \cite{DobrinevskiLeDoussalWiese2013}. Let us, as usual, start from a ``Middleton'' metastable state at $t=0$. On time scales $t \approx \tau_m \ll \tau$, one can neglect the $h$ on the right-hand-side of the second equation in \eqref{eq:BFMRetEOMExp}, and obtains
\bea
\label{eq:BFMRetSlowRelH}
\tau h(t) = \int_0^t \du(t_1)\rmd t_1 + \mO(t/\tau) = u(t)-u(0) + \mO(t/\tau)
\eea
Inserting this into the first equation of \eqref{eq:BFMRetEOMExp}, we obtain 
\bea
\eta \du(t) = F\big(u(t)\big) + m^2 \left[w(t) - u(t)\right] - a \left[u(t)-u(0)\right]+....
\eea
Combining the terms linear in $u$, we see that the mass is modified from $m^2 \to m^2+a$.
It is less trivial to see that the driving velocity $\dw = v$ entering the stationary velocity distribution \eqref{eq:BFMRetStatVel} is also rescaled by $v_m^a = \sigma/(\eta(m^2+a))$ instead of $v_m = \sigma/(\eta m^2)$; for this, the full boundary-layer calculation in \cite{DobrinevskiLeDoussalWiese2013} is necessary.

\begin{figure}
         \centering
         \begin{subfigure}[t]{0.45\textwidth}
                 \centering
                 \includegraphics[width=\textwidth]{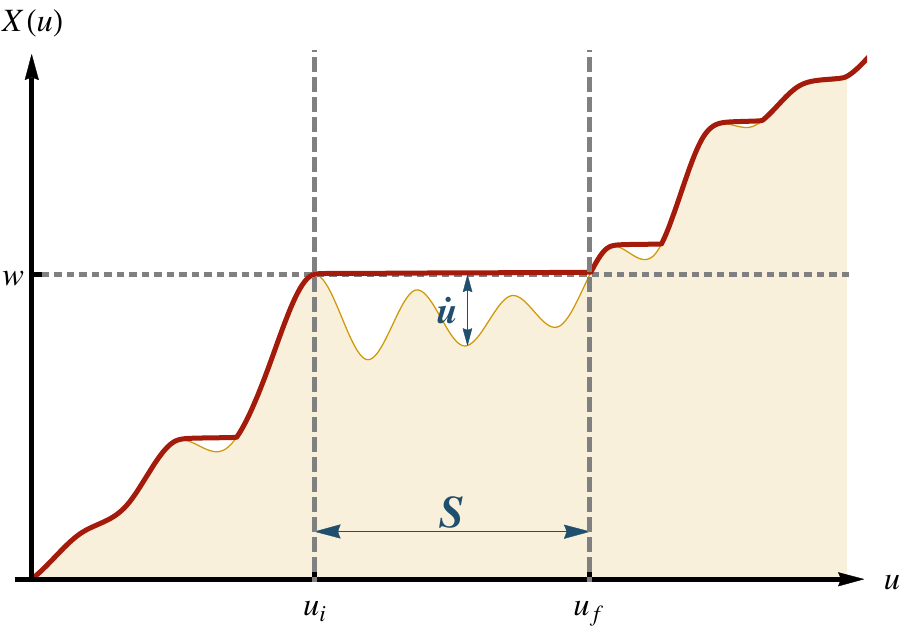}
                 \caption{Avalanches in the standard driven particle model with first-order dynamics, \eqref{eq:BFMABBM}}
                 \label{fig:BFMRetQSAvStd}
         \end{subfigure}%
         ~ 
         \begin{subfigure}[t]{0.45\textwidth}
                 \centering
                 \includegraphics[width=\textwidth]{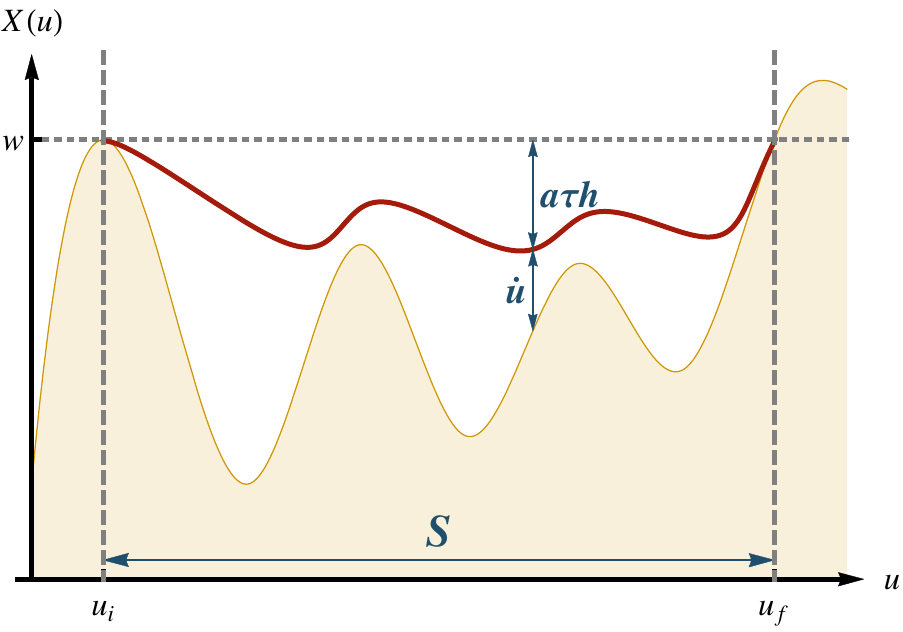}
                 \caption{Avalanches for a driven particle with retardation, \eqref{eq:BFMRetEOMU}, for $a=1, \tau/\tau_m = 1$.}
                 \label{fig:BFMRetQSAvTau1}
         \end{subfigure}
				\\
         \begin{subfigure}[t]{0.45\textwidth}
                 \centering
                 \includegraphics[width=\textwidth]{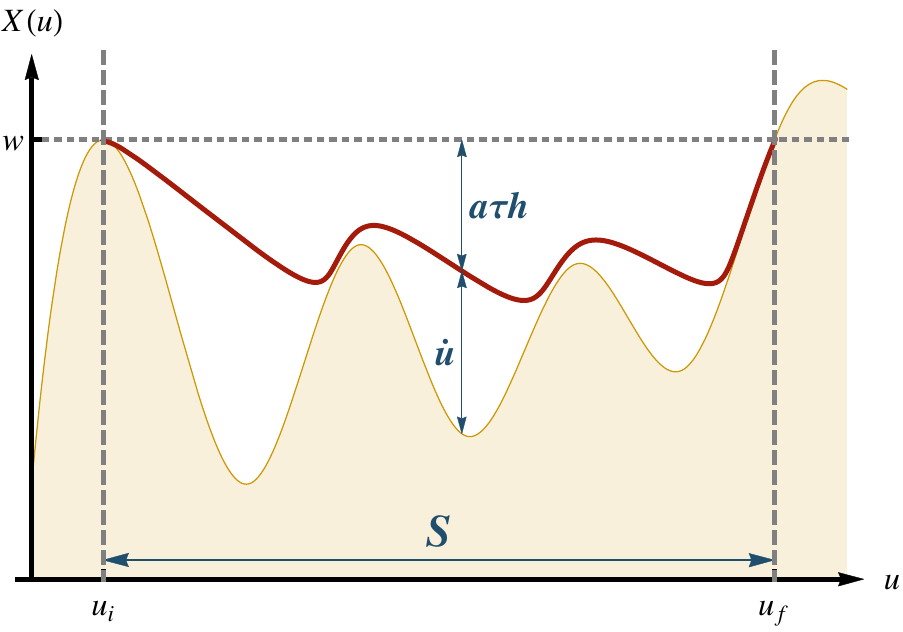}
                 \caption{Avalanches for a driven particle with retardation, \eqref{eq:BFMRetEOMU}, for $a=1, \tau/\tau_m = 3$.}
                 \label{fig:BFMRetQSAvTau3}
         \end{subfigure}
				~
         \begin{subfigure}[t]{0.45\textwidth}
                 \centering
                 \includegraphics[width=\textwidth]{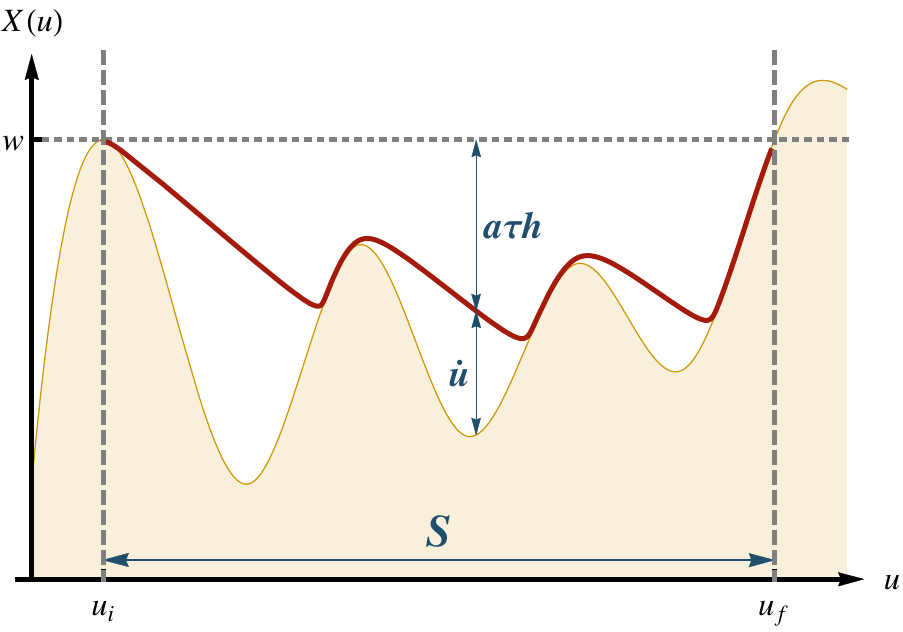}
                 \caption{Avalanches for a driven particle with retardation, \eqref{eq:BFMRetEOMU}, for $a=1, \tau/\tau_m = 10$.}
                 \label{fig:BFMRetQSAvTau10}
         \end{subfigure}
         \caption{Construction of quasi-static avalanches and subavalanches without retardation and with retardation, for a microscopically smooth random force landscape $F(u)$. The Brownian Force Landscape \eqref{eq:DisorderABBM} is microscopically rough; it has local maxima on all scales and thus an arbitrarily small step in $w$ will always trigger an avalanche }\label{fig:BFMRetQSAv}
\end{figure}

This modification of the effective mass also impacts the construction of quasi-static avalanches \cite{DobrinevskiLeDoussalWiese2013}. Again, let us take any disorder $F(u)$. In the standard driven particle model \eqref{eq:BFMABBM}, the quasi-static position $u(w)$ is obtained as the first passage of the tilted landscape $X(u):=m^2 u - F(u)$, at level $X(u) = m^2 w$. When $w$ passes over a local maximum of the landscape, this triggers an avalanche going from $u_i$ (the location of the maximum) to $u_f$, the next passage of $X(u)$ at the same level. Its size is $S=u_f-u_i$. During the avalanche, $w$ does not change and hence by \eqref{eq:BFMABBM} the velocity $\du(u)$ is given by $m^2 w - X(u)$. This is shown graphically in figure \ref{fig:BFMRetQSAvStd}.

In the model with retardation, \eqref{eq:BFMRetEOMU}, the velocity $\du(u)$ is decreased by the additional memory term $a \tau h(t)$, which by \eqref{eq:BFMRetSlowRelH} initially grows linearly $\sim a u$ (assuming $\tau \gg \tau_m$). Thus, initially the particle moves as if the landscape was tilted by an additional slope $-a u$ (corresponding to the change of the effective mass discussed above). At some point $u_1$ in the interval $]u_i;u_f[$, the velocity $\du(u_1)$ becomes essentially zero, since the ``additional tilt'' line with slope $-a u$  (corresponding to the energy stored in the memory term), starting from $m^2 w$ at $u_i$, hits $X(u)$. This constitutes the first \textit{subavalanche}. Then, the relaxation of the memory term leads to slow motion (with $\du \sim 1/\tau$) along the potential, as long as the slope $X'(u) > -a$. If there are no more local maxima of $X(u)$ between $u_1$ and $u_f$, this slow motion proceeds until a stable state is reached with $h = 0$ and $\du =0$ at $u=u_f$. On the other hand, if $X(u)$ has additional local maxima in that interval, additional sub-avalanches (starting from a finite value of $h$) will be triggered when the potential slope becomes $X'(u) < -a$. They consist, just as the first subavalanche, in short periods of rapid motion $\du \sim 1$ with an effective mass $a+m^2$. During each subavalanche, $h(u)$ grows linearly $\sim a u$, and each subavalanche terminates when $\du(u) \sim 1/\tau$, i.e. when the slope $-a u$, starting from the initial point of the subavalanche, hits $X(u)$. This is visualized graphically in figure \ref{fig:BFMRetQSAvTau10}. The intermediate figures  \ref{fig:BFMRetQSAvTau1} and  \ref{fig:BFMRetQSAvTau3} show how the subavalanche structure arises and becomes sharper when $\tau$ is increased from $\tau \approx 0$ (standard driven particle) to $\tau \gg \tau_m$. 

In agreement with the Middleton theorem, and the discussion in section \ref{sec:BFMRetSize}, we see that the total avalanche size $S=u(t=\infty)-u(t=0) = u_f -u_i$ is the same as in the standard ABBM model. However, the avalanche is split in a series of sub-avalanches, which are sharply delimited in the limit $\tau \to \infty$. Between them there are long periods $\sim \tau$ of slow motion $\du \sim 1/\tau$. The number of sub-avalanches is random, and depends on the properties of the potential landscape. In the Brownian Force landscape \eqref{eq:DisorderABBM} considered in the ABBM model, there are local maxima on all scales and hence there are infinitely many subavalanches. Each has a finite size $S_i \sim S_m$ and a finite duration $T_i \sim \tau_m$. The total time it takes the particle to reach the final point $u_f$ is infinite\footnote{At least for the case of exponential relaxation, or any other memory kernel which ranges infinitely far in time.}. As indicated above, this qualitatively new effect of slow retardation -- the splitting of one avalanche in several subavalanches -- appears for any random force landscape $F(u)$.

This response to external loading through a series of correlated sub-avalanches resembles aftershocks observed in earthquake dynamics (see section \ref{sec:Earthquakes} and \cite{Omori1894,UtsuEtAl1995,Scholz2002}). In the present model, the typical subavalanche sizes become smaller from one subavalanche to the next one. However, there are fluctuations and in some cases the first sub-avalanche will not be the largest one. This is again similar to earthquake aftershocks, where the main (largest) shock is not necessarily the first one, still typically the aftershock activity decays. The celebrated Omori-Utsu law \cite{Omori1894,UtsuEtAl1995,Scholz2002} states that the average decay of aftershock activity (measured e.g. by the mean rate of events $n(t)$) is a power law with an exponent close to 1:
\bea
n(t) = \frac{K}{(c+t)^p},\quad\quad\quad p\approx 0.7-1.5
\eea
Thus, attempting a description of earthquake aftershocks using the ABBM model with retardation would require understanding \eqref{eq:BFMRetEOMU} with a memory kernel $f$ decaying in time as a power-law of this form. This is certainly an interesting question, and it is known that some aftershocks can indeed be attributed to geophysical relaxation processes. However it is not entirely clear which of the various (and physically very different) mechanisms for triggering aftershocks\footnote{\cite{Freed2005} lists, among others, viscous relaxation, poroelastic rebound, afterslip, nonlinearities in constitutive laws for friction, and seismic waves. See also section \ref{sec:Earthquakes}.} would give rise to a memory term of the special form considered in \eqref{eq:BFMRetEOMU} (in particular, linear in $\du$). It is also not clear, if this model \eqref{eq:BFMRetEOMU} is sufficiently detailed to predict any additional information on aftershock correlations or distributions, beyond what is put in (see also the discussion in section \ref{sec:Earthquakes}).

\section{Non-monotonous motion\label{sec:BFMNonMon}}
In the previous sections we considered avalanches under not necessarily stationary, but always monotonous driving. Especially for magnetic domain walls, it is also interesting to consider non-monotonous driving, leading to hysteresis phenomena \cite{Bertotti1998,DurinZapperi2006b}. To some extent, the appearance of hysteresis can be explained by the simple model of a particle driven on a rough landscape \eqref{eq:BFMABBM} (see \cite{Neel1942,MagniBeatriceDurinBertotti1999,DurinZapperi2006b}). When moving monotonously in one direction, the particle position $u(t)$ (corresponding to the total magnetization of the sample) typically lags behind the position of the driving spring $w(t)$. When the driving direction is reversed, the particle moves less for a certain time, until the spring starts pulling backward sufficiently strongly. Then it moves again in a monotonous fashion, but lagging in the other direction. This yields a hysteresis loop such as shown in figure \ref{fig:IntroHystLoopSizes}.

For the simple case of the ABBM model \eqref{eq:BFMABBM}, a particle in a Brownian landscape \eqref{eq:DisorderABBM}, the hysteresis loop can be computed exactly for the case of quasi-static driving. I will explain this calculation in the following section \ref{sec:HystABBMQS}. The average shape of the hysteresis loop of the ABBM model has already been obtained in \cite{MagniBeatriceDurinBertotti1999,BertottiMayergoyzBassoMagni1999}. Here I extend this and give joint probability distributions of points on the hysteresis curve, in particular the exact avalanche size distribution at any distance from the turning point. 
Lastly, in section \ref{sec:HystFiniteV} I briefly discuss the more general case of non-monotonous motion at finite frequency.

\subsection{Quasi-static hysteresis loop of the ABBM model\label{sec:HystABBMQS}}
Let us consider a particle at position $u$ in a Brownian force landscape $F(u)$, moving according to \eqref{eq:BFMABBM}. Since here we will be interested in the actual particle position (and not just position differences), it does not suffice to fix the increments of $F$ as in \eqref{eq:DisorderABBM}. We choose $F(u)$ to be a random walk starting at $F(u=0)=0$, so that $F(u)$ is Gaussian with correlations $\overline{F(u)F(u')} = 2 \sigma \min u,u'$.\footnote{Note this means that $F$ is nonstationary. It nevertheless has independent increments, $\partial_u F(u) = \xi(u)$ where $\xi(u)$ is white noise. See also section \ref{sec:BFMPositionTheory}.} 
For the entire section \ref{sec:HystABBMQS} we will use dimensionless units, $m=\sigma=1$.

Let us first increase $w$ very slowly starting from $u=0, F(0)=0$ and $w=0$. On this \textit{forward branch} of the hysteresis loop, the quasi-static $u(w)$ will be the leftmost metastable state, i.e. the smallest $u$ for which $0 = F(u) - (u-w)$. In other words, on the forward branch $u(w)$ is the first-passage time of the Brownian Motion with drift $X(u) := u - F(u)$ at level $X=w$. The distribution of $u(w)$ is just the quasi-static avalanche size distribution \eqref{eq:BFMPwofS} (see also \cite{DobrinevskiLeDoussalWiese2012})
\beq
\label{eq:Hyst1PtForward}
P(u_w) = \frac{1}{\sqrt{4\pi}u_w^{3/2}}\exp\left[-\frac{(u_w-w)^2}{4u_w}\right] \Leftrightarrow \overline{e^{\lambda u_w}} = \exp\left[\frac{w}{2}\left(1-\sqrt{1-4\lambda}\right)\right].
\eeq
This has mean $\overline{u_w} = w$, a special property of the nonstationary Brownian.
Now let us consider the backward branch of the hysteresis loop. 

\subsubsection{Single-point probability density on backward branch \label{sec:HystABBMQS1Pt}}
Let us move the harmonic well up to position $w_m$, and then return it to a position $0 < w_1 < w_m$ (which ensures that $u(w_1)>0$ and we stay inside the range of definition of the Brownian). 
In this section we want to compute $P(u(w_1))$. 

\begin{figure}
         \centering
         \begin{subfigure}[t]{0.45\textwidth}
                 \centering
                 \includegraphics[width=\textwidth]{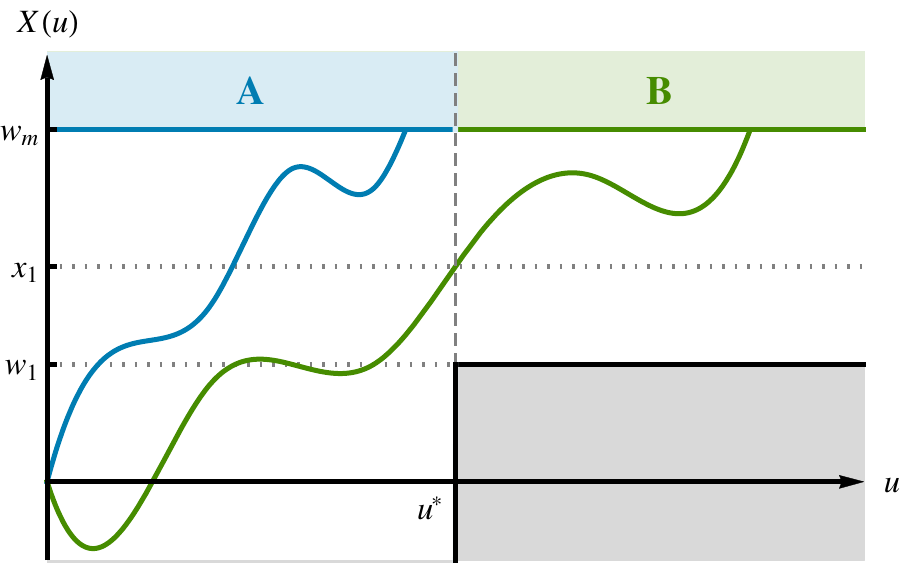}
                 \caption{Contributions $F^{(1)}_A(u^*)$ and $F^{(1)}_B(u^*)$ to the cumulative distribution of a single point on the backward branch of the quasi-static ABBM hysteresis loop. See text in section \eqref{sec:HystABBMQS1Pt}.}
                 \label{fig:HystABBM1Pt}
         \end{subfigure}%
         ~ 
         \begin{subfigure}[t]{0.45\textwidth}
                 \centering
                 \includegraphics[width=\textwidth]{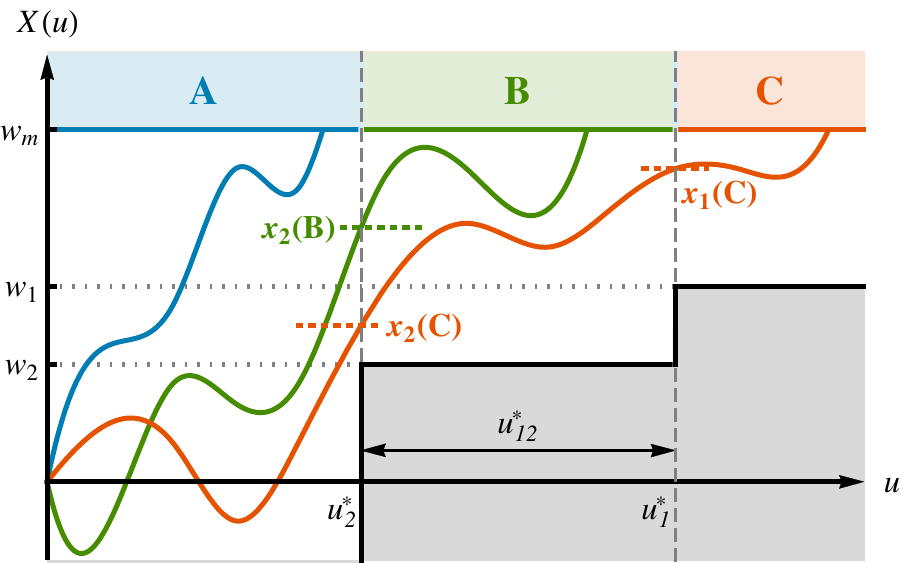}
                 \caption{Contributions $F^{(2)}_A(u_2^*)$,$F^{(2)}_B(u_2^*,u_{12}^*)$ and $F^{(2)}_C(u_2^*,u_{12}^*)$ to the cumulative distribution of two points on the backward branch of the quasi-static ABBM hysteresis loop. See text in section \eqref{sec:HystABBMQS2Pt}.}
                 \label{fig:HystABBM2Pt}
         \end{subfigure}
         \caption{Contributions to cumulative distributions of one and two points on the backward branch of the quasi-static ABBM hysteresis loop. }\label{fig:HystABBMPlots}
\end{figure}

As discussed above, at the turning point $w_m$ $u$ is given by the first passage time of $X(u) := u - F(u)$ at level $X=w_m$. This means that $u(w_1)$ is the \textit{last} passage time of $X(u)$ at level $w_1$, before reaching level $w_m$.
We thus have $u_1:=u(w_1) \leq u^*$ if and only if one of the following conditions is satisfied (see figure  \ref{fig:HystABBM1Pt}):
\begin{itemize}
	\item The first passage of $X(u)$ at $w_m$ is for $0 \leq u \leq u^*$. Let us call the probability of this event $F^{(1)}_A(u^*)$.
	\item From $0 \leq u \leq u^*$, $X(u) < w_m$, $w_m \geq X(u^*) \geq w_1$, and for $ u^* \leq u$, $X(u)$ touches $w_m$ before $w_1$. Let us call this contribution $F^{(1)}_B(u^*)$.
\end{itemize}
The cumulative distribution of $u_1$ is then
\bea
\label{eq:Hyst1PtCum}
F(u^*) := P(u_1 \leq u^*) = F^{(1)}_A(u^*) + F^{(1)}_B(u^*) .
\eea

To compute $F_A$, $F_B$, we will need the following propagators\footnote{As usual, by the propagator $\mathcal{P}_X (x_f,u_f | x_i,u_i)$ we mean the probability density of the final position $x_f$ of the stochastic process $X$ at time $u_f$, given that $X$ started at position $x_i$ at time $u_i < u_f$.}:
\bea
\label{eq:HystPropBM}
\mathcal{P}_X (x_f,u_f | x_i,u_i) = & \frac{1}{\sqrt{4\pi (u_f-u_i)}}\exp\left[-\frac{\left(x_f-x_i-(u_f-u_i)\right)^2}{4(u_f-u_i)}\right], \\
\label{eq:HystPropBM1b}
\mathcal{P}_{X < a} (x_f,u_f | x_i,u_i) = & \mathcal{P}_X (x_f,u_f | x_i,u_i) - e^{a-x_i} \mathcal{P}_X (x_f,u_f | 2a-x_i,u_i), \\
\nn
\mathcal{P}_{a_2 < X < a_1} (x_f,u_f | x_i,u_i) = & \sum_{n=-\infty}^\infty \left[e^{n(a_1-a_2)}\mathcal{P}_X (x_f,u_f | x_i+n(a_1-a_2),u_i) \right. \\
\label{eq:HystPropBM2b}
& \quad \left.- e^{n(a_1-a_2)+a_1-x_i} \mathcal{P}_X (x_f,u_f | 2n(a_1-a_2)+2a_1-x_i,u_i)\right].
\eea
The first line is just the standard propagator of a Brownian motion with drift. 
The last two lines are obtained using the method of images and the Bachelier-Levy formula (see e.g. \cite{Lerche1986} and the references therein). For $\mathcal{P}_{X < a}$, we suppose both $x_i$ and $x_f$ are less than $a$. For $\mathcal{P}_{a_2 < X < a_1}$, we suppose both $x_i$ and $x_f$ are between $a_2$ and $a_1$.

The probability of being absorbed at $w_m$ before $u^*$ is obtained by integrating the current $J_X$ of $\mathcal{P}_{X<w_m}$ at $x_f=w_m$, from $u=0$ to $u^*$:
\bea
\nn
F_A^{(1)}(u^*) =& \int_0^{u^*} \rmd u\, J_{X<w_m}(x_f= w_m, u|0,0) = \int_0^{u^*} \rmd u \left[-\partial_{x_f}\big|_{x_f=w_m}\mathcal{P}_{X < w_m} (x_f,u | 0,0)\right] \\
\label{eq:HystF1A}
=& \frac{1}{2} \left[\text{erf}\left(\frac{u^*-w_m}{2
   \sqrt{u^*}}\right)+e^{w_m}
   \text{erfc}\left(\frac{u^*+w_m}{2 \sqrt{u^*}}\right)+1\right].
\eea

The probability of being absorbed at $w_m$ before touching $w_1$, starting from $x_1$ at $u^*$, is obtained from $\mathcal{P}_{w_1 < X < w_m}$. We first compute the probability current of $\mathcal{P}_{w_1 < X < w_m}$ at $X=w_m$:
\bea
\nn
&J_{w_1 < X < w_m}(x_2=w_m,u^*+u_m) = -\partial_{x_2}\big|_{x_2 = w_m} \mathcal{P}_{w_1 < X < w_m} (x_2,u^*+u_m | x_1,u^*) = \\
\nn
&= \sum_{n=-\infty}^\infty \frac{1}{4\sqrt{\pi}u_m^{3/2}}e^{n (w_m-w_1)} \left\{e^{-\frac{[2 n (w_m-w_1)+u_m+w_m-x_1]^2}{4 u_m}+w_m-x_1} [2
   n (w_m- w_1)+u_m+w_m-x_1] \right. \\
\nn
	& \quad \quad \quad\quad\quad\quad\quad\quad \quad\quad\quad\quad
	\left.-e^{-\frac{[2 n(
   w_m- w_1)+u_m-w_m+x_1]^2}{4 u_m}} [2 n
   (w_m- w_1)+u_m-w_m+x_1]\right\}
\eea
Integrating this over the hitting time $u_m$ of level $w_m$ and performing the sum over $n$, we obtain the probability to hit $w_m$ at any time before $w_1$, starting from $x_1$:
\beq
\label{eq:HystBMSplit}
P_{hit}(w_m \, \text{before}\, w_1 | x_1, u^*) = \int_{0}^\infty \rmd u_f\,J_{w_1 < X < w_m}(x_2=w_m,u^*+u_f) = \frac{1-e^{w_1-x_1}}{1-e^{w_1-w_m}}
\eeq
This is the well-known \textit{splitting probability} of a Brownian with drift, between two boundaries \cite{Gardiner}. Through it, we can express the contribution B as
\bea
\label{eq:HystF1B}
F^{(1)}_B(u^*) =& \int_{w_1}^{w_m} \rmd x_1\, \mathcal{P}_{X < w_m} (x_1,u^* | 0,0) P_{hit}(w_m \, \text{before}\,w_1 | x_1, u^*)
\eea
So, finally we have
\bea
\nn
&P(u(w_1) \leq u^* ) = F_A^{(1)}(u^*) + F_B^{(1)}(u^*) \\
\nn
& \quad	= 1-{\textstyle\frac{e^{w_m} \left[\text{erfc}\left(\frac{u^*-w_1}{2
   \sqrt{u^*}}\right)+e^{w_1} \text{erfc}\left(\frac{u^*+w_1}{2
   \sqrt{u^*}}\right)\right]-e^{w_1} \text{erfc}\left(\frac{u^*+w_1-2
   w_m}{2 \sqrt{u^*}}\right)-e^{2 w_m}
   \text{erfc}\left(\frac{u^*-w_1+2 w_m}{2 \sqrt{u^*}}\right)}{2
   \left(e^{w_m}-e^{w_1}\right)}}.
\eea
Taking a derivative with respect to $u^*$, we obtain the distribution of $u_1:=u(w_1)$, on the backward branch:
\bea
\label{eq:NonMonSingleDist}
P(u_1) =& \partial_{u^*}\big|_{u^* = u_1} P(u(w_1) \leq u^* ) 
= \frac{e^{-\frac{(u_1-w_1)^2}{4
   u_1}}-e^{w_m-\frac{(u_1-w_1+2 w_m)^2}{4
   u_1}}}{2 \sqrt{\pi u_1 }
   \left(1-e^{w_1-w_m}\right)}.
\eea
Its Laplace transform is given by
\bea
\label{eq:Hyst1Pt}
\overline{\exp\left[ \lambda u(w_1) \right]} = \frac{\exp \left[\frac{w_m}{2}\left(1-\sqrt{1-4\lambda}\right)\right]}{\sqrt{1-4\lambda}}\frac{\sinh \frac{w_m-w_1}{2}\sqrt{1-4\lambda}}{\sinh \frac{w_m-w_1}{2}}.
\eea
This allows to determine easily the average shape of the backward branch,
\bea
\label{eq:Hyst1PtBackwardExact}
\overline{u(w_1)} = \partial_{\lambda}\big|_{\lambda=0}\overline{\exp\left[ \lambda u(w_1)\right]} = 2+w_1 + \frac{2(w_1-w_m)}{e^{w_m-w_1}-1}.
\eea
For $w_1 \ll w_m$ we have a constant value of the mean pinning force
\bea
\overline{u(w_1)-w_1} = 2,
\eea
which is also the result of a perturbative one-loop calculation, see appendix \ref{sec:AppABBMHystPert}. For $w_1 \approx w_m$, i.e. near the turning point, the average pinning force decays exponentially to its asymptotic value $2$.

\begin{figure}%
\centering
\begin{minipage}[c]{0.5\textwidth}
\includegraphics[width=\columnwidth]{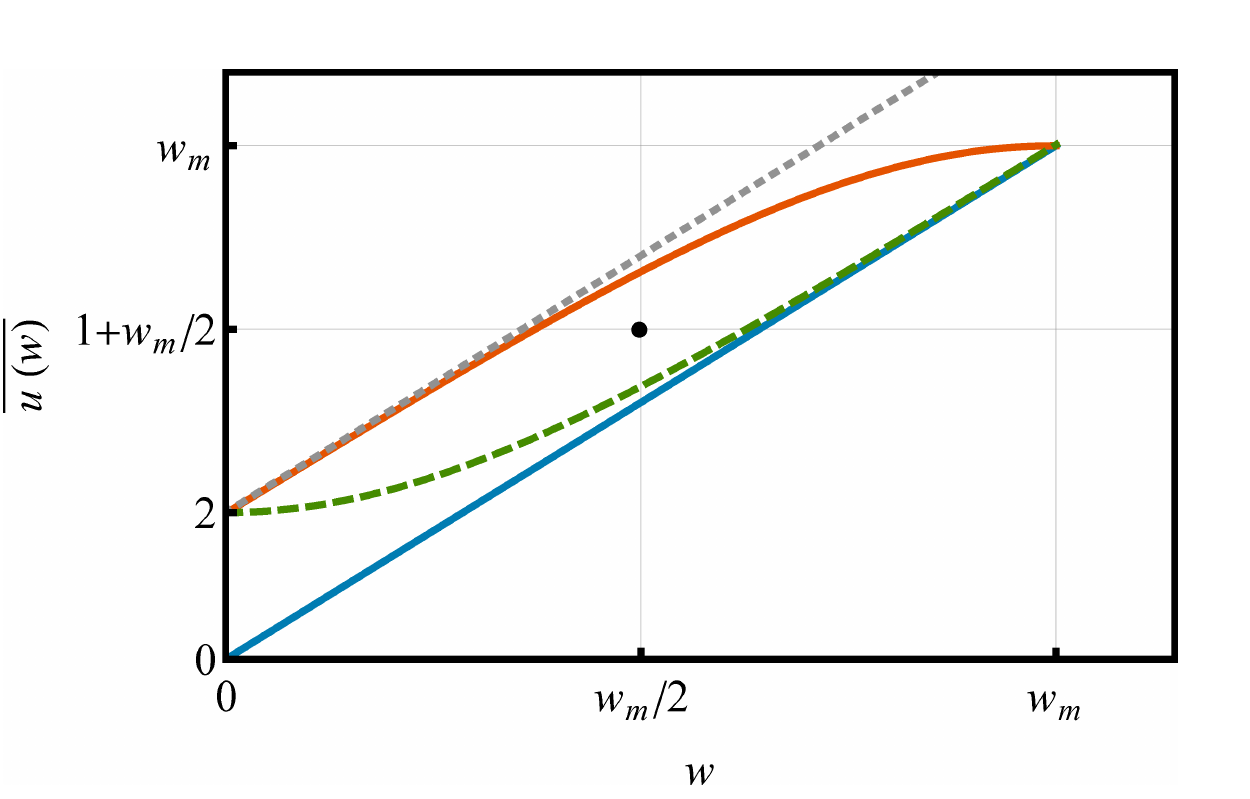}%
\end{minipage}~
\begin{minipage}[c]{0.45\textwidth}
\caption{Average hysteresis loop for a particle in the (nonstationary) Brownian force landscape. Blue line: Forward branch of the hysteresis loop, \eqref{eq:Hyst1PtForward}, $\overline{u(w)}=w$. Orange line: Backward branch of the hysteresis loop, \eqref{eq:Hyst1PtBackwardExact}, after turning at $w_m=7$. Dotted gray line: asymptotics of backward branch, $\overline{u(w)}=w+2$. Dashed green line: ``symmetrized'' forward branch after moving up to $w_m$, then down to $w=0$, and back up. Dot: Center of hysteresis loop $\left(w=\frac{w_m}{2},u=\frac{w_m}{2} + 1\right)$ (see text).\label{fig:ABBMHysteresisAvg}}
\end{minipage}
\end{figure}

If the harmonic confinement is moved up from $w=0$ to $w_m \gg 1$, then down from $w_m$ to $w=0$, and then up again, we expect to see a ``symmetrized'' version where the pinning force decays exponentially from its value $\approx 2$ at $w=0$ (since we are on the backward branch) to $0$ for large $w$. This will give a symmetric hysteresis loop, centered around $\left(\frac{w_m}{2},\frac{w_m}{2} + 1\right)$ (cf.~figure \ref{fig:ABBMHysteresisAvg}). This off-center position is due to our choice of a non-stationary Brownian for $F(u)$, whose variance increses with $u$. In a more realistic situation, which is still microscopically a Brownian (e.g. periodized on a large scale, or $F(u)$ an Ornstein-Uhlenbeck process with a large but finite correlation length), the center of the hysteresis loop will be shifted back to the origin but we expect the shape to remain the same.

Finally, we can consider the stationary regime of the backward branch, when the turning point $w_m \to \infty$. The distribution \eqref{eq:NonMonSingleDist} then simplifies to:
\beq
\label{eq:HystBMLPT}
P(u_1) = \frac{1}{2 \sqrt{\pi u_1}}\exp\left[-\frac{\left(u_1-w_1\right)^2}{4 u_1}\right].
\eeq
In this limit, $u_1=u(w_1)$ is just the last passage time of the Brownian with drift $X$ at level $w_1$, without any further constraints (since $w_m \to \infty$). \eqref{eq:HystBMLPT} then coincides with known results \cite{Durrett1996,ChoiNam2003}.

\subsubsection{Two-point probability density on the backward branch\label{sec:HystABBMQS2Pt}}
Now let us consider the joint distribution $P(u(w_1),u(w_2))$ on the backward branch, i.e. for $0<w_2<w_1<w_m$ after turning at $w_m$.
Let us set $u_1 := u(w_1)$, $u_2 := u(w_2)$. By monotonicity we have $u_1 \geq u_2$, hence we will set $u_1 = u_2 + u_{12}$ and determine the distributions of $u_2, u_{12}$.

There are three contributions to $P(u_2\geq u_2^*, u_{12} \geq u_{12}^*)$, with $u_2^* \leq u_1^*$ (see figure  \ref{fig:HystABBM2Pt}):
\begin{enumerate}
	\item When $w_m$ is hit before $u_2^*$. The probability of this event is $F^{(2)}_A(u_2^*) = F^{(1)}_A(u_2^*) $ derived in \eqref{eq:HystF1A}.
	\item When $w_m$ is hit at some $u_2^* \leq u \leq u_1^*$, without hitting $w_2$ after $u_2^*$. Let us denote the probability of this event by $F^{(2)}_B(u_2^*, u_{12}^*)$.
	\item When $w_m$ is hit after $u_1^*$, without hitting $w_2$ after $u_2^*$ and without hitting $w_1$ after $u_1^*$. Let us denote the probability of this event by $F^{(2)}_C(u_2^*, u_{12}^*)$.
\end{enumerate}
The cumulative distribution of $u_2, u_{12}$ is then
\bea
\label{eq:Hyst2PtCum}
F^{(2)}(u_2^*,u_{12}^*) := P(u_2\geq u_2^*, u_{12} \geq u_{12}^*) = F^{(2)}_A(u_2^*) + F^{(2)}_B(u_2^*,u_{12}^*) +F^{(2)}_C(u_2^*,u_{12}^*).
\eea

Each of these contributions can be computed as above. We have:
\bea
\nn
F^{(2)}_A(u_2^*) = & \int_0^{u_2^*} \rmd u_m\, J_{X < w_m} (x = w_m,u_m|0,0), \\
\nn
F^{(2)}_B(u_2^*,u_{12}^*) = & \int_{w_2}^{w_m}\rmd x_2 \mathcal{P}_{X<w_m}(x_2,u_2^*|0,0)\int_0^{u_{12}^*} \rmd u_m\, J_{w_2<X < w_m} (x = w_m,u_2^* + u_m|x_2,u_2^*), \\
\nn
F^{(2)}_C(u_2^*,u_{12}^*) = & \int_{w_2}^{w_m}\rmd x_2 \mathcal{P}_{X<w_m}(x_2,u_2^*|0,0)\int_{w_1}^{w_m}\rmd x_1\, \mathcal{P}_{w_2<X<w_m}(x_1,u_2^*+u_{12}^*|x_2,u_2^*) \times\\
\label{eq:Hyst2PtFABC1}
& \quad\quad\quad \times \int_0^{\infty} \rmd u_m\, J_{w_2<X < w_m} (x = w_m,u_1^* + u_m|x_1,u_1^*).
\eea

To simplify the calculation, here we will take a Laplace transform from $u$ to $\lambda$ straight away. The Laplace-transformed propagator \eqref{eq:HystPropBM} is:
\bea
\nn
\mathcal{\hP} (x_f| x_i; \lambda) := &\int_{u_i}^\infty \rmd u_f\, e^{\lambda(u_f-u_i)}\mathcal{P} (x_f,u_f | x_i,u_i) = \frac{e^{\frac{1}{2} \left(-\sqrt{1-4 \lambda } \left| x_f-x_i \right| +x_f-x_i \right)}}{\sqrt{1-4 \lambda }}  \\
=& \frac{e^{\frac{1}{2} \left[-b(\lambda ) \left| x_f-x_i \right| +x_f-x_i \right]}}{b(\lambda )},
\eea
where we introduced $b(\lambda) := \sqrt{1-4\lambda}$. This also allows to simplify the propagators with one or two absorbing boundaries \eqref{eq:HystPropBM1b}, \eqref{eq:HystPropBM2b}:
\bea
\nn
\mathcal{\hP}_{X < a} (x_f| x_i; \lambda) = & \frac{e^{\frac{x_f-x_i }{2}} \left[e^{-\frac{1}{2} b(\lambda ) \left| x_f-x_i \right| }-e^{-\frac{1}{2} b(\lambda ) (2 a-x_f-x_i
   )}\right]}{b(\lambda )}, \\
\nn
\mathcal{\hP}_{a_2 < X < a_1} (x_f| x_i; \lambda) = & \frac{1}{b(\lambda )}e^{\frac{x_f-x_i }{2}} \left\{e^{-\frac{1}{2}
   b(\lambda ) \left| x_f-x_i \right| }+ \right. \\
	\nn
	&\quad \left. + {\textstyle\frac{e^{-\frac{1}{2} b(\lambda ) (x_f+x_i )} \left[-e^{(a_1+a_2) b(\lambda )}+e^{(a_2+x_f)
   b(\lambda )}+e^{(a_2+x_i ) b(\lambda )}-e^{b(\lambda ) (x_f+x_i )}\right]}{e^{a_1 b(\lambda )}-e^{a_2 b(\lambda )}}}\right\}.
\eea
Similarly, one obtains simple expressions for the Laplace-transformed probability currents with one or two absorbing boundaries:
\bea
\nn
\mathcal{\hJ}_{X < a} (x_f=a| x_i; \lambda) = & e^{-\frac{1}{2} (a-x_i ) [b(\lambda )-1]}, \\
\nn
\mathcal{\hJ}_{a_2 < X < a_1} (x_f=a_1| x_i; \lambda) = & \frac{e^{\frac{1}{2} (a_1-x_i ) [b(\lambda )+1]} \left[e^{x_i  b(\lambda )}-e^{a_2 b(\lambda )}\right]}{e^{a_1 b(\lambda )}-e^{a_2 b(\lambda
   )}}.
\eea
This allows to compute the Laplace-transformed functions $F^{(2)}_{A,B,C}$ in \eqref{eq:Hyst2PtFABC1}:
\bea
\nn
\hF^{(2)}_{A}(\lambda_2) :=&\int_0^\infty \rmd u_2^* \, e^{\lambda_2 u_2^*}F^{(2)}_A(u_2^*) =  \left(-\frac{1}{\lambda_2}\right) \mathcal{\hJ}_{X < w_m} (x_f=w_m| x_i=0; \lambda_2) \\
\label{eq:Hyst2PtFinA}
=&  -\frac{1}{\lambda_2}e^{-\frac{1}{2} w_m (b(\lambda_2 )-1)}, \\
\nn
\hF^{(2)}_{B}(\lambda_2,\lambda_{12}) :=& \int_0^\infty \rmd u_2^*\int_{u_2^*}^\infty \rmd u_1^*  \, e^{\lambda_2 u_2^*+\lambda_{12}(u_1^*-u_2^*)}F^{(2)}_B(u_2^*,u_1^*-u_2^*) =\\
\nn
=& \int_{w_2}^{w_m}\rmd x_2\, \mathcal{\hP}(x_f = x_2|x_i=0;\lambda_2)  \left(-\frac{1}{\lambda_{12}}\right) \mathcal{\hJ}_{w_2<X < w_m} (x_f=w_m| x_i=x_1; \lambda_{12}) \\
\label{eq:Hyst2PtFinB}
=& \frac{e^{\frac{1}{2} [1- b(\lambda_{2})]w_m}}{\lambda_{12} b(\lambda_{2}) \left(\lambda_2-\lambda_{12}\right)}
   \left\{b(\lambda_{12}) \frac{\sinh \left[\frac{w_2-w_m}{2} b(\lambda_{2})
   \right]}{\sinh \left[\frac{w_2-w_m}{2} b(\lambda_{12})
   \right] }-b(\lambda_{2})\right\},
	\eea
	\bea
	\nn
&\hF^{(2)}_{C}(\lambda_2,\lambda_{12}) := \int_0^\infty \rmd u_2^*\int_{u_2^*}^\infty \rmd u_1^*  \, e^{\lambda_2 u_2^*+\lambda_{12}(u_1^*-u_2^*)}F^{(2)}_C(u_2^*,u_1^*-u_2^*) =\\
\nn
&= \int_{w_2}^{w_m}\rmd x_2\, \mathcal{\hP}(x_f = x_2|x_i=0;\lambda_2) \int_{w_1}^{w_m}\rmd x_1\, \mathcal{\hP}(x_f = x_1|x_i=x_2;\lambda_{12}) \times \\
\nn
& \quad\quad  \times\mathcal{\hJ}_{w_1<X < w_m} (x_f=w_m| x_i=x_2; 0) \\
\nn
&= \frac{e^{\frac{1}{2} [1-b(\lambda_2)] w_m}}{b(\lambda_2) \lambda_{12} \lambda_2
   (\lambda_{12}-\lambda_2)} \left(\lambda_2 \frac{ \sinh \left[\frac{b(\lambda_2)
   (w_m-w_2)}{2}\right]}{\sinh\left[\frac{b(\lambda_{12}) (w_m-w_2)}{2}\right]} \left\{b(\lambda_{12})-
	\frac{ \sinh \left[\frac{b(\lambda_{12}) (w_m-w_1)}{2}\right]}{\sinh\left[\frac{w_m-w_1}{2}\right]}\right\}+ \right. \\
	\label{eq:Hyst2PtFinC}
	& \left. \quad\quad+
	\lambda_{12}\left\{
   \frac{ \sinh \left[\frac{b(\lambda_2) (w_m-w_1)}{2}\right]}{\sinh\left[\frac{w_m-w_1}{2}\right]}-b(\lambda_2) \right\}\right).
\eea
In the previous section \ref{sec:HystABBMQS1Pt} we computed in \eqref{eq:HystF1A}, \eqref{eq:HystF1B} the contributions $F^{(1)}_A$, $F^{(1)}_B$ to the cumulative distribution of one point on the backward branch \eqref{eq:Hyst1PtCum}. Now we can re-express their Laplace transforms simply as 
\bea
\nn
\hF^{(1)}(\lambda) =& \hF^{(1)}_A(\lambda) +\hF^{(1)}_B(\lambda), \\
\nn
\hF^{(1)}_A(\lambda) = & \hF^{(2)}_A(\lambda) = -\frac{1}{\lambda}e^{-\frac{1}{2} w_m [b(\lambda )-1]}, \\
\label{eq:Hyst1PtCum2}
\hF^{(1)}_B(\lambda) = & \lim_{\lambda_{12}\to 0, w_2\to w_1} -\lambda_{12} \hF^{(2)}_{B}(\lambda,\lambda_{12}) 
= \frac{e^{\frac{1}{2} w_m [1-b(\lambda )]}}{\lambda b(\lambda )  } \left[b(\lambda )  - \frac{\sinh \left[\frac{w_1-w_m}{2} b(\lambda )
  \right]}{\sinh\left(\frac{w_1-w_m}{2}\right)}\right].
\eea

Let us now express the Laplace transform $\hP$ of the actual distribution $P(u_2,u_{12})$ through the Laplace transforms of the cumulative distributions \eqref{eq:Hyst2PtCum}, \eqref{eq:Hyst1PtCum2}
\bea
\nn
\hP(\lambda_2,\lambda_{12}) :=& \int_0^\infty \rmd u_2 \int_{u_2}^\infty \rmd u_1  \, e^{\lambda_2 u_2+\lambda_{12}(u_1-u_2)}P(u_1,u_2) \\
\nn
=&
\int_0^\infty \rmd u_2 \,e^{(\lambda_2-\lambda_{12})u_2}\int_{u_2}^\infty \rmd u_1  \, e^{\lambda_{12}u_1} \partial_{u_2}\partial_{u_1} F^{(2)}(u_1,u_2) = \
\\
\nn =&
\int_0^\infty \rmd u_2 \,e^{(\lambda_2-\lambda_{12})u_2} \partial_{u_2}\int_{u_2}^\infty \rmd u_1  \, e^{\lambda_{12}u_1} \partial_{u_1} F^{(2)}(u_1,u_2) \\
& \quad + \int_0^\infty \rmd u_2 \,e^{(\lambda_2-\lambda_{12})u_2} \, e^{\lambda_{12}u_2} \partial_{u_1}\big|_{u_1=u_2} F^{(2)}(u_1,u_2) = \\
\nn =&
-(\lambda_2-\lambda_{12})\int_0^\infty \rmd u_2 \,e^{(\lambda_2-\lambda_{12})u_2} \int_{u_2}^\infty \rmd u_1  \, e^{\lambda_{12}u_1} \partial_{u_1} F^{(2)}(u_1,u_2)
 \\
\nn & \quad + \int_0^\infty \rmd u_2 \,e^{\lambda_2 u_2}  \partial_{u_1}\big|_{u_1=u_2} F^{(2)}(u_1,u_2) = \\
\nn =&
(\lambda_2-\lambda_{12})\lambda_{12}\int_0^\infty \rmd u_2 \,e^{(\lambda_2-\lambda_{12})u_2} \int_{u_2}^\infty \rmd u_1  \, e^{\lambda_{12}u_1} F^{(2)}(u_1,u_2) \\
\nn & \quad + (\lambda_2-\lambda_{12})\int_0^\infty \rmd u_2 \,e^{\lambda_2 u_2} F^{(2)}(u_2,u_2) + \int_0^\infty \rmd u_2 \,e^{\lambda_2 u_2}  \partial_{u_1}\big|_{u_1=u_2} F^{(2)}(u_1,u_2) .
\eea
We used several times that $F^{(2)}(0,u_2) = F^{(2)}(u_1,0)=0$.
Now note that for $u_1 = u_2$, $F^{(2)}(u_1,u_2) = F^{(1)}(u_2)$. We thus obtain the final result:
\bea
\label{eq:Hyst2PtPLT}
\hP(\lambda_2,\lambda_{12}) =& (\lambda_2-\lambda_{12})\lambda_{12} \hF^{(2)}(\lambda_2,\lambda_{12}) - \lambda_{12}\hF^{(1)}(\lambda_2),
\eea
where $\hF^{(1)}$ is given by \eqref{eq:Hyst1PtCum2}. We can further simplify this by using that $\hF^{(1)}_A(\lambda_2) = \hF^{(2)}_A(\lambda_2)$, and the Laplace transform of \eqref{eq:Hyst2PtCum} is
\bea
\nn
\hF^{(2)}(\lambda_2,\lambda_{12}) =& \left(-\frac{1}{\lambda_{12}}\right)\hF^{(1)}_A(\lambda_2) + \hF^{(2)}_B(\lambda_2,\lambda_{12})+\hF^{(2)}_C(\lambda_2,\lambda_{12}), \\
\nn
\hF^{(1)}(\lambda_2) =& \hF^{(1)}_A(\lambda_2) + \hF^{(1)}_B(\lambda_2).
\eea
Inserting this into \eqref{eq:Hyst2PtPLT}, we get
\bea
\hP(\lambda_2,\lambda_{12}) = (\lambda_2-\lambda_{12})\lambda_{12}\left[\hF^{(2)}_B(\lambda_2,\lambda_{12})+\hF^{(2)}_C(\lambda_2,\lambda_{12})\right] - \lambda_2 \hF^{(1)}_A(\lambda_2) - \lambda_{12} \hF^{(1)}_B(\lambda_2).
\eea
Combining the expressions \eqref{eq:Hyst2PtFinA}, \eqref{eq:Hyst2PtFinB}, \eqref{eq:Hyst2PtFinC}, \eqref{eq:Hyst1PtCum2}, one obtains the simple final result
\bea
\nn
\hP(\lambda_2,\lambda_{12}) = & \frac{e^{-\frac{1}{2} [b(\lambda_2)-1] w_m}}{b(\lambda_2)} \frac{\sinh \left[\frac{w_2-w_m}{2} b(\lambda_2) \right]\sinh \left[\frac{w_1-w_m}{2} b(\lambda_{12}) \right]}{ \sinh\left[\frac{w_1-w_m}{2}\right]
 \sinh\left[\frac{w_2-w_m}{2} b(\lambda_{12})\right] }\\
\label{eq:Hyst2Pt}
	=& \frac{e^{-\frac{1}{2} (\sqrt{1-4\lambda_2}-1) w_m}}{\sqrt{1-4\lambda_2}} \frac{\sinh \left[\frac{w_2-w_m}{2} \sqrt{1-4\lambda_2} \right]\sinh \left[\frac{w_1-w_m}{2} \sqrt{1-4\lambda_{12}} \right]}{ \sinh\left[\frac{w_1-w_m}{2}\right]
 \sinh\left[\frac{w_2-w_m}{2} \sqrt{1-4\lambda_{12}}\right] }
\eea
It would be very interesting to find a field-theory-based derivation for this formula, similar to the instanton method used in section \ref{sec:BFMABBM}.
\eqref{eq:Hyst2Pt} satisfies many non-trivial checks. For example, when the first point $u_1$ is ignored, it reduces to the one-point distribution \eqref{eq:Hyst1Pt}:
\bea
\nn
\hP(\lambda_2,0) =& \hP^{(1)}(\lambda_2).
\eea
Comparing \eqref{eq:Hyst1Pt} and \eqref{eq:Hyst2Pt}, we can guess the natural extension to more than two points:
\bea
\hP(\lambda_n,\lambda_{n-1,n},\cdots,\lambda_{12}) = \frac{e^{-\frac{1}{2} [b(\lambda_n)-1] w_m}}{b(\lambda_n)} \prod_{j=1}^{n} \frac{\sinh \left[\frac{w_j-w_m}{2} b(\lambda_{j,j+1}) \right]}{\sinh \left[\frac{w_j-w_m}{2} b(\lambda_{j-1,j}) \right]},
\eea
where $\lambda_{01}:=0$ and $\lambda_{n,n+1} := \lambda_n$.

\subsubsection{Avalanche size distribution near the turning point}
\begin{figure}%
\centering
\includegraphics[width=0.7\columnwidth]{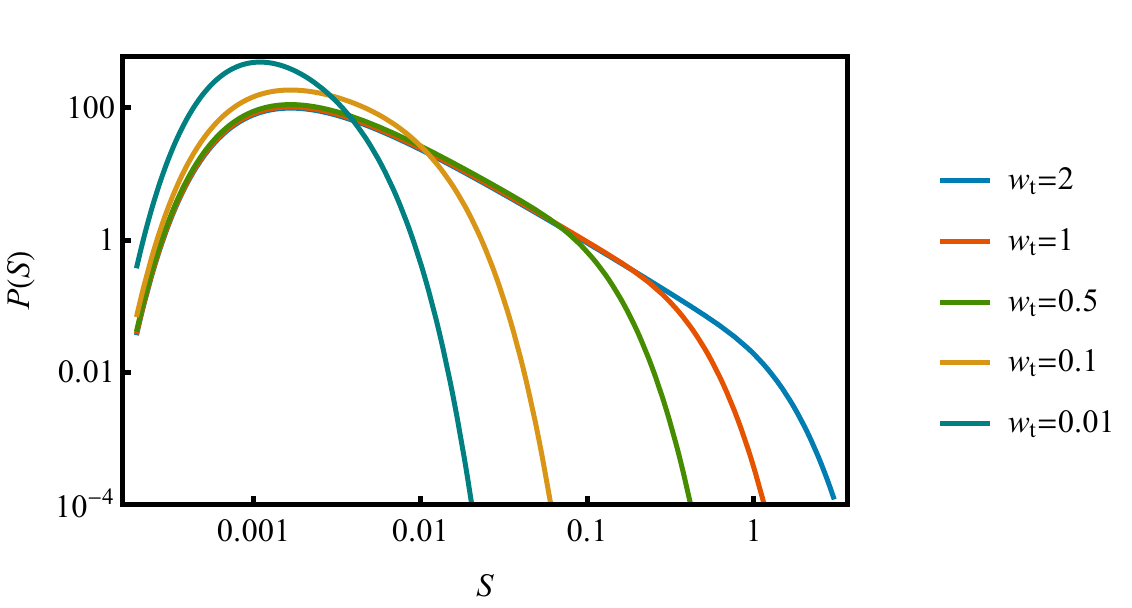}%
\caption{Avalanche size distrubution $P(S)$, for a step of size $w=0.1$, at various distances $w_t$ from the turning point of the hysteresis loop. The plots were obtained by numerical Laplace inversion of \eqref{eq:ABBMBackwardAvSize}. For large $w_t$, one observes a power-law distribution \eqref{eq:BFMPofS}, whereas for small $w_t$ one observes a regular distribution \eqref{eq:ABBMBackwardAvTurn}.}%
\label{fig:HystPofS}%
\end{figure}

We are now in a position to analyze the quasi-static avalanche size distribution in the ABBM model on the backward branch of the hysteresis loop. 
We define $S := u_1 - u_2$, and set $w_1 = w_2 + w$, $w_m = w_1 + w_t$. $w_t$ is the distance to the turning point, $w$ is the size of the ``kick'' controlling the avalanche size. We obtain:
\bea
\label{eq:ABBMBackwardAvSize}
\overline{e^{\lambda S}} = \frac{\sinh \left(\frac{w_t}{2} \sqrt{1-4 \lambda} \right) \sinh \left(\frac{w+w_t}{2}\right)}{ \sinh\left(\frac{w+w_t}{2} \sqrt{1-4 \lambda}\right)\sinh\left(\frac{w_t}{2}\right)}   .
\eea
In the limit $w_t \to \infty$ one recovers, as expected from \eqref{eq:BFMPwofS}, $\overline{e^{\lambda S}} = e^{\frac{w}{2}(1-\sqrt{1-4\lambda})}$. On the other hand, directly at the turning point $w_t=0$ one finds
\bea
\overline{e^{\lambda S}} = \frac{\sinh(w/2)}{\sinh(w/2\sqrt{1-4\lambda})}\sqrt{1-4\lambda}.
\eea 
This has a regular expansion around $w=0$:
\bea
\label{eq:ABBMBackwardAvTurn}
\overline{e^{\lambda S}} = \exp\left[\frac{\lambda w^2}{6}+\left(\frac{\lambda^2}{180}-\frac{\lambda}{360}\right) w^4+\mathcal{O}(w)^6\right].
\eea
We see that power-law avalanches are suppressed near the turning point; one has a regular avalanche size distribution (to leading order, Gaussian with mean $\overline{S} = \frac{w^2}{6}-\frac{w^4}{360}$ and variance $\overline{S^2}^c = \frac{w^4}{90}$). For finite $w_t$, the expression \eqref{eq:ABBMBackwardAvSize} interpolates between the power-law distribution \eqref{eq:BFMPofS} and the regular distribution \eqref{eq:ABBMBackwardAvTurn} (see figure \ref{fig:HystPofS}). 
Expanding \eqref{eq:ABBMBackwardAvSize} to order $w$, one gets the (unnormalized) density of avalanches:
\bea
\int (e^{\lambda S}-1)P(S) = \partial_w \big|_{w=0} \overline{e^{\lambda S}} = \frac{1}{2} \left[\coth \left(\frac{w_t}{2}\right)-\sqrt{1-4 \lambda} \coth \left(\frac{1}{2} w_t \sqrt{1-4 \lambda}\right)\right].
\eea
At large negative $\lambda$, this goes as $\sqrt{-\lambda}$ unless $w_t=0$.
Thus, the $\frac{1}{\sqrt{4\pi}}S^{-3/2}$ power-law is present for any $w_t > 0$ (with this precise prefactor); it only vanishes strictly at $w_t=0$.
It would now be interesting to compute the distribution $P(S)$ averaged over the entire hysteresis loop. As noticed in \cite{DurinZapperi2006}, this leads to an effective avalanche size exponent different from the stationary one. However, the averaged form of the distribution $P(S)$ is not known. This is left as a challenge for future work.

\subsection{Hysteresis loop at finite velocity\label{sec:HystFiniteV}}
The calculation of the quasi-static hysteresis loop in the previous section used the fact that the particle starts moving backward exactly at the same value of $w$ at which the driving turns around. This is no longer true when $w$ is varied with a finite velocity. 
This makes exact calculations much more difficult. For the ABBM model with an absorbing boundary at $\du=0$, one can obtain the statistics of the velocity after the driving has turned around, but while the particle is still moving forward (for example, using the instanton method as in \cite{DobrinevskiLeDoussalWiese2013}). In particular, the distribution of the delay between the turnaround of the driving and the turnaround of the particle can be computed \cite{Dobrinevski2012unpubl}. However, the full hysteresis loop for the ABBM model at finite driving velocity seems out of reach with the current methods.

The hysteresis loop of an elastic interface \eqref{eq:InterfaceEOM}, in short-range-correlated disorder, at a finite driving frequency, has been considered perturbatively in \cite{GlatzNattermannPokrovsky2003,SchutzeNattermann2011}. In \cite{GlatzNattermannPokrovsky2003}, it was discussed how the depinning transition of a finite-dimensional elastic interface, in short-range-correlated disorder, is smeared by a finite driving frequency. However, as seen in appendix \ref{sec:AppABBMHystPert}, already for the quasi-static ABBM model a perturbative calculation only gives some the asymptotic part of the hysteresis loop, and not all features near the turning point. Thus, the validity of the results in \cite{GlatzNattermannPokrovsky2003} is not entirely clear. These are interesting questions for further research.


\section{Short-range-correlated models based on ABBM}
As already suggested in \cite{AlessandroBeatriceBertottiMontorsi1990}, the BFM model can be generalized to a model with physical, short-range-correlated disorder. A natural way to do that is a pinning force $F(u,x)$, which is not a Brownian motion in the $u$ direction as in \eqref{eq:BFM}, but an Ornstein-Uhlenbeck process \cite{Gardiner}. More precisely, let us set
\bea
\label{eq:DisorderOU}
\partial_u F(u,x) = \xi(u,x) - \mu F(u,x),
\eea
where $\xi(u,x)$ is Gaussian white noise as in \eqref{eq:BFMCorrWhiteNoise}, $\mu>0$, and $\mu^{-1}$ is the finite correlation length of the disorder in the $u$ direction.
The Brownian case \eqref{eq:BFM} of the BFM, with infinite-ranged correlations, is recovered for $\mu=0$. For monotonous motion, we can proceed as in section \ref{sec:BFMVelocity}, and obtain an equation-of-motion for the pinning force $F$ with multiplicative, annealed, noise:
\bea
\label{eq:DisorderOU2}
\partial_t F(x,t) = \sqrt{\du_{xt}}\xi(x,t) - \mu \du_{xt} F(x,t).
\eea
The equation-of-motion \eqref{eq:IntVelEOM} remains the same:
\bea
\label{eq:IntVelEOM3}
\eta\partial_t \du_{xt} = - m^2(\du_{xt}-\dw_{xt}) + \nabla^2_x \du_{xt} + \partial_t F_{xt}.
\eea
Together, \eqref{eq:IntVelEOM3} and \eqref{eq:DisorderOU2} provide a closed theory for the interface velocity $\du_{xt}$ and the pinning force $F(x,t)$. This is particularly useful for numerical simulations, since it allows one to simulate short-range-correlated disorder without actually generating and storing a random potential.

In $d=0$, i.e. for a particle, the equations \eqref{eq:DisorderOU2} and \eqref{eq:IntVelEOM3} in the $m=0$ limit can be treated analytically. They reduce to the known Regge quantum mechanics, or Reggeon field theory in zero dimensions \cite{BronzanShapiroSugar1976,BondarenkoEtAl2007}. One observes non-singular avalanche size and duration distributions (i.e. $\tau=0$ and $\alpha=0$ in \eqref{eq:BFMSizeDurExp}), consistent with the results for discrete short-range-correlated disorder \cite{LeDoussalWiese2009}.

However, for an analytical treatment in finite dimensions \eqref{eq:DisorderOU2} and \eqref{eq:IntVelEOM3} are rather cumbersome: Although they can be turned into an MSR field theory as in section \ref{sec:BFMSolDeriv}, it has four fields $\du,\tu,F,\tilde{F}$ whose propagators mix. It is not clear if it can be analyzed. In fact, the easiest approach may be recognizing that $F$ given by \eqref{eq:DisorderOU} is Gaussian with correlator
\bea
\overline{F(u,x)F(u',x')} = \Delta(u-u')\delta(x-x'),\quad\quad\quad \Delta(u-u') = \frac{\sigma}{\mu}\exp\left(-\mu |u-u'|\right).
\eea
Thus, the theory which will be developed in the following chapter \ref{sec:OneLoop} for Gaussian disorder with a general short-ranged $\Delta$ is applicable.

On the other hand, by setting $E_{xt} := \eta \du_{xt} - F_{xt}$, and considering $E_{xt},\du_{xt}$ as independent fields, one can write \eqref{eq:DisorderOU2}, \eqref{eq:IntVelEOM3} as
\bea
\nn
\partial_t \du_{xt} =& - m^2(\du_{xt}-\dw_{xt}) + \nabla^2_x \du_{xt} + \sqrt{\du_{xt}}\xi_{xt} - \mu \eta \du_{xt}^2 + \mu \du_{xt} E_{xt}, \\
\nn
\partial_t E_{xt} = & \nabla^2_x \du_{xt} - m^2 (\du_{xt}-\dw_{xt}).
\eea
This looks strikingly similar to the reaction-diffusion equation of the \textit{conserved directed percolation} (C-DP) class (see eq.~(3) in \cite{BonachelaAlavaMunoz2009}). The only difference is the mass term in the equation for $E$. It has indeed been conjectured \cite{BonachelaChateDornicMunoz2007}, that the universality classes of elastic interfaces in short-ranged disorder and C-DP are one and the same; this mapping may provide a way of understanding this correspondence better.

%% file: OneLoop.tex
\chapter{Avalanches beyond mean-field\label{sec:OneLoop}}
Let us recall the first-order equation of motion \eqref{eq:InterfaceEOM},
\bea
\label{eq:InterfaceEOM2}
\eta\partial_t u_{xt} = - m^2(u_{xt}-w_{xt}) + \nabla^2_x u_{xt} + F(u_{xt},x).
\eea
We suppose $F$ to be Gaussian, with a general correlation function $\Delta$ in the $u$ direction
\bea
\overline{F(u,x)F(u',x')} = \delta^{(d)}(x-x') \Delta(u-u').
\eea
In chapter \ref{sec:BFM} we considered the Brownian Force Model, a linear $\Delta$, corresponding to forces with independent increments and infinite-ranged correlations. As already mentioned, the physical disorder, e.g. in a ferromagnet, has a finite correlation length; thus, the physical (microscopic) $\Delta$ is expected to decay to $0$ for large arguments. 
The basic picture for avalanches with such a short-ranged $\Delta$ is the following (cf.~also the discussion in section \ref{sec:IntroInterfaceModel}): 
The spatial structure of the interface moving according to \eqref{eq:InterfaceEOM2} has a single correlation length along the internal direction(s) $x$ of the interface given by $\xi \sim m^{-1}$. On distances $x \gg m^{-1}$, the harmonic confinement $-m^2 u$ is the dominant force. The system (of total size $L\gg m^{-1}$) splits into pieces of linear size $m^{-1}$, which are essentially independent. On length scales (again, along the internal direction(s) of the interface) $x \ll m^{-1}$, pinning by disorder is important. The interface becomes rough, and its local displacement $u_{xt}$ grows as $u_{xt} \sim x^{\zeta}$. This defines the roughness exponent $\zeta$. Since this happens up to a distance $x \sim \xi \sim m^{-1}$ (where the growth of the displacement is stopped by the harmonic confinement), the local displacement (again, for an infinitely large system, $L \gg m^{-1}$) scales like $u_{xt} \sim \xi^{\zeta} \sim m^{-\zeta}$. The typical pinning time scale, i.e. the time it takes an interface to make a jump forward, scales with an independent exponent $t \sim m^{-z}$. These exponents $\zeta$ and $z$ are linked to avalanche  exponents, such as the exponent $\tau$ of $P(S)$, $\alpha$ of $P(T)$, and $a$ of $P(\du)$, by scaling relations (see section \ref{sec:FRGScalingRelations}).

Let us assume, for the moment, that $\zeta>0$. As $m \to 0$, the typical displacements of the interface become larger. 
The interface settles down into a local energy minimum, chosen from a larger and larger set of possible states as $m \to 0$. Its center-of-mass feels an \textit{effective disorder}, corresponding to the energy of this local minimum. As usual in extreme value statistics, as the set of sampled states becomes larger, microscopic details matter less and less. We assume that the theory for the variables rescaled by appropriate powers of $m$ (in particular, the rescaled effective disorder correlator $\Delta$) has a finite limit as $m \to 0$, independent of microscopic details. This scale-invariant limiting theory 
should depend only on global features of the microscopic disorder, like the tail of its distribution, existence of long-ranged correlations  etc., and the driving protocol (zero-temperature static ground state vs. quasi-static depinning). It is this universal $m\to 0$ theory in rescaled variables, that we attempt to construct.

In section \eqref{sec:FRGReview}, I review how the universal effective disorder distribution is determined using the well-known functional renormalization group (FRG) \cite{Fisher1986a,NarayanFisher1993,NattermannStepanowTangLeschhorn1992,LeschhornNattermannStepanow1997,WieseLeDoussal2006}. Basically, it considers perturbatively the feedback of the disorder on smaller scales into the effective disorder on larger scales. The condition that both are given, up to scaling, by a Gaussian distribution with the same effective disorder correlator $\Delta$ then allows determining its shape and the exponents $\zeta$ and $z$. This FRG approach was first used for understanding the zero-temperature static ground state of an elastic interface in a random medium \cite{Fisher1986a}, and later for the quasi-static dynamics/depinning considered here \cite{NattermannStepanowTangLeschhorn1992,LeschhornNattermannStepanow1997,ChauveGiamarchiLeDoussal2000}. This will be discussed in detail in section \ref{sec:FRGReviewOneLoop}. I will then derive scaling relations (some of them new) in section \ref{sec:FRGScalingRelations}, which express the exponents characterizing avalanche statistics in terms of the roughness and dynamical exponents $\zeta$ and $z$.

As the internal dimension $d$ of the interface is increased, the roughness exponent $\zeta$ decreases. At a critical dimension $d_c$, the interface becomes flat and the roughness exponent $\zeta = 0$. For the short-ranged elastic interaction $\nabla^2$ in eq.~\eqref{eq:InterfaceEOMi}, $d_c=4$ \cite{BruinsmaAeppli1984}, as we shall see in section \ref{sec:FRGReview}. For long-ranged power-law elastic interactions in eq.~\eqref{eq:InterfaceEOMLRi} $d_c=2\mu$ \cite{CizeauZapperiDurinStanley1997}, as will be discussed in section \ref{sec:OneLoopLongRange}. The perturbative expansion for the universal effective disorder distribution is controlled by the dimensionless parameter $\epsilon = d_c-d$, and it will turn out that the roughness exponent $\zeta \propto \epsilon$ for small $\epsilon$.

The main quantitative results that will be discussed in the following sections (and presented in a forthcoming publication \cite{DobrinevskiLeDoussalWiese2013inpr,LeDoussal2013tobepublished,Wiese2013tobepublished}) are:
\begin{itemize}
	\item In section \ref{sec:FRGAvalanches}, I will show how the general FRG theory reviewed in section \ref{sec:FRGReview} can be applied to the study of avalanches. This was first discussed in \cite{LeDoussalWiese2008c} for static avalanche sizes of an interface, and later extended in \cite{LeDoussalWiese2011b,LeDoussalWiese2011,LeDoussalWiese2013,DobrinevskiLeDoussalWiese2013inpr} to the dynamics (in particular the instantaneous velocity distribution). The basic result is that as $d \to d_c$, the interface becomes flat and  typical avalanche scales become smaller. Thus, for $\epsilon \to 0$, only the behaviour of $\Delta(u)$ near $u=0$ is important: At order $\epsilon=0$, i.e.~in mean-field theory, only the linear cusp $\Delta'(0^+)$ is important for small avalanches, and the theory reduces to the BFM discussed in the previous chapter \ref{sec:BFM}. At order $\epsilon$, one needs to take into account $\Delta''(0^+)$ which can be done perturbatively. The general FRG theory (section \ref{sec:FRGReview}) determines the universal value of $\Delta''(0^+)$ in the limit of small $m$, and hence in this limit the perturbative corrections of $\mO(\epsilon)$ to general avalanche observables are also universal. The main result of section \ref{sec:FRGAvalanches} is formula \eqref{eq:OneLoopGenObsFinal}, which gives explicitly the correction of order $\epsilon$ to any avalanche observable.
		\item In section \ref{sec:OneLoopDurations}, I use the results of section \ref{sec:FRGAvalanches} to compute the correction of order $\epsilon$ to the form of the avalanche duration distribution $P(T)$, introduced in section \ref{sec:BFMDuration}. Its explicit form is given in \eqref{eq:OneLoopDurPofTCorr}, \eqref{eq:OneLoopDurDeltaF1}. In particular, this calculation shows that the correction to the avalanche duration exponent $\alpha$ is 
		\bea
		\alpha=2-\frac{2}{9}\epsilon + \mO(\epsilon)^2.
		\eea
		The plots in figure \ref{fig:OneLoopDurPlots} also show, that the corrections of order $\mO(\epsilon)$ add a characteristic bump to the shape of $P(T)$, which provides a clear sign of non-mean-field behaviour. 
	\item In section \ref{sec:OneLoopShapeTime}, I use the results of section \ref{sec:FRGAvalanches} to compute the correction of order $\epsilon$ to the average avalanche shape, at fixed avalanche duration $T$, $\mfs(\tm,T)$ as introduced in section \ref{sec:BFMShape}. Its explicit form is given in \eqref{eq:OneLoopShpDurFinal}. This new result yields the following conclusions:
	\begin{itemize}
		\item Near the beginning and the end of the avalanche, the linear shape of the BFM $\mfs(\tm,T)= 2\tm + \mO(\tm)^2$, $\mfs(T-\tmp,T)= 2\tmp+ \mO(\tmp)^2$ (cf. eq.~\eqref{eq:BFMShapeFixedT}) is modified to a steeper power law $\tm^{1-\frac{1}{9}\epsilon}$. The power law is the same for both ends of the avalanche, however the prefactor is different on the left and on the right (see \eqref{eq:OneLoopShpLeftEdgeTot} and \eqref{eq:OneLoopShpRightEdgeTot}). In both cases the leading behaviour near the avalanche beginning and the avalanche end is independent of the total duration $T$.
		\item For short avalanches, the average avalanche shape is, up to a global prefactor, a function of $x:=\tm/T$ only. This function, which is a parabola \eqref{eq:ABBMShapeShort} for the BFM, is corrected to $\mO(\epsilon)$. The correction can be expressed in closed form \eqref{eq:OneLoopShapeDurShort}, is universal (i.e. independent of the large-scale cutoff), and leads to a rightward asymmetry of the shape (shown in figure \ref{fig:OneLoopShapePlotEps}).
	\end{itemize}
	\item In section \ref{sec:OneLoopShapeSize} I discuss similarly the corrections of order $\epsilon$ to the avalanche size distribution $P(S)$ (which is known since \cite{LeDoussalWiese2008c,LeDoussalWiese2011b}) and the average avalanche shape at fixed avalanche size (which is new). In particular, I recall the modification of the avalanche size exponent $\tau$,
	\bea
	\tau = \frac{3}{2} - \frac{1}{12}\epsilon + \mO(\epsilon)^2.
	\eea
	The leading behaviour of the average avalanche shape at fixed size near the avalanche beginning (eq.~\eqref{eq:OneLoopShapeSizeLeft}) turns out to be identical to the leading-order of the average avalanche shape at fixed duration. This includes both the steeper-than-linear power law $t^{1-\frac{1}{9}\epsilon}$, and the numerical prefactor. A closed expression for the average avalanche shape for any fixed size $S$ is given in \eqref{eq:OneLoopShapeSizeCorrFinal}.
	\item In section \ref{sec:OneLoopLongRange} I give the generalizations of the results above to the case of a long-ranged, power-law elastic kernel \eqref{eq:InterfaceEOMLRi}. In this case, the critical dimension $d_c=2\mu$ and $\epsilon = d_c-d$. I obtain scaling relations for the avalanche exponents for arbitrary $\mu$. For $\mu=1$, I derive the corrections to the avalanche duration distribution and the average avalanche size, as in the previous sections for short-ranged elasticity. In particular, the corrections of $\mO(\epsilon)$ to the exponents $\alpha$ and $\tau$ are 
	\bea
	\alpha = 2-\frac{4}{9}\epsilon + \mO(\epsilon)^2,\quad\quad\quad \tau = \frac{3}{2} - \frac{1}{6}\epsilon+ \mO(\epsilon)^2.
	\eea
	Similarly, the corrections to the avalanche shape exponents near the left and right edges are doubled, the power laws become $t^{1-\frac{2}{9}\epsilon}$.
	The form of the correction to the universal avalanche shape for short avalanches (and, in particular, the rightward asymmetry) remains unchanged, up to a rescaling of the amplitude.
\end{itemize}

\section{A brief review: Functional Renormalization Group for depinning\label{sec:FRGReview}}
As in section \ref{sec:BFMSolDeriv} for the BFM, we apply the Martin-Siggia-Rose formalism in order to translate \eqref{eq:InterfaceEOM2} into a field theory. This yields a path integral formula for the generating functional (as in \eqref{eq:DefGSources}), averaged over realizations of the disorder
\bea
\overline{\exp\left(\int_{xt}\hat{\lambda}_{xt}u_{xt}\right)}^{w_{xt}}=:G[\hat{\lambda},w] = \int \mD[u,\hu] e^{-S[u,\hu] + \int_{xt}\hat{\lambda}_{xt}u_{xt} + \int_{xt} f_{xt}\hu_{xt}},
\eea
where the driving force is $f_{xt}:=m^2 w_{xt}$ and the Martin-Siggia-Rose action for the equation of motion \eqref{eq:InterfaceEOM} is
\bea
\label{eq:InterfaceAction}
S[u,\hu] = \int_{x,t} \hu_{xt}\left[\eta \partial_t u_{xt} + m^2 u_{xt} - \nabla_x^2 u_{xt} \right] - \frac{1}{2}\int_{x,t,t'} \hu_{xt}\hu_{xt'}\Delta(u_{xt}-u_{xt'}).
\eea
Note that the quenched random force generates a coupling in $S$ which is non-local in time. 
As in the section \ref{sec:BFMAvalanches}, for analyzing avalanches it will be useful to consider the interface velocity $\du$. Taking a derivative of \eqref{eq:InterfaceEOM2}, we obtain an equation-of-motion for $\du$, 
\bea
\label{eq:IntVelEOM2}
\eta\partial_t \du_{xt} = - m^2(\du_{xt}-\dw_{xt}) + \nabla^2_x \du_{xt} + \partial_t F(u_{xt},x).
\eea
The corresponding Martin-Siggia-Rose action is 
\bea
\label{eq:IntVelAction}
S[\du,\tu] = \int_{x,t} \tu_{xt}\left[\eta \partial_t \du_{xt} + m^2 \du_{xt} - \nabla_x^2 \du_{xt} \right] - \frac{1}{2}\int_{x,t,t'} \tu_{xt}\tu_{xt'}\partial_t \partial_{t'}\Delta(u_{xt}-u_{xt'}).
\eea
It can likewise be obtained from \eqref{eq:InterfaceAction} by setting $\hu_{xt} =: -\partial_t \tu_{xt}$. The generating functional for interface velocities is then given by \eqref{eq:DefGSources}, with the action \eqref{eq:IntVelAction}.

\subsection{Statistical Translational Symmetry\label{sec:InterfaceSTS}}
The action \eqref{eq:InterfaceAction} possesses a nontrivial symmetry: Mapping $u_{xt} \to u_{xt} + g_x$, one observes
\bea
\nn
S[u_{xt}+g_x,\hu_{xt}] = &S[u_{xt},\hu_{xt}]+ \int_{x,t} \hu_{xt}\left( m^2 g_{x} - \nabla_x^2 g_{x} \right)=S[u_{xt},\hu_{xt}]+ \int_{x,t} g_x \left( m^2 - \nabla_x^2 \right) \hu_{xt}.
\eea
Since the path integration measure $\mD[u,\hu]$ is invariant under the mapping $u_{xt} \to u_{xt} + g_x$, this means, that for any observable $O[u,\hu]$ one has
\bea
\nn
\overline{O[u_{xt}-g_x,\hu_{xt}]}^S =& \int \mD[u,\hu]\,e^{-S[u,\hu]}O[u_{xt}-g_x,\hu_{xt}]  \\
\nn
= &\int \mD[u,\hu]\,e^{-S[u,\hu] - \int_{x,t} g_x \left( m^2 - \nabla_x^2 \right) \hu_{xt}}O[u_{xt},\hu_{xt}] \\
\nn
=& \overline{e^{-\int_{x,t} g_x \left( m^2 - \nabla_x^2 \right)\hu_{xt}}O[u_{xt},\hu_{xt}]}^S .
\eea
This \textit{statistical translational symmetry}, first observed in \cite{HwaFisher1994}, holds for any choice of $g$. Taking a functional derivative with respect to $g_x$, we obtain that for any fixed value of $x$,
\bea
\label{eq:STS}
-\int_t \overline{\frac{\delta }{\delta u_{xt}}O[u,\hu]} = -\overline{O[u,\hu] \int_t \left( m^2 - \nabla_x^2 \right) \hu_{xt} }.
\eea
 In particular, let us consider the (renormalized) response function
\bea
R(\xf-\xxi,\tf-\ti) := \overline{u_{\xf,\tf}\hu_{\xxi,\ti}}.
\eea
Using \eqref{eq:STS} for $O[u,\hu] := u_{\xf,\tf}$, we obtain from \eqref{eq:STS} that $R$ satisfies \cite{Chauve2000}
\bea
\int_{\ti} \left( m^2 - \nabla_{\xxi}^2 \right)R(\xf-\xxi,\tf-\ti) = \delta(\xf-\xxi).
\eea
Fourier transforming from $x:=\xf-\xxi$ to $q$ and from $t:=\tf-\ti$ to $\omega$, we obtain
\bea
R(q,\omega=0):= \int_x e^{-i q x}\rmd x\int_0^\infty \rmd t\, R(x,t)= \frac{1}{q^2+m^2}.
\eea
This is the same result, as obtained without disorder. This means that the $\omega=0$ mode of the response function is not renormalized. In particular, there are no loop corrections to the mass term $\int_{xt} m^2 u_{xt} \hu_{xt}$, and the elastic term $\int_{xt} \hu_{xt} \nabla^2 u_{xt}$ in the action \eqref{eq:InterfaceAction}. This makes the mass $m$ a natural control parameter for the renormalization-group flow.

\subsection{Units and dimensions}
For small $m$, one observes that the interface driven according to \eqref{eq:InterfaceEOM} becomes rough, i.e. the typical local displacements $u_{xt}$ grow as $m^{-\zeta}$ where $\zeta$ is the roughness exponent. This motivates the following rescaling:
\bea
\label{eq:OneLoopScaling1}
x = & m^{-1} \underline{x}, \quad\quad\quad t = m^{-z} \underline{t}, \quad\quad\quad u_{xt} = m^{-\zeta} \uu_{\underline{xt}}.
\eea
$\zeta$ is the roughness exponent, and $z$ is the dynamical exponent. Both are, so far, arbitrary. Here and in the following, we shall always denote fields rescaled by the corresponding power of $m$ (the scaling dimension of the field) by underlined letters. Note that the scaling $x\sim m^{-1}$ supposes that the prefactor of the elastic term $\nabla^2$ is not renormalized. This is a consequence of the \textit{statistical translational symmetry} discussed in section \ref{sec:InterfaceSTS}. 
In the limit $m \to 0$ we expect criticality and scale invariance. Assuming that this limiting $m\to 0$ theory is still of the form \eqref{eq:InterfaceAction}, this means that after suitable rescalings of all fields with powers of $m$, the action \eqref{eq:InterfaceAction} should become independent of $m$. 
Inserting \eqref{eq:OneLoopScaling1} into \eqref{eq:InterfaceAction} we observe that this scale invariance holds, if and only if we also scale
\bea
\label{eq:OneLoopScaling2}
\eta = & m^{2-z} \underline{\eta}, \quad\quad\quad \hu_{xt} = m^{d+z-2+\zeta} \underline{\hu}_{\underline{xt}}, \quad\quad\quad \Delta = m^{4-d-2\zeta} \uD.
\eea
For the fields $\du,\tu$ in the velocity theory \eqref{eq:IntVelAction}, the scaling \eqref{eq:OneLoopScaling1}, \eqref{eq:OneLoopScaling2} implies
\bea
\label{eq:OneLoopScaling3}
\tu_{xt} = m^{d-2+\zeta}\underline{\tu}_{\underline{xt}}, \quad\quad\quad \du_{xt} = m^{z-\zeta} \underline{\du}_{\underline{xt}}.
\eea
Comparing \eqref{eq:OneLoopScaling3} and \eqref{eq:OneLoopScaling1} to the rescaling for the BFM performed in \eqref{eq:BFMRescaling}, we see that the BFM exponent values are
\bea
\label{eq:ScalingBFMExp}
\zeta = 4-d = \epsilon,\quad\quad\quad z=2.
\eea

\subsubsection{Exponent identities from scaling\label{sec:FRGScalingRelations}}
The assumption of a scale-invariant theory of the form \eqref{eq:InterfaceAction} in the limit $m\to 0$, with the scaling \eqref{eq:OneLoopScaling1}, \eqref{eq:OneLoopScaling2}, \eqref{eq:OneLoopScaling3}, also restricts the power-law exponents for avalanche size and duration distributions, as well as the instantaneous velocity distribution. 

\paragraph{Avalanche size exponent}
The total avalanche size $S := \int_{xt} \du_{xt}$ has the scaling $S = m^{-d-\zeta}\underline{S}$.
The definition \eqref{eq:DefPofS} of the density $P(S)$ implies that the scaling dimension of the Laplace transform of $P(S)$, and of the right-hand side of \eqref{eq:DefPofS} must be the same\footnote{This makes the assumption that the force required to trigger an avalanche scales with $m$ in the same way as all other forces according to \eqref{eq:OneLoopScaling1}, \eqref{eq:OneLoopScaling2}. This appears natural, but could conceivably be violated. In section \ref{sec:OneLoopShapeSize} an explicit calculation shows that to leading order in $\epsilon$, no such violations occur.}. The existence of a thermodynamic limit implies that for $L \to \infty$, $\frac{1}{L^d}\int_x \tu_{x,\ti}$ converges and scales as $\tu_{x,\ti}$. Hence, we have
\bea
\label{eq:OneLoopScalingTauTemp}
\int \rmd S \,\left(e^{\lambda S} - 1\right)P(S) \sim \tu_{x,\ti} \sim m^{d-2+\zeta},
\eea
where $\sim$ means here ``scales with the same power of $m$ as $m\to 0$''. 
This has to hold, in principle, for any $\lambda$, and hence in particular for large negative $\lambda$ which effectively singles out the contribution from small $S$. Supposing that near $S=0$, $P(S) = S^{-\tau}$, the left-hand side scales as $\lambda^{\tau-1}\sim S^{1-\tau} \sim m^{-(d+\zeta)(1-\tau)}$. Comparing the scaling of the left- and the right-hand sides gives the relation
\bea
\label{eq:OneLoopScalingTau}
(d+\zeta)(1-\tau) = 2-d-\zeta \Leftrightarrow \tau = 2-\frac{2}{d+\zeta}.
\eea
This scaling relation for the avalanche size exponent was first conjectured by \cite{NarayanFisher1993}. For the BFM, where $\zeta = \epsilon = 4-d$ (see eq.~\eqref{eq:ScalingBFMExp}), we recover the mean-field value $\tau=3/2$ as in section \ref{sec:BFMSize}. For short-range correlated disorder below the critical dimension, \eqref{eq:OneLoopScalingTau} can be verified to $\mO(\epsilon)$ through a one-loop calculation \cite{LeDoussalWiese2008c,LeDoussalWiese2011b,LeDoussalWiese2013}. I will briefly review this in section \ref{sec:OneLoopShapeSize}.

\paragraph{Avalanche duration exponent}
The definition \eqref{eq:DefPofT} of the distribution $P(T)$ implies that the scaling dimension of the cumulative density $F(\tf)$, and of the right-hand side of \eqref{eq:DefPofT} must be the same. This has to hold for any $\tf$, in particular for small $\tf$ where $P(T)\sim T^{-\alpha}$, and $F(\tf) \sim \tf^{1-\alpha}\sim m^{-z(1-\alpha)}$. The right-hand side of \eqref{eq:DefPofT}, as in the previous section, scales as $\tu_{x,\ti}$.   Hence, we have
\bea
\label{eq:OneLoopScalingAlpha}
-z(1-\alpha) = d-2+\zeta \Leftrightarrow \alpha = 1+\frac{d-2+\zeta}{z}.
\eea
This scaling relation for the avalanche duration exponent $\alpha$ was first found, using similar arguments, in \cite{ZapperiCizeauDurinStanley1998}. For the BFM, where $\zeta = \epsilon = 4-d$ -- see eq.~\eqref{eq:ScalingBFMExp} -- it gives the mean-field value $\alpha=2$ as in section \ref{sec:BFMDuration}. For short-range correlated disorder below the critical dimension, I will show in section \ref{sec:OneLoopDurations} how \eqref{eq:OneLoopScalingAlpha} is confirmed to order $\mO(\epsilon)$ by an explicit one-loop calculation of $P(T)$.

\paragraph{Stationary velocity distribution exponent}
The definition \eqref{eq:DefPofUdot} of the density $P(\du)$ implies that the scaling dimension of the Laplace transform of the density $P(\du)$, and of the right-hand side of \eqref{eq:DefPofUdot} must be the same. As in the previous sections, for $L \to \infty$, $\frac{1}{L^d}\int_x \tu_{x,\ti}$ scales as $\tu_{x,\ti}$. Hence, we have
\bea
\label{eq:OneLoopScalingATemp}
\int \rmd \du \,\left(e^{\lambda \du} - 1\right)P(\du) \sim \int_{\ti}\tu_{x,\ti} \sim m^{d-2+\zeta-z},
\eea
Note that in contrast to \eqref{eq:OneLoopScalingTauTemp}, there is now an additional time integral, due to the fact that the initial time $\ti$, at which the avalanche is triggered, is not fixed.
\eqref{eq:OneLoopScalingATemp} has to hold, in principle, for any $\lambda$, and hence in particular for large negative $\lambda$, which effectively singles out the contribution from small $\du$. Supposing that near $\du=0$, $P(\du) = \du^{-a}$, the left-hand side scales as $\lambda^{a-1}\sim \du^{1-a} \sim m^{-(d+\zeta-z)(1-a)}$. Comparing the scaling of the left- and the right-hand sides of \eqref{eq:OneLoopScalingATemp} gives the relation
\bea
\label{eq:OneLoopScalingA}
(d+\zeta-z)(1-a) = 2-d-\zeta + z \Leftrightarrow a = 2-\frac{2}{d+\zeta-z}.
\eea
This scaling relation is new, and clarifies the question, posed in \cite{LeDoussalWiese2011,LeDoussalWiese2013}, whether the exponent $a$ is independent. For the BFM, where $\zeta = \epsilon = 4-d$ and $z=2$ (see eq.~\eqref{eq:ScalingBFMExp}), it yields $a=1$ in agreement with the results of section \ref{sec:BFMABBMStat}. For short-range correlated disorder below the critical dimension, \eqref{eq:OneLoopScalingTau} agrees to $\mO(\epsilon)$ with the results of a one-loop calculation \cite{LeDoussalWiese2011,LeDoussalWiese2013}.

\paragraph{Local avalanche sizes, and local velocities}
The arguments in the preceding sections can be extended straightforwardly to local avalanche sizes, and local interface velocities.
The local avalanche size $S_{\phi}$ on a hyperplane $\phi$ of dimension $d_\phi \leq d$ of the interface (with internal dimension $d$) was defined in section \ref{sec:BFMLocalAv}, eq. \eqref{eq:DefLocSD}. Assuming its density $P(S_{\phi})$ (as defined in \eqref{eq:DefPofLocS}) diverges near $S_{\phi} =0$ as $P(S_{\phi}) \sim S_{\phi}^{\tau_{\phi}}$, a straightforward generalization of \eqref{eq:OneLoopScalingTauTemp} gives the relation
\bea
\label{eq:OneLoopScalingTauLocal}
\tau_{\phi} = 2-\frac{2}{d_\phi +\zeta}.
\eea
This relation was first conjectured in \cite{LeDoussalWiese2008c}. Here we show that it is, in fact, very closely related to the Narayan-Fisher conjecture \eqref{eq:OneLoopScalingTau} for the global avalanche size exponent $\tau$. For $d_\phi=d$, \eqref{eq:OneLoopScalingTauLocal} reduces to \eqref{eq:OneLoopScalingTau}. For the BFM case $\zeta = \epsilon = 4-d$, $\tau_\phi$ given by \eqref{eq:OneLoopScalingTauLocal} agrees with a scaling analysis of the local instanton solution generalizing \eqref{eq:BFMLocalODE}, \eqref{eq:BFMLocalInstSol} in arbitrary dimensions \cite{WiesePrivate}.

Similarly to \eqref{eq:DefLocSD}, we can define local velocities $\du_\phi$ on a hyperplane $\phi$, 
\bea
\label{eq:DefLocUdotD}
\du_{\phi,t} = \int_{x_1}\cdots \int_{x_{d_\phi}} \du_{(x_1,...x_{d_\phi},0...0),t}\quad.
\eea
Following the same type of arguments as above, we obtain a scaling relation for the exponent $a_\phi$ of the stationary local velocity distribution $P(\du_\phi)\sim \du_\phi^{a_\phi}$:
\bea
\label{eq:OneLoopScalingALocal}
a_\phi = 2-\frac{2}{d_\phi+\zeta-z}.
\eea
For $d_\phi = d$ this reduces to \eqref{eq:OneLoopScalingA}.


To summarize, we observe that the exponents $\zeta$ and $z$ characterizing the interface roughness and pinning dynamics also determine the scaling of avalanches. We thus expect that all these exponents, and the universal scaling forms of avalanche observables, are determined by the universal shape of the renormalized disorder. Let us make this more precise.

\subsection{One-loop FRG equation\label{sec:FRGReviewOneLoop}}
For clarity, let us denote the disorder correlator in the bare action \eqref{eq:InterfaceAction} by $\Delta_0$. The renormalized (effective) disorder correlator $\Delta$ is defined through a two-point function of the average interface displacement $u_t = \frac{1}{L^d}\int_x u_{xt}$, at various positions $w$ of the harmonic driving well \cite{LeDoussal2006,LeDoussalWiese2006a,LeDoussal2009,LeDoussalWiese2013}:
\bea
\label{eq:OneLoopDefDelta}
\Delta(w-w') := L^{d} m^4 \overline{\left[u(w)-w\right]\left[u(w')-w'\right]}^c.
\eea
This makes sense for quasi-static dynamics at a constant driving velocity $\dw_{xt}=v=0^+$, since then $u$ is a function of $w$ only (see section \ref{sec:InterfaceMonot}). 
Note that $\Delta(w)$, defined in this way, is an observable easily accessible in numerics \cite{LeDoussalWiese2006a,MiddletonLeDoussalWiese2007,RossoLeDoussalWiese2007} or experiments \cite{LeDoussalWieseMoulinetRolley2009}.
$\Delta$ can be computed perturbatively in $\Delta_0$, for example using a diagrammatic expansion \cite{LeDoussalWieseChauve2002}. At quasi-static driving, $\dw(x,t)=v=0^+$, the diagrams of order $n+1$ in the bare disorder $\Delta_0$ are exactly the diagrams with $n$ momentum loops. One obtains an expansion of the form 
\bea
\label{eq:OneLoopPertDelta}
\Delta(u) = \Delta_0(u) + \delta^{(1)}(\Delta_0, \Delta_0)  + \delta^{(2)}(\Delta_0,\Delta_0,\Delta_0) + \mO(\Delta_0)^3.
\eea
Here $\delta^{(1)}(\Delta_0, \Delta_0)$ is the one-loop perturbative correction \cite{LeDoussalWieseChauve2002}
\bea
\delta^{(1)}(\Delta_0, \Delta_0) = - \frac{1}{2}\frac{\rmd^2}{\rmd u^2}\left[\Delta_0(u)-\Delta_0(0)\right]^2 I_1.
\eea
$I_1$ is the one-loop integral given by \cite{LeDoussalWieseChauve2002} 
\bea
\label{eq:OneLoopIntegral}
I_1 := \int \frac{\rmd^d q}{(2\pi)^d} \frac{1}{(q^2+m^2)^2} = m^{d-4} \frac{\Gamma(2-d/2)}{(4\pi)^{d/2}} =: m^{d-4} \tilde{I}_1.
\eea
$\delta^{(2)}(\Delta_0,\Delta_0,\Delta_0)$ is the two-loop perturbative correction, first computed in \cite{LeDoussalWieseChauve2002}, and consists of several contributions of the form $I_2 \frac{\rmd^4}{\rmd u^4} \Delta_0^3$, with $I_2 \sim m^{-2\epsilon}$ representing different types of two-loop integrals. Higher-loop contributions keep the same structure but quickly become more complicated.

We observe that the expansion \eqref{eq:OneLoopPertDelta} (to any order) is homogeneous in $m$  with the rescaling \eqref{eq:OneLoopScaling1}, \eqref{eq:OneLoopScaling2}: all terms are of order $m^{4-d-2\zeta}$. This is the scaling dimension of the disorder correlator $\Delta$ as seen previously from \eqref{eq:OneLoopScaling2}; the rescaled correlator $\uD$ is dimensionless. The expansion \eqref{eq:OneLoopPertDelta} is controlled under the assumption that the dimensionless $\uD$ is small.

%

Let us now consider the flow of the dimensionless disorder $\uD(\uu)$ with the infrared cutoff $m$, keeping the bare (dimensionful) disorder $\Delta_0$ fixed. This is the $\beta$ function defined as
\bea
\nn
\beta(\uD) :=& -m\partial_m\big|_{\Delta_0} \uD(\uu) = -m\partial_m\big|_{\Delta_0} \left[m^{2\zeta-\epsilon}\Delta\left(m^{-\zeta}\uu\right)\right] \\
\label{eq:OneLoopBeta1}
=& (\epsilon-2\zeta) \uD(\uu) + \zeta \uu \uD'(\uu) + m^{2\zeta-\epsilon}\left(-m\partial_m\big|_{\Delta_0}\Delta\right)\left(\uu\right).
\eea
The first two terms correspond to the contribution from the rescaling according to \eqref{eq:OneLoopScaling1}, \eqref{eq:OneLoopScaling2}. The last term is the flow due to the modification of the infrared cutoff in the loop integrals $I_1, I_2, ...$. Using \eqref{eq:OneLoopPertDelta}, and the fact that the $n$-loop integral $I_n$ is proportional to $m^{-n\epsilon}$ (and otherwise independent of $m$), we obtain
\bea
\label{eq:OneLoopBeta2}
-m\partial_m\big|_{\Delta_0}\Delta = \epsilon \delta^{(1)}(\Delta_0, \Delta_0) + 2\epsilon \delta^{(2)}(\Delta_0,\Delta_0,\Delta_0)+...
\eea
The one-loop term (of order $\Delta_0^2$) can be easily re-expressed in terms of $\uD$. Inverting \eqref{eq:OneLoopPertDelta} order by order in $\uD$, and using the scaling relation \eqref{eq:OneLoopScaling1}, we obtain
\bea
\label{eq:OneLoopInvDelta}
\Delta_0(u) = & m^{\epsilon-2\zeta}\left\{\uD(\uu) + \tilde{I}_1 \frac{1}{2}\frac{\rmd^2}{\rmd \uu^2} \left[\uD(\uu)-\uD(0)\right]^2  + \mO(\uD)^3 \right\}.
\eea
Inserting the leading order in \eqref{eq:OneLoopBeta2}, one finds
\bea
\delta^{(1)}(\Delta_0, \Delta_0) I_1 =- \frac{1}{2}\frac{\rmd^2}{\rmd u^2}\left[\Delta_0(u)-\Delta_0(0)\right]^2 m^{-\epsilon}\tilde{I}_1 = -m^{\epsilon-2\zeta} \frac{1}{2}\frac{\rmd^2}{\rmd \uu^2} \left[\uD(\uu)-\uD(0)\right]^2  \tilde{I}_1 + \mO(\uD)^3.
\eea
Inserting this into \eqref{eq:OneLoopBeta1}, one obtains a closed expression for the $\beta$ function to one-loop order,
\bea
\label{eq:FRGOneLoopTemp}
-m\partial_m \uD(\uu) = (\epsilon-2\zeta) \uD(\uu) + \zeta \uu \uD'(\uu) - \epsilon \tilde{I}_1 \frac{1}{2}\frac{\rmd^2}{\rmd \uu^2} \left[\uD(\uu)-\uD(0)\right]^2  + \mO(\uD)^3.
\eea

Let us first assume that $\epsilon > 0$ and $\zeta > 0$, i.e. that we are below the critical dimension $d_c = 4$.
As $m\to 0$, one expects $\uD$ to tend to a fixed-point solution $\uD^*$ of \eqref{eq:FRGOneLoopTemp}, where both left- and right-hand sides of \eqref{eq:FRGOneLoopTemp} vanish. 
Comparing the terms in \eqref{eq:FRGOneLoopTemp} which are linear, respectively quadratic, in  $\uD$  we see that the fixed point must scale as $\zeta \sim \epsilon$, $\Delta \sim \epsilon$. Note using \eqref{eq:OneLoopIntegral} that the prefactor $\epsilon \tilde{I}_1 \to const.$ as $\epsilon \to 0$.
The two-loop terms we neglected are of the form \cite{LeDoussalWieseChauve2002}
\bea
\left(\frac{\rmd}{\rmd \uu}\right)^4 \uD(\uu)^3 \epsilon\left(2\tilde{I}_2- \tilde{I}_1^2 \right).
\eea
The combination $\epsilon\left(2\tilde{I}_2- \tilde{I}_1^2 \right) \to const.$ as $\epsilon \to 0$. Hence, the two- and higher-loop contributions in \eqref{eq:FRGOneLoopTemp} are of order $\epsilon^3$, i.e. a factor $\epsilon$ smaller than the one-loop and rescaling contributions in \eqref{eq:FRGOneLoopTemp}. This shows that $\epsilon$ is a small parameter which controls our expansion. Note that this requires an exact cancellation of the terms of order $\epsilon^{-2}$ between the repeated one-loop counterterms ($\tilde{I}_1^2$) and true two-loop terms ($2\tilde{I}_2$) above. While for a standard field theory (e.g. $\phi^4$) this is a simple consequence of factoring diagrams, here it is quite nontrivial and requires a correct evaluation of anomalous two-loop diagrams \cite{LeDoussalWieseChauve2002}.
In the limit of small $\epsilon$, the one-loop terms and the linear rescaling terms in \eqref{eq:FRGOneLoopTemp} are of the same order $\mO(\epsilon)^2$ and provide the leading order, self-consistent approximation to the behaviour of our theory at small $m$.\footnote{The self-consistency assumption is best seen on \eqref{eq:OneLoopInvDelta}. This seemingly formal manipulation is actually quite profound. For very small $m$, when $\uD$ is near the fixed point $\uD_*$, the $\Delta_0$ given by \eqref{eq:OneLoopInvDelta} is not really the nonuniversal, arbitrary microscopic disorder which we put into the model. Rather, the $\Delta_0$ given by \eqref{eq:OneLoopInvDelta} is a consistent continuation of the large-scale behaviour (the universal $\uD$) down to any microscopic scale. In other words, we suppose $\Delta$ to be given exactly by the leading power-law behaviour $\Delta \sim m^{\epsilon-2\zeta}$, and neglect terms which decay faster as $m\to 0$. 
} Terms of two- and higher-loop order in \eqref{eq:FRGOneLoopTemp} are $\mO(\epsilon)^3$ for small $\epsilon$, and give perturbative corrections to the one-loop result.

On the other hand, for $\epsilon < 0$, i.e. $d > d_c = 4$, \eqref{eq:FRGOneLoopTemp} shows that the rescaled disorder $\uD$ flows to $0$ as $m\to 0$. This means that the interface is only pinned microscopically, and flat on large scales \cite{NattermannStepanowTangLeschhorn1992}. In particular, $\zeta=0$. In this case, there is less universality: the renormalized $\Delta$ depends on the microscopic properties of the disorder (see \cite{LeDoussalWiese2008c}, appendix B). At the critical dimension $d=d_c$, we still have $\zeta=0$, but the universal power-laws in $d<d_c$ are replaced by universal logarithms. Higher-loop corrections correspond to logarithms which are subdominant on large scales. This is discussed in detail in \cite{FedorenkoStepanow2003}.

To simplify \eqref{eq:FRGOneLoopTemp}, we still have the freedom to rescale $\uD$ and $\uu$ by constant ($m$-independent) factors. We set $\uD(\uu) =: \tD(\uu)/(\epsilon \tilde{I}_1)$,
or equivalently, in terms of the original disorder,
\bea
\label{eq:RescDelta}
\Delta(u) =& m^{\epsilon - 2\zeta}\uD(m^{\zeta}u) = \frac{1}{\epsilon \tilde{I}_1}m^{\epsilon - 2\zeta}\tD(m^{\zeta}u).
\eea
We obtain the one-loop FRG equation 
\bea
\label{eq:FRGOneLoop}
-m\partial_m \tD(\uu) = (\epsilon-2\zeta)\tD(\uu) + \zeta \uu \tD'(\uu) - \frac{1}{2}\frac{\rmd^2}{\rmd \uu^2}\left[\tD(\uu)-\tD(0)\right]^2.
\eea
This equation was first found for the effective disorder for the static ground state of an elastic interface \cite{Fisher1986a}. In this context, $u(w)$ and $u(w')$ appearing in the definition \eqref{eq:OneLoopDefDelta} of $\Delta$ are the center-of-mass positions of the zero-temperature ground state, for two positions $w$, $w'$ of the harmonic well, but in the same disorder. In the depinning context considered here, $u(w)$ and $u(w')$ are the leftmost metastable states for monotonous, quasi-static driving (``Middleton'' states as discussed in section \ref{sec:InterfaceMonot}). Interestingly, the one-loop equation \eqref{eq:FRGOneLoop} is identical in both situations \cite{NattermannStepanowTangLeschhorn1992,LeschhornNattermannStepanow1997,ChauveGiamarchiLeDoussal2000}. However, starting from  two-loop order the $\beta$ functions for $\Delta$ in the statics and quasi-static depinning become different \cite{ChauveLeDoussalWiese2000a,LeDoussalWieseChauve2002}. Thus the effective disorder (and roughness exponents) in the two situations are similar, but not identical.

\begin{figure}%
\centering
\includegraphics[width=0.6\columnwidth]{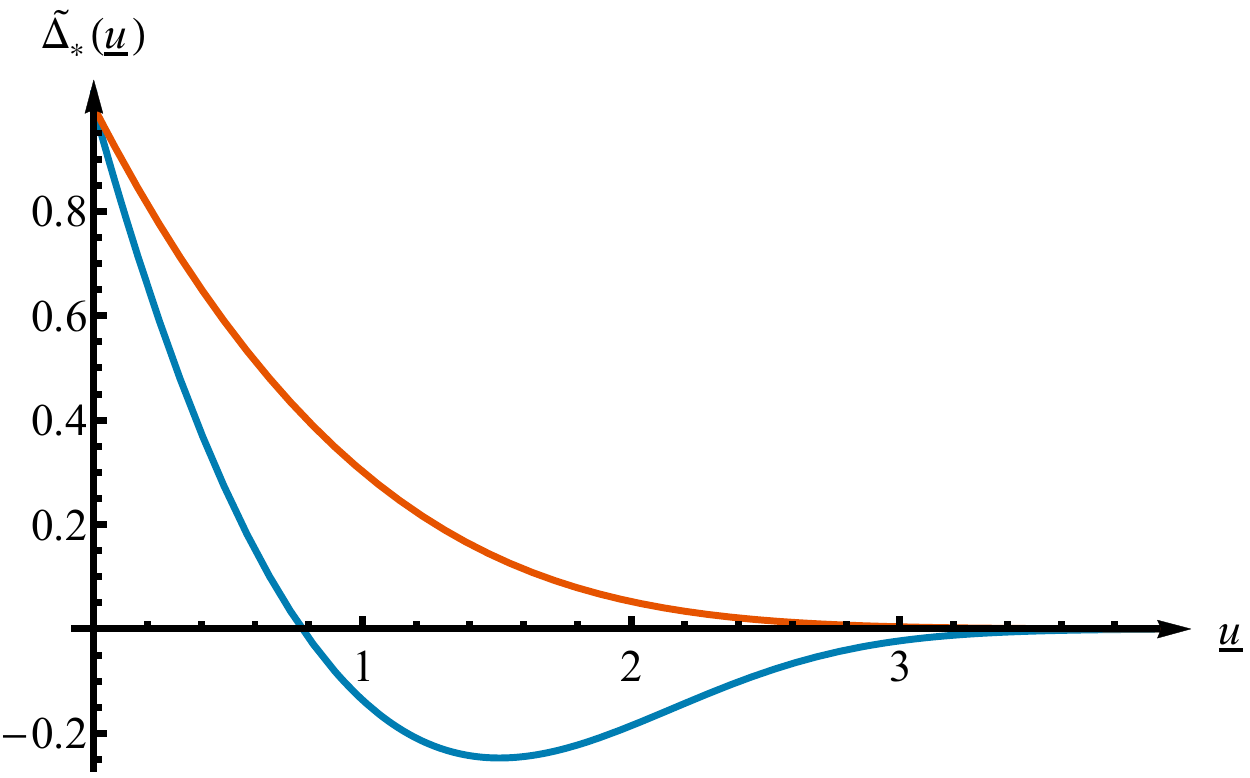}%
\caption{Random-field (top, orange) and random-bond (bottom, blue) fixed points of the one-loop FRG equation \eqref{eq:FRGOneLoop}, normalized to $\Delta(0)=1$.}%
\label{fig:DeltaFPs}%
\end{figure}

As $m\to 0$, one expects $\tD$ to tend to a fixed-point solution $\tD_*$ of \eqref{eq:FRGOneLoop}, where both left- and right-hand sides vanish. 
Such fixed-point solutions with reasonable (i.e. decaying) behaviour of $\tD$ at $\uu\to\infty$ only exist for a few specific values of $\zeta$. They correspond to distinct universality classes \cite{LeDoussalWieseChauve2002,WieseLeDoussal2006}:
\begin{itemize}
	\item A \textit{random-periodic} fixed point for $\zeta=0$. This describes physical systems with periodic disorder, e.g. sliding charge-density waves \cite{NarayanFisher1992}.
	\item A \textit{random-bond} fixed point $\zeta = 0.2083...\epsilon$ \cite{Fisher1986a,NarayanFisher1993} (see figure \ref{fig:DeltaFPs} for a plot). This describes physical systems where the random potential $V(u)$ affecting the interface is short-range correlated, i.e. $\overline{V(u)V(u')}$ decays quickly with $|u-u'|$. Experimentally, this is realized e.g. in diluted Ising antiferromagnets (see \cite{Fisher1986a} and references therein).
	\item A \textit{random-field} fixed point $\zeta = \epsilon/3$ \cite{Fisher1986a,NarayanFisher1993} (see figure \ref{fig:DeltaFPs} for a plot). This describes physical systems where the random force $F(u) = -V'(u)$ is short-range correlated, i.e. $\overline{F(u)F(u')}$ decays quickly with $|u-u'|$. The disorder $V(u)$ itself then has correlations growing linearly in $|u-u'|$. This can be realized, like the random-bond case, in diluted antiferromagnets by applying an additional random magnetic field (see \cite{Fisher1986a} and references therein).
	\item There is also the \textit{BFM} fixed point $\zeta=\epsilon$, and $\Delta''(u)=\sigma \delta(u)$. This is actually an exact fixed point of the FRG flow to any order in $\epsilon$ and for arbitrary driving velocity (due to the exact solution discussed in the previous chapter \ref{sec:BFM}, cf.~also \cite{LeDoussalWiese2011b,DobrinevskiLeDoussalWiese2012}). However, it can only be realized in $d<d_c$ if the bare disorder has infinite-ranged correlations as in the BFM.
\end{itemize}
The striking prediction of the FRG is that for small $m$ (i.e. large systems) near $\epsilon =0$, the effective disorder felt by 
the elastic interface becomes universal and is given by one of these universality classes. This fixes the roughness exponent $\zeta$ and the \textit{universal} shape of the renormalized disorder correlator $\uD$.

For the problem of the zero-temperature equilibrium ground state of an interface, all three universality classes are stable and can be seen e.g. in numerical simulations \cite{Middleton1995}. However, for interface depinning, the random-bond fixed point is unstable and flows to the random-field fixed point. This can be seen numerically \cite{RossoLeDoussalWiese2007}, through a two-loop calculation \cite{ChauveLeDoussalWiese2000a,LeDoussalWieseChauve2002,RossoLeDoussalWiese2007}, or by analyzing the impact of a small but nonzero driving velocity \cite{ChauveGiamarchiLeDoussal2000}.

Here we are interested in non-periodic quasi-static depinning. The relevant fixed point is thus the random-field one\footnote{Actually, there is a one-parameter family of fixed points linked by the rescaling $\tD(x) \to \frac{1}{q^2}\tD(q x)$ which leaves \eqref{eq:FRGOneLoop} invariant. We use this freedom to rescale $u$ and $\tD$, so that $y(0)=1$ in \eqref{eq:OneLoopDeltaResc}.}, given to one-loop order by \cite{Fisher1986a,ChauveGiamarchiLeDoussal2000,ChauveLeDoussalWiese2000a,LeDoussalWieseChauve2002} 
\bea
\nn
&\zeta=\epsilon/3,\quad\quad\quad \tD(\uu) = \frac{\epsilon}{3}y(\uu), \\
\label{eq:OneLoopDeltaResc}
&y(\uu) - \log y(\uu) = 1+\frac{1}{2}\uu^2 \quad\quad
\Leftrightarrow \quad\quad y(\uu) = -W_0\left[-\exp\left(-1-\frac{1}{2}\uu^2\right) \right].
\eea
$W_0$ is the principal branch of the Lambert W function \cite{CorlessEtAl1996}. The resulting $\uD$ is plotted in figure \ref{fig:DeltaFPs}. Around $\uu=0$, we have the expansion
\bea
\nn
y(\uu) = & 1-|\uu|+\frac{\uu^2}{3}-\frac{|\uu|^3}{36}-\frac{\uu^4}{270}+\mO\left(\uu\right)^5, \\
\label{eq:OneLoopDeltaFP}
\tD(\uu) = & \frac{\epsilon}{3} - \frac{\epsilon}{3}|\uu| + \frac{\epsilon}{9}\uu^2 - \mO(\uu)^3.
\eea
Two-loop corrections are computed in \cite{ChauveLeDoussalWiese2000a,LeDoussalWieseChauve2002}. In particular, one observes that the roughness exponent $\zeta$ becomes different at two-loop order. For the statics with random-field disorder, it is still given by $\zeta_{\text{stat}}=\epsilon/3$ (this likely holds to any order \cite{LeDoussalWieseChauve2002}), whereas for quasi-static depinning (in either random-field or random-bond disorder) it becomes $\zeta_{\text{dep}} = \epsilon/3(1+0.143313\epsilon)$ \cite{ChauveLeDoussalWiese2000a,LeDoussalWieseChauve2002}. The universal form of the effective disorder correlator, to one-loop order \eqref{eq:OneLoopDeltaResc} and including two-loop corrections \cite{ChauveLeDoussalWiese2000a,LeDoussalWieseChauve2002} has been confirmed numerically with great accuracy \cite{MiddletonLeDoussalWiese2007,RossoLeDoussalWiese2007}.

\subsubsection{The cusp}
All of the FRG fixed points mentioned above, including the explicit expression \eqref{eq:OneLoopDeltaFP} for the RF fixed point, exhibit a linear cusp at $u=0$.
This feature is shared by the Brownian force model discussed in  chapter \ref{sec:BFM}. The cusp indicates that the effective disorder is microscopically rough (non-differentiable).\footnote{On the other hand, for any smooth (differentiable) effective disorder, $\Delta$ would start as $\Delta(u) = d_1 + d_2 u^2 + \mO(u^4)$ near $u=0$.}

One of the important and amazing results of the functional-renormalization-group approach is the appearance of this non-analyticity at a finite length scale, even when starting from a microscopically smooth disorder \cite{WieseLeDoussal2006}. Starting from a $\Delta$ with $\tD'(0)=0$ and $\tD''(0)$ finite, \eqref{eq:FRGOneLoop} gives a closed flow equation for $\tD''(0)$ \cite{WieseLeDoussal2006}:
\bea
-m\partial_m \tD''(0) = \epsilon \tD''(0) - 3 \left[\tD''(0)\right]^2.
\eea
Its solution $\tD''_m(0)$ diverges at a finite value of $m$ \cite{WieseLeDoussal2006}. This is the \textit{Larkin scale}, at which the interface becomes pinned. Above this scale, $\Delta$ becomes nonanalytic; the blow-up of $\tD''(0)$ translates into a cusp singularity at $u=0$, so that $\tD'(0^+) \neq 0$.

The cusp is closely related to the intermittent avalanche dynamics \cite{BalentsBouchaudMezard1996,ChauveGiamarchiLeDoussal2000,WieseLeDoussal2006,LeDoussalWiese2008c}. When moving the harmonic confinement from $w$ to $w + \delta w$, there is a probability $\approx p \delta w$ (for $\delta w$ small) of triggering an avalanche. It has some typical size $S_t$ independent of $\delta w$. Thus, the leading-order fluctuations of the pinning force increments are \cite{ChauveGiamarchiLeDoussal2000}
\bea
\overline{\left[F(u)-F(u')\right]^2} \propto S_t^2 p \delta w + \mO(\delta w)^2.
\eea
Using \eqref{eq:OneLoopDefDelta} this yields a linear cusp for $\Delta$, $\Delta(\delta w)-\Delta(0) \propto S_t^2 p \delta w$. Thus, the slope of the cusp is proportional to the density of avalanches per unit interval. For smooth motion, we would have $\Delta(\delta w)-\Delta(0) \propto \delta w^2$ and no cusp.

\subsection{One-loop correction to the friction\label{sec:OneLoopEta}}
Analogously to the expansion \eqref{eq:OneLoopPertDelta} for the effective disorder in terms of the bare one, one obtains an expansion for the effective friction coefficient $\eta$ in terms of the bare friction coefficient $\eta_0$ \cite{LeDoussalWieseChauve2002}, 
\bea
\label{eq:OneLoopPertEta}
\eta = \eta_0\left[1 - \Delta''_0(0^+) I_1 + \mO(\Delta_0)^2\right],
\eea
where $I_1$ is given by \eqref{eq:OneLoopIntegral}. As for \eqref{eq:OneLoopPertDelta}, this expansion assumes quasi-static driving $\dw_{x,t}=v=0^+$. Proceeding in the same way as above for $\uD$, let us consider the dimensionless friction coefficient $\ueta$ defined in \eqref{eq:OneLoopScaling2}. We obtain the following RG flow for $\ueta$, at fixed (dimensionful) $\Delta_0, \eta_0$:
\bea
\nn
-m\partial_m \big|_{\Delta_0,\eta_0} \ueta =& -m\partial_m \big|_{\Delta_0,\eta_0} m^{z-2}\eta \\
\nn
=& (2-z)\ueta +m^{z-2}\left\{-m\partial_m \bigg|_{\Delta_0,\eta_0}\eta_0\left[1 - \Delta''_0(0^+) m^{-\epsilon}\tilde{I}_1 + \mO(\Delta_0)^2\right]\right\} \\
\label{eq:OneLoopFlowEtaTemp}
= & \left(2-z\right)\ueta -\epsilon \tilde{I}_1 m^{z-2-\epsilon}\Delta''_0(0^+)\eta_0  + \mO(\Delta_0)^2.
\eea
We can now use \eqref{eq:OneLoopInvDelta} to express $\Delta_0$ in terms of $\uD$. Similarly, we can invert \eqref{eq:OneLoopPertEta} to obtain a formula for $\eta_0$ in terms of $\ueta$:
\bea
\label{eq:OneLoopInvEta}
\eta_0 = & m^{2-z}\ueta \left[1+\tilde{I}_1 \frac{\rmd^2}{\rmd \uu^2}\bigg|_{\uu=0^+} \uD(\uu) + \mO(\uD)^2\right].
\eea
Inserting this and \eqref{eq:OneLoopInvDelta} into \eqref{eq:OneLoopFlowEtaTemp}, we obtain
\bea
\label{eq:OneLoopFlowEta}
-m\partial_m \ueta = & \left[2-z-\epsilon \tilde{I}_1 \uD''(0^+)  + \mO(\uD)^2\right]\ueta = \left[2-z-\tilde{\Delta}''(0^+)+ \mO(\tD)^2\right]\ueta.
\eea

In the limit $m\to 0$, $\tilde{\Delta}$ tends to a fixed point $\tilde{\Delta}_*$ of the FRG equation \eqref{eq:FRGOneLoop} as discussed above, and hence $\tilde{\Delta}''(0^+)$ tends to a constant. Recall that the rescaling \eqref{eq:OneLoopScaling1}, \eqref{eq:OneLoopScaling2} was chosen so that the rescaled dimensionless observables $\uu$, $\underline{t}$, $\ueta$, etc. are all finite and non-zero as $m\to 0$. Considering the flow \eqref{eq:OneLoopFlowEta}, this means that we must choose the dynamical exponent $z$ as
\bea
\label{eq:OneLoopZ}
z = 2-\tilde{\Delta}''_*(0^+),
\eea
since else according to \eqref{eq:OneLoopFlowEta} the rescaled dimensionless friction $\ueta$ will either flow to 0 or to $\infty$ as $m\to 0$. Let us now take $z$ given by \eqref{eq:OneLoopZ} from now on. The actual value of $\ueta$ is not fixed (since \eqref{eq:OneLoopFlowEta}, in contrast to \eqref{eq:FRGOneLoop}, is linear). It depends on the microscopic friction $\eta_0$, but also on the details of the flow of $\tilde{\Delta}$ while it approaches the fixed point $\tilde{\Delta}_*$.

For the random-field fixed point \eqref{eq:OneLoopDeltaFP} relevant at depinning, we have to order $\epsilon$
\bea
\label{eq:OneLoopZEps}
\tilde{\Delta}''_*(0^+) = \frac{2}{9}\epsilon + \mO(\epsilon)^2 \Rightarrow z = 2 - \frac{2}{9}\epsilon + \mO(\epsilon)^2.
\eea

\section{Applying the FRG to avalanches\label{sec:FRGAvalanches}}
One way to extend the discussion above are calculations to higher order, allowing to determine more precisely the roughness exponent and the disorder correlator $\Delta$. Here I will stick to the leading order in $\epsilon$, but compute more general observables than just $\zeta, z$ and $\Delta$. Since the effective rescaled disorder $\uD$ assumes a universal limit as $m \to 0$, we expect similar universality to hold, in principle, for any (rescaled) generating functional $G$ as defined in \eqref{eq:GenFct}. Here I will focus on avalanche observables such as defined in section \ref{sec:BFMAvalanches}.

\subsection{Avalanche size and time scales}
Avalanches, either at steady driving or following a step in the driving force, correspond to small displacements $\delta u_x := u_{x,t=\infty}-u_{x,t=\ti}$ from a quasi-static ``Middleton'' state at $t=\ti$. As discussed in section \ref{sec:BFMUnits} for the BFM, there is a natural scale $S_m$ for the total displacement $S := \int_x \delta u_x $, at which the harmonic confinement becomes as strong as the disorder.\footnote{There are also other ways to define the avalanche size scale $S_m$ \cite{RossoLeDoussalWiese2009,LeDoussalWiese2008c,LeDoussalWiese2011b}. To $\mO(\epsilon)$ they are equivalent to the present discussion.} This constitutes the cutoff for sizes of large avalanches \cite{LeDoussalWiese2008c,LeDoussalWiese2011b,LeDoussalWiese2013}. For an interface in a general effective disorder $\Delta$, this scale is obtained by matching
\bea
\nn
F_{harm} \sim F_{dis} \quad\quad\quad &\Leftrightarrow \quad\quad\quad 
m^2 S \sim \left(\overline{\left[F(u_{\ti}+S)-F(u_{\ti})\right]^2}\right)^{1/2} \\
\nn
& \Leftrightarrow \quad\quad\quad  m^4 S^2  \sim \Delta(S)-\Delta(0).
\eea
Now note that $S$ scales as $S = \int_x \delta u_x = m^{-d-\zeta} \int_{\ulx}\delta \uu_{\ulx} = m^{-d-\zeta}\underline{S}$. Thus, the dimensionless avalanche scale $\underline{S}_m$ satisfies
\bea
\label{eq:OneLoopSmDef}
\underline{S}_m^2 \sim \uD(\underline{S}_m)-\uD(0).
\eea
As discussed in \ref{sec:FRGReview}, in the limit $m\to 0$ the dimensionless $\uD$ tends to a fixed point $\uD_*$ of order $\epsilon$. This means that in the limit of small $\epsilon$, $\underline{S}_m$ shrinks to zero. To obtain its leading-order behaviour in $\epsilon$, one only needs to consider the small-$\uu$ behaviour of $\uD_*(\uu)$ above. We obtain from \eqref{eq:OneLoopSmDef}, that in the limit $m\to 0$, and for small $\epsilon$,
\bea
\nn
\underline{S}_m = -\uD_*'(0^+)  =:  \usigma_* \quad\quad\Leftrightarrow \quad\quad S_m = -m^{-4}\Delta_*'(0^+) =: \sigma_* / m^4.
\eea 
Here we defined $\underline{\sigma}_* := |\uD'_*(0^+)|$.
Note that the presence of the cusp in $\uD_*$ ensures that at scales $S \ll S_m$, the disorder force is dominant\footnote{For microscopically smooth disorder, this holds only for $S \gg S_0$, where $S_0$ is some microscopic cutoff related to the Larkin scale. Below this scale, the elasticity dominates over disorder. This is also the scale at which the cusp appears during the RG flow.}, whereas for $S \gg S_m$, the harmonic confinement is dominant. When including higher orders in $\epsilon$, higher-order terms in $\uD$ (i.e. its curvature) start contributing and a more precise definition of $\underline{S}_m$ is needed.
From $\underline{S}_m$ we also obtain a natural scale for \textit{local} displacements $\delta u_x \sim u_{xm}$, where the \textit{local} harmonic confinement becomes as strong as the \textit{local} disorder. It is given by 
\bea
u_{xm} = m^{-\epsilon} |\Delta'_*(0^+)| \Leftrightarrow \uu_{xm} = \usigma_* = |\uD'_*(0^+)|,
\eea
consistent with $\int_x u_{xm} = S_m$. As in section \ref{sec:BFMUnits}, it is natural to express total and local displacements in units of the corresponding scales. We denote, as in section \ref{sec:BFMUnits}, the resulting fields by primes. We have
\bea
\label{eq:RescUSM}
\uu_{xt} =: \usigma_* \cdot u'_{x't'} \Leftrightarrow \underline{S} = \usigma_* \cdot S' \Leftrightarrow S = \usigma_* m^{-d-\zeta} S' = \sigma_* m^{-4} \cdot S'.
\eea

The dynamical term in the action \eqref{eq:InterfaceAction} is given by
\bea
\int_{xt} \hu_{xt} \eta \partial_t u_{xt} = \int_{\underline{x}\underline{t}} \underline{\hu}_{xt} \ueta \partial_{\underline{t}} \uu_{\underline{x}\underline{t}}.
\eea
We saw in section \ref{sec:OneLoopEta} that for small $m$, the rescaled $\ueta$ tends to a constant $\ueta_*$ (which is, however, non-universal). In analogy to the discussion in section \ref{sec:BFMUnits}, this leads us to the definition of an avalanche time scale
\bea
\tau_m := \eta/m^2 = m^{-z}\ueta_*.
\eea
Correpondingly, expressing times in terms of the scale $\tau_m$ gives
\bea
\label{eq:RescTSM}
\ult = \ueta_*\cdot t' \quad\quad \Leftrightarrow \quad\quad t =  m^{-z} \ueta_* \cdot t' = \tau_m \cdot t'.
\eea

Combining \eqref{eq:RescUSM} and \eqref{eq:RescTSM}, we deduce the corresponding rescaling of the local velocities $\udu_{\ulx \ult}$ and of the response fields $\utu_{\ulx \ult}$ and $\uhu_{\ulx \ult} = -\partial_{\ult} \utu_{\ulx \ult}$
\bea
\label{eq:RescUHSM}
\udu_{\ulx \ult} =: \frac{\usigma_*}{\ueta_*} \cdot \du'_{x't'},\quad\quad
\uhu_{\ulx \ult} =: \frac{1}{\usigma_*\ueta_*} \cdot \hu'_{x't'},\quad\quad 
\utu_{\ulx \ult} =: \frac{1}{\usigma_*}  \cdot \tu'_{x't'}.
\eea
Computing in terms of $\hu'$, $u'$ effectively sets $\tau_m = S_m = 1$, or equivalently $\ueta_* = \usigma_* = 1$. The primed fields here play the same role as the primed fields in the discussion of the BFM in \ref{sec:BFMUnits}.

\subsection{The rescaled action for avalanches to one loop\label{sec:AvalancheAction}}
Inserting the scaling \eqref{eq:OneLoopScaling1}, \eqref{eq:OneLoopScaling2}, \eqref{eq:OneLoopScaling3} into \eqref{eq:IntVelAction}, we obtain the rescaled action without any explicit $m$ dependence
\bea
\label{eq:IntVelActionResc}
S[\du,\tu] = \int_{x,t} \utu_{xt}\left[\ueta_0 \partial_t \udu_{xt} + \udu_{xt} - \nabla_x^2 \udu_{xt} \right] - \frac{1}{2}\int_{x,t,t'} \utu_{xt}\utu_{xt'}\partial_t \partial_{t'}\uD_0(u_{xt}-u_{xt'}).
\eea
Here we wrote $\ueta_0$, $\uD_0$ to stress that we are considering bare parameters. They are given in terms of the rescaled, renormalized parameters through the inversion formulae \eqref{eq:OneLoopInvDelta}, \eqref{eq:OneLoopInvEta}. In the limit $m\to 0$, $\uD$ tends to a cuspy fixed-point as discussed in section \ref{sec:FRGReviewOneLoop}, hence we can also assume the expansion $\uD_0(u)= \uD_0(0) + \uD_0'(0^+)|u| + \frac{1}{2}\uD_0''(0^+)u^2 + ...$. 
 Let us now introduce the avalanche size and time scales given by \eqref{eq:RescUSM}, \eqref{eq:RescTSM}, \eqref{eq:RescUHSM}. 
Remembering that $\uD'_*(0^+) = \usigma_* \sim \epsilon$ is a small parameter, this permits us to expand the action \eqref{eq:IntVelActionResc} as 
\bea
\nn
S[\du,\tu] =& \int_{x',t'} \left\{\tu'_{x't'}\left[\frac{\ueta_0}{\ueta_*} \partial_{t'} \du'_{x't'} + \du'_{x't'} - \nabla_{x'}^2 \du'_{x't'} \right] + \frac{1}{\usigma_*} \left[\frac{\rmd}{\rmd \uu}\big|_{\uu=0^+}\uD_0(\uu)\right]\tu_{x't'}^2\du'_{x't'} \right\} \\
\label{eq:IntVelActionResc2temp}
&  + \frac{1}{2}\left[\frac{\rmd^2}{\rmd \uu^2}\big|_{\uu=0^+}\uD_0(\uu)\right]\int_{x',t'_1,t'_2} \tu'_{x't'_1}\tu'_{x't'_2}\du'_{x't'_1}\du'_{x't'_2} + ...
\eea
From now on we will drop the primes on the fields, and always use these rescaled units tacitly. This also permits us to write again $\uD_0''(0^+)=\frac{\rmd^2}{\rmd \uu^2}\big|_{\uu=0^+}\uD_0(\uu)$, $\uD_0'(0^+) = \frac{\rmd}{\rmd \uu}\big|_{\uu=0^+}\uD_0(\uu)$ without risk of confusion (the primes here are derivatives and unrelated to the primes in \eqref{eq:RescUSM}, \eqref{eq:RescTSM}, \eqref{eq:RescUHSM}):
\bea
\nn
S[\du,\tu] =& \int_{x,t} \left\{\tu_{xt}\left[\frac{\ueta_0}{\ueta_*} \partial_t \du_{xt} + \du_{xt} - \nabla_x^2 \du_{xt} \right] + \frac{\uD'_0(0^+)}{\usigma_*} \tu_{xt}^2\du_{xt} \right\} \\
\label{eq:IntVelActionResc2}
 & + \frac{\uD''_0(0^+)}{2}\int_{x,t_1,t_2} \tu_{xt_1}\tu_{xt_2}\du_{xt_1}\du_{xt_2} + ...
\eea
The first integral in \eqref{eq:IntVelActionResc2} is $\mO(\epsilon)^0$, and corresponds to the action of the BFM \eqref{eq:BFMAction}, which we discussed in the previous chapter. The second one is $\mO(\epsilon)$, and provides the leading correction below the upper critical dimension. Higher terms in the expansion of $\Delta$ around $u=0$ correspond to higher orders in $\epsilon$ (since the scaling \eqref{eq:RescUSM} gives $u\sim \usigma_* \sim \epsilon$), and will be neglected here. Note again, that this holds only due to the scaling \eqref{eq:RescUSM}, \eqref{eq:RescTSM}, which reflects that the typical displacement during an avalanche is small, in the sense $\mO(\epsilon)$. If one considers instead a fixed interface displacement (like for the RG flow of $\Delta$ discussed in section \ref{sec:FRGReview}), this no longer holds, the entire function $\Delta$ is of the same order and needs to be considered.

Finally, we now re-express the bare quantities $\ueta_0$, $\uD'_0(0^+)$ through the renormalized ones, $\ueta$, $\uD$. This splits the action \eqref{eq:IntVelActionResc} into a renormalized action (where the bare couplings are replaced by the renormalized ones), and a series of counterterms.
Explicitly, recall the inversion formulae \eqref{eq:OneLoopInvDelta}, \eqref{eq:OneLoopInvEta} to leading order in $\uD$:
\bea
\nn
\Delta_0(u) = & m^{\epsilon-2\zeta}\left[\uD(\uu) + \tilde{I}_1 \frac{1}{2}\frac{\rmd^2}{\rmd \uu^2} \left[\uD(\uu)-\uD(0)\right]^2  + \mO(\uD)^3 \right],\\
\label{eq:OneLoopInvDeltaEta}
\eta_0 = & m^{2-z}\ueta \left[1+\tilde{I}_1 \frac{\rmd^2}{\rmd \uu^2}\bigg|_{\uu=0^+} \uD(\uu) + \mO(\uD)^2\right].
\eea
In the limit $m\to 0$, following the discussion in section \ref{sec:FRGReviewOneLoop}, we can assume $\uD$ to be given by a (universal) fixed-point function $\uD_*$ and $\ueta$ by a constant (non-universal) value $\ueta_*$. As noted above, with equation \eqref{eq:OneLoopInvDeltaEta} we assume the microscopic theory to be given self-consistently by the universal large-scale behaviour. 

As above, when introducing the avalanche size and time scales given by \eqref{eq:RescUSM}, \eqref{eq:RescTSM}, \eqref{eq:RescUHSM}, only the behaviour of \eqref{eq:OneLoopInvDeltaEta} near $u=0$ is important. We obtain
\bea
\nn
\uD_0(u) = &\uD_*(0) + \tilde{I}_1 \uD'_*(0^+) + \left[\uD'_*(0^+) + 3 \tilde{I}_1 \uD'(_*0^+)\uD''_*(0^+)\right]|\uu| \\
\label{eq:OneLoopInvDelta2}
&+ \left[\frac{1}{2}\uD''_*(0^+) + \frac{3}{2}\uD''_*(0^+)^2 + 2\uD'_*(0^+)\uD'''_*(0^+)\right]|\uu|^2 + \mO(\uu)^3.
\eea
Inserting this into \eqref{eq:IntVelActionResc2} we obtain a splitting of the action $S$ into a tree-level action (with renormalized parameters, which simplify in our choice of scaling), of order $\mO(\epsilon)^0$, and terms of order $\epsilon$ (1-loop terms):
\bea
\nn
S[\du,\tu] =& S_{\tree}[\du,\tu] + S_{\onel}[\du,\tu] + \mO(\epsilon)^2, \\
\label{eq:IntVelActionCT1}
S_{\tree}[\du,\tu] = & \int_{x,t}\left\{ \tu_{xt}\left[\partial_t \du_{xt} + \du_{xt} - \nabla_x^2 \du_{xt} \right] + \tu_{xt}^2 \du_{xt} \right\},\\
\nn
S_{\onel}[\du,\tu] = &\frac{1}{2}\uD''_*(0^+)\int_{x,t_1,t_2}\tu_{xt_1}\tu_{xt_2}\du_{xt_1}\du_{xt_2} +\int_{x,t} \left[\tilde{I}_1 \uD''_*(0^+)\tu_{xt}\partial_t \du_{xt} + 3 \tilde{I}_1 \uD''_*(0^+)\tu_{xt}^2 \du_{xt}\right].
\eea 
\textbf{We note the striking feature that the tree-level action $S_{\tree}$ (or equivalently, the action $S$ to leading order $\mO(\epsilon)^0$), is identical to the BFM action \eqref{eq:BFMAction}, but with rescaled and renormalized parameters}, as first observed in \cite{LeDoussalWiese2011,LeDoussalWiese2013}. In this sense, the mean-field limit of avalanches in short-range disorder is the BFM discussed in chapter \ref{sec:BFM}. However, the avalanche size and time scales are now determined by the effective, renormalized parameters $\ueta_*$, $\usigma_*$ and not by the microscopic ones. Recall that this holds only for displacements $\uu \sim \usigma_* \sim \epsilon$, i.e. only on the scale of a single avalanche. The discussion here makes the intuitive argument in section \ref{sec:ReviewABBM} more precise, and provides a basis for a perturbative expansion. At the upper critical dimension $d=d_c$, $\epsilon=0$, universal power laws translate into universal logarithms. The BFM action then gives the dominant logarithmic behaviour on large scales (again, keeping in mind that the avalanche size and time scales are now determined by the renormalized parameters). One-loop and higher-loop terms give subdominant logarithms (see \cite{FedorenkoStepanow2003,LeDoussalWiese2013} for details).

The one-loop terms are all proportional to $\uD''_*(0^+) \sim \epsilon $. They include both the term of order $\uD''_*(0^+) \sim \epsilon$ in \eqref{eq:IntVelActionResc2}, and counterterms (also of order $\uD''_*(0^+)$) in the inversions \eqref{eq:OneLoopInvDeltaEta}, \eqref{eq:OneLoopInvDelta2}.
According to the usual RG paradigm, order by order in $\uD_0$, the counterterms (and repeated counterterms) cancel the leading divergences in $\epsilon$, giving a \textit{convergent} $\epsilon$ expansion for observables like correlation functions. Physically, they assure that when computing the renormalized $\uD$ using \eqref{eq:IntVelActionCT1}, one obtains indeed $\uD_*$ 
order by order in $\uD$; they adjust the microscopic action so that the effective observables are fixed. In the absence of counterterms, this would only hold at tree level, and loop corrections would modify them.

\subsection{Avalanche observables to one loop\label{sec:FRGAvalanchesObsOneLoop}}
\begin{figure}
         \centering
         \begin{subfigure}[t]{0.9\textwidth}
                 \centering
                 \includegraphics[width=0.2\textwidth]{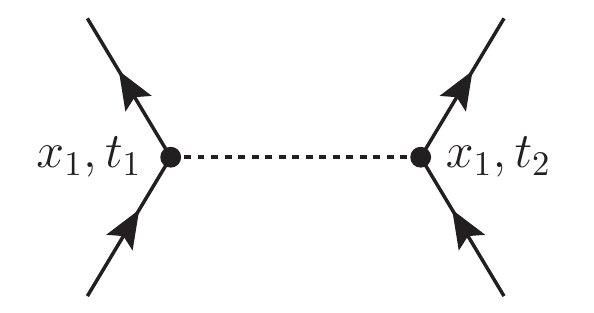}
                 \caption{Additional vertex in one-loop correction to general observables \eqref{eq:OneLoopDeltaG}. Note that it is local in space, i.e. the two points are at the same $x_1$. In Fourier space this gives rise to a loop integral over one free momentum.}
                 \label{fig:OneLoopVertex}
         \end{subfigure}
					\\
         \begin{subfigure}[t]{0.9\textwidth}
                 \centering
                 \includegraphics[width=\textwidth]{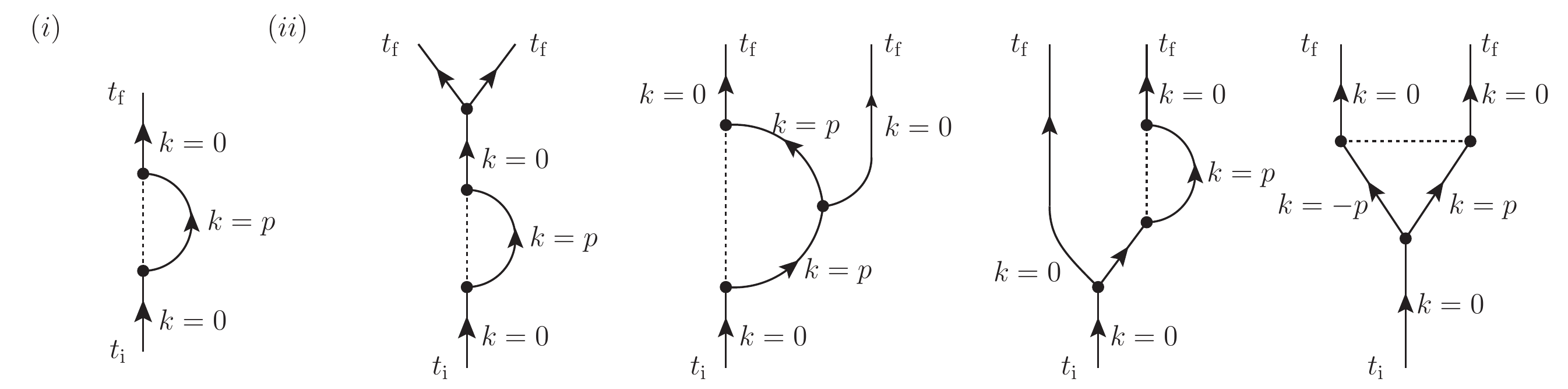}
                 \caption{Examples of one-loop corrections $\delta_1 O$ given by  \eqref{eq:OneLoopDeltaG}, for the correlation functions (i) $O[\du,\tu]=\tu_{\ti}\du_{\tf}$, (ii) $O[\du,\tu]=\tu_{\ti}\du_{\tf}^2$. These are the tree-level diagrams for the corresponding observables (cf.~figure \ref{fig:BFMDiagrammatics}), with one additional quartic $\Delta''$ vertex inserted (cf.~figure \ref{fig:OneLoopVertex}). The locality of the quartic vertex in space gives rise to a loop integral over one free momentum $p$. For the next order $\tu \du^3$ one finds a total of 11 diagrams, see \cite{LeDoussalWiese2013} appendix M.}
                 \label{fig:OneLoopDiags}
         \end{subfigure}%
\caption{Diagrammatics of the one-loop correction \eqref{eq:OneLoopDeltaG}.\label{fig:OneLoopDiagrammatics}}
\end{figure}

Let us now consider a general observable $O[\du,\tu]$ averaged with the action $S$, given to one-loop order by \eqref{eq:IntVelActionCT1}. 
We define the tree-level average of $\mO$ as a path integral with the tree-level action $S_{\tree}$ in \eqref{eq:IntVelActionCT1}
\bea
\overline{O[\du,\tu]}^{\tree} := \int \mD[\du,\tu]e^{-S_{\tree}[\du,\tu]} O[\du,\tu].
\eea
Since the one-loop part of the action is small (of order $\epsilon$), we can expand in it to obtain the average of $O$ to order $\epsilon$:
\bea
\nn
\overline{O[\du,\tu]}^{S} =& \int \mD[\du,\tu]e^{-S_{\tree}[\du,\tu]-S_{\onel}[\du,\tu] + \mO(\epsilon)^2} O[\du,\tu] \\
\nn
=& \overline{O[\du,\tu]}^{\tree} - \int \mD[\du,\tu]e^{-S_{\tree}[\du,\tu]} O[\du,\tu] S_{\onel}[\du,\tu] + \mO(\epsilon)^2 \\
\label{eq:OneLoopGenObsTemp}
= & \overline{O[\du,\tu]}^{\tree} - \uD''_*(0^+) \left[\delta_1 \overline{O[\du,\tu]} + c_\eta (O) + c_\sigma (O) \right] + \mO(\epsilon)^2.
\eea
In the last line we defined the one-loop correction $\delta_1 \overline{O[\du,\tu]}$ and the counterterms $c_\eta$, $c_\sigma$ as
\bea
\label{eq:OneLoopDeltaG}
\delta_1 \overline{O[\du,\tu]} :=& \frac{1}{2}\int_{x,t_1,t_2}\overline{\tu_{xt_1}\tu_{xt_2}\du_{xt_1}\du_{xt_2} O[\du,\tu]}^{\tree}, \\
\nn
c_\eta (O) := & \tilde{I}_1 \int_{x,t} \overline{\left(\tu_{xt}\partial_t \du_{xt}\right) O[\du,\tu]}^{\tree}, \\
\nn
c_\sigma (O) := & 3 \tilde{I}_1 \int_{x,t} \overline{\left(\tu_{xt}^2 \du_{xt}\right) O[\du,\tu]}^{\tree}.
\eea
Diagrammatically, the correction \eqref{eq:OneLoopDeltaG} corresponds to all tree diagrams (with an arbitrary number of cubic $\sigma$ vertices, as discussed in section \ref{sec:BFMSolDeriv}), with an additional quartic $\tu^2\du^2$ vertex inserted (cf.~figure \ref{fig:OneLoopDiagrammatics}).

We can simplify the counterterm $c_\eta$ using the fact that $\int_{x,t}\tu_{xt}\partial_t \du_{xt} = \partial_\eta\big|_{\eta=\sigma=m=1} S_{\dtree}$, where $S_{\dtree}$ is the tree-level (BFM) action \eqref{eq:BFMAction}, prior to scaling $\eta, \sigma, m$ to 1 as in \eqref{eq:IntVelActionCT1}. Similarly, $\int_{x,t}\tu_{xt}^2\du_{xt} = \partial_\sigma\big|_{\eta=\sigma=m=1} S_{\dtree}$. Introducing the \textit{dimensionful} average of $O$
\bea
\overline{O[\du,\tu]}^{\dtree} := \int \mD[\du,\tu]e^{-S_{\dtree}[\du,\tu]} O[\du,\tu],
\eea
we can express the counterterms as
\bea
\label{eq:OneLoopCEta}
c_\eta (O) =& - \tilde{I}_1 \partial_{\eta}\bigg|_{
\begin{subarray}{l} \eta=1 \\ \sigma=1 \\ m =1
\end{subarray}
} \overline{O[\du,\tu]}^{\dtree}, \\
\label{eq:OneLoopCSigma}
c_\sigma (O) =& - 3 \tilde{I}_1 \partial_{\sigma}\bigg|_{
\begin{subarray}{l} \eta=1 \\ \sigma=1 \\ m =1
\end{subarray}
} \overline{O[\du,\tu]}^{\dtree}.
\eea
The simplest way to obtain these counterterms is thus to compute the dimensionless tree-level result $G_{\tree}$, then to restore units using the scaling \eqref{eq:RescTSM}, \eqref{eq:RescUHSM}, \eqref{eq:RescUSM}, then to take the derivatives given above, and finally to set back all units to 1. The interpretation of this procedure is that we are subtracting the corrections leading to a renormalization of $\usigma$, $\ueta$, since in our rescaled units these are fixed at $1$. Renormalizability of our theory means that all divergences as $\epsilon \to 0$ are absorbed in the renormalized couplings $\sigma, \eta$; the expression $\delta_1 \overline{O} + c_\eta + c_\sigma$ in \eqref{eq:OneLoopGenObsTemp} should be finite as $\epsilon \to 0$.

\subsubsection{Mass counterterms}
However, following the procedure outlined above for the computation of the generating functional to one-loop order, one finds that some divergences remain. More precisely, the loop integrals arising in the calculation of \eqref{eq:OneLoopDeltaG} have contributions scaling as $\int_p p^{-2}$, which are strongly divergent near $d=d_c=4$. These correspond to an unphysical correction to the mass (which is actually forbidden in the original action \eqref{eq:InterfaceAction} due to the STS symmetry discussed in section \ref{sec:InterfaceSTS}). There are several equivalent ways to see why they arise and how this apparent non-renormalizability can be ``cured''.
\begin{enumerate}
	\item By considering some diagrams involving $\Delta'''(u)$, which we neglected in \eqref{eq:IntVelActionCT1} since they are subleading in $u$. However, they can become important when the distance $u$ can become arbitrarily large (and not only of the order $u\sim S_m \sim \epsilon$). By ``partial integration'' they give a term $~\sim \Delta''(0^+)$ which cancels precisely these additional divergences (see \cite{LeDoussalWiese2013}, section IV F 2 for details).
	\item By turning on the disorder adiabatically, the divergences are cured by ``boundary terms'' from $t\to-\infty$ (see \cite{LeDoussalWiese2013}, appendix P).
	\item In the simplified velocity theory \eqref{eq:IntVelActionCT1}, we effectively replace $\Delta$ by a ``simplified'' $\Delta_{\text{simpl}}$, the expansion to order $2$ around $0$:
	\bea
	\Delta(u) \to \Delta_{\text{simpl}}(u) := \Delta(0) + \Delta'(0^+)|u| + \frac{1}{2}\Delta''(0^+)u^2.
	\eea
	However, this means that $\Delta_{\text{simpl}}$ grows for large $u$ as $\Delta_{\text{simpl}}(u)\sim u^2$ and the typical pinning force grows as
	\bea
	F(u) \sim \sqrt{F(u)^2} \sim \sqrt{\Delta_{\text{simpl}}(u)} \sim u.
	\eea
	Thus, the pinning force in the simplified theory \eqref{eq:IntVelActionCT1} is (at large distances $u\to\infty$) as strong as the harmonic confinement $-m^2 u$, and gives a correction of order $\epsilon$ to the mass. This is not present in the original theory \eqref{eq:InterfaceAction}, since there the $\Delta$ decays to zero as $u\to\infty$, as it should in any physical model.
\end{enumerate}
The upshot of this discussion is that this quadratic divergence is related to the break-down of the approximations made in \eqref{eq:IntVelActionCT1} at extremely large distances, much larger than the distances covered in an avalanche and hence irrelevant to the discussion here. One finds that this quadratic divergence is removed by some of the terms we dropped when going to \eqref{eq:IntVelActionCT1}. They can be re-introduced in the form of a simple ``mass counterterm'' \cite{LeDoussalWiese2013} analogous to the counterterms $c_\sigma$, $c_\eta$ in \eqref{eq:OneLoopCSigma}, \eqref{eq:OneLoopCEta}
\bea
\label{eq:OneLoopCM}
c_m(O) = \int_p \frac{1}{1+p^2} \partial_{m^2}\bigg|_{
\begin{subarray}{l} \eta=1 \\ \sigma=1 \\ m =1
\end{subarray}
} \overline{O[\du,\tu]}^{\dtree}.
\eea

\subsubsection{Simplifying the one-loop integrals}
To summarize, any avalanche observable $O$, computed 
to order $\epsilon$, is given by adding the mass counterterm \eqref{eq:OneLoopCM} to  \eqref{eq:OneLoopGenObsTemp}
\bea
\label{eq:OneLoopGenObs}
\overline{O[\du,\tu]}^{S} = & \overline{O[\du,\tu]}^{\tree} - \uD''_*(0^+) \left[\delta_1 \overline{O[\du,\tu]} + c_\eta (O) + c_\sigma (O) + c_m(O) \right] + \mO(\epsilon)^2.
\eea
with $\delta_1 \overline{O[\du,\tu]}$ given in \eqref{eq:OneLoopDeltaG}, and the counterterms given in \eqref{eq:OneLoopCEta}, \eqref{eq:OneLoopCSigma}, \eqref{eq:OneLoopCM}.

As an explicit calculation will confirm later for several examples (see e.g. sections \ref{sec:OneLoopDurationsDiag} and \ref{sec:OneLoopShapeTimeDiag}), the correction $\delta_1 \overline{O[\du,\tu]}$ in \eqref{eq:OneLoopGenObs}, defined by \eqref{eq:OneLoopDeltaG}, corresponds to diagrams with one loop, i.e. integrated over one free momentum $p$. It can thus be written as\footnote{From here on we will frequently denote diagrams, counterterms, etc. prior to integration over momenta with hats.}
\bea
\label{eq:OneLoopDeltaGH}
\delta_1 \overline{O[\du,\tu]} = \int_p \hat{\delta}_1 \overline{O[\du,\tu]}(p^2),\quad\quad\quad \int_p := \int\frac{\rmd^d p}{(2\pi)^d}.
\eea
The same holds for the counterterms, where we can rewrite \eqref{eq:OneLoopCEta}, \eqref{eq:OneLoopCSigma}, \eqref{eq:OneLoopCM} explicitly as 
\bea
\label{eq:OneLoopCEtaH}
c_\eta(O) =& \int_p \hat{c}_\eta(O,p^2),\quad\quad\quad \hat{c}_\eta(O,p^2) = -\frac{1}{(1+p^2)^2} \partial_{\eta}\bigg|_{\eta=1} \overline{O[\du,\tu]}^{\dtree}, \\
\label{eq:OneLoopCSigmaH}
c_\sigma(O) =& \int_p \hat{c}_\sigma(O,p^2),\quad\quad\quad \hat{c}_\sigma(O,p^2) = -3\frac{1}{(1+p^2)^2} \partial_{\sigma}\bigg|_{\sigma=1} \overline{O[\du,\tu]}^{\dtree}, \\
\label{eq:OneLoopCMH}
c_m(O) =& \int_p \hat{c}_m(O,p^2),\quad\quad\quad \hat{c}_m(O,p^2) = \frac{1}{1+p^2} \partial_{m^2}\bigg|_{m=1}  \overline{O[\du,\tu]}^{\dtree}.
\eea
Since $\uD''_*(0^+) \sim \epsilon$ in \eqref{eq:OneLoopGenObs}, to leading order in $\epsilon$ it suffices to consider the correction $\delta_1 \overline{O[\du,\tu]} + c_\eta (O) + c_\sigma (O) + c_m(O)$ at the critical dimension, $\epsilon=0$ or $d=4$. Substituting $x := p^2$, we can write
\beq
\label{eq:OneLoopSRLoopInt}
\int_p = \int \frac{\rmd^4 p}{(2\pi)^4} =  \frac{S_4}{16 \pi^4} \frac{1}{2}\int_0^\infty x\, \rmd x =  \frac{1}{8\pi^2}\frac{1}{2}\int_0^\infty x\, \rmd x,
\eeq
where $S_4 = 2\pi^2$ is the volume of the $3$-dimensional hypersphere in $4$ dimensions\footnote{For long-ranged elastic interactions the critical dimension is different and the results will be discussed in section \ref{sec:OneLoopLongRange}.}.

Thus, to leading order in $\epsilon$ we can rewrite \eqref{eq:OneLoopGenObs} as
\bea
\nn
\overline{O[\du,\tu]}^{S} = & \overline{O[\du,\tu]}^{\tree}  \\
\label{eq:OneLoopGenObs2}
& - \frac{1}{16\pi^2}\uD''_*(0^+) \int_0^\infty \rmd x\, x\, \left[\hat{\delta}_1 \overline{O[\du,\tu]}(x) + \hat{c}_\eta (O,x) + \hat{c}_\sigma (O,x) + \hat{c}_m(O,x) \right] + \mO(\epsilon)^2.
\eea
Note from \eqref{eq:OneLoopIntegral}, that for $\epsilon \to 0$ we also have
\bea
 \label{eq:OneLoopSRLoopInt1}
\tilde{I}_1 = \frac{1}{8\pi^2} \frac{1}{\epsilon} + \mO(\epsilon)^0 \quad\quad\quad\Leftrightarrow \quad\quad\quad \lim_{\epsilon \to 0} \epsilon \tilde{I}_1 = \frac{1}{8\pi^2}.
\eea
Inserting this into the definition of the rescaled $\tD$ \eqref{eq:RescDelta}, we see that to leading order in $\epsilon$
\bea
\label{eq:OneLoopValueD1}
\frac{1}{8\pi^2}\uD''_*(0^+) = \tD''_*(0^+),
\eea
where $\tD_*$ is the relevant fixed-point of the one-loop FRG equation \eqref{eq:FRGOneLoop}. Since here we are considering quasi-static depinning, $\tD_*$ is the random-field fixed point given to one-loop order by \eqref{eq:OneLoopDeltaResc}. Its expansion \eqref{eq:OneLoopDeltaFP} gives the universal value
\bea
\label{eq:OneLoopValueD}
\tD''_*(0^+) = \frac{2}{9}\epsilon.
\eea
Inserting \eqref{eq:OneLoopValueD1} into \eqref{eq:OneLoopGenObs2}, we obtain the final simplified expression for one-loop avalanche observables:
\bea
\nn
\overline{O[\du,\tu]}^{S} = & \overline{O[\du,\tu]}^{\tree}  \\
\label{eq:OneLoopGenObsFinal}
&- \tD''_*(0^+)\, \frac{1}{2}\int_0^\infty \rmd x\, x\, \left[\hat{\delta}_1 \overline{O[\du,\tu]}(x) + \hat{c}_\eta (O,x) + \hat{c}_\sigma (O,x) + \hat{c}_m(O,x) \right] + \mO(\epsilon)^2.
\eea
Recall that $\tD''_*(0^+)$ is given by \eqref{eq:OneLoopValueD}, that $\hat{\delta}_1\overline{O}$ is given by \eqref{eq:OneLoopDeltaG} and \eqref{eq:OneLoopDeltaGH}, that the counterterms $\hat{c}$ are given by \eqref{eq:OneLoopCEtaH}, \eqref{eq:OneLoopCSigmaH}, \eqref{eq:OneLoopCMH}, and that $x=p^2$.

\eqref{eq:OneLoopGenObsFinal} permits us to compute, in principle, the correction of leading order in $\epsilon$, for short-range disorder below the critical dimension,  for any observable $O$ expressible through the generating functional $G$. 

This idea of computing loop corrections to avalanche observables, shown to be universal via the FRG, was first applied for the stationary velocity distribution $P(\du)$ discussed in section \ref{sec:BFMABBMStat} (see \cite{LeDoussalWiese2011,LeDoussalWiese2013}). This yielded interesting results, including the correction of the power-law exponent $P(\du)\sim \du^{-a}$. While the mean-field (BFM) result in section \ref{sec:BFMABBMStat} gives $a = 1$ for quasi-static, stationary driving ($v=0^+$), the corrections of order $\epsilon$ in \eqref{eq:OneLoopGenObsFinal} modify this to \cite{LeDoussalWiese2011,LeDoussalWiese2013} 
\bea
a = 1 - \tD''_*(0^+) +\mO(\epsilon)^2 = 1-\frac{2}{9}\epsilon +\mO(\epsilon)^2,
\eea
in accordance with the scaling relation \eqref{eq:OneLoopScalingA}. The correction of $\mO(\epsilon)$ to the scaling form of $P(\du)$ was also obtained \cite{LeDoussalWiese2011,LeDoussalWiese2013}.



Let us now apply the more general formula \eqref{eq:OneLoopGenObsFinal} for a few new, interesting and more complicated examples, which are to be published soon \cite{DobrinevskiLeDoussalWiese2013inpr}. 

\section{Avalanche durations\label{sec:OneLoopDurations}}
\subsection{Diagrammatics\label{sec:OneLoopDurationsDiag}}
Let us consider avalanches following a step in the driving field, at $t=\ti$, $\dw_{xt} = w \delta(t-\ti)$, with total size $w^\rmt = L^d w$. The density of avalanche durations has been defined in section \ref{sec:BFMDuration}, equation \eqref{eq:DefPofT} by taking the limit $w^\rmt \to 0$:
\bea
\nn
P(T) =& \partial_{\tf}\bigg|_{\tf=T} F(\tf),\quad\quad\quad F(\tf) = \lim_{\lambda \to -\infty} F_{\lambda}(\tf), \\
\label{eq:OneLoopDefDur}
F_\lambda(\tf) := & \int \mD[\du,\tu] e^{-S[\du,\tu]} \frac{1}{L^d}\int_{\xxi}\tu_{\xxi,\ti}\exp\left(\lambda \int_{\xf} \du_{\xf,\tf}\right).
\eea
Applying \eqref{eq:OneLoopGenObsFinal}, the expansion of $F_\lambda$ to $\mO(\epsilon)$ below the upper critical dimension is given by
\bea
\label{eq:OneLoopObsDur}
F_\lambda(\tf) = F_\lambda^{\tree}(\tf) - \tD''_*(0^+)\, \frac{1}{2}\int_0^\infty \rmd x\, x\,  \left[\hat{\delta}_1 \overline{O_T[\du,\tu]}(x) + \hc_\eta (O_T,x) + \hc_\sigma (O_T,x) + \hc_m(O_T,x) \right].
\eea
Here we defined the avalanche duration observable $O_T$ by
\bea
\label{eq:OneLoopObsDurFT}
 O_T[\du,\tu] := \frac{1}{L^d}\int_{\xxi}\tu_{\xxi,\ti}\exp\left(\lambda \int_{\xf} \du_{\xf,\tf}\right).
\eea
The mean-field contribution $F_\lambda^{\tree}(\tf)$ is given by the BFM solution discussed in section \ref{sec:BFMDuration}:
\beq
\label{eq:OneLoopDurTree}
F_\lambda^{\tree}(\tf) = \int \mD[\du,\tu] e^{-S_{\tree}[\du,\tu]} \frac{1}{L^d}\int_{\xxi}\tu_{\xxi,\ti}\exp\left(\lambda \int_{\xf} \du_{\xf,\tf}\right) = \tu^{(00)}_{\ti},
\eeq
where $\tu^{(00)}$ is the solution of the BFM instanton equation \eqref{eq:BFMInstanton} with a source constant in space, $\lambda_{xt} = \lambda \delta(t-\tf)$, given by \eqref{eq:ABBMOneTimeInstanton2}:
\beq
\label{eq:OneLoopDurInst0}
\tu_{xt}^{(00)} = \frac{\lambda}{\lambda + (1-\lambda)e^{-(t-\tf)}}.
\eeq
In the limit $\lambda \to -\infty$ \eqref{eq:OneLoopDurTree} then reduces to \eqref{eq:BFMFofT}, as expected.

\begin{figure}%
\includegraphics[width=0.95\columnwidth]{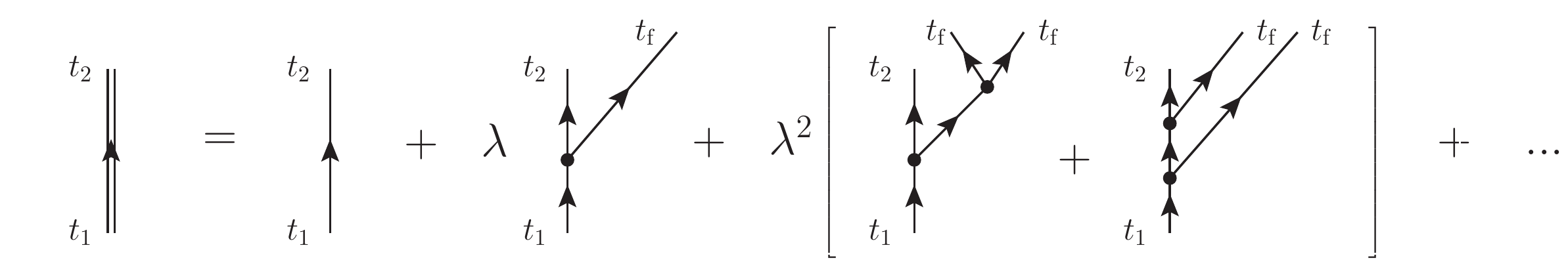}%
\caption{The dressed propagator $\mathbb{R}$ (indicated by a double line), given in \eqref{eq:OneLoopRdr} resums all tree diagrams branching off from a propagator. For the avalanche duration, i.e. for $\tu^{(00)}$ in \eqref{eq:OneLoopDurInst0}, these are all trees terminating at $\tf$ weighted by $\lambda^{\text{\# of vertices at }\tf}$. \label{fig:OneLoopDurRdr}}
\end{figure}

\begin{figure}
         \centering
					\includegraphics[width=0.5\textwidth]{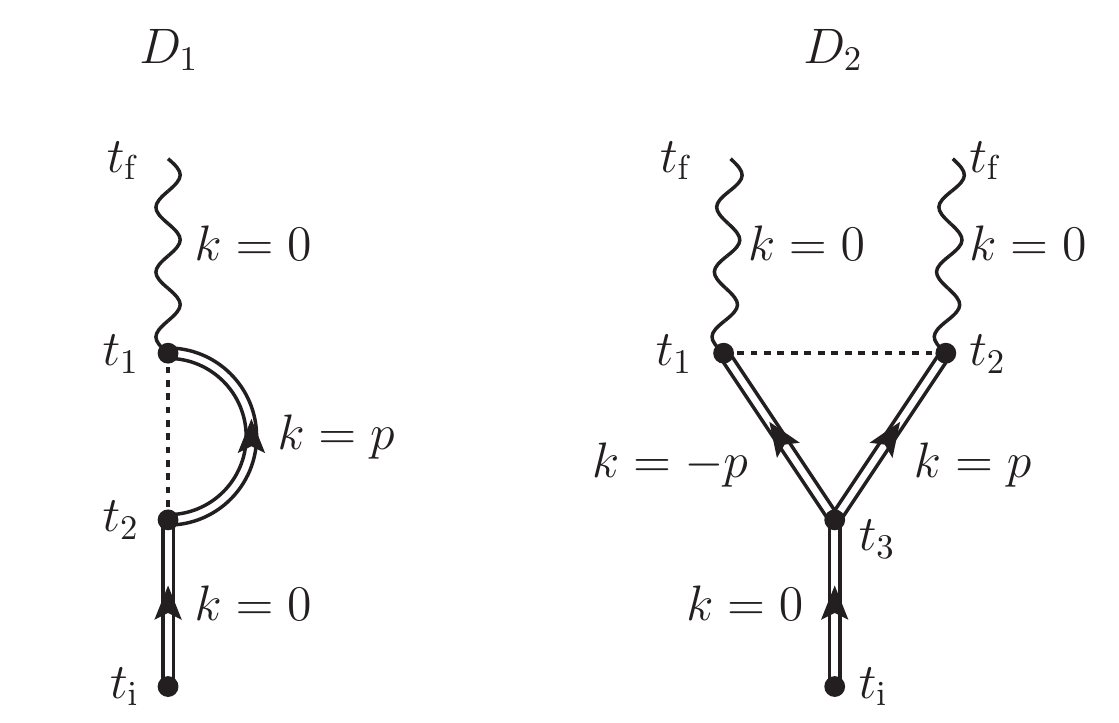}
         \caption{Diagrammatic expression of the one-loop correction to the avalanche duration distribution.}\label{fig:OneLoopDuration}
\end{figure}

The terms of order $\epsilon$ in \eqref{eq:OneLoopObsDur} are the counterterms and the one-loop contribution $\delta_1 \overline{O_T}$.
We will discuss the counterterms later, in section \ref{sec:OneLoopDurationsCT}. For now, let us focus on the one-loop contribution (we set $\ti=0$ without loss of generality):
\bea
\label{eq:OneLoopDurCorr0}
\delta_1 \overline{O_T} = \frac{1}{2}\frac{1}{L^d}\int_{\xxi}\int_{x_1,t_1,t_2}\overline{\tu_{\xxi,\ti=0}\du_{x_1,t_1}\du_{x_1,t_2}\tu_{x_1,t_1}\tu_{x_1,t_2}\exp\left(\int_{\xf}\lambda \du_{\xf,\tf}\right)}^{\tree}.
\eea
The integrand is now again an observable in the tree theory, but a more complicated one than $O_T$. It can be computed using \eqref{eq:BFMSolGenFct}, by introducing the modified source term
\beq
\lambda_{xt} := \lambda \delta(t-\tf) + \mu_1 \delta(t-t_1)\delta^d(x-x_1) + \mu_2 \delta(t-t_2)\delta^d(x-x_1).
\eeq
To compute \eqref{eq:OneLoopDurCorr0}, we now need the solution $\tu_{xt}[\lambda]$ of the instanton equation \eqref{eq:BFMInstanton}
\beq
\label{eq:OneLoopInstanton}
\partial_t \tu_{xt} + \nabla^2 \tu_{xt} - \tu_{xt} + \tu_{xt}^2 = -\lambda_{xt},
\eeq
to order $\mu_1\mu_2$, while keeping all orders in $\lambda$. For this, let us expand $\tu$ in $\mu_1, \mu_2$:
\beq
\label{eq:OneLoopPertAnsatz}
\tu_{xt} = \tu_{xt}^{(00)} + \mu_1 \tu_{xt}^{(10)} + \mu_2 \tu_{xt}^{(01)} + \mu_1\mu_2 \tu_{xt}^{(11)} + \text{higher orders}.
\eeq
Inserting this into \eqref{eq:OneLoopInstanton}, one obtains an hierarchy of linear equations:
\bea
\nn
\partial_t \tu_{xt}^{(00)} + \nabla^2 \tu_{xt}^{(00)} - \tu_{xt}^{(00)} + \left[\tu_{xt}^{(00)}\right]^2 =& -\lambda\delta(t-\tf), \\
\nn
\partial_t \tu_{xt}^{(10)} + \nabla^2 \tu_{xt}^{(10)} - \tu_{xt}^{(10)} + 2\tu_{xt}^{(00)}\tu_{xt}^{(10)} =& -\delta(t-t_1)\delta^d(x-x_1), \\
\nn
\partial_t \tu_{xt}^{(01)} + \nabla^2 \tu_{xt}^{(01)} - \tu_{xt}^{(01)} + 2\tu_{xt}^{(00)}\tu_{xt}^{(01)} =& -\delta(t-t_2)\delta^d(x-x_1), \\
\label{eq:OneLoopPertSeries}
\partial_t \tu_{xt}^{(11)} + \nabla^2 \tu_{xt}^{(11)} - \tu_{xt}^{(11)} + 2\tu_{xt}^{(00)}\tu_{xt}^{(11)} + 2\tu_{xt}^{(10)}\tu_{xt}^{(01)}  =& 0.
\eea
By causality, as discussed in section \ref{sec:BFMSol}, the boundary conditions are
\bea
\label{eq:OneLoopPertBC}
\tu_{x,t > \tf}^{(00)} = 0,\quad\quad\quad \tu_{x,t > t_1}^{(10)} = 0,\quad\quad\quad \tu_{x,t > t_2}^{(01)} = 0, \quad\quad\quad \tu_{x,t > \min t_1,t_2}^{(11)} = 0.
\eea
Using this expansion, the correction $\delta_1 \overline{O_T}$ is expressed as
\bea
\nn
\delta_1 \overline{O_T} =& \frac{1}{2}\partial_{\mu_1}\partial_{\mu_2}\bigg|_{\mu_1=\mu_2=0} \int_{\xxi,x_1,t_1,t_2}\tu_{\xxi,t=0}[\lambda]\tu_{x_1,t_1}[\lambda]\tu_{x_1,t_2}[\lambda]\\
\nn
=& \int_{\xxi,x_1,0<t_1,t_2<\tf}\tu_{\xxi,t=0}^{(11)}\tu_{x_1,t_1}^{(00)}\tu_{x_1,t_2}^{(00)} \\
\label{eq:OneLoopPertRes}
& \quad + \frac{1}{2}\int_{\xxi,x_1,0<t_1<t_2<\tf}\tu_{\xxi,t=0}^{(10)}\tu_{x_1,t_1}^{(01)}\tu_{x_1,t_2}^{(00)} + \frac{1}{2}\int_{\xxi,x_1,0<t_2<t_1<\tf}\tu_{\xxi,t=0}^{(01)}\tu_{x_1,t_2}^{(10)}\tu_{x_1,t_1}^{(00)} ,
\eea
where $\tu[\lambda]$ is the solution of the instanton equation \eqref{eq:OneLoopInstanton}.
The factor 2 in the expansion for $\tu^{(11)}$ is cancelled by the global prefactor $1/2$.
Those configurations not listed above do not contribute, due to the boundary conditions \eqref{eq:OneLoopPertBC} enforced by causality. 
Clearly, the two last terms in \eqref{eq:OneLoopPertRes} are the same and one can write the final result as 
\beq
\label{eq:OneLoopPertRes2}
\delta_1 \overline{O_T} = \int_{x_i,x_1,0<t_1,t_2<\tf}\tu_{x_i,t=0}^{(11)}\tu_{x_1,t_1}^{(00)}\tu_{x_1,t_2}^{(00)} + \int_{x_i,x_1,0<t_1<t_2<\tf}\tu_{x_i,t=0}^{(10)}\tu_{x_1,t_1}^{(01)}\tu_{x_1,t_2}^{(00)}.
\eeq
Let us now proceed to evaluating the functions $\tu^{(..)}$ explicitly. 
\eqref{eq:OneLoopPertSeries} allows us to identify $\tu^{(00)}$ as the space-independent mean-field instanton solution \eqref{eq:OneLoopDurInst0}. 
The three remaining equations are linear, and are easier to treat in Fourier space. Following the conventions in appendix \ref{sec:AppendixNotations}, they read
\bea
\nn
\partial_t \tu_{kt}^{(10)} -\left(1+k^2-2\tu_{t}^{(00)}\right) \tu_{kt}^{(10)} =& -\delta(t-t_1), \\
\nn
\partial_t \tu_{kt}^{(01)} -\left(1+k^2-2\tu_{t}^{(00)}\right) \tu_{kt}^{(01)} =& -\delta(t-t_2), \\
\label{eq:OneLoopInstPertFourier}
\partial_t \tu_{kt}^{(11)} -\left(1+k^2-2\tu_{t}^{(00)}\right) \tu_{kt}^{(11)} + 2\int_p\tu_{pt}^{(10)}\tu_{(k-p),t}^{(01)}  =& 0.
\eea
Note that here we used the fact that $\tu_{xt}^{(00)} = \tu_{t}^{(00)}$ is constant in space. Let us now define the dressed propagator as the Green's function for \eqref{eq:OneLoopInstPertFourier}
\bea
\nn
\left[\partial_t-\left(1+k^2-2\tu_{t}^{(00)}\right) \right] \mathbb{R}_{k,t_1,t} = -\delta(t_1-t), \\
\label{eq:OneLoopRdr}
\Rightarrow \mathbb{R}_{k,t_1,t} = \exp\left[\int_{t_1}^t \rmd s \left(1+k^2-2\tu_{s}^{(00)}\right) \right]\theta(t_1-t).
\eea
In terms of the dressed propagator $\mathbb{R}$, the perturbative expansion \eqref{eq:OneLoopPertSeries} can be expressed succinctly:
\bea
\nn
\tu_{kt}^{(10)}& = \mathbb{R}_{k,t_1,t}, \quad\quad\quad \tu_{kt}^{(01)} = \mathbb{R}_{k,t_2,t}, \\
\tu_{kt}^{(11)}& = 2\int_{p,t_3}\mathbb{R}_{k,t_3,t}\tu_{pt_3}^{(10)}\tu_{(k-p),t_3}^{(01)} = 2\int_{p,t_1,t_2,t_3}\mathbb{R}_{k,t_3,t}\mathbb{R}_{p,t_1,t_3}\mathbb{R}_{k-p,t_2,t_3}.
\eea
The final integral over $x_i$ in \eqref{eq:OneLoopPertRes2} picks out the $k=0$ contribution of $\tu^{(11)}$. 
We thus obtain the following simple expression for the one-loop correction $\hat{\delta}_1 \overline{O_T}$ in \eqref{eq:OneLoopGenObsFinal}:
\bea
\label{eq:OneLoopPertRes3}
\hat{\delta}_1 \overline{O_T}(p^2) = & \hD_1(p^2)+\hD_2(p^2),\\
\label{eq:OneLoopPertResD1}
\hD_1(p^2) := & \int_{0<t_2<t_1<\tf}\mathbb{R}_{k=0,t_2,t=0}\mathbb{R}_{p,t_1,t_2}\tu_{t_1}^{(00)}, \\
\label{eq:OneLoopPertResD2}
\hD_2(p^2) := & \int_{0<t_3<t_1,t_2<\tf}\mathbb{R}_{k=0,t_3,t=0}\mathbb{R}_{p,t_1,t_3}\mathbb{R}_{-p,t_2,t_3}\tu_{t_1}^{(00)}\tu_{t_2}^{(00)}
\eea
Here we introduced a natural splitting in Feynman diagrams $D_1$ and $D_2$, shown in figure \ref{fig:OneLoopDuration}. In these Feynman diagrams, 
the dressed propagator $\mathbb{R}$ is indicated by a double line. It effectively resums all tree diagrams branching off from a propagator (see figure \ref{fig:OneLoopDurRdr}). The tree-level response functions $\tu$ are indicated by wiggly lines, and the disorder vertex $\Delta''(0^+)$, which is local in $x$ and hence leads to an integration in momentum space, is indicated by a dashed line. Note that this dashed line is just a guid to the eye, it does not appear in the formulae \eqref{eq:OneLoopPertResD1}, \eqref{eq:OneLoopPertResD1}.
For the particular case of the avalanche durations distribution, $\tu_s^{(00)}$ is given by \eqref{eq:OneLoopDurInst0}. One can then compute the dressed propagator explicitly:
\bea
\nn
\int_{t_1}^t \rmd s \, \tu_{s}^{(00)} = \ln \frac{(1-\lambda) + e^{t-\tf}\lambda}{(1-\lambda) + e^{t_1-\tf}\lambda}
\Rightarrow \mathbb{R}_{k,t_1,t} = e^{(1+k^2)(t-t_1)}\frac{\left[(1-\lambda) + e^{t_1-\tf}\lambda\right]^2}{\left[(1-\lambda) + e^{t-\tf}\lambda\right]^2}\theta(t_1-t).
\eea
Note again that by \eqref{eq:OneLoopPertRes} it suffices to consider $0<t_1,t<\tf$. From now on we will only consider the limit $\lambda \to -\infty$, as relevant for \eqref{eq:OneLoopDefDur}. In this limit, $\mathbb{R}$ simplifies to
\bea
\mathbb{R}_{k,t_1,t} = e^{(1+k^2)(t-t_1)}\frac{\left(1-e^{t_1-\tf}\right)^2}{\left(1 - e^{t-\tf}\right)^2}\theta(t_1-t).
\eea

Let us now compute the $x$ integral (i.e. momentum integral) in \eqref{eq:OneLoopGenObsFinal}, with the one-loop correction $\hat{\delta}_1 \overline{O_T}$ given in \eqref{eq:OneLoopPertRes3}. We compute the diagrams $D_1$ and $D_2$ individually, in order to have tractable expressions. In order to make the $x$ integrals convergent for each diagram, we will add artifical counterterms diagram by diagram. In section \ref{sec:OneLoopDurationsCT} we will check that these coincide with the physical counterterms in \eqref{eq:OneLoopGenObsFinal}.

\subsubsection{Diagram $D_1$}

Let us first perform the $t_1$ and $t_2$ integral in \eqref{eq:OneLoopPertResD1}. Without loss of generality we set $\ti=0$.
\bea
\nn
\hD_1(p^2) & = \int_{0<t_2<t_1<\tf}\mathbb{R}_{k=0,t_2,t=0}\mathbb{R}_{p,t_1,t_2}\tu_{t_1}^{(00)} 
= \int_{0<t_2<t_1<\tf} \frac{e^{\left(p^2+1\right) \left(t_2-t_1\right)-t_2}
   \left(e^{t_1-\tf}-1\right){}^2}{\left(e^{-\tf}-1\right)^2
   \left(1-e^{\tf-t_1}\right)} \\
\label{eq:OneLoopDurDiag1a}
& = \frac{-p^4+e^{\tf-p^2 \tf}-\left(p^2-1\right) e^{\tf}
   \left[p^2 (\tf-1)-1\right]}{p^4 \left(p^2-1\right)
   \left(e^{\tf}-1\right)^2}.
\eea
From this expression we can see that for large $p$, $\hD_1(p^2) \sim p^{-2}$. This means that the $x$ integral in \eqref{eq:OneLoopGenObsFinal}, $D_1 := \frac{1}{2}\int_0^\infty \rmd x\, x\, \hD_1(p^2=x)$, is divergent. In order to make it convergent, one needs to remove the contributions $\sim p^{-2}, \sim p^{-4}$ at $p\to\infty$ in \eqref{eq:OneLoopDurDiag1a}. We do this by matching them with two counterterms $\sim \frac{1}{1+p^2}, \frac{1}{(1+p^2)^2}$, since these are the forms of loop integrals arising in the physical counterterms \eqref{eq:OneLoopCEtaH}, \eqref{eq:OneLoopCSigmaH}, \eqref{eq:OneLoopCMH}:
\beq
\label{eq:OneLoopDurD1CT}
\hc_1(p^2) := \frac{e^{\tf} (\tf-1)+1}{\left(e^{\tf}-1\right)^2}\frac{1}{1+p^2}+ \frac{e^{\tf} (\tf-2)+2}{\left(e^{\tf}-1\right)^2} \frac{1}{(1+p^2)^2}.
\eeq
These counterterms can also be interpreted diagrammatically \cite{WiesePrivate}: They corresponds to the subtraction of the contribution of $\hD_1$ to the renormalization of $m$, $\sigma$, $\eta$
\bea
\nn
\hc_1 =& \hc_1^{(m)} + 2 \hc_1^{(\sigma)} + \hc_1^{(\eta)} \\
\nn
\hc_1^{(m)}(p^2) = & -\frac{1}{1+p^2}\int_{\ti=0}^{\tf} \rmd t_1\, \Rdr_{k=0,t_1,\ti=0} \tu_{t_1}^{(00)} = \frac{1}{1+p^2}\frac{e^{\tf} (\tf-1)+1}{\left(e^{\tf}-1\right)^2}, \\
\nn
\hc_1^{(\sigma)}(p^2) = & -\frac{1}{(1+p^2)^2} \int_{\ti=0}^{\tf} \rmd t_1\, \Rdr_{k=0,t_1,\ti=0} \left(\tu_{t_1}^{(00)}\right)^2 = \frac{1}{(1+p^2)^2}\frac{1}{1-e^{\tf}},\\
\nn
\hc_1^{(\eta)}(p^2) = & -\frac{1}{(1+p^2)^2} \int_{\ti=0}^{\tf} \rmd t_1\, \Rdr_{k=0,t_1,\ti=0} \partial_{t_1}\tu_{t_1}^{(00)} = \frac{1}{(1+p^2)^2}\frac{e^{\tf} \tf}{\left(e^{\tf}-1\right)^2}.
\eea
We will check later in section \ref{sec:OneLoopDurationsCT} that the counterterms introduced in this way in order to render the diagrams convergent actually coincide with the physical counterterms in \eqref{eq:OneLoopGenObsFinal}. 

We can now evaluate the integral over $x$ in \eqref{eq:OneLoopGenObsFinal}, for diagram 1:
\bea
\nn
D_1 + c_1 & := \frac{1}{2}\int_{0}^\infty \rmd x\, x\, \left[\hD_1(p^2=x) + \hc_1(p^2=x) \right] = \frac{e^\tf}{(e^\tf-1)^2}d_1(\tf), \\
\label{eq:OneLoopDurDiag1b}
	d_1(t) & := \frac{1}{2} 
   \left[2 +\gamma + \log (t)-2 e^{-t}-e^{-t}\text{Ei}(t)-
   t\right].
\eea

\subsubsection{Diagram $D_2$}
For diagram $D_2$ computing all time integrals first gives complicated expressions. We thus first integrate over $t_1$ and $t_2$ in \eqref{eq:OneLoopPertResD2} (again setting without loss of generality $\ti=0$), but leave the $t_3$ integral for later:
\bea
\nn
& \hD_2(p^2) = \int_{0<t_3<t_2,t_1<\tf}\mathbb{R}_{k=0,t_3,t=0}\mathbb{R}_{p,t_1,t_3}\tu_{t_1}^{(00)}\mathbb{R}_{p,t_2,t_3}\tu_{t_2}^{(00)} \\
\nn
& = \int_{0<t_3<t_2,t_1<\tf} \frac{\left(e^{t_1-\tf}-1\right)^2
   \left(e^{t_2-\tf}-1\right)^2 \exp
   \left[\left(p^2+1\right) (t_3-t_1)+\left(p^2+1\right)
   (t_3-t_2)-t_3\right]}{\left(e^{-\tf}-1\right)
   ^2 \left(1-e^{\tf-t_1}\right)
   \left(1-e^{\tf-t_2}\right)
   \left(e^{t_3-\tf}-1\right)^2}\\
\label{eq:OneLoopDurDiag2a}
& = \int_{0<t_3<\tf} \frac{e^{t_3-2 p^2 \tf} \left\{e^{p^2 \tf}
   \left[\left(p^2-1\right) e^{\tf}-p^2
   e^{t_3}\right]+e^{p^2 t_3+\tf}\right\}^2}{p^4
   \left(p^2-1\right)^2 \left(e^{\tf}-1\right)^2
   \left(e^{t_3}-e^{\tf}\right)^2}.
\eea
From this expression we can see that for large $p$, $\hD_2(p^2) \sim p^{-4}$. As for diagram $D_1$, this means that the $x$ integral in \eqref{eq:OneLoopGenObsFinal}, $D_2 := \frac{1}{2}\int_0^\infty \rmd x\, x\, \hD_2(p^2=x)$, is divergent. Now, in order to make it convergent, it suffices to remove the contribution $\sim p^{-4}$ at $p\to\infty$ in \eqref{eq:OneLoopDurDiag2a}. For this we can take a single counterterm $\frac{1}{(1+p^2)^2}$ (again, motivated by the form of the physical counterterms \eqref{eq:OneLoopCEtaH}, \eqref{eq:OneLoopCSigmaH}):
\beq
\label{eq:OneLoopDurD2CT}
\hc_2(p^2) := -\int_{0<t_3<\tf}\frac{e^{t_3}}{\left(e^{\tf}-1\right)^2} \frac{1}{(1+p^2)^2}.
\eeq
Just as for diagram 1, this counterterm can be interpreted diagrammatically \cite{WiesePrivate}: It corresponds to the subtraction of the contribution of $D_2$ to the renormalization of $\sigma$
\bea
\nn
\hc_2(p^2) =& \hc_2^{(\sigma)}(p^2) = -\frac{1}{(1+p^2)^2} \int_{\ti=0}^{\tf} \rmd t_3\, \Rdr_{k=0,t_3,\ti=0} \left(\tu_{t_3}^{(00)}\right)^2 .
\eea

We can now evaluate the integral over $x$ in \eqref{eq:OneLoopGenObsFinal}, for diagram 2:
\bea
\nn
& D_2 + c_2  := \frac{1}{2}\int_{0}^\infty \rmd x\, x\, \left[\hD_2(p^2=x) + \hc_2^{(\sigma)}(x)\right] \\
\nn
	& = \int_{0<t_3<\tf}{\textstyle\frac{2  \left[e^{t_3}
   (\tf-t_3)+e^{\tf}\right]
   \text{Ei}(\tf-t_3)+e^{ t_3} (2 t_3-2
   \tf-1) \text{Ei}(2 \tf-2 t_3)-e^{2
   \tf-t_3} \left[\log
   \left(\frac{\tf-t_3}{2}\right)+\gamma \right]}{2
   \left(e^{\tf}-1\right)^2
   \left(1-e^{\tf-t_3}\right)^2} }\\
	\nn
	& = \frac{e^\tf}{(e^\tf-1)^2} \int_{0}^{\tf}\rmd t'_3\, d_2(t'_3),\\
\label{eq:OneLoopDurDiag2b}
	& d_2(t) := \frac{e^{t}}{2 \left(e^{t}-1\right)^2
  } \left[2 \left(t+e^{t}\right)e^{-2 t}
   \text{Ei}(t)-(2 t+1)e^{-2 t} \text{Ei}(2 t)- \log
   \left(\frac{t}{2}\right)-\gamma \right].
\eea
where in the last line we introduced $t'_3 := \tf-t_3$.
It does not seem possible to evaluate the remaining $t'_3$ integral in closed form. 

\subsection{Counterterms revisited\label{sec:OneLoopDurationsCT}}
For evaluating the momentum integrals in the diagrams $\hD_1$ \eqref{eq:OneLoopDurDiag1a}, $\hD_2$ \eqref{eq:OneLoopDurDiag2a}, we introduced the counterterms $\hc_1$, $\hc_2$ given by \eqref{eq:OneLoopDurD1CT}, \eqref{eq:OneLoopDurD2CT}. Here I will check that they are equal to the physical counterterms appearing in \eqref{eq:OneLoopGenObsFinal}. The latter ones correspond to a subtraction of the loop corrections to the effective $\eta, \sigma, m$ since we are computing at $\eta=\sigma=m=1$. They are given in \eqref{eq:OneLoopCEtaH}, \eqref{eq:OneLoopCSigmaH}, \eqref{eq:OneLoopCMH} in terms of the dimensionful tree-level result. Restoring the dimensions in \eqref{eq:OneLoopDurTree}, \eqref{eq:OneLoopDurInst0} (recall that for the durations distribution we only need the limit $\lambda \to -\infty$ as in \eqref{eq:OneLoopDefDur}) we obtain:
\bea
\label{eq:DurMFwUnits}
F^{\dtree}(\tf) = F_{\lambda=-\infty}^{\dtree}(\tf) =  \frac{m^2/\sigma}{1-e^{(\tf-\ti)m^2/\eta}}.
\eea
Computing the derivatives with respect to $\eta,\sigma,m$, we obtain the counterterms given in \eqref{eq:OneLoopCEtaH}, \eqref{eq:OneLoopCSigmaH}, \eqref{eq:OneLoopCMH}:
\bea
\nn
\hc_\eta(p^2) & = - \frac{1}{(1+p^2)^2} \partial_\eta\bigg|_{m=\sigma=\eta=1}F^{\dtree}(\tf)  = \frac{1}{(1+p^2)^2}  \frac{e^{\ti-\tf} (\tf-\ti)}{\left(1-e^{\ti-\tf}\right)^2}, \\
\nn
\hc_\sigma(p^2) & = - 3 \frac{1}{(1+p^2)^2}  \partial_\sigma\bigg|_{m=\sigma=\eta=1}F^{\dtree}(\tf)  =  3\frac{1}{(1+p^2)^2}  \frac{1}{1-e^{\tf-\ti}}, \\
\label{eq:OneLoopDurwCT}
\hc_{m}(p^2) & = \frac{1}{1+p^2} \frac{1}{2}\partial_m\bigg|_{m=\sigma=\eta=1}F^{\dtree}(\tf)  =  \frac{1}{1+p^2} \frac{e^{\ti-\tf}
   \left(e^{\ti-\tf}+\tf-\ti-1\right)}{\left(1-e^{\ti-\tf}.
   \right)^2}
\eea
Comparing this to the per-diagram counterterms $\hc_1(p^2)$, $\hc_2(p^2)$ we introduced in \eqref{eq:OneLoopDurD1CT}, \eqref{eq:OneLoopDurD2CT} (note that there $\ti=0$), we observe
\bea
\hc_1(p^2) = \hc_{m}(p^2) + \hc_\eta(p^2) + \frac{2}{3}\hc_\sigma(p^2),\quad\quad\quad \hc_2(p^2) = \frac{1}{3}\hc_\sigma(p^2),
\eea
so that in total
\bea
\label{eq:OneLoopDurCTMatch}
\hc_1(p^2) + \hc_2(p^2) = \hc_\eta(p^2)+\hc_\sigma(p^2)+\hc_{m}(p^2),
\eea
as required. This nontrivial check shows that the physical counterterms \eqref{eq:OneLoopCEtaH}, \eqref{eq:OneLoopCSigmaH}, \eqref{eq:OneLoopCMH} indeed cancel all the divergences in the diagrams $D_1$ and $D_2$. In other words, a renormalization of these coupling constants is sufficient to render the theory finite.

\subsection{Final result for avalanche duration distribution\label{sec:OneLoopDurFinal}}
Let us now insert the results \eqref{eq:OneLoopDurDiag1b}, \eqref{eq:OneLoopDurDiag2b} into \eqref{eq:OneLoopPertRes3} and \eqref{eq:OneLoopObsDur}. Due to the matching of the counterterms \eqref{eq:OneLoopDurCTMatch}, one can write the ``cumulative'' one-loop correction \eqref{eq:OneLoopDefDur} to the avalanche durations density as 
\bea
\nn
F(\tf) =&  F^{\tree}(\tf)  - \tD''_*(0^+) \delta_c F(\tf) + \mO(\epsilon)^2, \\
\label{eq:OneLoopDurDeltaF1}
 \delta_c F(\tf) =& (D_1 + c_1) + (D_2 + c_2) = \frac{e^\tf}{(e^\tf-1)^2}\left[d_1(\tf) + \int_0^{\tf} \rmd t'\,d_2(t')\right],
\eea
with $d_1$ and $d_2$ defined in \eqref{eq:OneLoopDurDiag1b}, \eqref{eq:OneLoopDurDiag2b}. We wrote $\delta_c$ here to stress that this is the correction including counterterms. As mentioned above, it does not seem possible to evaluate the remaining $t'$ integral in closed form. 
However, one can simplify the integrand by using $\int \frac{e^{t'}}{(e^{t'}-1)^2}\rmd t'= \frac{1}{e^{t'}-1}$ and performing partial integration. After some additional steps, evaluating parts of the integral, one obtains
\bea
\nn
 \delta_c F(\tf) &= \frac{e^\tf}{(e^\tf-1)^2}\left[d_1(\tf) + \int_0^{\tf} \rmd t'\,d_2(t')\right] \\
\nn
& = \frac{1}{2 \left(e^{\tf}-1\right)^2}\left\{-\text{Ei}(\tf)-e^{\tf} \tf+2
   e^{\tf}+\gamma  e^{\tf}+e^{\tf} \log
   (\tf)-2 \right. \\
	& \quad\quad\left. - 2\int_{0<t'<\tf} \frac{t' e^{\tf-2 t'} \left[e^{t'}
   \text{Ei}(t') \left(e^{t'}-2 e^{\tf}+1\right)-2
   \text{Ei}(2 t')
   \left(e^{t'}-e^{\tf}\right)\right]}{\left(e^{t'}-1
   \right) \left(e^{\tf}-1\right)}\right\}.
\eea
In contrast to \eqref{eq:OneLoopDurDeltaF1}, the integrand is now regular near $t'=0$, and hence the integral is easier to evaluate numerically. We can expand it for small $\tf$:
\bea
\nn
 \delta_c F(\tf) &= -\frac{-4 \log (\tf)-4 \gamma +1+\log (16)}{4 \tf}+\frac{1}{6} \left[-3 \log (\tf)-3 \gamma
   -1+\log (16)\right]\\
\label{eq:OneLoopDurSerF1}
	 & \quad +\frac{1}{288} \tf
   \left[24 \log (\tf)+24 \gamma +27-64 \log (2)\right]+\frac{1}{480} \tf^2 \left[24 \log (2)-7\right] + \mathcal{O}(\tf)^3.
\eea
The series expansion becomes nicer if one factors the mean-field contribution, $F^{\tree}(\tf) = 1/(1-e^{\tf})$:
\bea
\label{eq:OneLoopDurSerF2}
 \delta_c F(\tf) &= \frac{1}{e^\tf-1}\left\{\log \left(\frac{\tf}{2}\right)+\gamma -\frac{1}{4}+\frac{1}{24}
   \tf \left[4 \log (2)-7\right]+\tf^2 \left[-\frac{1}{32}-\frac{\log (2)}{18}\right]+ \mathcal{O}(\tf)^3\right\}.
\eea
In particular, one sees that the contributions with a logarithmic singularity at $\tf=0$ are resummed exactly by the first term, $\frac{1}{e^\tf-1}\log \tf$ (this holds to all orders).

\begin{figure}%
         \centering
         \begin{subfigure}[t]{0.4\textwidth}
                 \centering
                 \includegraphics[width=\textwidth]{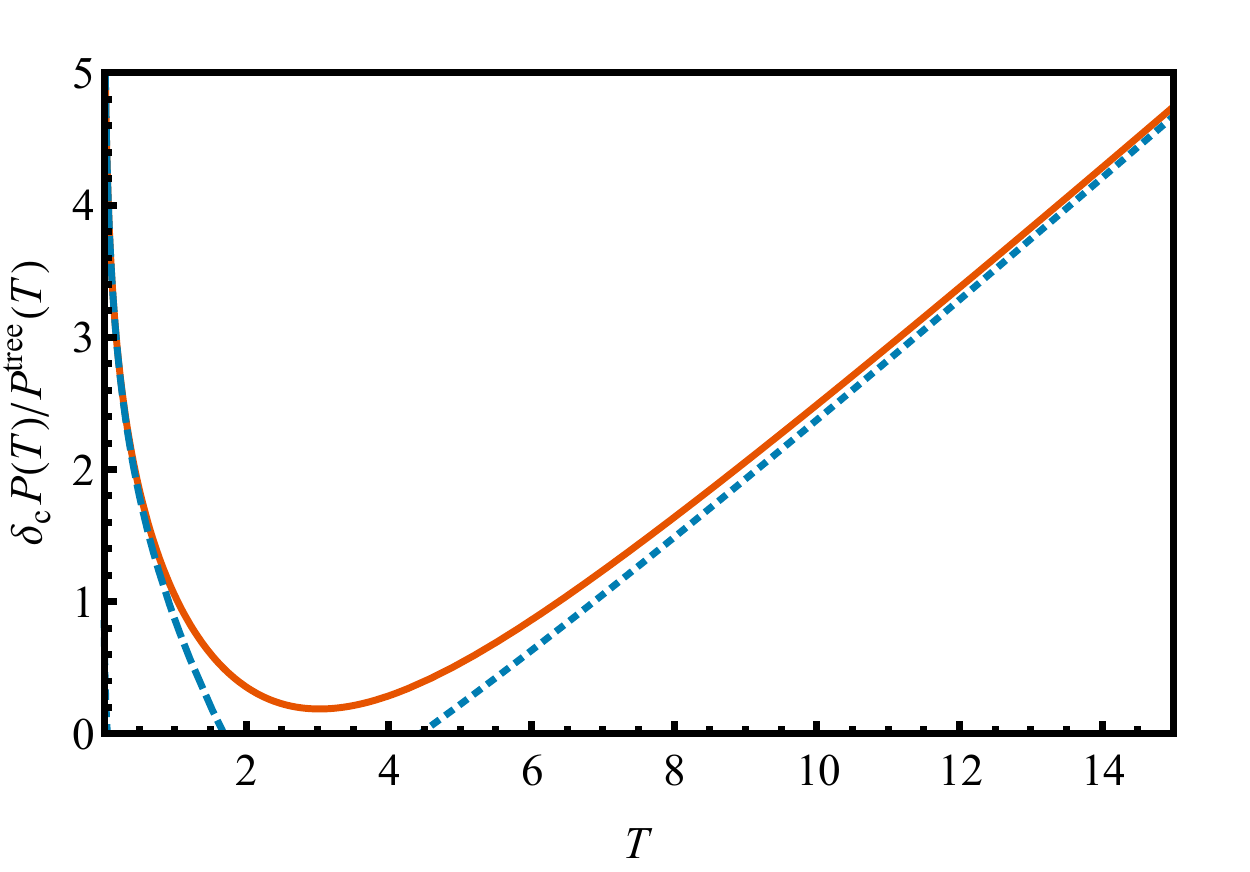}
                 \caption{One-loop correction $\delta_c P(T) / P^{\tree}(T)$ to the avalanche duration distribution, given by \eqref{eq:OneLoopDurPofTCorr}, \eqref{eq:OneLoopDurDeltaF1}. The dashed and dotted lines are the asymptotics for small $T$ (given by \eqref{eq:OneLoopDurPofTCorrSmallT} up to $\mO(T)^2$) and large $T$ (given by \eqref{eq:OneLoopDurPofTCorrLargeT}).}
                 \label{fig:OneLoopDurCorrPlot}
         \end{subfigure}%
         ~ 
         \begin{subfigure}[t]{0.5\textwidth}
                 \centering
                 \includegraphics[width=\textwidth]{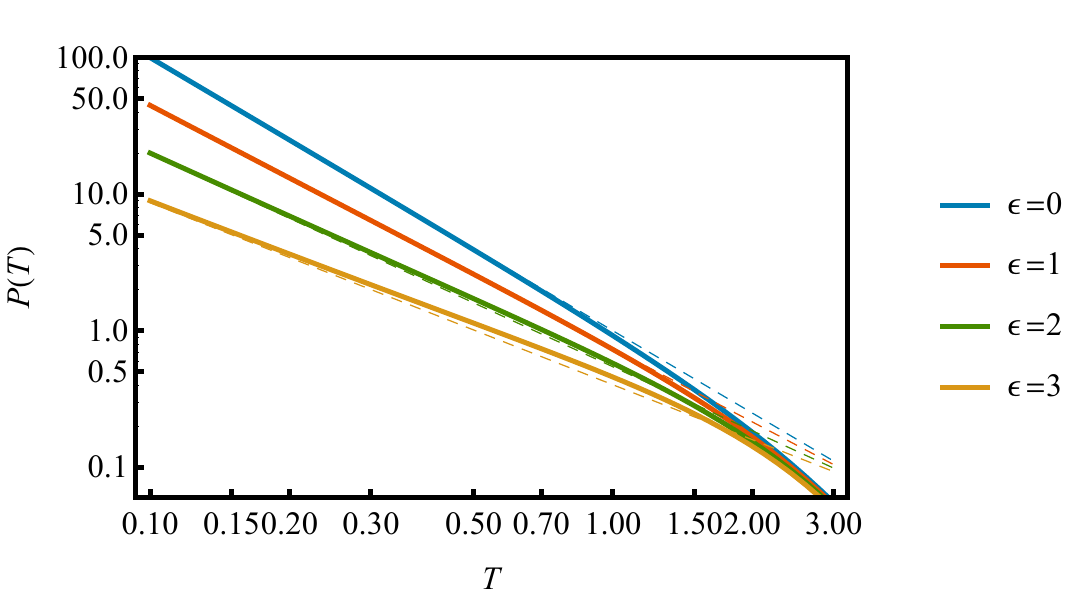}
                 \caption{Avalanche duration $P(T)$ to one-loop order, obtained by ``resumming'' \eqref{eq:OneLoopDurPofTCorr} as $P(T) = P^{\tree}(T) \exp\left[-\frac{2}{9}\epsilon \frac{\delta_c P(T)}{P^{\tree}(T)}\right]$, for various values of $\epsilon$. Dashed lines are the leading power-law terms at $T=0$, with exponents (from top to bottom) $-2$, $-16/9$, $-14/9$,$-4/3$. Observe that the distribution for $\epsilon > 0$ obtains a characteristic bump around $T=1$, which is similar to the one observed for the distribution of avalanche sizes (cf. \cite{LeDoussalWiese2008c,LeDoussalMiddletonWiese2009} and section \ref{sec:OneLoopSize}). }
                 \label{fig:OneLoopDurPlotEps}
         \end{subfigure}%
				                \caption{Avalanche duration distribution $P(T)$ to one-loop order. \label{fig:OneLoopDurPlots}}
\end{figure}

The correction of order $\epsilon$ to the avalanche durations density is given from \eqref{eq:OneLoopDefDur} as a derivative of $\delta_c F$ given in \eqref{eq:OneLoopDurDeltaF1}:
\bea
\label{eq:OneLoopDurPofTCorr}
P(T) = \partial_\tf \bigg|_{\tf = T} F(\tf) = P^{\tree}(T) - \tD''_*(0^+) \delta_c P(T)+ \mO(\epsilon)^2,\quad\quad \delta_c P(T) := \partial_\tf \big|_{\tf = T}\delta_c F(\tf).
\eea
As discussed in \eqref{eq:OneLoopValueD}, \eqref{eq:OneLoopGenObsFinal}, the fixed-point value $\tD''_*(0^+) = \frac{2}{9}\epsilon$ to leading order in $\epsilon$. The tree-level result is $P^{\tree}(T) = \partial_\tf \big|_{\tf = T} F^{\tree}(\tf)$ given by \eqref{eq:BFMPofT}.
The correction $\delta_c P(T)$, as well as the distribution $P(T)$ for several values of $\epsilon$, are shown in figure \ref{fig:OneLoopDurPlots}. One observes that around $T \approx 1$, the distribution $P(T)$ including the loop corrections rises above the leading power-law for small $T$, before the cutoff makes it fall quickly. This ``bump'' is a characteristic sign of non-mean-field behaviour and has also been observed for the avalanche size distribution, analytically and numerically \cite{LeDoussalWiese2008c,LeDoussalWiese2008c,RossoLeDoussalWiese2009,LeDoussalMiddletonWiese2009}.

By taking a derivative of the series \eqref{eq:OneLoopDurSerF1}, \eqref{eq:OneLoopDurSerF2}, one obtains the expansion of $\delta_c P(T)$ for small $T$:
\bea
\nn
\delta_c P(T) = &\frac{-\log (T)-\gamma +\frac{5}{4}+\log (2)}{T^2}-\frac{1}{2 T}+\frac{1}{288} \left[24 \log (T)+24
   \gamma +51-64 \log (2)\right]+\mO\left(T\right)^2 \\
	\label{eq:OneLoopDurPofTCorrSmallT}
	=& \frac{e^T}{(e^T-1)^2}\left\{\left[-\log (T)-\gamma +\frac{5}{4}+\log (2)\right]-\frac{T}{2}+T^2 \left[\frac{9}{32}-\frac{5
   \log (2)}{36}\right]+\mO(T)^3 \right\}.
\eea
Again, in the second series the only term with a logarithmic singularity at $T=0$ is $-\frac{e^T}{(e^T-1)^2}\log (T)$.
Combining this with the tree-level contribution in \eqref{eq:OneLoopDurPofTCorr}, we obtain the leading term for small $T$ in $P(T)$,
\bea
\label{eq:OneLoopDurPofTCorr2}
P(T) = \frac{e^T}{(e^T-1)^2}\left[1 + \tD''_*(0^+) \left(\log (T)+\gamma -\frac{5}{4}-\log (2)\right)  + \mO(\epsilon)^2, \mO(T) \right].
\eea
Interpreting the logarithm in \eqref{eq:OneLoopDurPofTCorr2} as a modification to the exponent, the order-$\epsilon$ correction to the power-law exponent $\alpha$ of the avalanche duration distribution is:
\bea
\label{eq:OneLoopDurPofTExp}
P(T) = T^{-\alpha},\quad\quad \alpha = 2 - \frac{2}{9}\epsilon +\mO(\epsilon)^2.
\eea
This is consistent, to order $\epsilon$, with the scaling relation \eqref{eq:OneLoopScalingAlpha} given the values of $\zeta$ and $z$ in \eqref{eq:OneLoopDeltaFP}, \eqref{eq:OneLoopZEps}. This correction to the exponent should be completely universal for large systems, i.e. independent of the details of the infrared cutoff. The correction \eqref{eq:OneLoopDurPofTCorr} to the entire function $P(T)$ is also universal as $m\to 0$, but assumes that the infrared cutoff is a harmonic well. If the cutoff is instead e.g.~a finite system size, the shape of the distribution for $T \gtrapprox 1$ may be different.

Similarly to the small-$T$ behaviour in \eqref{eq:OneLoopDurPofTExp}, from \eqref{eq:OneLoopDurPofTCorr} we can also obtain the asymptotics of $\delta_c P(T)$ for large $T$:
\bea
\nn
\frac{\delta_c P(T)}{P^{\tree}(T)} =& \frac{1}{2}\left(T - \log(T) - 3-\gamma\right) - \int_0^\infty \rmd t'\, d_2(t'), \\
\label{eq:OneLoopDurPofTCorrLargeT}
\int_0^\infty \rmd t'\, d_2(t') =& -0.31555221566...
\eea

\section{Average avalanche shape, at fixed avalanche duration\label{sec:OneLoopShapeTime}}
Let us trigger, as above, an avalanche by a step of the driving field at $\ti=0$, $\dw(t) = w \delta(t)$, with total size $w^\rmt = L^d w$. The average avalanche shape, at fixed avalanche duration, has been defined in section \ref{sec:BFMShape}, equation \eqref{eq:DefShapeFixedT} by taking the limit $w^\rmt \to 0$:
\bea
\nn
\mfs(\tm,T) =& \frac{1}{P(T)}\partial_{\tf}\bigg|_{\tf=T} F^\mfs(\tm,\tf),\quad\quad\quad F^\mfs(\tm,\tf) = \lim_{\lambda\to-\infty} F^\mfs_\lambda(\tm,\tf), \\
\label{eq:OneLoopDefShp}
F^\mfs_\lambda(\tm,\tf) := & \int \mD[\du,\tu] e^{-S[\du,\tu]} \frac{1}{L^d}\int_{\xxi}\tu_{\xxi,\ti}\int_{\tm}\du_{\xm,\tm}\exp\left(\lambda \int_{\xf} \du_{\xf,\tf}\right).
\eea
Applying \eqref{eq:OneLoopGenObsFinal}, the expansion of $F^\mfs$ to $\mO(\epsilon)$ below the upper critical dimension is given by
\bea
\nn
F^\mfs(\tm,\tf) =& \left(F^\mfs\right)^{\tree}(\tm,\tf) - \tD''_*(0^+)\,\delta_c F^\mfs(\tm,\tf), \\
\label{eq:OneLoopObsShp}
 \delta_c F^\mfs(\tm,\tf) = & \lim_{\lambda \to -\infty} \frac{1}{2}\int_0^\infty \rmd x\, x\,  \left[\hat{\delta}_1 \overline{O_\mfs[\du,\tu]}(x) + \hc_\eta (O_\mfs,x) + \hc_\sigma (O_\mfs,x) + \hc_m(O_\mfs,x) \right].
\eea
Here we defined the avalanche shape observable $O_\mfs$ by
\bea
\label{eq:OneLoopObsShpFT}
 O_\mfs[\du,\tu] := \frac{1}{L^d}\int_{\xxi}\tu_{\xxi,\ti}\int_{\xm}\du_{\xm,\tm}\exp\left(\lambda \int_{\xf} \du_{\xf,\tf}\right).
\eea
The mean-field contribution $\left(F^\mfs\right)^{\tree}(\tm,\tf)$ is given by the BFM solution discussed in section \ref{sec:BFMShape}:
\beq
\label{eq:OneLoopShpTree}
\left(F^\mfs\right)^{\tree}(\tm,\tf) = \tu^{(1)}_{t=0} = \frac{e^{\tm}(e^{\tf-\tm}-1)^2}{(e^{\tf}-1)^2},
\eeq
where $\tu^{(1)}$ is given by \eqref{eq:ABBMShapeLinPertSol}. Applying \eqref{eq:OneLoopDefShp}, i.e. taking a derivative with respect to $\tf$, and dividing by the tree-level durations density $P(T)$ in \eqref{eq:BFMPofT}, one recovers the tree-level shape $\mfs^{\tree}(\tm,T)$ given in \eqref{eq:ABBMShapewFixedT}, for $w^\rmt=0^+$:
\bea
\label{eq:OneLoopShpTTree}
\mfs^{\tree}(\tm,T) = \frac{4\sinh \left(\frac{\tm}{2}\right)\sinh \left(\frac{T-\tm}{2}\right)}{\sinh \left(\frac{T}{2}\right)}.
\eea
In order to obtain the average avalanche shape to order $\epsilon$ using \eqref{eq:OneLoopDefShp}, \eqref{eq:OneLoopObsShp}, we need the contributions of order $\epsilon$ both to $F^\mfs$ and to $P(T)$. 

The durations density $P(T)$ was computed to order $\epsilon$ in \eqref{eq:OneLoopDurPofTCorr}:
\beq
P(T) = P^{\tree}(T)-\tD''_*(0^+) \delta_c P(T) + \mO(\epsilon)^2,
\eeq
where $\delta_c P(T)$ is given in \eqref{eq:OneLoopDurPofTCorr}, \eqref{eq:OneLoopDurDeltaF1}.

To order $\epsilon$, the average avalanche shape \eqref{eq:OneLoopDefShp} is thus given as 
\bea
\nn
\mfs(\tm,T) = & \mfs^{\tree}(\tm,T) - \tD''_*(0^+) \delta_c \mfs(\tm,T) + \mO(\epsilon)^2, \\
\label{eq:OneLoopShapeCorr}
\delta_c \mfs(\tm,T) = & \frac{\partial_{\tf}\big|_{\tf=T} \delta_c F^\mfs(\tm,\tf)}{P^{\tree}(T)} - \mfs^{\tree}(\tm,T)\frac{\delta_c P(T)}{P^{\tree}(T)}.
\eea

\subsection{Diagrammatics\label{sec:OneLoopShapeTimeDiag}}
\begin{figure}
         \centering
					\includegraphics[width=\textwidth]{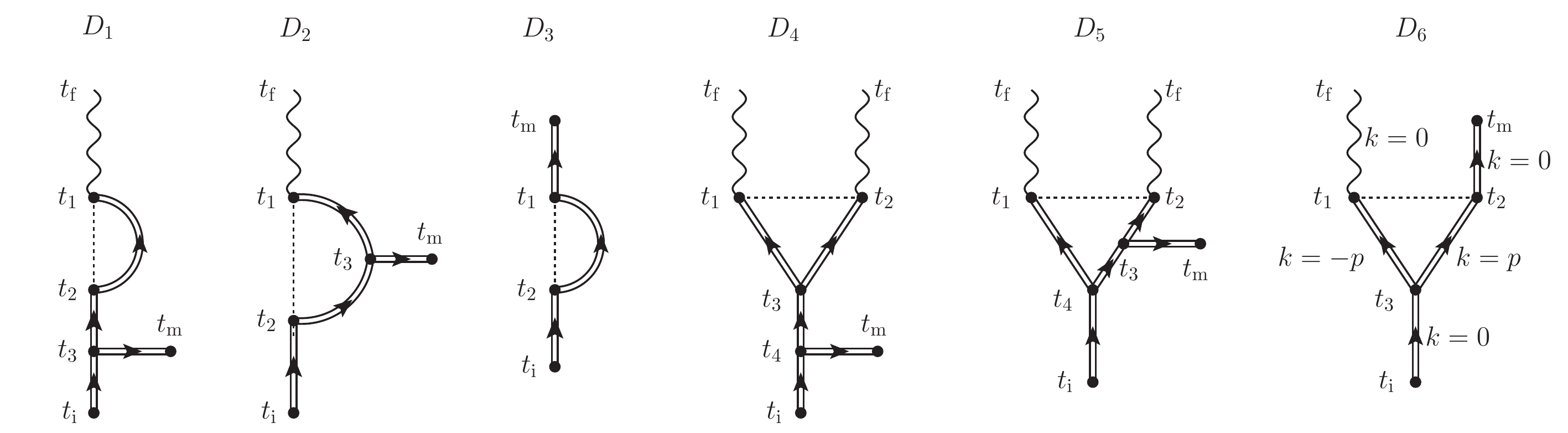}
         \caption{Diagrams contributing to the one-loop correction to the mean temporal avalanche shape. For clarity, the internal momenta are indicated only on diagram $D_6$. \label{fig:OneLoopShapeDiags}}
\end{figure}

The one-loop contribution $\hat{\delta}_1 \overline{O_\mfs}$ in \eqref{eq:OneLoopObsShp} is defined through \eqref{eq:OneLoopDeltaG}, \eqref{eq:OneLoopDeltaGH}, with the avalanche shape observable $O_\mfs$ given in \eqref{eq:OneLoopObsShpFT}.
For computing $\hat{\delta}_1 \overline{O_\mfs}$ we can build a perturbative treatment similar to what was done in section \ref{sec:OneLoopDurationsDiag} for the avalanche duration. However, it is more efficient to use a diagrammatic technique here from the start. Comparing to the avalanche duration observable $O_T$ in \eqref{eq:OneLoopObsDurFT}, the shape observable $O_\mfs$ in \eqref{eq:OneLoopObsShpFT} has an additional factor $\du_{\tm}$. This means, that one-loop diagrams contributing to the correction \eqref{eq:OneLoopDeltaG} for the shape contain one additional vertex at $t=\tm$ (apart from the arbitrary number of vertices at $t=\tf$ generated by the $e^{\lambda \du_\tf}$ part of $O_\mfs$), where one of the response functions can terminate. Since there is still only one $\tu_\ti$ factor in $O_\mfs$, just as in $O_T$, all diagrams still start at a single vertex at $\ti$. 
 Overall, we obtain the following diagrams (see also figure \ref{fig:OneLoopShapeDiags}; the diagrammatic conventions are as in section \ref{sec:OneLoopDurationsDiag}):
\bea
\nonumber
\hD_1(p^2) &= \int_{\ti}^{\tm} \rmd t_3 \, \Rdr_{k=0,t_3,\ti} \Rdr_{k=0,\tm,t_3}  \int_{t_3}^{\tf} \rmd t_2 \, \Rdr_{k=0,t_2,t_3} \int_{t_2}^{\tf} \rmd t_1 \, \Rdr_{p,t_1,t_2} \tu_{t_1}^{(00)} \\
\label{eq:OneLoopShpHD1} 
&= \int_{\ti}^{\tm} \rmd t_3 \, \Rdr_{k=0,\tm,t_3}  \int_{t_3}^{\tf} \rmd t_2 \, \Rdr_{k=0,t_2,\ti} \int_{t_2}^{\tf} \rmd t_1 \, \Rdr_{p,t_1,t_2} \tu_{t_1}^{(00)} \\
\nonumber
\hD_2(p^2) &= \int_{\ti}^{\tm} \rmd t_2 \, \Rdr_{k=0,t_2,\ti} \int_{t_2}^{\tm} \rmd t_3 \, \Rdr_{p,t_3,t_2} \Rdr_{k=0,\tm,t_3} \int_{t_3}^{\tf} \rmd t_1 \, \Rdr_{p,t_1,t_3} \tu_{t_1}^{(00)} \\
\label{eq:OneLoopShpHD2} 
 &= \int_{\ti}^{\tm} \rmd t_2 \, \Rdr_{k=0,t_2,\ti} \int_{t_2}^{\tm} \rmd t_3 \, \Rdr_{k=0,\tm,t_3} \int_{t_3}^{\tf} \rmd t_1 \, \Rdr_{p,t_1,t_2} \tu_{t_1}^{(00)} \\
\label{eq:OneLoopShpHD3} 
\hD_3(p^2) &= \int_{\ti}^{\tm} \rmd t_2 \, \Rdr_{k=0,t_2,\ti} \int_{t_2}^{\tm} \rmd t_1 \, \Rdr_{k=0,\tm,t_1} \Rdr_{p,t_1,t_2} \\
\label{eq:OneLoopShpHD4} 
\hD_4(p^2) &= \int_{\ti}^{\tm} \rmd t_4 \, \Rdr_{k=0,t_4,\ti}  \Rdr_{k=0,\tm,t_4}  \int_{t_4}^{\tf} \rmd t_3 \, \Rdr_{k=0,t_3,t_4} \int_{t_3}^{\tf} \rmd t_2 \, \Rdr_{p,t_2,t_3} \tu_{t_2}^{(00)} \int_{t_3}^{\tf} \rmd t_1 \, \Rdr_{p,t_1,t_3}\tu_{t_1}^{(00)} \\
\label{eq:OneLoopShpHD5} 
\hD_5(p^2) &= \int_{\ti}^{\tm} \rmd t_4 \, \Rdr_{k=0,t_4,\ti} \int_{t_4}^{t} \rmd t_3 \, \Rdr_{p,t_3,t_4} \Rdr_{k=0,\tm,t_3}  \int_{t_3}^{\tf} \rmd t_2 \,  \Rdr_{p,t_2,t_3}\tu_{t_2}^{(00)} \int_{t_4}^{\tf} \rmd t_1 \, \Rdr_{p,t_1,t_4}\tu_{t_1}^{(00)} \\
\label{eq:OneLoopShpHD6} 
\hD_6(p^2) &= \int_{\ti}^{\tm} \rmd t_4 \, \Rdr_{k=0,t_4,\ti} \int_{t_4}^{\tf} \rmd t_1 \, \Rdr_{p,t_1,t_4} \tu_{t_1}^{(00)} \int_{t_4}^{t} \rmd t_2 \, \Rdr_{p,t_2,t_4} \Rdr_{k=0,\tm,t_2}
\eea

To simplify the diagrams $\hD_1$ and $\hD_2$ from the first to the second line, respectively, we used the relationship
\beq
\label{eq:RdrProduct}
\Rdr_{k,t_3,t_2}\Rdr_{k,t_2,t_1} = \Rdr_{k,t_3,t_1},
\eeq
which follows easily from the definition of the dressed response function \eqref{eq:OneLoopRdr}.
In contrast to the calculation for the avalanche duration distribution in section \ref{sec:OneLoopDurationsDiag}, these diagrams now contribute with different combinatoric prefactors. There is a global prefactor $1/2$, coming from the expansion in \eqref{eq:OneLoopDeltaG}. For each diagram, one has in addition the following factors
\begin{itemize}
\item Diagram 1: The $\sigma$ vertex at $t_3$ gives a factor $2$. Since the ordering of $t_1$ and $t_2$ was arbitrary in \eqref{eq:OneLoopDeltaG}, but is fixed in \eqref{eq:OneLoopShpHD1}, one must add an additional factor $2$.
\item Diagram 2: The $\sigma$ vertex at $t_3$ gives a factor $2$. Due to the ordering of $t_1$ and $t_2$ (cf. diagram 1), one has an additional factor $2$.
\item Diagram 3: Due to the ordering of $t_1$ and $t_2$ (cf. diagram 1), one has an additional factor $2$.
\item Diagram 4: Each of the $\sigma$ vertices (at $t_3$ and $t_4$) gives a factor $2$.
\item Diagram 5: Each of the $\sigma$ vertices (at $t_3$ and $t_4$)  gives a factor $2$. Since the $t_3$ vertex can be either on the $t_1-t_4$ leg or on the $t_2-t_4$ leg, one has an additional factor of $2$. 
\item Diagram 6: The $\sigma$ vertex at $t_4$ gives a factor $2$. Since the dressed response going to $\tm$ can connect to either $t_1$ or $t_2$, there is an additional factor $2$.
\end{itemize}
Thus, the total one-loop correction $\hat{\delta}_1 \overline{O_\mfs}$ in \eqref{eq:OneLoopObsShp} can be expressed as
\bea
\label{eq:OneLoopShpDiagSum}
\hat{\delta}_1 \overline{O_\mfs}(p^2) = 2 \hD_1(p^2) + 2 \hD_2(p^2) + \hD_3(p^2) + 2\hD_4(p^2) + 4 \hD_5(p^2) + 2 \hD_6(p^2).
\eea
Note that after applying the relation \eqref{eq:RdrProduct}, the integrand in the expressions for $\hD_1$ and $\hD_2$ \eqref{eq:OneLoopShpHD1}, \eqref{eq:OneLoopShpHD2}  is the same. Thus, their sum simplifies to
\beq
\label{eq:OneLoopShpDiag12a}
\hD_1(p^2) + \hD_2(p^2) = \int_{\ti}^{\tm} \rmd t_3 \, \Rdr_{k=0,t,t_3}  \int_{t_3}^{\tf} \rmd t_1 \, \int_{\ti}^{t_1} \rmd t_2 \, \Rdr_{k=0,t_2,\ti} \Rdr_{p,t_1,t_2} \tu_{t_1}^{(00)}.
\eeq

From now on, we set $\ti=0$ without loss of generality, and take the limit $\lambda\to-\infty$. We will now perform the momentum integral, i.e. the integral over $x=p^2$ in \eqref{eq:OneLoopObsShp}. This will be done diagram by diagram, as for the avalanche durations distribution, in order to keep the expressions manageable. This will require adding diagramwise counterterms in order to make the $x$ integral convergent. We will check in section \ref{sec:OneLoopShapeTimeCT} that these diagramwise counterterms are identical to the physical counterterms in \eqref{eq:OneLoopObsShp}.

\subsubsection{Diagrams $D_1 + D_2$}
We now proceed by evaluating diagrams $D_1 + D_2$. Let us first evaluate the integrals over $t_2$ and $t_1$ in \eqref{eq:OneLoopShpDiag12a}.
\bea
\nn
&\hD_1(p^2) + \hD_2(p^2) = \int_{\ti}^{\tm} \rmd t_3 \int_{t_3}^{\tf} \rmd t_1 \int_{\ti}^{t_1} \rmd t_2 \, \frac{\left(e^\tm-e^{\tf}\right)^2
   \left(e^{t_1}-e^{\tf}\right) e^{p^2
   (t_2-t_1)-\tm+t_3}}{\left(e^{\tf}-1\right)^2
   \left(e^{t_3}-e^{\tf}\right)^2} \\
\label{eq:OneLoopShpDiag12b}
	& = \int_{\ti}^{\tm} \rmd t_3 {\textstyle\frac{e^{t_3-\tm} \left(e^\tm-e^{\tf}\right)^2}{p^2 \left(e^{\tf}-1\right)^2
   \left(e^{t_3}-e^{\tf}\right)^2} \left[e^{-p^2
   t_3}
   \left(\frac{e^{\tf}}{p^2}-\frac{e^{t_3}}{p^2-1}\right)+\frac{e^{\tf-p^2 \tf}}{p^2 \left(p^2-1\right)}+t_3
   e^{\tf}-e^{t_3}+e^{\tf}-e^{\tf}
   \tf\right]}
\eea
As in section \ref{sec:OneLoopDurationsDiag}, we introduce artificial counterterms $\sim \frac{1}{1+p^2}, \frac{1}{(1+p^2)^2}$ in order to remove the contributions $\sim p^{-2}$, $\sim p^{-4}$ for large $p$ in \eqref{eq:OneLoopShpDiag12b}:
\bea
\nn
\hc_{12}(p^2) = &\frac{e^{t_3-\tm} \left(e^\tm-e^{\tf}\right)^2
   \left[e^{\tf}
   (-t_3+\tf-1)+e^{t_3}\right]}{\left(e^{\tf}-1\right)^2 \left(e^{t_3}-e^{\tf}\right)^2}\frac{1}{1+p^2} \\
\label{eq:OneLoopShpD12CT}
	&
	+ \frac{e^{t_3-\tm} \left(e^\tm-e^{\tf}\right)^2
   \left[e^{\tf}
   (-t_3+\tf-1)+e^{t_3}\right]}{\left(e^{\tf}-1\right)^2 \left(e^{t_3}-e^{\tf}\right)^2}\frac{1}{(1+p^2)^2} 
\eea
These counterterms can also be interpreted diagrammatically \cite{WiesePrivate}, as for the case of the avalanche durations distribution in section \ref{sec:OneLoopDurationsDiag}: They corresponds to the subtraction of the contribution of $\hD_1$, $\hD_2$ to the renormalization of $m$, $\sigma$, $\eta$
\bea
\nn
\hc_{12} =& \hc_1^{(m)} + 2 \hc_1^{(\sigma)} + \hc_1^{(\eta)} + \hc_2^{(\sigma)} \\
\nn
\hc_1^{(m)}(p^2) = & -\frac{1}{1+p^2}\int_{0}^{\tm} \rmd t_3\, \int_{t_3}^\tf \rmd t_2\, \tu_{t_2}^{(00)}\Rdr_{k=0,t_2,t_3} \Rdr_{k=0,\tm,t_3}  \Rdr_{k=0,t_3,\ti=0} , \\
\nn
\hc_1^{(\sigma)}(p^2) = & -\frac{1}{(1+p^2)^2} \int_{0}^{\tm} \rmd t_3\, \int_{t_3}^\tf \rmd t_2\, \left(\tu_{t_2}^{(00)}\right)^2\Rdr_{k=0,t_2,t_3} \Rdr_{k=0,\tm,t_3}  \Rdr_{k=0,t_3,\ti=0},\\
\nn
\hc_1^{(\eta)}(p^2) = & -\frac{1}{(1+p^2)^2}  \int_{0}^{\tm} \rmd t_3\, \int_{t_3}^\tf \rmd t_2\, \left(\partial_{t_2}\tu_{t_2}^{(00)}\right)\Rdr_{k=0,t_2,t_3} \Rdr_{k=0,\tm,t_3}  \Rdr_{k=0,t_3,\ti=0}, \\
\nn 
\hc_2^{(\sigma)}(p^2) = & -\frac{1}{(1+p^2)^2}\int_{0}^{\tm} \rmd t_3\, \tu_{t_3}^{(00)} \Rdr_{k=0,\tm,t_3}  \Rdr_{k=0,t_3,\ti=0}.
\eea
Adding these counterterms, the momentum integral over $x=p^2$ in \eqref{eq:OneLoopObsShp} converges, giving:
\bea
\nn
& D_1 + D_2 + c_{12} :=  \frac{1}{2}\int_0^\infty \rmd x\,x\,\left[\hD_1(p^2=x) + \hD_2(x)+\hc_{12}(x)\right]\\
\nn
&= \int_0^\tm \rmd t_3 \,{\textstyle \frac{e^{t_3-\tm} \left(e^\tm-e^{\tf}\right)^2
   \left[\text{Ei}(t_3)-\text{Ei}(\tf)+t_3
   e^{\tf}-e^{\tf} \log
   (t_3)-e^{t_3}+e^{\tf}-e^{\tf}
   \tf+e^{\tf} \log (\tf)\right]}{2
   \left(e^{\tf}-1\right)^2
   \left(e^{t_3}-e^{\tf}\right)^2}} \\
\nn
	&= \frac{e^{-\tm} \left(e^\tm-e^{\tf}\right)^2
   \left[\text{Ei}(\tf)+e^{\tf} \tf-e^{\tf} \log
   (\tf)-\gamma \right]}{2
   \left(e^{\tf}-1\right)^3} \\
\label{eq:OneLoopShpDiag12Fin}
	& \quad -\frac{e^{-\tm}
   \left(e^\tm-e^{\tf}\right)
   \left[\text{Ei}(\tm)-\text{Ei}(\tf)+e^\tm \tm-e^\tm \log
   (\tm)-e^{\tf} \tf+e^{\tf} \log (\tf)\right]}{2
   \left(e^{\tf}-1\right)^2}.
\eea

\subsubsection{Diagram $D_3$}
We first evaluate the integrals over $t_1$ and $t_2$ in \eqref{eq:OneLoopShpHD3}. This gives
\bea
\label{eq:OneLoopShpDiag3a}
\hD_3(p^2) &= \int_{0}^\tm \rmd t_2 \int_{t_2}^{\tm}\rmd t_1 \frac{\left(e^\tm-e^{\tf}\right)^2 e^{p^2
   (t_2-t_1)-\tm}}{\left(e^{\tf}-1\right)^2} = \frac{e^{-\tm} \left(p^2 \tm+e^{-p^2 \tm}-1\right)
   \left(e^\tm-e^{\tf}\right)^2}{p^4 \left(e^{\tf}-1\right)^2}.
\eea
As above, we introduce artificial counterterms $\sim \frac{1}{1+p^2}, \frac{1}{(1+p^2)^2}$ in order to remove the contributions $\sim p^{-2}$, $\sim p^{-4}$ for large $p$ in \eqref{eq:OneLoopShpDiag3a}:
\bea
\label{eq:OneLoopShpD3CT}
\hc_3(p^2) & = -\frac{e^{-t} t
   \left(e^t-e^{\tf}\right)^2}{\left(e^{\tf}-1\right)^2
  }\frac{1}{1+p^2}  -\frac{e^{-t} (t-1)
   \left(e^t-e^{\tf}\right)^2}{\left(e^{\tf}-1\right)^2
   }\frac{1}{(1+p^2)^2}.
\eea
These counterterms can also be interpreted diagrammatically \cite{WiesePrivate}, as for the previous diagram: They corresponds to the subtraction of the contribution of $\hD_3$ to the renormalization of $m$, $\sigma$, $\eta$
\bea
\nn
\hc_{3} =& \hc_3^{(m)} + 2 \hc_3^{(\sigma)} + \hc_3^{(\eta)} \\
\nn
\hc_3^{(m)}(p^2) = & -\frac{1}{1+p^2}\int_{0}^{\tm} \rmd t_2\, \Rdr_{k=0,\tm,t_2}\Rdr_{k=0,t_2,\ti=0} , \\
\nn
\hc_3^{(\sigma)}(p^2) = & -\frac{1}{(1+p^2)^2} \int_{0}^{\tm} \rmd t_2\, \tu_{t_2}^{(00)} \Rdr_{k=0,\tm,t_2}\Rdr_{k=0,t_2,\ti=0},\\
\nn
\hc_3^{(\eta)}(p^2) = & -\frac{1}{(1+p^2)^2}  \left[-\Rdr_{k=0,\tm,\ti=0}+\int_{0}^{\tm} \rmd t_2\, \left(\partial_{t_2} \Rdr_{k=0,\tm,t_2}\right)\Rdr_{k=0,t_2,\ti=0} \right].
\eea
Adding these counterterms, the momentum integral over $x=p^2$ in \eqref{eq:OneLoopObsShp} converges, giving:
\bea
\label{eq:OneLoopShpDiag3Fin}
D_3 + c_3  :=  \frac{1}{2}\int_0^\infty \rmd x\,x\,\left[\hD_3(x) +\hc_{3}(x)\right]=
-\frac{e^{-\tm} \left(e^\tm-e^{\tf}\right)^2 [\tm+\log (\tm)+\gamma -1]}{2
   \left(e^{\tf}-1\right)^2}.
\eea

\subsubsection{Diagram $D_4$}
We first perform the integrals over $t_1$ and $t_2$ in \eqref{eq:OneLoopShpHD4} :
\bea
\nn
&\hD_4(p^2) = \int_{\ti}^{\tm}\rmd t_4\,\int_{t_4}^{\tf}\rmd t_3\,\int_{t_3}^{\tf}\rmd t_1\,\int_{t_3}^{\tf}\rmd t_2\,{\textstyle\frac{\left(e^\tm-e^{\tf}\right)^2
   \left(e^{\tf}-e^{t_1}\right)
   \left(e^{\tf}-e^{t_2}\right) e^{-p^2(t_1+t_2-2
   t_3)-\tm+t_3+t_4}}{\left(e^{\tf}-1\right)^2
   \left(e^{t_3}-e^{\tf}\right)^2
   \left(e^{t_4}-e^{\tf}\right)^2}} \\
\label{eq:OneLoopShpDiag4a}
	&=\int_{\ti}^{\tm}\rmd t_4\,\int_{t_4}^{\tf}\rmd t_3\,\frac{\left(e^\tm-e^{\tf}\right)^2 \left(-p^2 e^{p^2
   \tf+t_3}+e^{p^2
   t_3+\tf}+\left(p^2-1\right) e^{p^2
   \tf+\tf}\right)^2 e^{-2 p^2
   \tf-\tm+t_3+t_4}}{p^4 \left(p^2-1\right)^2
   \left(e^{\tf}-1\right)^2
   \left(e^{t_3}-e^{\tf}\right)^2
   \left(e^{t_4}-e^{\tf}\right)^2}.
\eea
Now, it suffices to introduce a single counterterm $\sim \frac{1}{(1+p^2)^2}$ in order to remove the contribution $\sim p^{-4}$ for large $p$ in \eqref{eq:OneLoopShpDiag4a}
 (there is no contribution $\sim p^{-2}$):
\beq
\label{eq:OneLoopShpD4CT}
\hc_4(p^2) := -\int_0^\tm\rmd t_4\int_{t_4}^{\tf}\rmd t_3\, \frac{\left(e^\tm-e^{\tf}\right)^2
   e^{-\tm+t_3+t_4}}{\left(e^{\tf}-1\right)^2 
   \left(e^{t_4}-e^{\tf}\right)^2}\frac{1}{(1+p^2)^2}.
\eeq
As for the preceding diagrams, this counterterm can be interpreted diagrammatically \cite{WiesePrivate}: It corresponds to the subtraction of the contribution of $D_4$ to the renormalization of $\sigma$
\bea
\nn
\hc_4(p^2) =& \hc_4^{(\sigma)}(p^2) = -\frac{1}{(1+p^2)^2} \int_{0}^{\tf} \rmd t_3 \int_{0}^{t_3} \rmd t_4\, \left(\tu_{t_3}^{(00)}\right)^2 \Rdr_{k=0,t_3,t_4} \Rdr_{k=0,\tm,t_4}\Rdr_{k=0,t_4,\ti=0}.
\eea
Adding this counterterm, the momentum integral over $x=p^2$ in \eqref{eq:OneLoopObsShp} converges, giving:
\bea
\nn
&D_4 + c_4 := \frac{1}{2}\int_0^\infty \rmd x\,x\,\left[\hD_4(x) +\hc_{4}(x)\right]\\
\nn
&=  \int_{\ti}^{\tm}\rmd t_4\,\int_{t_4}^{\tf}\rmd t_3\,\frac{\left(e^\tm-e^{\tf}\right)^2 e^{-\tm+t_3+t_4}}{4
   \left(e^{\tf}-1\right)^2
   \left(e^{t_3}-e^{\tf}\right)^2
   \left(e^{t_4}-e^{\tf}\right)^2}
   \left\{-4 e^{t_3} \left[e^{t_3}
   (\tf-t_3)+e^{\tf}\right]
   \text{Ei}(\tf-t_3) \right. \\
\label{eq:OneLoopShpDiag4b}
	& \quad \left.-2 e^{2 t_3} (2 t_3-2
   \tf-1) \text{Ei}(2 \tf-2 t_3)+e^{2 \tf} \left[2
   \log (\tf-t_3)+2 \gamma -2 \log (2)\right]\right\}.
\eea
It seems difficult to find a closed-form expression for the primitive of the integrand with respect to $t_3$. However, its primitive with respect to $t_4$ is much simpler. We will thus evaluate the integral over $t_4$ first and the integral over $t_3$ later. $t_4$ runs from $\ti=0$ to the minimum of $t_3$ and $\tm$. We thus split the $t_3$ integral as follows:
\beq
\int_{\ti}^{\tm}\rmd t_4\,\int_{t_4}^{\tf}\rmd t_3 = \int_0^\tf \rmd t_3 \int_0^\tm \rmd t_4 - \int_0^\tm \rmd t_3 \int_{t_3}^\tm \rmd t_4.
\eeq
We denote the first piece as diagram $D_4^a$, and the second piece as diagram $D_4^b$.
It turns out that both have simple expressions through the mean-field shape \eqref{eq:OneLoopShpTTree}, and the function $d_2$ arising in the diagram $D_2$ for the durations distribution \eqref{eq:OneLoopDurDiag2b}:
\bea
\nn
D_4 + c_4 =& D_4^{a} + D_4^b, \\
\nn
D_4^a =& \frac{1}{2}  \frac{e^\tf}{(e^\tf-1)^2} \mathfrak{s}^{\tree}(\tm,\tf) \int_0^{\tf} \rmd t \, d_2(\tf-t) = \frac{1}{2}  \frac{e^\tf}{(e^\tf-1)^2} \mathfrak{s}^{\tree}(\tm,\tf) \int_0^{\tf} \rmd t' \, d_2(t'),\\
\nn
 D_4^b =& -\frac{1}{2} \frac{e^\tf}{(e^\tf-1)^2}\int_0^{\tm} \rmd t \,\mathfrak{s}^{\tree}(\tm-t,\tf-t)\, d_2(\tf-t) \\
\label{eq:OneLoopShpDiag4bFin}
 =& -\frac{1}{2} \frac{e^\tf}{(e^\tf-1)^2}\int_{\tmp}^{\tf} \rmd t' \,\mathfrak{s}^{\tree}(\tmp,t')\, d_2(t'),
\eea
where we set $t := t_3$, $t' := \tf-t_3$.

\subsubsection{Diagram $D_5$}
We first perform the integrals over $t_1$, $t_2$, and $t_3$ in \eqref{eq:OneLoopShpHD5} :
\bea
\nn
\hD_5(p^2) &= \int_0^\tm \rmd t_4 \int_{t_4}^{\tm}\rmd t_3 \int_{t_3}^{\tf}\rmd t_2 \int_{t_4}^{\tf} \rmd t_1 \,{\textstyle \frac{\left(e^\tm-e^{\tf}\right)^2
   \left(e^{\tf}-e^{t_1}\right)
   \left(e^{\tf}-e^{t_2}\right) e^{-p^2
   (t_1+t_2-2
   t_4)-\tm+t_3+t_4}}{\left(e^{\tf}-1\right)^2
   \left(e^{t_3}-e^{\tf}\right)^2
   \left(e^{t_4}-e^{\tf}\right)^2}} \\
\nn
	& = \int_0^\tm \rmd t_4 \frac{\left(e^\tm-e^{\tf}\right)^2 \left[-p^2 e^{p^2
   \tf+t_4}+e^{p^2
   t_4+\tf}+\left(p^2-1\right) e^{p^2
   \tf+\tf}\right] e^{-2 p^2 \tf-\tm+t_4}}{p^4
   \left(p^2-1\right)^2 \left(e^{\tf}-1\right)^2
   \left(e^{t_4}-e^{\tf}\right)^2}\times \\
	\label{eq:OneLoopShpDiag5a}
	& \quad\quad \times
   \left[\frac{e^{p^2 t_4} \left(e^{p^2
   (\tf-\tm)+\tm}-e^{\tf}\right)}{e^\tm-e^{\tf}}+\frac{e^{p^2
   t_4+\tf}-e^{p^2
   \tf+t_4}}{e^{t_4}-e^{\tf}}\right].
\eea
The momentum integral over $x=p^2$ in \eqref{eq:OneLoopObsShp} is now convergent without any counterterms, and one obtains
\bea
\nn
&D_5 :=\frac{1}{2}\int_0^\infty \rmd x\,x\, \hD_5(x) = D_5^a + D_5^b,\quad\quad\quad D_5^a = -D_4^b, \\
\nn
& D_5^b =  \frac{e^\tf}{(e^\tf-1)^2}\int_{0}^{\tm} \rmd t\,f_5(\tf-t,\tf-\tm)= \frac{e^\tf}{(e^\tf-1)^2}\int_{\tmp}^{\tf} \rmd t'\,f_5(t',\tmp), \\
\nn
 &f_5(t',\tmp) :=  \frac{\left(e^{\tmp}-1\right)
  e^{-t'-\tmp} }{2   \left(e^{t'}-1\right)^2 } 
   \left\{\text{Ei}(t') \left(t'
   e^{\tmp}+e^{t'}\right) \right. \\
	\nn
	& \quad\quad \left. +e^{\tmp}
   \left[\left(t'+e^{t'}-\tmp\right) \text{Ei}(t'-\tmp)+(-2
   t'+\tmp-1) \text{Ei}(2
   t'-\tmp)\right]\right. \\
\label{eq:OneLoopShpDiag5bFin}
	& \quad\quad \left.-e^{2 t'}
   \left[\log (t'-\tmp)-\log (2
   t'-\tmp)+\log
   (t')+\gamma \right]\right\}.
	\eea
Here we defined $t:=t_4, t':=\tf-t_4, \tmp:=\tf-\tm$, and simplified by extracting a part $D_5^a$ which is the opposite of diagram $D_4^b$ computed in \eqref{eq:OneLoopShpDiag4bFin}.

\subsubsection{Diagram $D_6$}
We first perform the integrals over $t_1$ and $t_2$ in \eqref{eq:OneLoopShpHD6} :
\bea
\nn
\hD_6(p^2) = & \int_0^\tm \rmd t_4 \int_{t_4}^\tm \rmd t_2 \int_{t_4}^\tf \rmd t_1 \frac{\left(e^\tm-e^{\tf}\right)^2
   \left(e^{t_1}-e^{\tf}\right) e^{-p^2(t_1+t_2-2
   t_4)-\tm+t_4}}{\left(e^{\tf}-1\right)^2
   \left(e^{t_4}-e^{\tf}\right)^2} \\
\label{eq:OneLoopShpDiag6a}
	= & \int_0^\tm\rmd t_4 \,{\textstyle \frac{e^{t_4-\tm} \left(e^\tm-e^{\tf}\right)^2 \left\{e^{p^2
   \tf} \left[p^2 e^{t_4}-\left(p^2-1\right)
   e^{\tf}\right]-e^{p^2 t_4+\tf}\right\} \left[e^{-p^2
   \tf}-e^{-p^2 (\tm-t_4+\tf)}\right]}{p^4
   \left(p^2-1\right) \left(e^{\tf}-1\right)^2
   \left(e^{t_4}-e^{\tf}\right)^2}}.
\eea
To remove the contribution $\sim p^{-4}$, we add the counterterm
\beq
\label{eq:OneLoopShpD6CT}
\hc_6(p^2) := -\int_0^\tm \rmd t_4\, \frac{e^{t_4-\tm}
   \left(e^\tm-e^{\tf}\right)^2}{\left(e^{\tf}-1\right)^2
  \left(e^{t_4}-e^{\tf}\right)}\frac{1}{(1+p^2)^2}.
\eeq
As for the preceding diagrams, this counterterm can be interpreted diagrammatically \cite{WiesePrivate}: It corresponds to the subtraction of the contribution of $D_4$ to the renormalization of $\sigma$
\bea
\nn
\hc_6(p^2) =& \hc_6^{(\sigma)}(p^2) = -\frac{1}{(1+p^2)^2} \int_{0}^{\tf} \rmd t_4 \, \tu_{t_4}^{(00)} \Rdr_{k=0,\tm,t_4} \Rdr_{k=0,t_4,\ti=0}.
\eea
Adding this counterterm, the momentum integral over $x=p^2$ in \eqref{eq:OneLoopObsShp} converges, giving:
\bea
\nn
&D_6+c_6 :=  \frac{1}{2}\int_0^\infty \rmd x\,x\,\left[\hD_6(x) +\hc_{6}(x)\right]\\
\nn
&= -\int_0^\tm \rmd t_4 \frac{e^{t_4-2 \tm} \left(e^\tm-e^{\tf}\right)^2}{2
   \left(e^{\tf}-1\right)^2
   \left(e^{t_4}-e^{\tf}\right)^2} \left\{e^{2
   t_4} \left[\text{Ei}(\tm-2 t_4+\tf)- \text{Ei}(\tm-t_4)\right]+ \right. \\
	\nn
	& \quad\left.+e^\tm \left[-e^{t_4}
   \text{Ei}(\tf-t_4)+e^{\tf} \log
   (\tm-t_4)-e^{\tf} \log (\tm-2
   t_4+\tf) \right. \right. \\
	\nn
	& \quad \quad \left.\left.+e^{\tf} \log
   (\tf-t_4)-e^{t_4}+e^{\tf}+\gamma 
   e^{\tf}\right] \right\} \\
	\nn
	& = \frac{e^\tf}{(e^\tf-1)^2}\int_{0}^{\tm} \rmd t\,f_6(\tf-t,\tf-\tm) = \frac{e^\tf}{(e^\tf-1)^2}\int_{\tmp}^{\tf} \rmd t'\,f_6(t',\tmp), \\
	\nn
	&f_6(t',\tmp):= \frac{\left(e^{\tmp}-1\right)^2
   e^{-t'-\tmp}}{2
   \left(e^{t'}-1\right)^2}
   \left\{e^{\tmp}
   \left[\text{Ei}(t'-\tmp)-\text{Ei}(2
   t'-\tmp)\right]+e^{t'}
   (\text{Ei}(t')+1) \right. \\
\label{eq:OneLoopShpDiag6Fin}
	& \quad\quad \left. -e^{2 t'}
   \left[\log (t'-\tmp)-\log (2
   t'-\tmp)+\log
   (t')+\gamma +1\right]\right\},
\eea
where as above we defined $t:= t_4, t':=\tf-t_4, \tmp:=\tf-\tm$.

\subsubsection{Counterterms\label{sec:OneLoopShapeTimeCT}}
Let us revisit the counterterms $\hc_1 \cdots \hc_6$ we added above for computing the diagrams $D_1 \cdots D_6$, contributing to the one-loop correction $\hat{\delta}_c F^\mfs$ to the cumulative shape function $F^\mfs$. First, note that after performing the $t_3$ integral in \eqref{eq:OneLoopShpD4CT}, and comparing it to \eqref{eq:OneLoopShpD6CT},
\beq
\hc_4(p^2)+\hc_6(p^2) = 0.
\eeq
Since diagram 5 is convergent without any counterterms, the only remaining counterterms come from diagrams $1,2$ and $3$.
Evaluating the integrals over $t_3$ in the counterterms for diagrams 1 and 2, eq.~\eqref{eq:OneLoopShpD12CT}, we obtain
\bea
\nn
& 2\hc_{12}(p^2)+\hc_3(p^2) = \frac{e^{-\tm} \left(e^{\tm}-e^{\tf}\right) }{\left(e^{\tf}-1\right)^3}\left[e^{\tf} (2
   \tf-\tm)+e^{2 \tf} \tm+e^{\tf+\tm} (\tm-2
   \tf)-e^{\tm} \tm\right]\frac{1}{1+p^2} \\
	\label{eq:OneLoopShpTCTdiags}
	& {\textstyle+\frac{e^{-\tm} \left(e^{\tm}-e^{\tf}\right) }{\left(e^{\tf}-1\right)^3
   }\left[e^{\tf} (2
   \tf-\tm+1)+e^{2 \tf} (\tm-1)+e^{\tf+\tm} (-2
   \tf+\tm+1)-e^{\tm} (\tm+1)\right]\frac{1}{(1+p^2)^2}}.
\eea
Note that the factor $2$ for $\hc_{12}(p^2)$ comes from the combinatoric prefactor $2$ for the diagrams $D_1$ and $D_2$ in \eqref{eq:OneLoopShpDiagSum}.

On the other hand, the physical counterterms, corresponding to the subtraction of loop corrections to the renormalized parameters $\eta,\sigma,m$, are given in \eqref{eq:OneLoopObsShp}, \eqref{eq:OneLoopCEtaH}, \eqref{eq:OneLoopCSigmaH}, \eqref{eq:OneLoopCMH} in terms of the dimensionful tree-level  cumulative shape $F_\mfs^{\dtree}$. Restoring the units in the tree-level expression \eqref{eq:OneLoopShpTree}, we obtain
\bea
F_\mfs^{\dtree} = \frac{e^{(m^2/\eta)t_m}\left[e^{(m^2/\eta)\left(\tf-t_m\right)}-1\right]^2}{\eta \left[e^{(m^2/\eta)\tf}-1\right]^2}.
\eea
Thus, the counterterms given by \eqref{eq:OneLoopCEtaH}, \eqref{eq:OneLoopCSigmaH}, \eqref{eq:OneLoopCMH} are
\bea
\nn
\hc_\eta(p^2)& = - \frac{1}{(1+p^2)^2} \partial_\eta\bigg|_{m=\sigma=\eta=1}F_\mfs^{\dtree}  \\
\nn
&= \frac{1}{(1+p^2)^2}  \frac{e^{-\tm} \left(e^{\tm}-e^{\tf}\right)}{\left(e^{\tf}-1\right)^3} \left[  e^{\tf} (2
   \tf-\tm+1)+e^{2 \tf} (\tm-1) \right. \\
	\nn
	& \quad\quad\quad\left.+e^{\tf+\tm} (-2
   \tf+\tm+1)-e^{\tm} (\tm+1)\right], \\
	\nn
\hc_\sigma(p^2) &= - 3 \frac{1}{(1+p^2)^2}  \partial_\sigma\bigg|_{m=\sigma=\eta=1}F_\mfs^{\dtree}  =  0,  \\
\nn
\hc_{m}(p^2)& = \frac{1}{1+p^2} \frac{1}{2}\partial_m\bigg|_{m=\sigma=\eta=1}F_\mfs^{\dtree}  \\
\label{eq:OneLoopShpTCT}
&=  \frac{1}{1+p^2} \frac{e^{-\tm} \left(e^{\tm}-e^{\tf}\right)}{\left(e^{\tf}-1\right)^3} \left[e^{\tf} (2
   \tf-\tm)+e^{2 \tf} \tm+e^{\tf+\tm} (\tm-2
   \tf)-e^{\tm} \tm\right].
\eea
Comparing this to the per-diagram counterterms 	\eqref{eq:OneLoopShpTCTdiags}, we observe
\bea
\label{eq:OneLoopShpTCTMatch}
2\hc_{12}(p^2)+\hc_3(p^2) = \hc_\eta(p^2) + \hc_\sigma(p^2) + \hc_{m}(p^2),
\eea
as required. As in section \ref{sec:OneLoopDurationsCT} for the avalanche durations, this nontrivial check shows that the physical counterterms \eqref{eq:OneLoopCEtaH}, \eqref{eq:OneLoopCSigmaH}, \eqref{eq:OneLoopCMH} appearing in \eqref{eq:OneLoopObsShp} indeed cancel all the divergences in the diagrams $D_1 \cdots D_6$ contributing to $\hat{\delta}_c F^\mfs$, with the correct combinatoric prefactors. In other words, a renormalization of these coupling constants is sufficient to render the theory finite.

\subsubsection{Final result for the shape}
When the diagrams above are combined in order to evaluate \eqref{eq:OneLoopObsShp}, they simplify. 
In particular, just as $D_4^{a}$ is expressible in terms of the function $d_2$ introduced for diagram 2 of the durations distribution \eqref{eq:OneLoopDurDiag2b}, diagrams 1, 2 and 3 given in \eqref{eq:OneLoopShpDiag12Fin}, \eqref{eq:OneLoopShpDiag3Fin}
 simplify when introducing $d_1$ from \eqref{eq:OneLoopDurDiag1b}:
\bea
\nn
& 2(D_1 + D_2 + c_{12}) + (D_3 + c_3) = \frac{e^\tf}{(e^\tf-1)^2}\left[\mathfrak{s}^{\tree}(\tm,\tf) d_1(\tf) + d^\mfs_1(\tm,\tf)\right], \\
\label{eq:OneLoopShpTimeD123}
& d^\mfs_1(\tm,\tf) :=  -\left(e^\tmp-e^{-\tmp}\right) d_1(\tm)- \left(e^{\tmp}+e^{-\tmp}-2\right) d'_1(\tm),\\
\nn
&  = \frac{1}{2} 
   \left(e^{-\tf}-e^{-\tm}\right) \left[-2
   \text{Ei}(\tm)+e^{\tf} (-\tm+\gamma
   +1)+\left(e^{\tf}+e^{\tm}\right) \log
   (\tm)+e^{\tm} (-\tm+\gamma +3)-4\right],
\eea
where as above $\tmp := \tf-\tm$.
Finally, combining diagrams $4,5,6$ from \eqref{eq:OneLoopShpDiag4bFin}, \eqref{eq:OneLoopShpDiag5bFin}, \eqref{eq:OneLoopShpDiag6Fin}
 there are also some simplifications, giving
\bea
\nn
2 D_4^a =& \mfs^{\tree}(\tm,\tf) \frac{e^\tf}{(e^\tf-1)^2} \int_0^\tf \rmd t'\, d_2(t'), \\
\label{eq:OneLoopShpTimeD456}
2 D_4^b + 4 D_5 + 2 (D_6+c_6) =& \frac{e^\tf}{(e^\tf-1)^2}\left[ \int_{\tmp}^\tf \rmd t'\,   d^\mfs_2(t',\tmp) \right],
\eea
where  
\bea
d^\mfs_2 (t',\tmp) := \mathfrak{s}^{\tree}(\tmp,t')d_2(t')  + 4 f_5(t',\tmp) + 2 f_6(t',\tmp).
\eea
$f_5, f_6$ are given in \eqref{eq:OneLoopShpDiag5bFin}, \eqref{eq:OneLoopShpDiag6Fin} and $d_2$ in \eqref{eq:OneLoopDurDiag2b}. Explicitly, 
\bea
\nn
d^\mfs_2 (t',\tmp) =& \mathfrak{s}^{\tree}(\tmp,t')d_2(t')+ \\
\nn
& + \frac{\left(e^{\tmp}-1\right)}{\left(e^{t'}-1\right)^2} \left\{ 
\text{Ei}(t') \left(2   e^{-t'} t'+e^{-\tmp}+1\right) \right. \\
\nn & \left.\quad+
	e^{-t'}   \left(2 t'+2 e^{t'}+e^{\tmp}-2   \tmp-1\right) \text{Ei}(t'-\tmp) \right. \\
	\nn
	& \quad\left. -e^{-t'}   \left(4 t'+e^{\tmp}-2 \tmp+1\right) \text{Ei}(2   t'-\tmp)+
	\left(e^{t'}-1\right)
   \left(e^{-\tmp}-1\right) \right. \\
\label{eq:OneLoopShpDurDMFS}
	& \quad\left. - e^{t'}\left(1+e^{\tmp}\right)\left[\gamma   + \log   (t')+  \log   (t'-\tmp) - \log (2
   t'-\tmp) \right]\right\}.
\eea
Altogether we can write the correction to the cumulative shape $\delta_c F^\mfs$ in \eqref{eq:OneLoopObsShp} as
\bea
\label{eq:OneLoopShpDiagSum2}
&\delta_c F^\mfs (\tm,\tf) =  2(D_1 + D_2 + c_{12}) + (D_3 + c_3) + 2 (D_4+c_4) + 4 D_5 + 2 (D_6+c_6) \\
\label{eq:OneLoopShpDurCumFinal}
&= \mfs^{\tree}(\tm,\tf) \delta_c F(\tf) + \frac{e^\tf}{(e^\tf-1)^2}\left[ d^\mfs_1(\tm,\tf) +  \int_{\tmp}^\tf \rmd t'\,   d^\mfs_2(t',\tmp) \right],
\eea
where $\delta_c F$ is the correction to the cumulative duration distribution computed in \eqref{eq:OneLoopDurDeltaF1}, $d_\mfs$ is given by \eqref{eq:OneLoopShpDurDMFS} and $d_1$, $d_2$ are given by \eqref{eq:OneLoopDurDiag1b}, \eqref{eq:OneLoopDurDiag2b}. Inserting this into \eqref{eq:OneLoopShapeCorr}, the first term simplifies (since $\partial_{\tf}\big|_{\tf=T} \delta_c F(\tf) = \delta_c P(T)$, as defined in \eqref{eq:OneLoopDurPofTCorr}, cancels the second term in \eqref{eq:OneLoopShapeCorr}) and we obtain
\bea
\nn
\delta_c \mfs(\tm,T) =& \left[\partial_\tf\bigg|_{\tf=T} \mfs^{\tree}(\tm,\tf)\right]\left[d_1(T) + \int_0^T \rmd t'\, d_2(t')\right] + \\
\label{eq:OneLoopShpDurFinal}
 &+ \left[\frac{1+e^\tf}{1-e^\tf}+\partial_\tf\right]\bigg|_{\tf=T}\left[d^\mfs_1(\tm,\tf)  + \int_{\tmp}^\tf \rmd t'\,   d^\mfs_2(t',\tmp) \right].
\eea
\begin{figure}%
\centering
                 \includegraphics[width=0.7\textwidth]{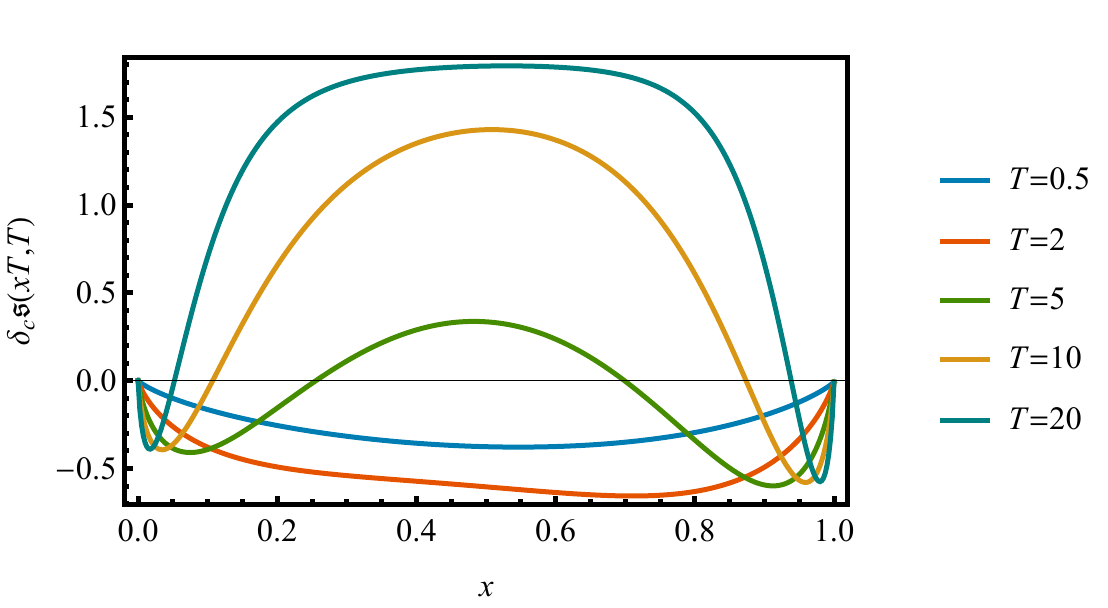}
                 \caption{One-loop correction to the average avalanche shape, eq.~\eqref{eq:OneLoopShpDurFinal}, for various $T$. For small $T$, this reduces to the asymptotic shape \eqref{eq:OneLoopShapeDurShortCorr} discussed in section \ref{sec:OneLoopShapeTimeShort}. For any $T$, the leading behaviour near the left and the right edge is given by \eqref{eq:OneLoopShpLeftEdge} and \eqref{eq:OneLoopShpRightEdge}, respectively.}
                 \label{fig:OneLoopShapePlotLong}
\end{figure}
In order to take the $\tf$ derivative, we can set $t' = \tf- t$ and rewrite the integrals as
\bea
\nn
&\delta_c \mfs(\tm,T) = \left[\partial_\tf\bigg|_{\tf=T} \mfs^{\tree}(\tm,\tf)\right]\left[d_1(T) + \int_0^T \rmd t\, d_2(T-t)\right] + \\
\label{eq:OneLoopShpDurFinalIntRewr}
& + \left[\frac{1+e^\tf}{1-e^\tf}+\partial_\tf\right]\bigg|_{\tf=T}\left[d^\mfs_1(\tm,\tf)\right] +  \int_{0}^{\tm} \rmd t\, \left[\frac{1+e^\tf}{1-e^\tf}+\partial_\tf\right]\bigg|_{\tf=T}\left[ d^\mfs_2(\tf-t,\tf-\tm) \right].
\eea
Some plots of this expression, for various total duration $T$, are shown in figure \ref{fig:OneLoopShapePlotLong}. For large $T$, the correction $\delta_c \mfs$ saturates to a constant value, just like the mean-field shape $\mfs^{\tree}$ given by \eqref{eq:OneLoopShpTTree} and shown in figure \ref{fig:ABBMShapeAnal}. For small $T$, the correction $\delta_c \mfs \propto T$ and has a simple analytical expression which will be computed in section \ref{sec:OneLoopShapeTimeShort}. For any $T$, the leading power-law behaviour of $\delta_c \mfs$ near the left and right edge will be obtained in sections \ref{sec:OneLoopShapeTimeBeg} and \ref{sec:OneLoopShapeTimeEnd}.

\subsection{Asymptotics near the beginning of an avalanche\label{sec:OneLoopShapeTimeBeg}}
Using \eqref{eq:OneLoopShpDurCumFinal}, \eqref{eq:OneLoopShpDurFinal} we can now compute the asymptotics near the beginning and the end of an avalanche.
We expand each term in \eqref{eq:OneLoopShpDurFinal} individually for small $\tm$, keeping $T$ fixed. 
The tree-level shape $\mathfrak{s}^{\tree}(\tm,\tf) = 2\tm + \mO(\tm)^2$ is independent of $\tf$ to order $\tm$. Hence, the term 
\bea
\label{eq:OneLoopShpSerLeft1}
\partial_\tf\bigg|_{\tf=T} \mfs^{\tree}(\tm,\tf) = \mO(\tm)^2,
\eea
and only the second line in \eqref{eq:OneLoopShpDurFinal} contributes. It consists in
\bea
\label{eq:OneLoopShpSerLeft2}
\left[\frac{1+e^\tf}{1-e^\tf}+\partial_\tf\right]\bigg|_{\tf=T}\left[d^{\mfs}_1(\tm,\tf)\right] =& -\tm + \mO(\tm)^2 .
\eea
For the second term involving the integral over $d^{\mfs}_2$, there is a boundary layer of size $\tm$ near $t=\tm$. Hence, it does not suffice to take the (pointwise) leading order of the integrand for small $\tm$ in \eqref{eq:OneLoopShpDurFinal} as is. Instead, we substitute $t = x \tm$ and consider $x$ of order $1$; we then expand for small $\tm$. The result is
\bea
\nn
&\left[\frac{1+e^\tf}{1-e^\tf}+\partial_\tf\right]\bigg|_{\tf=T}\left[\int_{\tmp}^\tf \rmd t'\,   d^{\mfs}_2(t',\tmp)\right] = \\
\nn
& = \left[\frac{1+e^\tf}{1-e^\tf}+\partial_\tf\right]\bigg|_{\tf=T}\left\{-\tm (1-e^{-\tf})\int_{0}^1 \rmd x\,\left[1+\gamma+\log\tm + \log(1-x)\right] +\mO(\tm)^2\right\} . \\
\label{eq:OneLoopShpSerLeft3}
&= \tm\left(\gamma + \log \tm\right)+\mO(\tm)^2
\eea
Inserting the series \eqref{eq:OneLoopShpSerLeft1},\eqref{eq:OneLoopShpSerLeft2},\eqref{eq:OneLoopShpSerLeft3} into \eqref{eq:OneLoopShpDurFinal}, we obtain:
\beq
\label{eq:OneLoopShpLeftEdge}
\delta_c\mathfrak{s}(\tm,T) = \tm(\gamma-1+\log \tm) + \mathcal{O}(\tm)^2 .
\eeq
Amazingly, all $T$-dependent terms drop out to this order. 
The expansion to next order becomes $T$-dependent and much more complicated, containing integrals without closed-form expressions. So, we restrict ourselves to $\mathcal{O}(\tm)$. 
Inserting \eqref{eq:OneLoopShpLeftEdge} into \eqref{eq:OneLoopShapeCorr} together with the mean-field contribution \eqref{eq:OneLoopShpTTree}, we can write the asymptotics near the left edge to order $\epsilon$ as
\bea
\nn
\mfs(\tm,T) =& 2\tm - \tD_*''(0^+)\tm(\gamma-1+\log \tm) + \mO(\tm)^2 \\
\label{eq:OneLoopShpLeftEdgeTot}
=& \tm^{1-\frac{1}{9}\epsilon}\left[2-\frac{2}{9}\epsilon(\gamma-1)+\mO(\epsilon)^2\right] + \mO(\tm)^2.
\eea
Here we used the universal value $\tD_*''(0^+) = \frac{2}{9}\epsilon + \mO(\epsilon)^2$ from \eqref{eq:OneLoopValueD}, valid at the depinning fixed point.
Thus, the power-law growth of the average velocity near the left edge of an avalanche is modified from an exponent $1\to 1-\frac{1}{9}\epsilon$; the shape becomes steeper. However, the leading order is still independent of the total duration $T$ just like in the mean-field case.

\subsection{Asymptotics near the end of an avalanche\label{sec:OneLoopShapeTimeEnd}}
Similarly to the above, we can compute the leading-order asymptotics near the end of an avalanche. 
We now first take the $\tf$ derivative in \eqref{eq:OneLoopShpDurFinalIntRewr} at $\tf=T$, and then expand in $\tmp := T-\tm$, keeping $T$ fixed.
Some cancellations occur:
\bea
\label{eq:OneLoopShpSerRight1}
&&\left[\partial_\tf\bigg|_{
\begin{subarray}{l} \tf=T \\ \tm = T-\tmp
\end{subarray}
} \mfs^{\tree}(\tm,\tf)\right] d_1(T) + \left[\frac{1+e^\tf}{1-e^\tf}+\partial_\tf\right]\bigg|_{
\begin{subarray}{l} \tf=T \\ \tm = T-\tmp
\end{subarray}
}\left[d^{\mfs}_1(\tm,\tf)\right] & = 0 + \mO(\tmp)^2, \\
\nn
&\lefteqn{
 \left[\partial_\tf\bigg|_{
\begin{subarray}{l} \tf=T \\ \tm = T-\tmp
\end{subarray}
} \mfs^{\tree}(\tm,\tf)\right] d_2(T-t)+}& & \\
&&\label{eq:OneLoopShpSerRight2}
 \quad\quad + \left[\frac{1+e^\tf}{1-e^\tf}+\partial_\tf\right]\bigg|_{
\begin{subarray}{l} \tf=T \\ \tm = T-\tmp
\end{subarray}
}\left[d^\mfs_2(\tf-t,\tf-\tm)\right] & = 0 + \mO(\tmp)^2,
\eea
where as usual $\tmp = \tf-\tm, t' = \tf-t$. So, taking the pointwise limit of the integrand in the $t$ integration in \eqref{eq:OneLoopShpDurFinalIntRewr}, the mean-field term (the first line in \eqref{eq:OneLoopShpDurFinalIntRewr}) cancels the contribution from $d^\mfs_2$ up to $\mO(\tmp)^2$. However, the limits of the integrals are different, so one additional term remains
\bea
\label{eq:OneLoopShpSerRight3}
-\int_{\tm}^T \rmd t\, \left[\frac{1+e^\tf}{1-e^\tf}+\partial_\tf\right]\bigg|_{
\begin{subarray}{l} \tf=T \\ \tm = T-\tmp
\end{subarray}
}d^\mfs_2(\tf-t,\tf-\tm) = \tmp\left(-\frac{1}{2}+\gamma  - 2\log 2 + \log \tmp \right) + \mO(\tmp)^2
\eea 
Furthermore, like in \eqref{eq:OneLoopShpSerLeft3}, there is a boundary-layer contribution near the limit $t = \tm$ of the $t$ integral in \eqref{eq:OneLoopShpDurFinalIntRewr}, for $\tm - t \sim \tmp$, which is not captured by the pointwise expansion \eqref{eq:OneLoopShpSerRight2}. To obtain it, we subtract the pointwise series \eqref{eq:OneLoopShpSerRight2} from the integrand, set $t' = T - t = (1+y)\tmp$, and $y$ of order $1$, and expand in $\tmp$. We obtain the additional term
\bea
\nn
&- \tmp \int_0^\infty \rmd y\,   \left[\frac{4 y^2 \log (y)+2 y-(8 y (y+1)+4) \log
   \left(y+\frac{1}{2}\right)+1}{2 (y+1)^2}+\log (y)+\log (y+1)\right] + \mO(\tmp)^2 \\
\label{eq:OneLoopShpSerRight4}
	& = \tmp\left[\frac{3}{2}-\frac{\pi ^2}{3}-2 \log ^2(2)+\log (16)\right] + \mO(\tmp)^2.
\eea
Combining \eqref{eq:OneLoopShpSerRight1}, \eqref{eq:OneLoopShpSerRight2}, \eqref{eq:OneLoopShpSerRight3} and \eqref{eq:OneLoopShpSerRight4}, we obtain the simple expansion for the shape near the right edge
\beq
\label{eq:OneLoopShpRightEdge}
\delta_c \mathfrak{s}(T-\tmp,T) = \tmp \left[2\ln 2 + \log (\tmp)-\frac{\pi ^2}{3}+\gamma +1-2 \log
   ^2(2)\right]+\mathcal{O}(\tmp)^2.
\eeq
Just like near the left edge, all $T$-dependent terms drop out to order $\mathcal{O}(\tmp)$. 
Inserting \eqref{eq:OneLoopShpRightEdge} into \eqref{eq:OneLoopShapeCorr} together with the mean-field contribution \eqref{eq:OneLoopShpTTree}, we can write the asymptotics near the right edge to order $\epsilon$ as
\bea
\label{eq:OneLoopShpRightEdgeTot}
\mfs(T-\tmp,T) =  \tmp^{1-\frac{1}{9}\epsilon}\left\{2-\frac{2}{9}\epsilon\left[2\ln 2 -\frac{\pi ^2}{3}+\gamma +1-2 \log
   ^2(2)\right]+\mO(\epsilon)^2\right\} + \mO(\tmp)^2.
\eea
Here, as in the previous section, we used the universal value $\tD_*''(0^+) = \frac{2}{9}\epsilon + \mO(\epsilon)^2$ from \eqref{eq:OneLoopValueD}, valid at the depinning fixed point.
We observe that the power-law exponent of the average velocity decay near the end of the avalanche is modified from $1\to 1-\frac{1}{9}\epsilon$, \textit{identically} to the exponent near the beginning of the avalanche. However, the correction to the linear term is different, which leads to 
 skewness of the average avalanche shape.

\subsection{Shape for short avalanches\label{sec:OneLoopShapeTimeShort}}
\begin{figure}%
         \centering
         \begin{subfigure}[t]{0.45\textwidth}
                 \centering
                 \includegraphics[width=\textwidth]{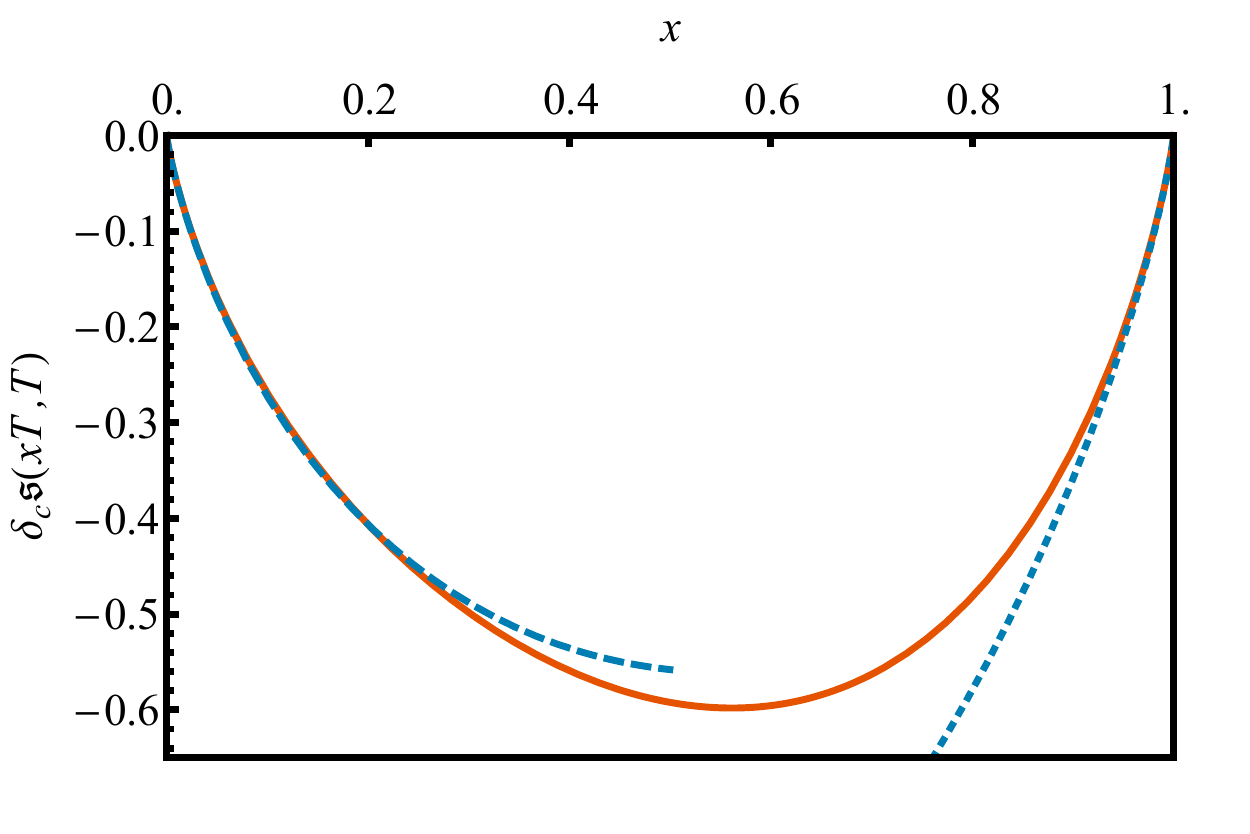}
                 \caption{One-loop correction to the average avalanche shape, for short avalanches, given by \eqref{eq:OneLoopShapeDurShortCorr}. The dashed and dotted lines are the asymptotics near the left and the right edge, given by \eqref{eq:OneLoopShpLeftEdge} and \eqref{eq:OneLoopShpRightEdge}, respectively.}
                 \label{fig:OneLoopShapePlotCorr}
         \end{subfigure}%
         ~ 
         \begin{subfigure}[t]{0.45\textwidth}
                 \centering
                 \includegraphics[width=\textwidth]{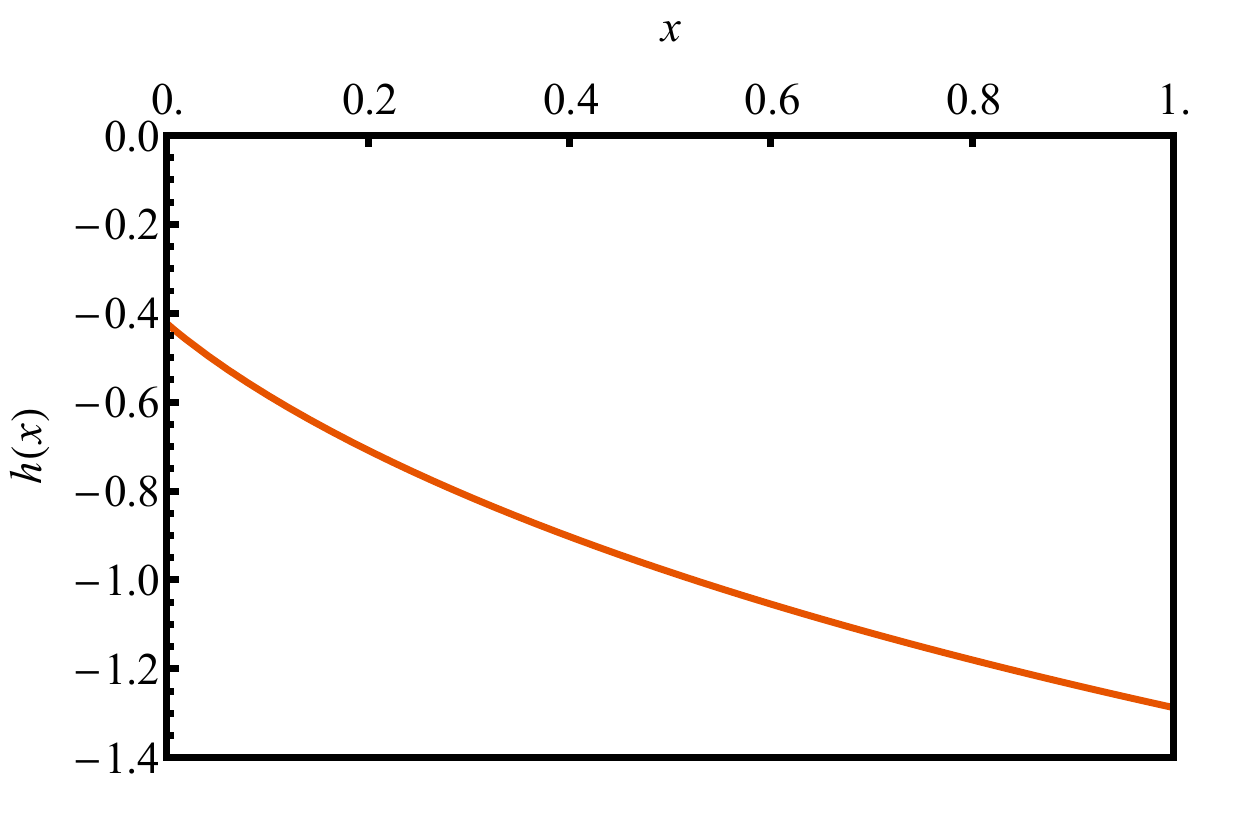}
                 \caption{Auxiliary function $h(x)$ appearing in eq.~\eqref{eq:OneLoopShapeDurShort} for the average shape of short avalanches to one-loop.}
                 \label{fig:OneLoopShapePlotH}
         \end{subfigure}%
				                \caption{ \label{fig:OneLoopShapeShort}}
\end{figure}

\begin{figure}%
                 \centering
                 \includegraphics[width=0.7\textwidth]{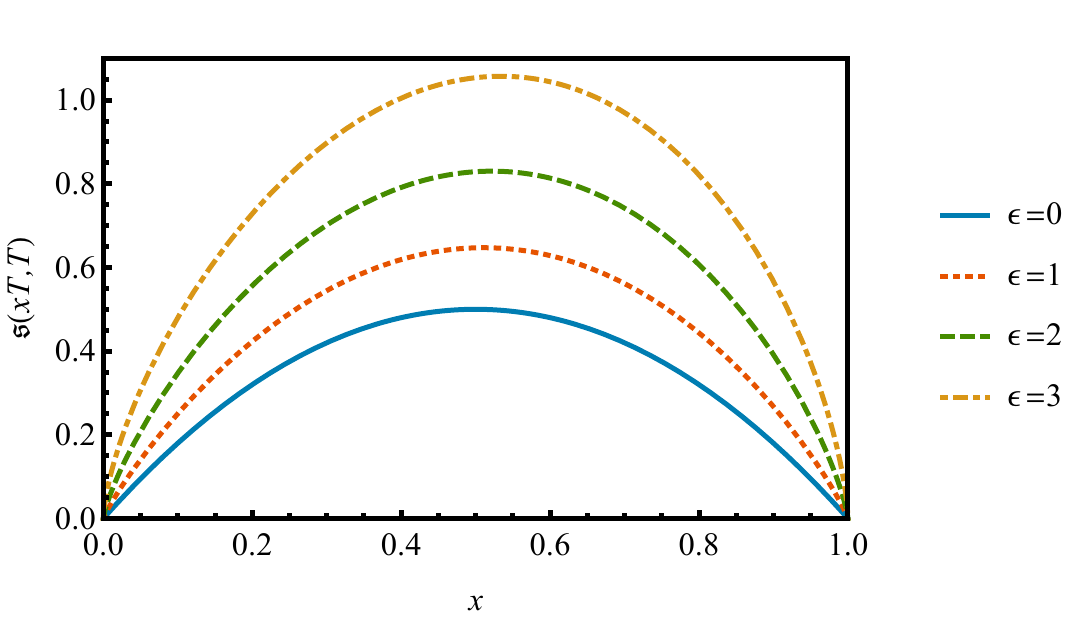}
                 \caption{Average avalanche shape, for short avalanches, including the one-loop correction correction, for various values of $\epsilon$. The analytical expression is given in \eqref{eq:OneLoopShapeDurShort} (the $T$ power-law prefactor was dropped).}
                 \label{fig:OneLoopShapePlotEps}
\end{figure}

Another interesting expansion is the shape for small total durations $T$. It is obtained by rescaling all times with $T$, i.e. setting $\tm=xT$.
We perform this for each term in \eqref{eq:OneLoopShpDurFinalIntRewr} individually:
\bea
\nn
& \left[\partial_\tf\bigg|_{
\begin{subarray}{l} \tf=T \\ \tm = xT
\end{subarray}
} \mfs^{\tree}(\tm,\tf)\right]d_1(T)+ \left[\frac{1+e^\tf}{1-e^\tf}+\partial_\tf\right]\bigg|_{
\begin{subarray}{l} \tf=T \\ \tm = xT
\end{subarray}
}d^\mfs_1(\tm,\tf) =  T x \left(x-1-x \log x\right) + \mO(T)^2, \\
\nn
& \left[\partial_\tf\bigg|_{
\begin{subarray}{l} \tf=T \\ \tm = xT
\end{subarray}
} \mfs^{\tree}(\tm,\tf)\right]\int_0^T \rmd t'\, d_2(t') = T x^2 \left[ \log (T)+ \gamma -\frac{1}{2}-2 \log (2)\right] + \mO(T)^2, \\
\nn
& 
\int_{0}^{\tm} \rmd t\, \left[\frac{1+e^\tf}{1-e^\tf}+\partial_\tf\right]\bigg|_{
\begin{subarray}{l} \tf=T \\ \tm = xT
\end{subarray}
}\left[ d^\mfs_2(\tf-t,\tf-\tm) \right] = T x \left\{4 (1-x)
   \left[\text{Li}_2(1-x)-\text{Li}_2\left(\frac{1-x}{2}\right)\right] \right. \\
	\nn
	& \quad\quad \left. +(1
   -2 x) \log (T)+\gamma  (1-2 x)+\frac{x}{2}+(1-x) \left[\log
   (1-x)-\frac{\pi ^2}{3}-2 \log ^2(2)\right] \right. \\
	\nn
	& \quad\quad \left. +\log (x)-2 x [2 \log
   (x)+\log (2)]+2 (x+1) \log (x+1)\right\} + \mO(T)^2.
\eea
Taking all these contributions together, one obtains the correction to the shape of short avalanches:
\bea
\label{eq:OneLoopShapeDurShortCorr}
\delta_c\mathfrak{s}(xT,T) =& Tx(1-x)\left[\log T + \log x + \log(1-x) + h(x)\right],\\
\nn
h(x) := & 4\left[
   \text{Li}_2(1-x)-  \text{Li}_2\left(\frac{1-x}{2}\right)\right]-1+ \gamma -\frac{\pi ^2}{3}-2 \log ^2(2) \\
	\label{eq:OneLoopShapeDurDefH}
	& \quad \quad -\frac{2}{1-x}\left[2 x \log (2 x)- (x+1)
   \log (x+1)\right].
\eea
The function $h(x)$ is smooth and non-singular at the edges $x=0$ and $x=1$, as shown in figure \ref{fig:OneLoopShapePlotH}.
The correction \eqref{eq:OneLoopShapeDurShortCorr} is plotted in figure \ref{fig:OneLoopShapePlotCorr}, and its expansion near $x=0$ and $x=1$ agrees with the results \eqref{eq:OneLoopShpLeftEdge}, \eqref{eq:OneLoopShpRightEdge}. Recall from \eqref{eq:OneLoopShpTTree} that the mean-field result is
\bea
\mfs^{\tree}(xT,T) = 2Tx(1-x) + \mO(T)^2.
\eea
Thus, the logarithms in \eqref{eq:OneLoopShapeDurShortCorr} can be interpreted as $\epsilon$-dependent modifications of the exponents for the duration, and near the left and right edge, as already observed in the previous sections.
Re-exponentiating this part, we can write the shape \eqref{eq:OneLoopShapeCorr} as
\bea
\label{eq:OneLoopShapeDurShort}
\mathfrak{s}(xT,T) =& \left[Tx(1-x)\right]^{1-\frac{1}{9}\epsilon} \left[2-\frac{2}{9}\epsilon h(x) + \mO(\epsilon)^2 \right],
\eea
where $h(x)$ is given in \eqref{eq:OneLoopShapeDurDefH}. 
This is plotted for various values of $\epsilon$ in figure \eqref{fig:OneLoopShapePlotEps}. 
We observe that the shape becomes asymmetric, and tilted to the right. Near the two edges it is not linear (as in the mean-field limit), but steeper with a power law $1 - \frac{\epsilon}{9} <1$, as discussed in the previous sections.  Preliminary numerical simulations show a comparable asymmetry in $d=2$ \cite{Laurson2013inpr}. However, for $d=1$ (which is easier to access numerically) the consensus is that the avalanche shape is skewed to the left. 

As discussed in section \ref{sec:OneLoopDurFinal} for the avalanche durations distribution $P(T)$, we expect the asymptotic average shape \eqref{eq:OneLoopShapeDurShort} for short avalanches to be more universal than the result \eqref{eq:OneLoopShpDurFinal} for avalanches of arbitrary duration. In fact, preliminary calculations \cite{LeDoussal2013unpublished} show that the shape for short avalanches \eqref{eq:OneLoopShapeDurShort} is recovered even when the infrared cutoff is a hard cutoff in momentum space, whereas the result \eqref{eq:OneLoopShpDurFinal} is only expected to be universal when the infrared cutoff is a harmonic well with $m\to 0$.

We can conclude that short-range correlated disorder leads to an asymmetry of the average avalanche shape in $d<d_c$. However, the effect is subtle and seems to depend strongly on the dimensionality of the space.
It would be interesting to see if this can be understood physically from the geometry of the avalanche motion. Perhaps, the asymmetry is related to splitting of an avalanche into several spatially separated subavalanches.

\section{Average avalanche shape, at fixed avalanche size\label{sec:OneLoopShapeSize}}
As we saw in section \ref{sec:BFMShape} for the BFM and later in section \ref{sec:BFMRetSize} for the ABBM model with retardation, another interesting and easily accessible observable is the average avalanche shape, as a function of time, averaged over avalanches of a fixed total size. In this section we will compute it for short-range correlated disorder below the critical dimension, to order $\mO(\epsilon)$. For this, we first need the avalanche size distribution to order $\mO(\epsilon)$. This is by now a well-known result \cite{LeDoussalWiese2008c,LeDoussalWiese2011b,LeDoussalWiese2013}.  Let us nevertheless briefly repeat its derivation in the present formalism.

\subsection{Avalanche size distribution to one-loop order\label{sec:OneLoopSize}}
The Laplace transform of the avalanche size distribution is defined in \eqref{eq:DefPofS} through the following MSR field theory observable:
\bea
\hP_S(\lambda) = \int \mD[\du,\tu] e^{-S[\du,\tu]} \frac{1}{L^d}\int_{\xxi}\tu_{\xxi,\ti}\exp\left(\lambda \int_{\xf,\tf} \du_{\xf,\tf}\right).
\eea
Applying \eqref{eq:OneLoopGenObsFinal}, we can write this to order $\epsilon$ as 
\bea
\label{eq:OneLoopObsSize}
\hP_S(\lambda) = \hP_S^{\tree}(\lambda) - \tD''_*(0^+)\, \frac{1}{2}\int_0^\infty \rmd x\, x\,  \left[\hat{\delta}_1 \overline{O_S[\du,\tu]}(x) + \hc_\eta (O_S,x) + \hc_\sigma (O_S,x) + \hc_m(O_S,x) \right].
\eea
Here we defined the avalanche shape observable $O_S$ by
\bea
\label{eq:OneLoopObsSizeFT}
 O_S[\du,\tu] := \frac{1}{L^d}\int_{\xxi}\tu_{\xxi,\ti}\exp\left(\lambda \int_{\xf,\tf} \du_{\xf,\tf}\right).
\eea
The mean-field contribution $\hP_S^{\tree}(\lambda)$ is given by the BFM solution \eqref{eq:ABBMSizeInstanton}:
\bea
\label{eq:OneLoopSizeInst00}
\hP_S^{\tree}(\lambda) = \tu_{x,t}^{(00)} = \frac{1}{2}\left(1-\sqrt{1-4\lambda}\right).
\eea
The diagrammatic structure and the prefactors of the diagrams contributing to \eqref{eq:OneLoopObsShp} are exactly the same as those of the avalanche duration diagrams in section \eqref{sec:OneLoopDurationsDiag}. The only change is that the $\tu_t^{(00)}$ wiggly external lines 
 are given by \eqref{eq:OneLoopSizeInst00}. They do not terminate at a fixed time $\tf$ (as in the case of avalanche durations), instead $\tf$ is integrated over.

The dressed response function \eqref{eq:OneLoopRdr} is now 
\bea
\label{eq:OneLoopRdrSize}
\mathbb{R}_{k,t_1,t} =& \exp\left[\int_{t_1}^t \rmd s \left(1+k^2-2\tu_{s}^{(00)}\right) \right]\theta(t_1-t) = \exp\left[-(k^2+\sqrt{1-4\lambda})(t_1-t)\right] \theta(t_1-t).
\eea
We obtain the following expressions for the size diagrams:
\bea
\nn
\hD_1(p^2) =& \int_\ti^\infty \rmd t_2 \Rdr_{k=0,t_2,t} \int_{t_2}^\infty \rmd t_1 \tu^{(00)} \Rdr_{p,t_1,t_2} \\
\nn
=& \frac{1}{2 \sqrt{1-4 \lambda } p^2-8 \lambda +2}-\frac{1}{2 \left(p^2+\sqrt{1-4 \lambda }\right)} = -\frac{b-1}{2 b \left(b+p^2\right)}, \\
\nn
\hD_2(p^2) =& \int_\ti^\infty \rmd t_3 \Rdr_{k=0,t_3,t} \int_{t_3}^\infty \rmd t_1 \tu^{(00)} \Rdr_{p,t_1,t_3} \int_{t_3}^\infty \rmd t_2 \tu^{(00)} \Rdr_{p,t_2,t_3}\\
\nn
=& \frac{\left(\sqrt{1-4 \lambda }-1\right)^2}{4 \sqrt{1-4 \lambda } \left(p^2+\sqrt{1-4 \lambda
   }\right)^2} = \frac{(b-1)^2}{4 b \left(b+p^2\right)^2},
\eea
where we defined $b:=\sqrt{1-4\lambda}$ in order to simplify notation.
The required counterterms can be identified from the $p^{-2}, p^{-4}$ divergences. We obtain:
\bea
\nn
\hc_1(p^2) = & -\frac{b^2-b \left(p^2+3\right)+p^2+2}{2 b \left(p^2+1\right)^2}, \\
\nn
\hc_2(p^2) = & -\frac{(b-1)^2}{4 b \left(p^2+1\right)^2}.
\eea
It can be checked, as in sections \ref{sec:OneLoopDurationsCT} and \ref{sec:OneLoopShapeTimeCT}, that the sum $\hc_1+\hc_2$ is identical to the physical counterterms written in \eqref{eq:OneLoopObsSize}.
Now, the integrals over the momentum $x=p^2$ in \eqref{eq:OneLoopObsSize} can be performed:
\bea
\nn
D_1 + c_1 := &\frac{1}{2}\int_0^\infty \rmd x\,\left[\hD_1(p^2=x)+\hc_1(x)\right] = -\frac{(b-1) [-b+b \log (b)+1]}{4 b}, \\
\nn
	D_2 + c_2 = &\frac{1}{2}\int_0^\infty \rmd x\,\left[\hD_2(p^2=x)+\hc_2(x)\right] = -\frac{(b-1)^2 \log (b)}{8 b}.
\eea
Finally, the correction to the size distribution \eqref{eq:OneLoopObsSize} is given by
\bea
\nn
\hP_S(\lambda) = & \hP_S^{\tree}(\lambda) - \tD''_*(0^+) \delta_c \hP_S(\lambda), \\
\nn
\delta_c \hP_S(\lambda) =& D_1 + c_1 + D_2 + c_2 = -\frac{(b-1) (-2 b+(3 b-1) \log (b)+2)}{8 b} =\\
\label{eq:OneLoopPofSCorr}
=& \frac{\lambda  \left[\left(12 \lambda +\sqrt{1-4 \lambda }-3\right) \log (1-4 \lambda )-4 \left(4
   \lambda +\sqrt{1-4 \lambda }-1\right)\right]}{4 \left(\sqrt{1-4 \lambda }+1\right) (4 \lambda
   -1)}.
\eea
This is identical to eq. (166) in \cite{LeDoussalWiese2008c}. The Laplace transform can be inverted, giving \cite{LeDoussalWiese2008c}  
\bea
\nn
P(S) =& P^{\tree}(S) - \tD''_*(0^+) \delta_c P(S) + \mO(\epsilon)^2, \\
\label{eq:OneLoopSizeDeltaPofS}
\delta_c P(S) = &  \frac{1}{16} P^{\tree}(S) \left[S \log S + \gamma S + 4 S - 8\sqrt{\pi S} - 6 \log S -6 \gamma + 4 \right].
\eea
The logarithmic terms can be interpreted as corrections to the power-law exponent and to the exponential tail \cite{LeDoussalWiese2008c}: To $\mO(\epsilon)^2$, 
\bea
\nn
P(S) =& \frac{A}{\sqrt{4\pi}S^{\tau}}\exp\left[C\sqrt{S} - \frac{B}{4}S^{\delta}\right],  &&  \\
\nn
\tau = & \frac{3}{2}+\frac{3}{8}\tD''_*(0^+),& \delta = & 1 + \frac{1}{4}\tD''_*(0^+) \\
\nn
A = & 1 - \frac{1}{8}(2-3\gamma)\tD''_*(0^+),& B =& 1 + \tD''_*(0^+)\left(1+\frac{\gamma}{4}\right), \quad\quad\quad C = \frac{1}{2}\sqrt{\pi} \tD''_*(0^+).
\eea
Taking into account that $\tD''_*(0^+) = \frac{2}{9}\epsilon+ \mO(\epsilon)^2$ by \eqref{eq:OneLoopValueD}, we see that the avalanche size exponent is modified as
\bea
\tau = \frac{3}{2}-\frac{1}{12}\epsilon.
\eea
This result is known since \cite{LeDoussalWiese2008c,LeDoussalWiese2011b}. It is also in agreement with the scaling relation discussed in section \ref{sec:FRGScalingRelations}, to order $\epsilon$, and with numerical simulations \cite{RossoLeDoussalWiese2009,LeDoussalMiddletonWiese2009}. In particular, there is a characteristic ``bump'' around $S \approx 1$ where $P(S)$ exceeds its leading power-law behaviour (just as for $P(T)$ discussed in section \ref{sec:OneLoopDurFinal}), and which has been observed in numerics \cite{RossoLeDoussalWiese2009,LeDoussalMiddletonWiese2009}.

Now let us proceed with the average avalanche shape at fixed avalanche size, which is a new result.

\subsection{Avalanche shape at fixed avalanche size -- Diagrams}
As we saw in section \ref{sec:BFMShape}, for obtaining the average avalanche shape at fixed avalanche size $S$ it is useful to first compute $\hsh$ defined in \eqref{eq:DefShapeFixedS}
\bea
\nn
\hsh(\tm,\lambda) := & \overline{e^{\lambda S}\mfs(\tm,S) P(S)} = \int \mD[\du,\tu] e^{-S[\du,\tu]} O_{\hsh}[\du,\tu], \\
O_{\hsh}[\du,\tu] :=&  \frac{1}{L^d}\int_{\xxi}\tu_{\xxi, \ti=0}\int_{\xm}\du_{\xm,\tm}\, \exp\left(-S[\du,\tu] + \lambda\int_{\xf,\tf}\du_{\xf,\tf} \right).
\eea
The original shape at fixed size $\mfs(\tm,S)$ can then be obtained by inverting the Laplace transform, and dividing by $P(S)$.
Using \eqref{eq:OneLoopGenObsFinal}, we can express $\hsh$ to $\mO(\epsilon)$ as 
\bea
\label{eq:OneLoopObsShpSize}
\hsh(\tm,\lambda) = \hsh^{\tree}(\tm,\lambda) - \frac{\tD''_*(0^+)}{2}\int_0^\infty \rmd x\, x\,  \left[\hat{\delta}_1 \overline{O_{\hsh}[\du,\tu]}(x) + \hc_\eta (O_{\hsh},x) + \hc_\sigma (O_{\hsh},x) + \hc_m(O_{\hsh},x) \right].
\eea
The tree-level contribution is obtained from the $w^\rmt \to 0$ limit of \eqref{eq:ABBMShapeSizeLTRes}
\bea
\hsh^{\tree}(\tm,\lambda) = \exp\left(-\sqrt{1-4\lambda}\tm\right).
\eea
The diagrams contributing to the one-loop correction in \eqref{eq:OneLoopObsShpSize} are again identical to the ones for the shape at fixed duration, discussed in section \ref{sec:OneLoopShapeTimeDiag} (except that the final time $\tf$ is integrated over). There are still $6$ diagrams $D_1\cdots D_6$ drawn in figure \ref{fig:OneLoopShapeDiags}, with the  expressions \eqref{eq:OneLoopShpHD1}...\eqref{eq:OneLoopShpHD6}, and with the combinatoric factors \eqref{eq:OneLoopShpDiagSum}. The only difference, similar to the previous section \ref{sec:OneLoopSize}, is that now the final time $\tf$ is integrated over, and the mean-field $\tu^{(00)}$ is given by \eqref{eq:OneLoopSizeInst00}. Similar to the previous section, let us introduce $b:=\sqrt{1-4\lambda}$. All time integrals can be evaluated explicitly, giving the following expressions for the diagrams:
\bea
\nn
\hD_1(p^2)+\hD_2(p^2) = & -\frac{(b-1) e^{-\tm \left(b+p^2\right)} \left\{b \left[e^{p^2 \tm} \left(p^2
   \tm-1\right)+1\right]+p^4 \tm e^{p^2 \tm}\right\}}{2 b p^4 \left(b+p^2\right)}, \\
\nn
	\hD_3(p^2) = & \frac{e^{-b \tm} \left(p^2 \tm+e^{-p^2 \tm}-1\right)}{p^4}, \\
\nn
	\hD_4(p^2) = & \frac{(b-1)^2 \tm e^{-b \tm}}{4 b \left(b+p^2\right)^2}, \\
\nn
	\hD_5(p^2) = & \frac{(b-1)^2 \left[e^{p^2 \tm} \left(p^2 \tm-1\right)+1\right] e^{-\tm
   \left(b+p^2\right)}}{4 p^4 \left(b+p^2\right)^2}, \\
\nn
	\hD_6(p^2) = & -\frac{(b-1) \left[e^{p^2 \tm} \left(p^2 \tm-1\right)+1\right] e^{-\tm
   \left(b+p^2\right)}}{2 p^4 \left(b+p^2\right)}.
\eea
The counterterms per diagram are:
\bea
\nn
\hc_{12}(p^2) = & \frac{(b-1) \left(p^2+2\right) \tm e^{-b \tm}}{2 b \left(p^2+1\right)^2}, \\
\nn
	\hc_3(p^2) = & -\frac{(\tm-1) e^{-b \tm}}{\left(p^2+1\right)^2}-\frac{\tm e^{-b \tm}}{p^2+1}, \\
\nn
	\hc_4(p^2) = & -\frac{(b-1)^2 \tm e^{-b \tm}}{4 b \left(p^2+1\right)^2}, \\
\nn
	\hc_5(p^2) = & 0, \\
\nn
	\hc_6(p^2) = & \frac{(b-1) \tm e^{-b \tm}}{2 \left(p^2+1\right)^2}.
\eea
Evaluating the momentum integrals (for short-ranged elasticity) one obtains
\bea
\nn
D_1+D_2+c_{12} = & \frac{(b-1) e^{-b \tm}}{4 b} \left[-e^{b \tm} \text{Ei}(-b \tm)+\log (b)-\tm+\log (\tm)+\gamma \right], \\
\nn
	D_3+c_3 = & -\frac{1}{2} e^{-b \tm} \left[-\tm+\log (\tm)+\gamma +1\right], \\
\nn
	D_4+c_4 = & -\frac{(b-1)^2 \tm e^{-b \tm} \log (b)}{8 b}, \\
\nn
	D_5 = &  -\frac{(b-1)^2 e^{-b \tm}}{8 b^2} \left[e^{b \tm} (b \tm-1) \text{Ei}(-b \tm)-b \tm+\log (b)+\log (\tm)+\gamma
   \right], \\
\nn
	D_6+c_6 = & \frac{(b-1) e^{-b \tm}}{4 b} \left[-e^{b \tm} \text{Ei}(-b \tm)-b \tm+b \tm \log (b)+\log (b)+\log (\tm)+\gamma
   \right].
\eea
Combining these with the correct combinatorical prefactors (see \eqref{eq:OneLoopShpDiagSum}), we obtain the correction to the Laplace-transformed shape:
\bea
\nn
\delta_c \hsh(\tm,\lambda) = & \frac{e^{-b \tm}}{4 b^2} \left\{\left(b^2-1\right) (b \tm+2) \log (b)-2 \left[b^2+(b-2) b \tm+\log (\tm)+\gamma
   \right] \right. \\
\label{eq:OneLoopShpSizeCorr}
	& \quad \left. -2 (b-1) e^{b \tm} \left[(b-1) b \tm+b+1\right] \text{Ei}(-b \tm)\right\}.
\eea

\subsection{Final result and asymptotics}
The correction of order $\epsilon$ to the shape at fixed size is
\bea
\nn
\mfs(\tm,S) =& \mfs^{\tree}(\tm,S) - \tD''_*(0^+) \delta_c \mfs(\tm,S),\\
\label{eq:OneLoopShapeSizeCorr}
\delta_c \mfs(\tm,S) = & \frac{1}{P^{\tree}(S)}\delta_c \left[\mfs(\tm,S)P(S)\right] - \mfs^{\tree}(\tm,S) \frac{1}{P^{\tree}(S)}\delta_c P(S),
\eea 
where $\mfs^{\tree}(\tm,S)$ is given by \eqref{eq:ABBMShapeSizeResSmallW}, and $\delta_c P(S)$ is given by \eqref{eq:OneLoopSizeDeltaPofS}. $\delta_c \left[\mfs(\tm,S)P(S)\right]$ is the inverse Laplace transform of \eqref{eq:OneLoopShpSizeCorr},
\bea
\int_0^\infty e^{\lambda S}\delta_c \left[\mfs(\tm,S)P(S)\right]\rmd S = \delta_c \hsh(\tm,\lambda).
\eea

We can use \eqref{eq:OneLoopShapeSizeCorr} together with \eqref{eq:OneLoopShpSizeCorr} to check e.g. that the normalization of the shape is preserved:
\bea
\nn
\int_0^\infty \rmd \tm \, \hsh(\tm,\lambda) = & \frac{b^2+\left(3 b^2-1\right) \log (b)-4 b+3}{4 b^3}\\
\nn
=&  \partial_\lambda\big|_{\lambda=0} \delta_c \hP_S(\lambda) = \int_0^\infty \rmd S \, e^{\lambda S}\,\int_0^\infty \rmd \tm \, \mfs^{\tree}(\tm,S) \delta_c P(S),
\eea
with $\delta_c \hP_S(\lambda) $ given in \eqref{eq:OneLoopPofSCorr}. Inserting this into \eqref{eq:OneLoopShapeSizeCorr}, we obtain
\bea
\int_0^\infty \rmd \tm \, \delta_c \mfs(\tm,S) = 0 \Rightarrow \int_0^\infty \rmd \tm \, \mfs(\tm,S) = \int_0^\infty \rmd \tm \, \mfs^{\tree}(\tm,S) = S.
\eea

\begin{figure}%
         \centering
         \begin{subfigure}[t]{0.45\textwidth}
                 \centering
                 \includegraphics[width=\textwidth]{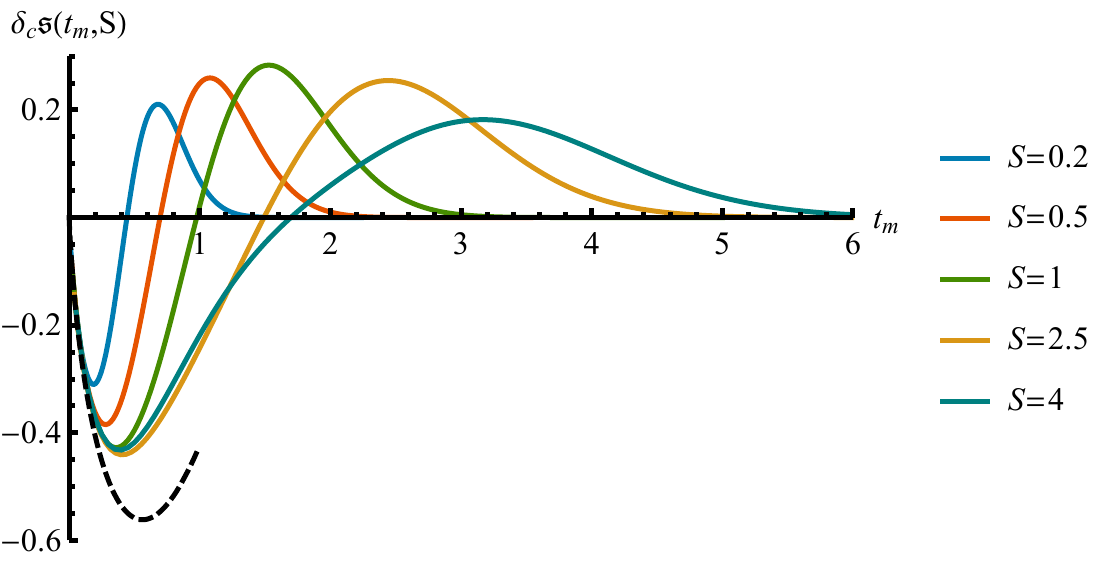}
                 \caption{One-loop correction to the average avalanche shape for various fixed avalanche sizes, given by \eqref{eq:OneLoopShapeSizeCorrFinal}. For small $S$, the shape is $S$-independent (up to rescaling), given by \eqref{eq:OneLoopShapeSizeDefH} and shown in figure \ref{fig:OneLoopShapeSizeCorrShort}. For any $S$, the asymptotics near the beginning is given by \eqref{eq:OneLoopShapeSizeLeft} (black dashed line).}
                 \label{fig:OneLoopShapeSizeCorrLong}
         \end{subfigure}%
         ~ 
         \begin{subfigure}[t]{0.45\textwidth}
                 \centering
                 \includegraphics[width=\textwidth]{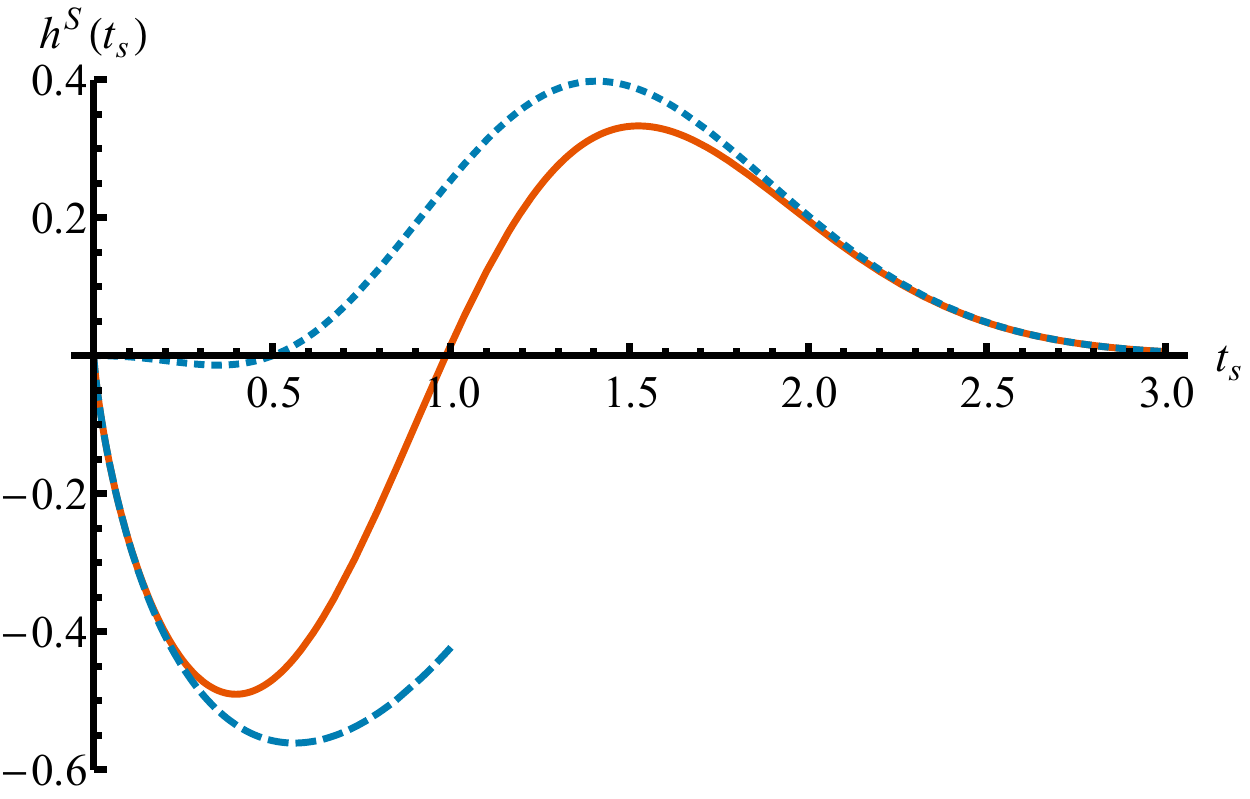}
                 \caption{Asymptotic correction to the shape at fixed size $h^S(t_s)$ for small avalanches, defined in eq.~\eqref{eq:OneLoopShapeSizeDefH}. The dashed and dotted lines are the asymptotics near the beginning and the end, given by \eqref{eq:OneLoopShapeSizeLeftH} and \eqref{eq:OneLoopShapeSizeRightH}, respectively.}
                 \label{fig:OneLoopShapeSizeCorrShort}
         \end{subfigure}%
				                \caption{Average avalanche shape at fixed avalanche size, to one loop.\label{fig:OneLoopShapeSizeCorr}}
\end{figure}

We can also invert the Laplace transform in \eqref{eq:OneLoopShpSizeCorr}, giving via \eqref{eq:OneLoopShapeSizeCorr} an explicit expression for $\delta_c \mfs(\tm,S)$:
\bea
\nn
\delta_c \mfs(\tm,S) =& \frac{e^{-t_s^2}}{8
   \sqrt{S}} \left(t_s \left\{4 S
   \left[(1-t_s^2)\log (S)-2 \sqrt{S} t_s+S+2 \sqrt{\pi } \sqrt{S}-\gamma  t_s^2+2
   t_s^2+\gamma \right] \right. \right. \\
\nn
	& \left.\left. \quad +\pi  S \text{erfi}(t_s) \left(-S+4 t_s^2+2\right)\right\}-S e^{t_s^2} \left\{4 S t_s \,
   _2F_2\left(\frac{1}{2},\frac{1}{2};\frac{3}{2},\frac{3}{2};-t_s^2\right) \right. \right. \\
\nn
	& \left.\left. \quad
	+S t_s\left[\partial_a\big|_{a=\frac{1}{2}} + 2\partial_a\big|_{a=\frac{3}{2}}\right]\left[
	\,_1F_1\left(a,\frac{1}{2},-t_s^2\right)-2 \,_1F_1\left(a,\frac{3}{2},-t_s^2\right) \right] \right.\right. \\
\nn
	& \left.\left. \quad
	+\sqrt{\pi } S \partial_a\big|_{a=0}\,_1F_1\left(a,\frac{1}{2},-t_s^2\right)
	-\sqrt{\pi}\left[8 \sqrt{S} t_s	+\gamma   S  +2 S \log
   (t_s)-4 \right]\text{erf}(t_s) \right.\right. \\
\label{eq:OneLoopShapeSizeCorrFinal}
	& \left.\left. \quad
+2t_s\left(S -2\right)\text{Ei}\left(-t_s^2\right)
   +4 \sqrt{\pi } t_s\left(2 \sqrt{S} +t_s\right)\right\}\right),
\eea
where $t_s := \tm/\sqrt{S}$. 
This is plotted as a function of $t_s$ for various total sizes $S$ in figure \ref{fig:OneLoopShapeSizeCorrLong}.

The expansion of $\delta_c \mfs(\tm,S)$ for small $\tm$ at fixed $S$ is
\bea
\label{eq:OneLoopShapeSizeLeft}
\delta_c \mfs(\tm,S) = \tm \left(-1 + \gamma + \log \tm \right) + \mO(\tm)^2.
\eea
This is independent of $S$ and identical to the expansion of the average avalanche shape at fixed duration, near the left edge, eq.~\eqref{eq:OneLoopShpLeftEdge}. Thus, the leading-order behaviour of the average interface velocity near the avalanche beginning is independent of how the avalanche is delimited: Not only does it not matter \textit{which} value of a fixed size $S$ or fixed duration $T$ we restrict ourselves to, we actually obtain the same behaviour no matter \textit{whether} we consider avalanches of a fixed size or a fixed duration.
Inserting \eqref{eq:OneLoopShapeSizeLeft} into \eqref{eq:OneLoopShapeSizeCorr}, and using that $\mfs^{\tree}(\tm,S) = 2\tm + \mO(\tm)^2$, we can resum the logarithm as a correction to the exponent. Using the universal value of $\tD''_*(0^+)$ in \eqref{eq:OneLoopValueD}, we obtain the average shape at fixed size, near the avalanche beginning
\bea
\label{eq:OneLoopShapeSizeLeftTot}
\mfs(\tm,S) = \tm^{1-\frac{1}{9}\epsilon}\left[2 - \frac{2}{9}\epsilon\left(-1+\gamma\right)\right] + \mO(\tm)^2.
\eea
Again, this is identical to the average avalanche shape at fixed duration, near the avalanche beginning \eqref{eq:OneLoopShpLeftEdgeTot}. The shape becomes steeper, with an exponent $1\to 1-\frac{1}{9}\epsilon$.

Another interesting limit is that of small avalanches, $S\to 0$. There the correction to the shape \eqref{eq:OneLoopShapeSizeCorrFinal} takes the form 
\bea
\label{eq:OneLoopShapeSizeShort}
\delta_c \mfs(\tm,S) =& \sqrt{S}\left[\frac{1}{2}e^{-t_s^2}t_s(1-t_s^2) \log (S) + h^S(t_s)\right], \\
\nn
 h^S(t_s) :=& \frac{1}{4} e^{-t_s^2} t_s \left\{\left(1-2
   t_s^2\right) t_s^2 \left[\,
   _2F_2\left(1,1;-\frac{1}{2},2;-t_s^2\right)-\,
   _2F_2\left(1,1;\frac{3}{2},2;-t_s^2\right)\right] \right. \\
\nn
	& \left. -2 \left(3-2
   t_s^2\right) t_s^2 \,
   _2F_2\left(1,1;\frac{1}{2},2;-t_s^2\right) \right. \\
	\nn
	& \left. +2
   e^{t_s^2} \left[\sqrt{\pi } \left(2 t_s^2
   F(t_s)+F(t_s)-t_s\right)
   \text{erfc}(t_s)+\text{Ei}\left(-t_s^2\right)\right] \right. \\
\label{eq:OneLoopShapeSizeDefH}
	& \left. +2
   \left(4 t_s^4+3\right) t_s F(t_s)+2 \left(-2
   t_s^4-\gamma  t_s^2+t_s^2+\gamma -2\right)\right\}.
\eea
Here, as above, $t_s := \tm/\sqrt{S}$ and $F(x) := e^{-x^2}\int_0^x \rmd y\,e^{y^2}$ is the Dawson $F$-function. The term $\sim \log S$ in \eqref{eq:OneLoopShapeSizeShort} arises from the correction \eqref{eq:OneLoopZEps} to the dynamical exponent 
\bea
z = 2 - \tD''_*(0^+) + \mO(\epsilon)^2 = 2-\frac{2}{9}\epsilon + \mO(\epsilon)^2,
\eea
and a correspondingly modified scaling variable $\tm/S^{1/z}$.
For small $t_s$, $h^S(t_s)$ reproduces the asymptotics \eqref{eq:OneLoopShapeSizeLeft}:
\bea
\label{eq:OneLoopShapeSizeLeftH}
h^S(t_s) = t_s \left(-1 + \gamma + \log t_s \right) + \mO(t_s)^2.
\eea
On the other hand, for large $t_s$,
\bea
\label{eq:OneLoopShapeSizeRightH}
h^S(t_s) = e^{-t_s^2}\left[t_s^3 \log (2t_s) + \mO(t_s)^2\right].
\eea
The function $h^S(t_s)$ in \eqref{eq:OneLoopShapeSizeDefH}, as well as its asymptotics in 
\eqref{eq:OneLoopShapeSizeLeftH} and \eqref{eq:OneLoopShapeSizeRightH}, are shown in figure \ref{fig:OneLoopShapeSizeCorrShort}. As for the average avalanche shape at fixed duration, we expect the scaling form $h^S$ for small avalanches to be universal and cutoff-independent.

\section{Long-ranged elastic interactions\label{sec:OneLoopLongRange}}
\subsection{Scaling\label{sec:OneLoopLRScaling}}
As we saw in section \ref{sec:IntroInterfaceModel}, for some applications, it is useful to consider \eqref{eq:InterfaceEOMi} with more general elasticity than the short-ranged kernel $\nabla^2$ discussed so far. One typical example, as discussed in section \ref{sec:InterfacePhenomenology}, is a nonlocal power-law elastic kernel \eqref{eq:InterfaceEOMLRi}. In Fourier space, it replaces the elastic term in \eqref{eq:InterfaceEOMi} by
\bea
\label{eq:OneLoopLRKernelRepl}
\left(-k^2-m^2\right) u_{k,t}u_{-k,t} \to -J(k) u_{k,t}u_{-k,t}.
\eea
The power-law decay of elastic forces in \eqref{eq:InterfaceEOMLRi} with exponent\footnote{Recall that as above, the exponent is denoted $\mu$ following \cite{DurinZapperi2000}. In the notations of \cite{LeDoussalWieseChauve2002}, $\mu=\alpha$ (but here we use $\alpha$ for the exponent of the avalanche duration distribution), in the notations of \cite{LeDoussalWiese2013}, $\mu=\gamma$.} $\mu$ yields a power-law behaviour of $J$ for large $|k|$,
\bea
J(|k|) \sim |k|^{\mu}.
\eea
The case $\mu=2$ corresponds to the universal behaviour of an interface with short-ranged elasticity, discussed in the previous sections. 
While \eqref{eq:OneLoopLRKernelRepl} does not modify the scaling dimensions of $x, t$ and $u$ in \eqref{eq:OneLoopScaling1}, the fact that now the elastic kernel scales $\sim m^{\mu}$ instead of $\sim m^2$ modifies the inferred scaling dimensions for $\tu, \Delta$. The generalization of \eqref{eq:OneLoopScaling1}, \eqref{eq:OneLoopScaling2}, \eqref{eq:OneLoopScaling3} is:
\bea
\nn
x = & m^{-1} \underline{x}, & t =& m^{-z} \underline{t}, & u_{xt} =& m^{-\zeta} \uu_{\underline{xt}}, \\
\nn
\eta = & m^{\mu-z} \underline{\eta}, & \hu_{xt} =& m^{d+z-\mu+\zeta} \underline{\hu}_{\underline{xt}}, & \Delta =& m^{2\mu-d-2\zeta} \uD, \\
\label{eq:OneLoopScalingLR}
\tu_{xt} =& m^{d-\mu+\zeta}\underline{\tu}_{\underline{xt}},& \du_{xt} =& m^{z-\zeta} \underline{\du}_{\underline{xt}}.
\eea
From the scaling of $\Delta$, we observe that the critical dimension $d_c$ is now modified to $d_c=2\mu$. As already observed in \cite{CizeauZapperiDurinStanley1997,ZapperiCizeauDurinStanley1998}, this shows that mean-field theory is applicable to domain walls in polycristalline ferromagnets, where $\mu=1$ and $d=d_c=2$. However, e.g.~crack fronts in fracture (see section \ref{sec:Fracture}) and fluid contact lines (see section \ref{sec:OtherExp}) have $\mu=1$ and $d=1<d_c$, thus it is interesting to go beyond mean-field and generalize the results of the previous sections to the case of long-ranged elasticity, too.
The modified scaling \eqref{eq:OneLoopScalingLR} also leads to modified scaling relations for the avalanche size and duration exponents, discussed in section \ref{sec:FRGScalingRelations}. The relations \eqref{eq:OneLoopScalingTau}, \eqref{eq:OneLoopScalingAlpha}, \eqref{eq:OneLoopScalingA}, \eqref{eq:OneLoopScalingTauLocal}, \eqref{eq:OneLoopScalingALocal} are modified as follows:
\bea
\nn
\tau =& 2-\frac{\mu}{d+\zeta},& \alpha =& 1+\frac{d-\mu+\zeta}{z}, & a =& 2-\frac{\mu}{d+\zeta-z}, \\
\label{eq:OneLoopScalingExpLR}
\tau_{\phi} =& 2-\frac{\mu}{d_\phi +\zeta}, & a_\phi =& 2-\frac{\mu}{d_\phi+\zeta-z}.
\eea
As in section \ref{sec:FRGScalingRelations}, these exponents reduce to their mean-field values for the BFM case $\zeta = \epsilon = 2\mu-d$, $z = \mu$.

Let us now specialize to the case $\mu=1$. This case is relevant for fracture and fluid contact lines (see section \ref{sec:IntroInterfaceModel}). In particular, we shall consider a kernel of the form 
\bea
\label{eq:OneLoopLRKernel}
J(k) = \sqrt{m^2+k^2}.
\eea

\subsection{Functional RG}
The generalization of the one-loop functional RG treatment in section \ref{sec:FRGReview} for determining the universal effective disorder $\Delta$ was discussed in \cite{ChauveLeDoussalWiese2000a,LeDoussalWieseChauve2002}. It turns out that to leading order in $\epsilon$, the one-loop FRG equation \eqref{eq:FRGOneLoop} is not modified (apart from the fact that now $\epsilon=2-d$), and the random-field fixed point $\tD_*$ relevant for depinning is still given by \eqref{eq:OneLoopDeltaResc}, with $\zeta=\epsilon/3 + \mO(\epsilon)^2$. In particular, we still have the universal value $\tD''_*(0^+) = \frac{2}{9}\epsilon$ as in \eqref{eq:OneLoopValueD}. Modifications appear starting from order $\epsilon^2$.

Let us now apply this to the avalanche field theory discussed in section \ref{sec:FRGAvalanches}. 
\eqref{eq:OneLoopLRKernel} means that the momentum $p^2$ appearing in the loop integrals has to be replaced by 
\bea
-1-p^2 \to -\sqrt{1+p^2} \Leftrightarrow p^2 \to \sqrt{1+p^2}-1.
\eea
In particular, the correction $\delta_1 \overline{O[\du,\tu]}$ with long-ranged elasticity can be written as
\bea
\label{eq:OneLoopDeltaGHLR}
\delta_{1,LR} \overline{O[\du,\tu]} = \int_p \hat{\delta}_1 \overline{O[\du,\tu]}(\sqrt{1+p^2}-1),\quad\quad\quad \int_p := \int\frac{\rmd^d p}{(2\pi)^d},
\eea
where $\hat{\delta}_1 \overline{O[\du,\tu]}$ is the same for long-ranged and short-ranged elasticity. This applies analogously to the counterterms \eqref{eq:OneLoopCEtaH}, \eqref{eq:OneLoopCSigmaH}, \eqref{eq:OneLoopCMH}.

To simplify \eqref{eq:OneLoopDeltaGHLR}, let us now substitute $x := \sqrt{1+p^2}-1$. Then the momentum shell integral \eqref{eq:OneLoopSRLoopInt} for a general integrand $f$ reads
\bea
\nn
\int_p f(\sqrt{1+p^2}-1)  =& \int \frac{\rmd^2 p}{(2\pi)^2} f(\sqrt{1+p^2}-1) =  \frac{S_2}{4 \pi^2} \int_0^\infty \rmd x\,(x+1)f(x)\\
\label{eq:OneLoopLRLoopInt}
 = & \frac{1}{2\pi}\int_0^\infty \rmd x\,(x+1)f(x).
\eea
Applying this similarly to \eqref{eq:OneLoopSRLoopInt}, the one-loop integral \eqref{eq:OneLoopIntegral} becomes
\bea
 \label{eq:OneLoopLRLoopInt1}
\tilde{I}_1 = \frac{1}{2\pi} \frac{1}{\epsilon} + \mO(\epsilon)^0 \quad\quad\quad\Leftrightarrow \quad\quad\quad \lim_{\epsilon \to 0} \epsilon \tilde{I}_1 = \frac{1}{2\pi}.
\eea
Combining this, the modification of eq.~\eqref{eq:OneLoopGenObsFinal} for the correction of $\mO(\epsilon)$ to a general observable becomes, for long-ranged elasticity,
\bea
\nn
\overline{O[\du,\tu]}^{S,LR} = & \overline{O[\du,\tu]}^{\tree} \\
\label{eq:OneLoopGenObsFinalLR}
& - \tD''_*(0^+)\, \int_0^\infty \rmd x\, (x+1)\, \left[\hat{\delta}_1 \overline{O[\du,\tu]}(x) + \hat{c}_\eta (O,x) + \hat{c}_\sigma (O,x) + \hat{c}_m(O,x) \right] + \mO(\epsilon)^2.
\eea
At order $\epsilon^0$, i.e. $d \geq d_c=2$, the observable $O$ is given by the tree-level (BFM) result, just like for $d \geq d_c =4$ in the case of short-ranged elasticity. 
At order $\epsilon^1$ below the critical dimension $d_c=2$, we still have $\tD''_*(0^+) = \frac{2}{9}\epsilon + \mO(\epsilon)^2$ as in \eqref{eq:OneLoopValueD}. The expressions for $\hat{\delta}_1 \overline{O[\du,\tu]}$, $\hat{c}_\eta$, $\hat{c}_\sigma$, $\hat{c}_m$ are identical to the case of short-ranged elasticity. 
The only change in \eqref{eq:OneLoopGenObsFinalLR} compared to \eqref{eq:OneLoopGenObsFinal} is the modified integration measure, 
\bea
\label{eq:OneLoopLRMeasure}
\frac{1}{2}\int_0^\infty x\, \rmd x  \to \int_0^\infty (x+1)\,\rmd x.
\eea
This can now be used to generalize the results of sections \ref{sec:OneLoopDurations}, \ref{sec:OneLoopShapeTime}, \ref{sec:OneLoopShapeSize} to the long-ranged elastic kernel \eqref{eq:OneLoopLRKernel}. This will be performed in the following sections.

\subsection{Avalanche durations with long-ranged elasticity}
Let us first discuss the distribution of avalanche durations $P(T)$ for long-ranged elasticity. $P(T)$ was defined and computed to tree level in section \ref{sec:BFMDuration}, and for short-ranged elasticity to $\mO(\epsilon)$ in section \ref{sec:OneLoopDurations}.
As discussed around eq.~\eqref{eq:OneLoopGenObsFinalLR}, the expressions for the non-integrated loop correction $\hat{\delta}_1 \overline{O[\du,\tu]}(x)$  remain unchanged for long-ranged elasticity. In \eqref{eq:OneLoopPertRes3} it is expressed through the diagrams $\hD_1$, $\hD_2$ in \eqref{eq:OneLoopDurDiag1a}, \eqref{eq:OneLoopDurDiag2a}, with the replacement $p^2 \to x$. The same holds for the counterterms \eqref{eq:OneLoopDurD1CT}, \eqref{eq:OneLoopDurD2CT}. Performing the momentum integral in \eqref{eq:OneLoopGenObsFinalLR} with the modified measure \eqref{eq:OneLoopLRMeasure}, one obtains the modification of 
\eqref{eq:OneLoopDurDiag1b} for long-ranged elasticity:
\beq
\label{eq:OneLoopDurDiag1bLR}
(D_1 + c_1)^{LR} = \frac{e^\tf}{(e^\tf-1)^2} d_1^{LR}(\tf),\quad\quad d_1^{LR}(\tf) = (1-\gamma)  \tf-(\tf-2) \log (\tf)+2 \gamma -2e^{-\tf}
   \text{Ei}(\tf).
\eeq
Similarly, we obtain the long-range modification of the expression \eqref{eq:OneLoopDurDiag2b} for diagram 2:
\bea
\nn
(D_2 + c_2)^{LR} =& \frac{e^\tf}{(e^\tf-1)^2} \int_0^\tf \rmd t'\, d_2^{LR}(t'), \\
\nn
d_2^{LR}(t') = &\frac{e^{t'}}{\left(e^{t'}-1\right)^2} \left\{-2 \left(2 t'+2 e^{t'}+1\right)
   e^{-2t'}\text{Ei}(t')+(4 t'+3)e^{-2t'} \text{Ei}(2 t') \right. \\
\label{eq:OneLoopDurDiag2bLR}
	& \left. + \left[t'
   \log (4)+\log \left(\frac{t'}{8}\right)+\gamma -2\right]+2 e^{-t'} \left[\log
   (t')+\gamma +2\right]-2e^{-2t'}\right\}.
\eea
The correction of order $\epsilon$ to the avalanche duration density with long-ranged elastic interactions is then given, analogously to \eqref{eq:OneLoopDurPofTCorr}, \eqref{eq:OneLoopDurDeltaF1}, as
\bea
\nn
P^{LR}(T) = & P^{\tree}(T) - \tD''_*(0^+) \delta_c P(T),\quad\quad \quad \delta_c P^{LR}(T) = \partial_{\tf}\bigg|_{\tf=T} \delta_c F^{LR}(\tf),\\
\label{eq:OneLoopDurDeltaF1LR}
\delta_c F^{LR}(\tf) = & (D_1 + c_1)^{LR} + (D_2 + c_2)^{LR} = \frac{e^\tf}{(e^\tf-1)^2}\left[d_1^{LR}(\tf) + \int_0^\tf \rmd t'\, d_2^{LR}(t') \right].
\eea
The tree-level result $P^{\tree}(T)$ is still given by \eqref{eq:BFMPofT}, and $d_1^{LR}$ and $d_2^{LR}$ are given in \eqref{eq:OneLoopDurDiag1bLR}
 and \eqref{eq:OneLoopDurDiag2bLR}.
Expanding $P^{LR}(T)$ for small $T$, and using the value $\tD''_*(0^+) = \frac{2}{9} \epsilon$ from \eqref{eq:OneLoopValueD}, we obtain the series
\bea
P^{LR}(T) = \frac{e^T}{(e^T-1)^2}\left\{1 +\tD''_*(0^+) \left[\left(2 \log (T)+2 \gamma -\frac{9}{2}-\log (4)\right)+\frac{3 T}{2}+O\left(T^2\right)\right] \right\}.
\eea
Interpreting the logarithm as a modification to the exponent, just like in \eqref{eq:OneLoopDurPofTCorr2} we obtain the order-$\epsilon$ correction to the power-law exponent $\alpha$ of the avalanche duration distribution, for long-ranged elasticity:
\bea
P(T) = T^{-\alpha},\quad\quad \alpha = 2 - \frac{4}{9}\epsilon +\mO(\epsilon)^2.
\eea
This is consistent, to order $\epsilon$, with the scaling relation \eqref{eq:OneLoopScalingExpLR}, given the values $\zeta = \epsilon/3 + \mO(\epsilon)^2$, $z = 1 - \frac{2}{9}\epsilon + \mO(\epsilon)^2$. 

\subsection{Average avalanche shape at fixed duration, for long-ranged elasticity}
Now let us discuss the average avalance shape $\mfs(\tm,T)$ at fixed duration $T$, as defined in section \ref{sec:BFMShape} and specifically \eqref{eq:DefShapeFixedT}. As for the avalanche duration distribution in the previous section, the tree-level result \eqref{eq:BFMShapeFixedT} is modified at large scales below the critical dimension; this correction was computed to $\mO(\epsilon)$ for short-ranged elasticity in section \ref{sec:OneLoopShapeTime}. Here I discuss the modifications to this calculation for the case of long-ranged elasticity.\footnote{Similarly, one can also generalize the computation of the average avalanche shape $\mfs(\tm,S)$ as a function of time $\tm$, at fixed avalanche size $S$ (discussed in section \ref{sec:OneLoopShapeSize}), to long-ranged elasticity. The explicit calculation is not difficult and will be left for future work.}

Recall from \eqref{eq:OneLoopDefShp} that the average avalance shape $\mfs(\tm,T)$ is the derivative of the cumulative shape $F^{\mfs}$, which is expressible as a field theory observable:
\bea
\mfs^{LR}(\tm,T) = \frac{1}{P^{LR}(T)}\partial_{\tf}\bigg|_{\tf=T}F^{\mfs,LR}(\tm,\tf),\quad\quad F^{\mfs,LR}(\tm,\tf) = \overline{O_\mfs},
\eea
where $O_\mfs$ is defined in \eqref{eq:OneLoopObsShpFT}. Here we added the superscript LR and took the $\lambda \to -\infty$ limit in \eqref{eq:OneLoopDefShp} implicitely.
For long-ranged elasticity, \eqref{eq:OneLoopGenObsFinalLR} allows us to express the cumulative shape $F^{\mfs,LR}$ to order $\epsilon = d_c-d = 2-d$ as
\bea
\nn
F^{\mfs,LR}(\tm,\tf) =& \left(F^\mfs\right)^{\tree}(\tm,\tf) - \tD''_*(0^+)\,\delta_c F^{\mfs,LR}(\tm,\tf), \\
\label{eq:OneLoopObsShpLR}
 \delta_c F^{\mfs,LR}(\tm,\tf) = & \lim_{\lambda \to -\infty} \int_0^\infty \rmd x\, (x+1)\,  \left[\hat{\delta}_1 \overline{O_\mfs[\du,\tu]}(x) + \hc_\eta (O_\mfs,x) + \hc_\sigma (O_\mfs,x) + \hc_m(O_\mfs,x) \right].
\eea
As for the distribution of avalanche durations, the only difference to the expression \eqref{eq:OneLoopObsShp} for short-ranged elasticity is the modified integration measure.
The non-integrated one-loop correction $\hat{\delta}_1 \overline{O_\mfs[\du,\tu]}(x)$ given in \eqref{eq:OneLoopShpDiagSum} is unchanged, just as the expressions for the diagrams $\hD_1\cdots \hD_6$ in \eqref{eq:OneLoopShpDiag12b}, \eqref{eq:OneLoopShpDiag3a}, \eqref{eq:OneLoopShpDiag4a}, \eqref{eq:OneLoopShpDiag5a}, \eqref{eq:OneLoopShpDiag6a} remain unchanged, with the replacement $p^2 \to x$. The same holds for the counterterms \eqref{eq:OneLoopShpD12CT}, \eqref{eq:OneLoopShpD3CT}, \eqref{eq:OneLoopShpD4CT}, \eqref{eq:OneLoopShpD6CT}. 

Let us now perform the $x$ integral with the integration measure for long-ranged elasticity in \eqref{eq:OneLoopObsShpLR}. 
In \eqref{eq:OneLoopShpTimeD123}, we obtained a simple expression for diagrams $1-3$ in terms of the function $d_1$ giving the diagram 1 for the avalanche duration distribution. Interestingly, this expression remains unchanged when going to long-ranged elasticity, and replacing $d_1$ by $d_1^{LR}$ as computed for the avalanche duration distribution diagram 1 in \eqref{eq:OneLoopDurDiag1bLR}:
\bea
\nn
& 2(D_1 + D_2 + c_{12}) + (D_3 + c_3) = \frac{e^\tf}{(e^\tf-1)^2}\left[\mathfrak{s}^{\tree}(\tm,\tf) d_1^{LR}(\tf) + d^{\mfs,LR}_1(\tm,\tf)\right], \\
\nn
& d^{\mfs,LR}_1(\tm,\tf) :=  -\left(e^\tmp-e^{-\tmp}\right) d_1^{LR}(\tm) - \left(e^{\tmp}+e^{-\tmp}-2\right) (d^{LR}_1)'(\tm),\\
\nn
&  = \left(e^{-\tm}-e^{-\tf}\right)
   \left\{4 \text{Ei}(\tm)+e^{\tf} \left[(\gamma -1)
   \tm+(\tm-1) \log (\tm)-\gamma \right] \right. \\
\label{eq:OneLoopShpTimeD123LR}
	& \quad\quad \left. +e^{\tm}
   \left[(\gamma -1) \tm+(\tm-3) \log (\tm)-3 \gamma
   \right]\right\}.
\eea
Also, the sum of diagrams $4-6$ simplifies like in \eqref{eq:OneLoopShpTimeD456}
\bea
\nn
& 2 D_4^a = \mfs^{\tree}(\tm,\tf) \frac{e^\tf}{(e^\tf-1)^2} \int_0^\tf \rmd t'\, d_2^{LR}(t'), \\
\label{eq:OneLoopShpTimeD456LR}
&2 D_4^b + 4 D_5 + 2 (D_6+c_6) = \frac{e^\tf}{(e^\tf-1)^2}\left[ \int_{\tmp}^\tf \rmd t'\,   d^{\mfs,LR}_2(t',\tmp) \right],
\eea
where  $d^{\mfs,LR}_2 (t',\tmp)$ is now a bit more complicated than in the short-range case \eqref{eq:OneLoopShpDurDMFS}
\bea
\nn
& d^{\mfs,LR}_2 (t',\tmp) =  \mathfrak{s}^{\tree}(\tmp,t')d_2^{LR}(t')  
+ \frac{\left(e^{\tmp}-1\right)}{\left(e^{t'}-1\right)^2} \left\{4 \text{Ei}(t')
   \left[e^{-\tmp}+e^{-t'} (2 t'+1)+1\right] \right. \\
\nn
	& \left. \quad +4
   e^{-t'} \left[2
   \left(-\tmp+e^{t'}+t'\right)+e^{\tmp}\right]
   \text{Ei}(t'-\tmp)-4 e^{-t'} \left[-2
   \tmp+e^{\tmp}+4 t'+2\right] \text{Ei}(2
   t'-\tmp) \right. \\
\nn
	& \left.\quad +8 \left(e^{t'}-1\right)
   e^{-\tmp-t'} \left(e^{t'}-e^{\tmp}\right)-2
   \gamma  e^{-\tmp} \left[e^{\tmp}
   \left(e^{t'}+3\right)+e^{t'}+1\right] \right. \\
\nn
	& \left. \quad + \left[2
   \left(e^{\tmp}+1\right) (t'-2)
   e^{t'-\tmp}-4\right] \log (t') \right. \\
\nn
	& \left. \quad -2 e^{-\tmp}
   \left[e^{\tmp+t'} (\tmp-t'+1)+e^{t'}
   (\tmp-t'+3)+e^{\tmp}+1\right] \log
   (t'-\tmp) \right. \\
\label{eq:OneLoopShpDurDMFSLR}
	& \left. \quad -2 e^{t'-\tmp} \left[-e^{\tmp}
   (\tmp-2 t'+2)-\tmp+2 t'-4\right] \log (2
   t'-\tmp)\right\}.
\eea
Altogether we can write the correction to the cumulative shape $\delta_c F^{\mfs,LR}$  (for long-ranged elasticity) in \eqref{eq:OneLoopObsShp} as
\bea
\label{eq:OneLoopShpDiagSum2LR}
&\delta_c F^{\mfs,LR} (\tm,\tf) =  2(D_1 + D_2 + c_{12}) + (D_3 + c_3) + 2 (D_4+c_4) + 4 D_5 + 2 (D_6+c_6) \\
\label{eq:OneLoopShpDurCumFinalLR}
&= \mfs^{\tree}(\tm,\tf) \delta_c F^{LR}(\tf) + \frac{e^\tf}{(e^\tf-1)^2}\left[d^\mfs_1(\tm,\tf) + \int_{\tmp}^{\tf} \rmd t'\,  d^{\mfs,LR}_2(t',\tmp) \right],
\eea
where $\delta_c F^{LR}$ is the correction to the cumulative duration distribution (for long-ranged elasticity) computed in \eqref{eq:OneLoopDurDeltaF1LR}, $ d^{\mfs,LR}_2$ is given by \eqref{eq:OneLoopShpDurDMFSLR} and $d_1^{LR}$, $d_2^{LR}$ are given by \eqref{eq:OneLoopDurDiag1b}, \eqref{eq:OneLoopDurDiag2b}. Inserting this into \eqref{eq:OneLoopShapeCorr}, the first term simplifies as in the short-range case and we obtain
\bea
\nn
 &\mfs^{LR}(\tm,T) = \mfs^{\tree}(\tm,T) - \tD''_*(0^+)\,\delta_c \mfs^{LR}(\tm,T), \\
\nn
&\delta_c \mfs^{LR}(\tm,T) = \left[\partial_\tf\bigg|_{\tf=T} \mfs^{\tree}(\tm,\tf)\right]\left[d_1^{LR}(T) + \int_0^T \rmd t'\, d_2^{LR}(t')\right] + \\
\nn
& \quad\quad\quad\quad\quad\quad\quad + \left[\frac{1+e^\tf}{1-e^\tf}+\partial_\tf\right]\bigg|_{\tf=T}\left[d^{\mfs,LR}_1(\tm,\tf)  + \int_{\tmp}^{\tf} \rmd t'\,  d^{\mfs,LR}_2(t',\tmp) \right] \\
\nn
&= \left[\partial_\tf\bigg|_{\tf=T} \mfs^{\tree}(\tm,\tf)\right]\left[d_1^{LR}(T) + \int_0^T \rmd t'\, d_2^{LR}(t')\right] + \left[\frac{1+e^\tf}{1-e^\tf}+\partial_\tf\right]\bigg|_{\tf=T}d^{\mfs,LR}_1(\tm,\tf)+\\
\label{eq:OneLoopShpDurFinalLR}
&\quad\quad + \int_{0}^{\tm} \rmd t\,  \left[\frac{1+e^\tf}{1-e^\tf}+\partial_\tf\right]\bigg|_{\tf=T} d^{\mfs,LR}_2(\tf-t,\tf-\tm).
\eea

\subsubsection{Asymptotics}
Let us now discuss the asymptotics of the average long-range avalanche shape \eqref{eq:OneLoopShpDurFinalLR}, near the beginning and the end of an avalanche, and for short avalanches.

In order to obtain the asymptotics near the avalanche beginning, we follow section \ref{sec:OneLoopShapeTimeBeg} and expand \eqref{eq:OneLoopShpDurFinalLR} in $\tm$ at fixed $T$. As in section \ref{sec:OneLoopShapeTimeBeg}, the term involving $\mfs^{\tree}$ does not contribute to order $\tm$. We have
\bea
\label{eq:OneLoopShpSerLeft2LR}
& \left[\frac{1+e^\tf}{1-e^\tf}+\partial_\tf\right]\bigg|_{\tf=T}\left[d^{\mfs,LR}_1(\tm,\tf)\right] = -2\tm + \mO(\tm)^2,\\
\nn
&\left[\frac{1+e^\tf}{1-e^\tf}+\partial_\tf\right]\bigg|_{\tf=T}\left[\int_{\tmp}^\tf \rmd t'\,   d^{\mfs,LR}_2(t',\tmp)\right] = \\
\nn
& = \left[\frac{1+e^\tf}{1-e^\tf}+\partial_\tf\right]\bigg|_{\tf=T}\left\{-\tm (1-e^{-\tf})\int_{0}^1 \rmd x\,2\left[\gamma+\log\tm + \log(1-x)\right] +\mO(\tm)^2\right\} \\
\label{eq:OneLoopShpSerLeft3LR}
&= 2\tm\left(-1+ \gamma + \log \tm\right)+\mO(\tm)^2.
\eea
Note that the last term contains a boundary-layer contribution near $t=\tm$ just as in \eqref{eq:OneLoopShpSerLeft3}.

Inserting the series \eqref{eq:OneLoopShpSerLeft2LR}, \eqref{eq:OneLoopShpSerLeft3LR} into \eqref{eq:OneLoopShpDurFinalLR}, we obtain:
\beq
\label{eq:OneLoopShpLeftEdgeLR}
\delta_c\mathfrak{s}^{LR}(\tm,T) = 2\tm(\gamma-2+\log \tm) + \mathcal{O}(\tm)^2 .
\eeq
Hence, the average avalanche shape (for long-ranged elasticity) takes the following form near the left edge:
\bea
\nn
\mfs^{LR}(\tm,T) =& 2\tm - \tD_*''(0^+)2\tm(\gamma-2+\log \tm) + \mO(\tm)^2 =\\
\label{eq:OneLoopShpLeftEdgeTotLR}
=& \tm^{1-\frac{2}{9}\epsilon}\left[2-\frac{4}{9}\epsilon(\gamma-2) + \mO(\epsilon)^2\right] + \mO(\tm)^2.
\eea
Here we used the universal value $\tD_*''(0^+) = \frac{2}{9}\epsilon + \mO(\epsilon)^2$ from \eqref{eq:OneLoopValueD}, valid at the depinning fixed point.
The power-law growth of the average velocity near the left edge of an avalanche is modified from an exponent $1\to 1-\frac{2}{9}\epsilon$, which is twice the correction to the exponent obtained in the case of short-ranged elasticity.

Now let us proceed with the expansion near the right edge, following section \ref{sec:OneLoopShapeTimeEnd}. We set $\tm = \tf - \tmp$, and find to leading order in $\tmp$
\bea
&& \label{eq:OneLoopShpSerRight1LR}
\left[\partial_\tf\bigg|_{
\begin{subarray}{l} \tf=T \\ \tm = T-\tmp
\end{subarray}
} \mfs^{\tree}(\tm,\tf)\right] d_1^{LR}(T) + \left[\frac{1+e^\tf}{1-e^\tf}+\partial_\tf\right]\bigg|_{
\begin{subarray}{l} \tf=T \\ \tm = T-\tmp
\end{subarray}
}\left[d^{\mfs,LR}_1(\tm,\tf)\right] & = 0 + \mO(\tmp)^2, \\
\nn
& \lefteqn{
 \left[\partial_\tf\bigg|_{
\begin{subarray}{l} \tf=T \\ \tm = T-\tmp
\end{subarray}
} \mfs^{\tree}(\tm,\tf)\right] d_2^{LR}(T-t) + } &&  \\
\label{eq:OneLoopShpSerRight2LR}
&&
 + \left[\frac{1+e^\tf}{1-e^\tf}+\partial_\tf\right]\bigg|_{
\begin{subarray}{l} \tf=T \\ \tm = T-\tmp
\end{subarray}
}\left[d^{\mfs,LR}_2(\tf-t,\tf-\tm)\right] &= 0 + \mO(\tmp)^2.
\eea
So, the cancellation of integrands in \eqref{eq:OneLoopShpDurFinalLR} up to $\mO(\tmp)^2$, which was observed in \eqref{eq:OneLoopShpSerRight1}, \eqref{eq:OneLoopShpSerRight2} for short-ranged elasticity, also occurs in the case of long-ranged elasticity. 
As discussed in section \ref{sec:OneLoopShapeTimeEnd}, there are two additional contributions. 
The first comes from the difference in integration boundaries in \eqref{eq:OneLoopShpDurFinalLR}
\bea
\nn
&\lefteqn{ -\int_{\tm}^T \rmd t\, \left[\frac{1+e^\tf}{1-e^\tf}+\partial_\tf\right]\bigg|_{
\begin{subarray}{l} \tf=T \\ \tm = T-\tmp
\end{subarray}
}\left[d^{\mfs,LR}_2(\tf-t,\tf-\tm)\right] = } & & \\
\label{eq:OneLoopShpSerRight3LR}
&& & = \tmp \left[2 \log (\tmp)+2 \gamma -3-4 \log (2)\right] + \mO(\tmp)^2
\eea 
The second comes from a boundary layer near $t=\tm$
\bea
\nn
&- \tmp \int_0^\infty \rmd y\,   \left[\frac{\left(6 y^2+4 y+2\right) \log (y)-\left(8 y^2+8 y+4\right) \log
   \left(y+\frac{1}{2}\right)+2 y+1}{(y+1)^2}+2 \log (y+1)\right] + \mO(\tmp)^2 \\
\label{eq:OneLoopShpSerRight4LR}
	& = \tmp\left[3-\frac{2 \pi ^2}{3}-4 \log ^2(2)+8\log (2)\right] + \mO(\tmp)^2
\eea
Combining \eqref{eq:OneLoopShpSerRight1LR}, \eqref{eq:OneLoopShpSerRight2LR}, \eqref{eq:OneLoopShpSerRight3LR} and \eqref{eq:OneLoopShpSerRight4LR}, we obtain the simple expansion for the shape near the right edge
\beq
\label{eq:OneLoopShpRightEdgeLR}
\delta_c \mathfrak{s}^{LR}(T-\tmp,T) = 2\tmp \left[\log (\tmp)-\frac{\pi ^2}{3}+\gamma -2 \log
   ^2(2)+2\log (2)\right]+\mathcal{O}(\tmp)^2.
\eeq
Just like near the left edge, all $T$-dependent terms drop out to order $\mathcal{O}(\tmp)$. 
Inserting \eqref{eq:OneLoopShpRightEdgeLR} into \eqref{eq:OneLoopShpDurFinalLR} together with the mean-field contribution \eqref{eq:OneLoopShpTTree}, we can write the asymptotics near the right edge to order $\epsilon$ as
\bea
\label{eq:OneLoopShpRightEdgeTotLR}
\mfs^{LR}(T-\tmp,T) =  \tmp^{1-\frac{2}{9}\epsilon}\left\{2-\frac{4}{9}\epsilon\left[2\ln 2 -\frac{\pi ^2}{3}+\gamma -2 \log
   ^2(2)\right] + \mO(\epsilon)^2\right\} + \mO(\tmp)^2.
\eea
Here, as in the previous section, we used the universal value $\tD_*''(0^+) = \frac{2}{9}\epsilon + \mO(\epsilon)^2$ from \eqref{eq:OneLoopValueD}, valid at the depinning fixed point.
We observe that the power-law exponent of the average velocity decay near the end of the avalanche is modified $1\to 1-\frac{2}{9}\epsilon$, again identical to the exponent near the beginning of the avalanche \eqref{eq:OneLoopShpLeftEdgeTotLR}.

Finally, let us give the expansion for the shape of short avalanches. We do this as usual term-by-term in \eqref{eq:OneLoopShpDurFinalLR}. Setting $x := \tm/T$, we have
\bea
\nn
&\textstyle \left[\partial_\tf\bigg|_{
\begin{subarray}{l} \tf=T \\ \tm = xT
\end{subarray}
} \mfs^{\tree}(\tm,\tf)\right]d_1^{LR}(T)+ \left[\frac{1+e^\tf}{1-e^\tf}+\partial_\tf\right]\bigg|_{
\begin{subarray}{l} \tf=T \\ \tm = xT
\end{subarray}
}d^{\mfs,LR}_1(\tm,\tf) =  2 T x \left(x-1-x \log x\right) + \mO(T)^2, \\
\nn
& \left[\partial_\tf\bigg|_{
\begin{subarray}{l} \tf=T \\ \tm = xT
\end{subarray}
} \mfs^{\tree}(\tm,\tf)\right]\int_0^T \rmd t'\, d_2^{LR}(t') = T x^2 \left[2 \log (T)+2 \gamma -3-4 \log (2)\right] + \mO(T)^2, \\
\nn
& 
\int_{0}^{\tm} \rmd t\, \left[\frac{1+e^\tf}{1-e^\tf}+\partial_\tf\right]\bigg|_{
\begin{subarray}{l} \tf=T \\ \tm = xT
\end{subarray}
}\left[ d^{\mfs,LR}_2(\tf-t,\tf-\tm) \right] = T x \left\{8 (1-x)
   \left[\text{Li}_2(1-x)-\text{Li}_2\left(\frac{1-x}{2}\right)\right] \right. \\
	\nn
	& \quad\quad \left. +2
   (1-2 x) [\log (T)+\gamma ]+5 x-(1-x) \left[\frac{2 \pi
   ^2}{3}+4 \log ^2(2)\right]+2 (1-x) \log (1-x) \right. \\
	\nn
	& \quad\quad \left. +2 (1-4 x) \log (x)+4
   (x+1) \log (x+1)-4 x \log (2)-2\right\} + \mO(T)^2.
\eea
Taking all these contributions together, one obtains the correction to the shape of short avalanches:
\bea
\label{eq:OneLoopShapeDurShortCorrLR}
\delta_c\mathfrak{s}^{LR}(xT,T) =& Tx(1-x)\left[2\log T + 2\log x + 2\log(1-x) + h^{LR}(x)\right],\\
\nn
h^{LR}(x) = & 8\left[ \text{Li}_2(1-x) - \text{Li}_2\left(\frac{1-x}{2}\right)\right] + 2 \gamma - 4 - \frac{2}{3}\pi^2 - 4 \log^2(2) \\
& 
	\label{eq:OneLoopShapeDurDefHLR}
-\frac{4}{1-x}\left[2x\log(2x) - (1+x)\log(1+x)\right].
\eea
As in the case of short-ranged elasticity, the first line corresponds to a modification of the exponent near the left and right edges. Re-exponentiating this part, we can write the shape as
\bea
\nn
\mfs^{LR}(xT,T) =& \left[Tx(1-x)\right]^{1-\frac{2}{9}\epsilon}\left[2-\frac{2}{9}\epsilon h^{LR}(x) + \mO(\epsilon)^2\right] 
\eea
Comparing \eqref{eq:OneLoopShapeDurDefHLR} to the corresponding expression \eqref{eq:OneLoopShapeDurShort} for short-ranged elasticity, we see that
\bea
h^{LR}(x) = 2 h(x) - 2.
\eea
Thus, the the only difference between the $\mO(\epsilon)$ corrections to the average avalanche shape in the case of LR elasticity, compared to SR elasticity, is the correction to the amplitude of the mean-field shape. The asymmetric parts of the correction are identical (up to the global factor $2$ in \eqref{eq:OneLoopShapeDurShortCorrLR} compared to \eqref{eq:OneLoopShapeDurShortCorr}).

%
%
%

%% file: KPZ.tex
\chapter{Directed Polymers and the Kardar-Parisi-Zhang equation\label{sec:KPZ}}
\section{Overview\label{sec:KPZOverview}}
So far we have exclusively considered an elastic interface with $d$ internal dimensions, in a $d+1$-dimensional embedding space. This interface was parametrized by a one-component height function $u(x)$, where $x\in \mathbb{R}^d$ is the internal coordinate. It is also interesting to discuss elastic interfaces with $d$ internal dimensions, in a $d+\dsf$-dimensional embedding space. This means that the height function $u(x) \in \mathbb{R}^\dsf$ now has $\dsf$ components.

One particularly interesting case is a \textit{directed polymer} (DP), where $d=1$. Its internal (longitudinal) direction $x \in \mathbb{R}$ can be naturally interpreted as a ``time'' direction; we will denote it $t$ from now on. The name \textit{directed} polymer comes from the absence of overhangs: For each time $t$, there is a unique transversal position $u(t) \in \mathbb{R}^\dsf$; the polymer never returns to the same $t$ twice.

In contrast to the previous sections, we shall be particularly interested in the equilibrium properties of the DP in $\dsf$ dimensions, at a temperature $T =: \beta^{-1}$, in a random medium. 
The partition sum $Z$ of a polymer of length $L$, with its ends fixed at $u_i \in \mathbb{R}^\dsf$ and $u_f \in \mathbb{R}^\dsf$, respectively, is given by
\bea
\label{eq:KPZPathInt}
Z_L(u_i,u_f) = \int_{u(0)=u_i}^{u(L)=u_f} \mD[u(t)]\,e^{-\beta H[u]},\quad\quad H[u] = \int_0^L \rmd t\, \left[\frac{1}{2}(\partial_t u(t))^2 + \eta(u(t),t) \right].
\eea
Here $\eta(u,t)$ is a random potential describing the disordered medium. We assume it to be a Gaussian, uncorrelated white noise with
\bea
\label{eq:KPZNoise}
\overline{\eta(u,t)} = 0,\quad\quad\quad \overline{\eta(u,t)\eta(u',t')} = 2 D \delta^{\dsf}(u-u') \delta(t-t').
\eea
This is a formal continuum definition; it will require some small-scale regularization
. For concreteness one may think of a lattice with $L^{\dsf+1}$ sites. On each site there is a random, independent potential $\eta$ chosen from a normal distribution. The partition sum is obtained as the sum over the Boltzmann weights of all directed paths between one corner of the lattice and the diagonally opposite one. 

Applying the Feynman-Kac formula, the partition sum $Z_t(u) := Z_{L=t}(u_i,u_f=u)$ as a function of the final point $u,t$ satisfies the following PDE \cite{HwaFisher1994}
\bea
\label{eq:KPZMNZ}
\partial_t Z_t(u) = \frac{1}{2\beta}\nabla^2_u Z_t(u)  - \beta \eta(u,t) Z_t(u).
\eea
This is the stochastic heat equation with multiplicative noise. Recall that $u \in \mathbb{R}^\dsf$ is the final transversal position of the polymer. \eqref{eq:KPZMNZ} holds independently of the initial condition, i.e. independently of whether the polymer end at $t=0$ is held fixed or is free.

Through the Cole-Hopf mapping $Z(u,t) =: e^{-\beta F_t(u)}$ one obtains a PDE for the free energy $F$ of the directed polymer, 
\bea
\label{eq:KPZKPZ}
\partial_t F_t(u) = \frac{T}{2}\nabla^2_u F_t(u) - \frac{1}{2}\left[\nabla_u F_t(u)\right]^2 + \eta(u,t).
\eea
This is the celebrated \textit{Kardar-Parisi-Zhang (KPZ) equation}. Its deterministic version was originally derived \cite{KardarParisiZhang1986} as a model for the growth of rough surfaces. Note that this formulation allows one to compute directly the free energy landscape at time $t+ \rmd t$ from the free energy landscape at time $t$ (which is less simple in the path integral formulation \eqref{eq:KPZPathInt}). Numerically, this suggests an efficient transfer matrix algorithm for the computation of $F$.

The free energy $F$ shows anomalous fluctuations with an exponent $\theta$,
\bea
\overline{\left[F(x,t)-\overline{F(x,t)}\right]^2} \sim t^{2\theta}.
\eea 
The anomalous fluctuations of the ground state of the directed polymer, $|u| \sim t^\zeta$, translate into free-energy fluctuations $\delta F \sim \int \rmd t\,\left[\partial_t u(t)\right]^2 \sim t^{2\zeta-1}$. 
Thus, the roughness exponent $\zeta$ and the free-energy fluctuation exponent $\theta$ are related via \cite{HuseHenley1985,HuseHenleyFisher1985,FisherHuse1991,HwaFisher1994}
\bea
\label{eq:KPZZetaTheta}
\theta = 2\zeta - 1.
\eea
Fluctuations of $F$ in space scale with a KPZ surface roughness exponent $\chi$, $F(u) \sim u^{\chi}$ (note this is called $\sigma$ in \cite{FisherHuse1991}). Since $u \sim t^{\zeta}$, we have \cite{FisherHuse1991}
\bea
\label{eq:KPZChi}
\chi = \frac{\theta}{\zeta} = 2 - \frac{1}{\zeta}.
\eea

The KPZ equation \eqref{eq:KPZKPZ} is equivalent to the \textit{Burgers equation} known from fluid mechanics. The latter is obtained by defining the velocity vector field $v(u,t) := \nabla_u F(u,t)$, and deriving \eqref{eq:KPZKPZ} with respect to $u$:
\bea
\partial_t v(u,t) = \nu \nabla_u^2 v(u,t) - v(u,t) \cdot \nabla_u v(u,t) + \eta'(u,t),
\eea
where $\nu := T/2$ is the Burgers viscosity. In the zero-temperature limit (frozen phase), it goes to zero and one obtains the \textit{inviscid Burgers equation}.

In the following sections I will review known results on the directed polymer problem and the KPZ equation. In particular, I will review known exact results in the mean-field limit (section \ref{sec:KPZMF}), and in $\dsf=1$ (section \ref{sec:KPZD1}). I will then discuss the challenges for $1<\dsf<\infty$, and in particular some new perspectives on what can be learned from the moments of $Z$ (section \ref{sec:KPZIntDimMoments}). Finally I will discuss the more general problem of a directed polymer with complex random weights in section \ref{sec:KPZComplex}. This model describes some aspects of quantum disordered systems, in particular variable-range-hopping in a disordered insulator. I will explain several new results obtained during this thesis, for this model and for a related solvable toy model, in sections \ref{sec:KPZCxRG} and \ref{sec:KPZCxToy}.

\section{Mean-field limit\label{sec:KPZMF}}
Analytical progress can be made in the mean-field limit. One way to implement it is a tree geometry, for which Derrida and Spohn obtained an exact solution in a landmark article \cite{DerridaSpohn1988,Derrida1991}. In a discrete setting, this means that instead of $\dsf$ neighboring sites on a finite lattice for the transversal coordinate at each time $t=1...L$, the polymer can choose one of $b$ independent branches at each time.
 The total number of paths is then equal to the number of sites, $b^L$. This tree geometry is supposed to describe well the situation in very high dimensions, $\dsf \to \infty$, where paths rarely cross and correlations disappear. 
Instead of the discrete tree, a similar geometry which is easier to treat analytically is a randomly branching tree in continuous time. In a time interval $[t;t+ \rmd t]$, the tree splits up in two independent branches with probability $\lambda\, \rmd t$. Otherwise, it accumulates a random energy $\int_t^{t+\rmd t}\rmd s\, \xi(s)$, where $\xi(s)$ is a white noise. This yields the following evolution equation for the partition sum $Z$:
\bea
\label{eq:KPZMFRecZ}
Z(t+\rmd t) = \begin{cases}
Z^{(1)}(t) + Z^{(2)}(t) & \text{with probability} \quad \lambda\,\rmd t, \\
Z(t)\cdot \exp\left[-\beta\int_t^{t+\rmd t}\rmd s\, \xi(s)\right] & \text{with probability} \quad 1- \lambda\,\rmd t .
\end{cases}
\eea 
Here, $Z^{(1)}$ and $Z^{(2)}$ are \textit{independent} sub-trees. This permits to write a simple recursion relation for the generating function \cite{DerridaSpohn1988}
\bea
\label{eq:KPZMFDefG}
G_t(x) := &\overline{\exp\left[-e^{-\beta x}Z(t)\right]},\\
\label{eq:KPZMFRecG}
G_{t+\rmd t}(x) = & \lambda\,\rmd t\,G_t(x)^2 + (1-\lambda\,\rmd t)\int_{-\infty}^{\infty}\,\frac{\rmd V}{\sqrt{2\pi \, \rmd t}}\,e^{-\frac{V^2}{2\,\rmd t}}G_t(x+V).
\eea
Here, $V := \int_t^{t+\rmd t}\rmd s\, \xi(s)$, and we assumed $\overline{\xi(s)\xi(s')}=\delta(s-s')$. In the continuum limit $\rmd t \to 0$, \eqref{eq:KPZMFRecG} gives a nonlinear PDE for $G$,
\bea
\label{eq:KPZMFPDEG}
\partial_t G_t(x) = \frac{1}{2}\partial_x^2 G_t(x) + \lambda G_t(x)\left[G_t(x)-1\right].
\eea
This is the well-known Fisher-KPP equation \cite{Fisher1999}. Its solution is a traveling wave, which goes from $G_t(-\infty)=0$ to $G_t(\infty)=1$ (this follows from the definition \eqref{eq:KPZMFDefG}). Inserting a traveling-wave ansatz $G_t(x) = w(x-ct)$ into \eqref{eq:KPZMFPDEG}, and using the asymptotics $G_t(x) = 1- e^{-\beta x}$ for large $x$, one finds that the velocity $c$ must satisfy
\bea
\label{eq:KPZMFValc}
c \beta = \frac{1}{2}\beta^2 + \lambda,\quad\quad\quad c = \frac{\lambda}{\beta} + \frac{1}{2}\beta.
\eea
$c$ is equal to the mean free-energy density of the polymer \cite{DerridaSpohn1988}
\bea
\label{eq:KPZMFDefc}
c = \lim_{L\to \infty}\frac{1}{ L} \overline{F(L)}, \quad\quad F(L) := -\frac{1}{\beta} \log Z(L).
\eea
From \eqref{eq:KPZMFValc} one sees that there is a minimum value of the free energy at $\beta=\beta_c = \sqrt{2\lambda}$. This is the critical temperature, at which the DP transitions between a high-temperature phase for $\beta < \beta_c$ and a low-temperature (frozen) phase for $\beta > \beta_c$. In the high-temperature phase, the mean free energy is given by \eqref{eq:KPZMFValc}, whereas in the low-temperature phase it is frozen at its minimal value $c(\beta_c) = 2\sqrt{\lambda}$. Let us set $\lambda=1$ from now on for simplicity.\footnote{Without loss of generality, by rescaling $x$ and the free energy of the DP accordingly.} 
From the shape of the traveling wave profile which solves \eqref{eq:KPZMFPDEG}, one can also obtain the distribution of the fluctuations of the free energy around its average
\bea
\label{eq:KPZMFDeff}
f : = F(L) - c L.
\eea
In the limit of large $L$, the distribution of $f$ is \cite{DerridaSpohn1988}
\begin{itemize}
	\item In the low-temperature phase, 
	\bea
	P(f) \sim \begin{cases} 
	-f\,\exp(\sqrt{2} f) &\mbox{for } f \to -\infty \\ 
 \exp\left[-(2-\sqrt{2})f\right] & \mbox{for } f \to \infty 
\end{cases}.
	\eea
	\item In the high-temperature phase, 
	\bea
	P(f) \sim \begin{cases} 
	\exp\left(\frac{2}{\beta} f\right) &\mbox{for } f \to -\infty \\ 
 \exp\left(-\alpha f\right) & \mbox{for } f \to \infty 
\end{cases}.
	\eea
	Here, $\alpha := \frac{1}{2}\left[\sqrt{\left(\beta + \frac{2}{\beta}\right)^2+8} - \left(\beta + \frac{2}{\beta}\right)\right]$ (see \cite{DerridaSpohn1988}, eq.~(5.13)).
\end{itemize}
	In particular, this implies that the moments $\overline{Z^n}$ cease to exist for $\beta > \beta_n := \sqrt{2/n}$. The true transition temperature is  $\beta_c = \sqrt{2}$ (which is formally $\beta_1 = \beta_c$), but moments of order $n \geq 2$ start to diverge already inside the high-temperature phase.

Let us now discuss the roughness exponent in the frozen state. For simplicity, consider the zero-temperature ground state. Since at each branching, the polymer will continue in one of the two branches with equal probability, its roughness is that of a random walk, i.e. the roughness exponent is $\zeta=1/2$. This is confirmed by a calculation of the free-energy fluctuations: for large $L$, the traveling wave profile in \eqref{eq:KPZMFPDEG} is stationary, hence the distribution of the free-energy fluctuations $f$ defined in \eqref{eq:KPZMFDeff} is also stationary. So, free energy fluctuations do not grow with the polymer length ($\theta=0$), and using the scaling law \eqref{eq:KPZZetaTheta} one finds again $\zeta=1/2$.

\section{Directed polymer in 1+1 dimensions\label{sec:KPZD1}}
\subsection{Scaling Exponents}
In $\dsf=1$, the steady state distribution of the DP free energy $F$, growing according to the KPZ equation \eqref{eq:KPZKPZ}, is known \cite{HuseHenleyFisher1985}. It is a Brownian in the $u$ direction, i.e. 
\bea
\label{eq:KPZD1StatMeas}
P[F] \propto \exp\left\{-\int \rmd u \left[\partial_u F(u)\right]^2 \right\}.
\eea
This is the same stationary distribution which $F$ would have in the absence of the nonlinearity in \eqref{eq:KPZKPZ}. This amazing fact is related to the existence of a fluctuation-dissipation theorem (see \cite{HuseHenleyFisher1985} and references therein).
In the steady state \eqref{eq:KPZD1StatMeas}, fluctuations of $F$ scale as $F(u) \sim u^{1/2}$, i.e. the KPZ surface roughness exponent is $\chi = 1/2$. Using \eqref{eq:KPZChi} and  \eqref{eq:KPZZetaTheta} one obtains the directed polymer roughness exponent $\zeta$ and the free-energy exponent $\theta$. Thus, the exact scaling exponents of directed polymers and the KPZ equation in $\dsf=1$ are \cite{HuseHenley1985,HuseHenleyFisher1985}
\bea
\label{eq:KPZD1Exp}
\zeta = \frac{2}{3}, \quad\quad \theta = \frac{1}{3}, \quad\quad \chi = \frac{1}{2}.
\eea

\subsection{Bethe Ansatz}
The stationary measure \eqref{eq:KPZD1StatMeas} is not a complete solution of the KPZ equation in $\dsf=1$, since it does not provide the distribution of the actual free energy, only of free energy differences between different points of the interface. Furthermore, it does not contain any information on initial and boundary conditions.
A more precise solution, which takes these points into account, is possible using the Replica Bethe Ansatz method. This is based on the observation that the multiplicative noise equation \eqref{eq:KPZMNZ} gives closed equations for the moments of $Z$. Denote $\Psi_n(u_1 \cdots u_n;t) := \overline{Z_t(u_1)\cdots Z_t(u_n)}$. Using \eqref{eq:KPZMNZ} and \eqref{eq:KPZNoise}, we obtain an $n$-body Schr\"odinger equation for $\Psi_n$
\bea
\partial_t \Psi_n(u_1 \cdots u_n;t) = \frac{1}{2\beta}\sum_{j=1}^n \nabla_{u_j}^2 \Psi_n + 2 \beta^2 D \sum_{j<k}^n \delta^\dsf(u_j-u_k) \Psi_n.
\eea
Thus, in any dimension $\dsf$ the $n$-th moment $\overline{Z^n}$ corresponds to the wave function of $n$ bosons with attractive pairwise $\delta$-function interactions.
It is useful to rescale time as $t \to 2 \beta t$, giving
\bea
\label{eq:KPZMoments}
\partial_t \Psi_n(u_1 \cdots u_n;t) = \sum_{j=1}^n \nabla_{u_j}^2 \Psi_n + 2 c \sum_{j<k}^n \delta^\dsf(u_j-u_k) \Psi_n,\quad\quad c = 2 \beta^3 D.
\eea
Here it is clear that there is one free parameter governing the physics, the attraction between bosons $c$ (which is proportional to the disorder strength $D$ in the original model).
In $\dsf=1$, this Bose gas with $\delta$-function interactions is the exactly solvable \textit{Lieb-Liniger model} \cite{LiebLiniger1963,Lieb1963} (see \cite{Sutherland2004} and \cite{Franchini2011} for a review): The ground state and all excited states of \eqref{eq:KPZMoments} have analytic expressions.
In particular, the $n$-boson ground state is given by
\bea
\label{eq:KPZGroundState1d}
\Psi_n^{(0)} \propto \exp\left(-\frac{c}{2}\sum_{j<k}|x_j-x_k| \right), \quad\quad  \partial_t \Psi_n^{(0)} = E_n \Psi_n^{(0)}, \quad\quad\quad E_n = \frac{c^2}{12} n(n^2-1).
\eea
This was used by Kardar \cite{Kardar1987} and Zhang \cite{Zhang1990} to obtain the tail of the free-energy distribution, and the exponent $\theta$. Their argument is that for large times,
\bea
\int_{-\infty}^\infty \rmd F\, P(F) e^{-n F} = \overline{e^{-n F}} = \overline{Z^n} \sim e^{E_n t} \sim e^{\frac{c^2}{12}n(n^2-1)t}.
\eea
The scaling $n^3 t$ for large $n,t$ of the exponent on the right-hand side is consistent with a tail of the form $P(F) \sim e^{-|F/F_*|^{3/2}}$ as $F \to -\infty$, where $F_* \sim t^{1/3}$. This can be seen by inserting this ansatz into the left-hand side, and approximating the integral using the saddle-point method. The scaling $F_* \sim t^{1/3}$ is, of course, consistent with $\theta=1/3$ as obtained in \eqref{eq:KPZD1Exp}.\footnote{This argument is not without fallacies. When applying the saddle-point argument for large $n$, one considers $P(F)$ for $F$ of order $n^2 F_*^3$, which is much larger than $F_*$ for large $t$ (and hence large $F_*$). Thus, the moments $Z^n$ are actually determined not by the bulk of the distribution $P(F)$, on scales $F \sim F_*$, but by its tail at atypically large scales. For the present case, it turns out that the ``typical tail'' of $P(F)$, for $F/F_*$ large but fixed, is identical to the ``far tail'' for $F \sim n^2 F_*^3$, including the precise prefactor \cite{KolokolovKorshunov2007,KolokolovKorshunov2008,KolokolovKorshunov2009}. The reason why this works in $\dsf=1$ is not entirely clear, but in section \ref{sec:KPZIntDimMoments} we shall see on an example that this is expected to fail in $\dsf \geq 2$.}

The entire free energy distribution is less universal than the scaling exponent. It depends on nonuniversal scales determined by the microscopic cutoff of the systems (e.g.~the lattice spacing, or the form of a short-ranged microscopic disorder correlation function regularizing the $\delta$-interaction in \eqref{eq:KPZMoments}). The description via the continuum equation \eqref{eq:KPZMoments} with $\delta$-correlated disorder is only applicable in the limit of high temperatures \cite{BustingorryLeDoussalRosso2010}. The coupling constants are effective, coarse-grained parameters. 
On a technical level, obtaining the complete free-energy distribution in this universal high-temperature limit requires taking into account all excited states of \eqref{eq:KPZMoments}. This has been first performed in \cite{BrunetDerrida2000} for the steady state of a directed polymer on a cylinder, i.e. with periodic boundaries in the $u$ direction. Later, the free-energy distribution for a DP of finite length in infinite space was computed for various initial conditions:
\begin{itemize}
	\item Initial condition with one end of the DP fixed \cite{SasamotoSpohn2010,SasamotoSpohn2010a,CalabreseLeDoussalRosso2010,AmirCorwinQuastel2011,Dotsenko2010,Dotsenko2010a,DotsenkoKlumov2010}. In terms of the free energy $F$, this corresponds to a ``narrow wedge'' initial condition. For long polymers, the distribution of free-energy fluctuations converges to a Tracy-Widom distribution which also describes the largest eigenvalue of a random matrix from the Gaussian Unitary Ensemble (GUE) \cite{TracyWidom1993}.
	\item Initial condition with one end of the DP free \cite{CalabreseLeDoussal2011,LeDoussalCalabrese2012}. In terms of the free energy $F$, this corresponds to a ``flat'' initial condition. For long polymers, the distribution of free-energy fluctuations converges to a Tracy-Widom distribution for the largest eigenvalue of a random matrix from the Gaussian Orthogonal Ensemble (GOE) \cite{TracyWidom1993}.
\end{itemize}
Recently this was generalized to joint distributions of the free energy at multiple points \cite{ProlhacSpohn2011,ProlhacSpohn2011a,DotsenkoIoffeGeschkenbeinKorshunovBlatter2008,Dotsenko2013a}, multiple times \cite{Dotsenko2013b}, to a half-space with a reflecting boundary \cite{GueudreLeDoussal2012} (which yields a Tracy-Widom distribution of the Gaussian Symplectic Ensemble (GSE) for long times), and to the distribution of the DP endpoint \cite{Dotsenko2013}.

Experimentally, KPZ growth in $\dsf=1$ can be observed in liquid-crystal turbulence \cite{TakeuchiSano2010,TakeuchiSanoSasamotoSpohn2011,TakeuchiSano2012}. These measurements show excellent agreement with the exact solution, including the different universality classes (GUE vs. GOE) of the long-time Tracy-Widom distribution, depending on the boundary conditions (fixed vs. flat).

\section{Directed polymer above 1+1 dimensions\label{sec:KPZHD}}
In the mean-field limit discussed in section \ref{sec:KPZMF}, we saw that the directed polymer model has two phases: a high-temperature phase, where disorder is weak, and a low-temperature, \textit{pinned} or \textit{frozen}, phase, where disorder is dominant. In $\dsf = 1$, discussed in the previous section \ref{sec:KPZD1}, disorder is always dominant and only the frozen phase appears. It is well-established\footnote{Rigorously \cite{ImbrieSpencer1988,CookDerrida1989b}, and using RG methods \cite{KardarParisiZhang1986,MedinaHwaKardarZhang1989,FreyTaeuber1994,Wiese1998}, see section \ref{sec:KPZRG} for details.} that the high-temperature phase and the phase transition appear above a lower critical dimension $\dsf = 2$.

Let us first discuss the pinned, or frozen, phase of the DP. From the results in the previous sections we know that in this phase the polymer is rough. Its transversal displacement $u$ scales as $|u| \sim t^\zeta$, with a roughness exponent $\zeta$ which is $\zeta=2/3$ in $\dsf =1$ and $\zeta = 1/2$ in $\dsf=\infty$ (mean-field). This suggests the natural question of what happens for intermediate $\dsf$. Two scenarios are envisageable:
\begin{itemize}
	\item Existence of an upper critical dimension $\dsf_c$. This would mean that for $\dsf \geq \dsf_c$, the roughness exponent is $\zeta=1/2$, and the behaviour is described by the mean-field model. 
	\item Non-existence of an upper critical dimension $\dsf_c$. This would mean that for any $\dsf < \infty$, the roughness exponent is $\zeta > 1/2$, and that $\zeta$ decreases smoothly towards $1/2$ as $\dsf \to \infty$.
\end{itemize}
Clarifying which of these two scenarios is correct is certainly one of the most important open questions in the theory of directed polymers. Other important challenges are 
\begin{itemize}
	\item Obtaining quantitative estimates for the scaling exponents in the low-temperature phase, which can be improved systematically.
	\item Understanding the nature of the phase transition between the high-temperature and the low-temperature phases in $\dsf > 2$.
\end{itemize}
In the following sections, I will review how these questions were approached in the past, and discuss some new results and ideas.

\subsection{Numerics \label{sec:KPZNum}}
\subsubsection{$\dsf=2$}
The case $\dsf = 2$ is, by now, well-studied numerically. A recent study by Halpin-Healy \cite{HalpinHealy2012} compares different ways of realizing KPZ growth -- through lattice growth models, via the mapping to directed polymers, or by direct integration of \eqref{eq:KPZKPZ}. All of these yield the same critical exponents, and, more importantly, a universal form of the free-energy distribution (in agreement with previous studies \cite{AaraoReis2004,MirandaReis2008}). This provides strong empirical evidence for the existence of a unique KPZ universality class in $\dsf=2$. As in the $\dsf=1$ case, the distributions of free energy fluctuations are universal but depend on the boundary conditions \cite{HalpinHealy2012,OliveiraAlvesFerreira2013}.

The most recent precision study in $\dsf=2$ \cite{KellingOdor2011} maps the KPZ growth model to a driven lattice gas of dimers. This mapping is an extension of the KPZ/ASEP mapping in $\dsf = 1$ and was originally proposed in \cite{OdorLiedkeHeinig2009}. Through an efficient GPU implementation, system sizes up to $L = 2^{17} \approx 10^5$ can be reached. The exponent estimates $\theta = 0.2415 \pm 0.0015$ and $\chi = 0.393 \pm 0.004$ were obtained, as well as a very precise result for the free-energy distribution. In agreement with previous numerical studies \cite{MarinariPagnaniParisi2000} on lattice growth models (restricted solid-on-solid (RSOS) models, see also \cite{HalpinHealyZhang1995} for a review), this seems to rule out the simple fractional value $\chi = 2/5$ (corresponding to $\theta = 1/4$ and $\zeta = 5/8$), which has been previously proposed \cite{KimKosterlitz1989,Laessig1998}. 

\subsubsection{$\dsf \geq 3$ and upper critical dimension}
Recent simulations in higher dimensions were performed e.g. in \cite{MarinariPagnaniParisiRacz2002} using an RSOS model. 
They obtain the estimates $\chi_{\dsf=2} = 0.393 \pm 0.003$, $\chi_{\dsf=3}=0.3135 \pm 0.0015$, $\chi_{\dsf=4}=0.255\pm 0.003$. 
This seems to exclude the rational values $\chi_{\dsf=2}=2/5, \chi_{\dsf=3}=1/3, \chi_{\dsf=4}=2/7$ proposed previously \cite{KimKosterlitz1989,Laessig1998}.

A recent simulation of directed polymers in $\dsf=4$ \cite{SchwartzPerlsman2012} shows a roughness exponent $\zeta = 0.57 > 1/2$ and $\theta = 0.14 > 0$. Via the scaling laws \eqref{eq:KPZZetaTheta}, \eqref{eq:KPZChi}, this is in very good agreement with the result $\chi_{\dsf=4}=0.255\pm 0.003$ from \cite{MarinariPagnaniParisiRacz2002}, obtained by a completely different method. These results were further confirmed on an RSOS model in $\dsf=4$ in \cite{PagnaniParisi2013}. A previous study \cite{MarinariPagnaniParisiRacz2002} also found smoothly varying distributions of the KPZ height profile for $\dsf = 1...5$. 
These studies provide very strong evidence against an upper critical dimension $\dsf_c = 4$, which appears in several theoretical arguments (most prominently, mode coupling, see \cite{FreyTaeuber1994,ColaioriMoore2001} and section \ref{sec:KPZRG}). This points to an upper critical dimension $\dsf_c \geq 5$, and possibly $\dsf_c = \infty$.

\subsection{Lattice approaches}
The Derrida-Spohn solution on a tree, discussed in section \ref{sec:KPZMF}, corresponds formally to the limit $\dsf \to \infty$. One attempt to perturb around it is by adding sections of a finite-dimensional lattice into the tree \cite{CookDerrida1989,CookDerrida1990,Derrida1991}, which yields a $1/\dsf$ expansion. It gives a quantitative estimate of the correction to the free-energy distribution, and predicts that the roughness exponent $\zeta= 1/2$ for $\dsf = \infty$ is not modified to order $1/\dsf$ \cite{CookDerrida1989,CookDerrida1990}. This could have different interpretations: 
\begin{itemize}
	\item The existence of a finite upper critical dimension, above which identically $\zeta= 1/2$.
	\item A decay of $\zeta$ with $\dsf$ towards $\zeta=1/2$ which is faster than $1/\dsf$ (e.g. exponential, stretched-exponential or power-law with exponent $>1$).
	\item Technical problems with non-commutation of limits as discussed in \cite{CookDerrida1989,CookDerrida1990}.
\end{itemize}

\begin{figure}%
\centering
\includegraphics[width=0.7\columnwidth]{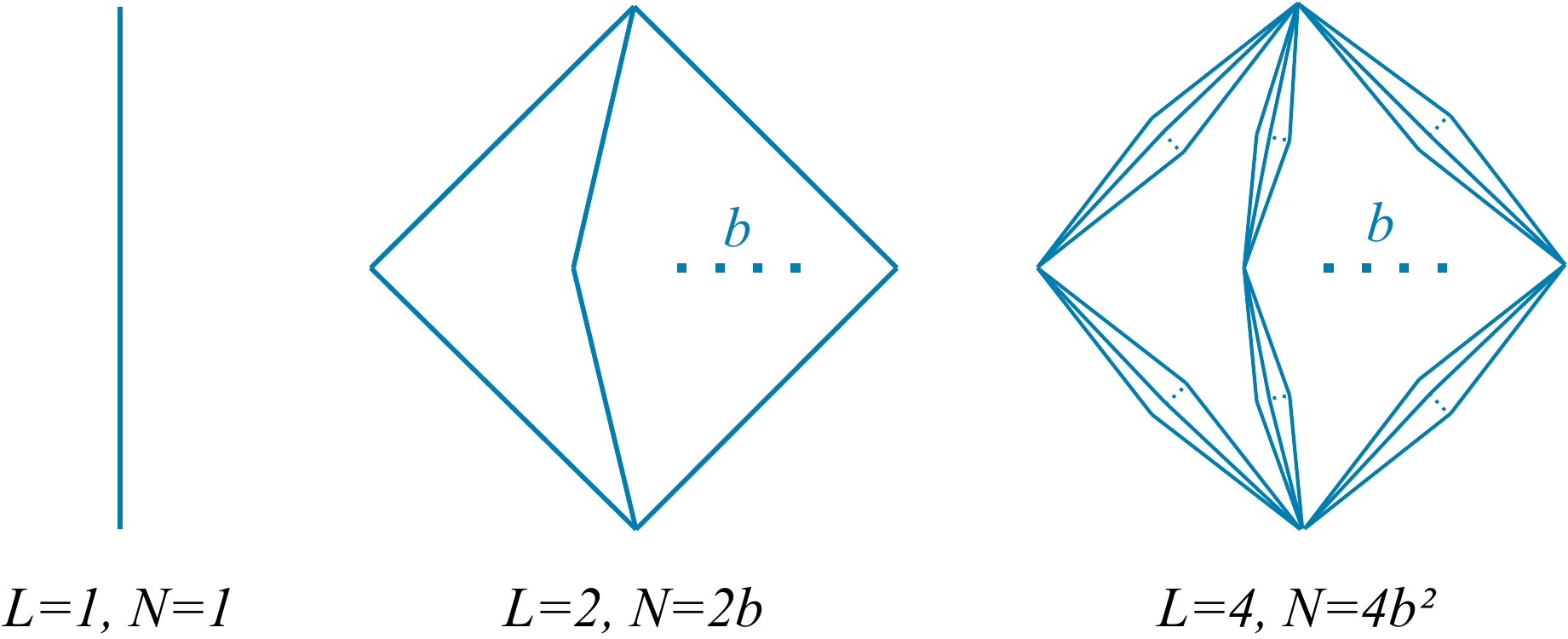}%
\caption{Recursive construction of hierarchical lattice with branching ratio $b$. $N$ is the total number of bonds in the system, the system size $L$ is the length of paths from top to bottom.\label{fig:HierarchLattice}}
\end{figure}

The tree geometry allowed a solution due to the simple form of the recurrence relation \eqref{eq:KPZMFRecZ} for $Z$ between successive iterations. Another geometry where this is possible is a \textit{hierarchical lattice}, or \textit{diamond lattice}, which is constructed by concatenating self-similar sublattices in a ``diamond'' structure with $b$ branches (see figure \ref{fig:HierarchLattice}). In contrast to the tree geometry, such hierarchical lattices contain loops on all scales and are thus non-mean-field models. Nevertheless, some analytical results are possible \cite{DerridaGriffiths1989,CookDerrida1989b,FisherHuse1991}. The branching ratio $b$ of the hierarchical lattice plays a role similar to the space dimension $\dsf$ in the original directed polymer model. Above a critical branching ratio $b_c = 2$, there is a high-temperature and a low-temperature phase. For $b \leq b_c = 2$, only the frozen, low-temperature phase persists \cite{CookDerrida1989b}. The roughness exponents and the free-energy distribution in the frozen phase were computed in an expansion around $b=1$ (which is an exactly solvable case), for zero \cite{DerridaGriffiths1989} and finite \cite{CookDerrida1989b} temperatures. Numerically, the roughness exponent has been computed up to $b \approx 30$; it was seen that $\theta$ decreases smoothly to $0$ and that there is no critical branching ratio \cite{FisherHuse1991}.

A similar approach is the real-space renormalization group of Perlsman and Schwartz \cite{PerlsmanSchwartz1996}. Although arguably one of the least sophisticated methods, it agrees very well with present numerical results. It predicts an exponent $\theta$ smoothly decreasing to $0$ as $\dsf \to \infty$, i.e. $\zeta$ smoothly decreasing to $1/2$ and no upper critical dimension. The quantitative predictions for the roughness exponents $\theta_{\dsf=2}=0.242$, $\theta_{\dsf=3}=0.188$ and $\theta_{\dsf=4}=0.153$ agree with the best available numerical estimates to $1\% $ accuracy in $\dsf =2$ and $\dsf = 3$, and to $5 \%$ accuracy in $\dsf = 4$.

\subsection{RG and mode-coupling\label{sec:KPZRG}}
Several renormalization-group studies of the KPZ equation were performed. The Martin-Siggia-Rose dynamical action corresponding to \eqref{eq:KPZKPZ} is
\bea
\label{eq:KPZRGMSR}
S = \int_{xt} \tF_{xt}\left[\partial_t F_{xt} - \nabla^2 F_{xt} + c \left(\nabla_x F_{xt}\right)^2 \right] - \frac{1}{2}\int_{xt}\tF_{xt}^2,
\eea
where we rescaled $F$ and $t$ so that only one coupling constant $c := 2\beta^3 D$ characterizing the strength of the disorder remains, as above in section \ref{sec:KPZD1}. The perturbative RG flow of $c$ was obtained in \cite{KardarParisiZhang1986,FreyTaeuber1994}. One finds to one-loop order \cite{FreyTaeuber1994}
\bea
\label{eq:KPZRGOneLoop}
-\mu \partial_\mu c(\mu) = -\epsilon c(\mu) + K_\dsf c(\mu)^2 + \mO(c)^3.
\eea
Here, $\mu$ is the RG scale; the universal behaviour for large systems is obtained from the limit $\mu \to 0$. $\epsilon := \dsf - 2$ is assumed to be a small parameter, and \eqref{eq:KPZRGOneLoop} is a controlled perturbative expansion to $\mO(\epsilon)^2$ assuming $c \sim \epsilon$.
 $K_\dsf$ is a geometrical factor which evaluates to $K_{\dsf=2} = 1/(2\pi)$ (the only value needed to this order in $\epsilon$).  
For $\dsf > 2$, we see that there is a stable high-temperature phase for $c < c_\text{crit.}$, where the disorder flows to zero on large scales (i.e. $\lim_{\mu \to 0} c(\mu) = 0$). $c_\text{crit.} := \epsilon/K_{\dsf}$ is an unstable fixed-point, which separates the high-temperature phase from the low-temperature phase. For $c>c_\text{crit.}$, the flow as $\mu \to 0$ goes to strong coupling, and the perturbative RG breaks down. Thus, while the existence of a high-temperature phase is confirmed for $\dsf > 2$, no predictions can be made on the scaling in the low-temperature phase. A two-loop perturbative RG calculation \cite{FreyTaeuber1994} does not change this picture qualitatively. Furthermore, the RG flow for the multiplicative noise equation \eqref{eq:KPZMNZ} can be computed exactly (i.e.~to all orders in perturbation theory) \cite{Wiese1998}. It likewise yields the same qualitative results, in particular, no strong-coupling fixed point is found and the coupling strength goes to infinity on large scales.

An approach for understanding the strongly-coupled, low-temperature phase is \textit{mode coupling}. This consists in resumming the diagrams renormalizing the propagator in the MSR field theory \eqref{eq:KPZRGMSR} of \eqref{eq:KPZKPZ}, but not renormalizing the nonlinear interaction vertex $\tF\left(\nabla F\right)^2$ \cite{FreyTaeuberHwa1996}. This approximation is motivated by the observation \cite{HwaFisher1994}, that the constraints imposed on this interaction vertex by the STS symmetry of the directed polymer discussed in section \ref{sec:InterfaceSTS} are most easily satisfied by assuming it to be of the same form as the bare one (up to scaling) \cite{HwaFisher1994}. Still, this is an uncontrolled approximation. Numerical solution of the resulting self-consistent equations yields exponent values $z\approx 1.62$ in $\dsf=2$, $z\approx 1.78$ in $\dsf=3$ \cite{ColaioriMoore2001}. Using the scaling law $\chi = 2-z$ \cite{ColaioriMoore2001}, this gives $\chi= 0.38$, $\chi=0.22$ in $\dsf=2,3$ which is in quite good agreement with the numerical simulation results discussed in section \ref{sec:KPZNum}. However, the mode-coupling approach predicts an upper critical dimension $\dsf_c=4$, above which $\chi=0$, which is inconsistent with recent numerical results (see section \ref{sec:KPZNum}). Furthermore, it is not clear whether it can be the starting point for a systematic expansion, which would allow to improve the predicted values of the exponents.

An approach using a non-perturbative RG scheme (NPRG) has been attempted in \cite{CanetEtAl2010,CanetEtAl2011}. The resulting predictions for exponents and scaling functions in $\dsf=1$ compare very well to exact results known using other methods. Around $\dsf=4$, the NPRG predicts a nontrivial change of behavior with the roughness exponent $\chi$ decreasing to a small value $\chi \approx 0.05$ in $\dsf=4$, but then increasing again to $\chi \approx 0.2$ in $\dsf=8$ \cite{CanetEtAl2011} (results for higher dimensions were not reported). This is intriguing, since this makes the NPRG one of the few theoretical methods which do not predict an upper critical dimension $\dsf=4$. However, it remains to be seen if the unexpected growth of $\chi$ with the dimension is confirmed using other methods (in particular, it does not agree well with numerical results in $\dsf=5$).

\subsection{Functional renormalization group}
In section \ref{sec:FRGReview} we reviewed how the functional renormalization group can be used to describe the roughening of elastic interfaces with $d$ internal dimensions and a $\dsf=1$-component height field. This was done by expanding around internal dimension $d=4$, where the interface becomes flat. This can be generalized to a manifold with $\dsf = N>1$ transversal directions, to one-loop \cite{BalentsFisher1993} and to higher-loop \cite{LeDoussalWiese2005,WieseLeDoussal2006} order. The leading-order result for the roughness exponent $\zeta$, for large $N$ and small $\epsilon$, is \cite{BalentsFisher1993}
\bea
\zeta \approx \epsilon / N,\quad\quad\quad \epsilon = 4-d.
\eea
In particular, for fixed $\epsilon$ and $\dsf=N\to\infty$, $\zeta \to 0$. Two-loop corrections and calculations for finite $N$ do not change this qualitative picture \cite{LeDoussalWiese2005,WieseLeDoussal2006}. This is in disagreement with the exact mean-field result, which predicts $\zeta \to 1/2$ for $\dsf \to \infty$, \textit{at fixed} $d=1$ ($\epsilon=3$). This is not very surprising: at or above internal dimension $d=2$, elasticity forces an interface to be flat at large scales in the absence of disorder\footnote{This is due to the metric constraint, that several paths from $x_1$ to $x_2$ are available which must lead to the same height difference $u(x_2)-u(x_1)$. In other words, the roughness allowed by the elastic term $\int \rmd^d x (\nabla_x u)^2$, $\zeta' = (2-d)/2$, vanishes at $d=2$.}. This is not the case for internal dimension $d=1$ relevant for the directed polymer: There, as the mean-field solution suggests, at each step from $x$ to $x+\delta x$ the interface wanders freely in order to minimize its energy (and it does that even at zero temperature; thus, the roughness we observe here is independent of the thermal roughness which incidentally leads to the same exponent $\zeta=1/2$). Thus, for $d<2$ (or $\epsilon>2$) one expects a nontrivial change of behaviour, which is nonanalytic in $\epsilon$ (as envisaged in \cite{LeDoussalWiese2005}) and cannot be obtained from a $4-\epsilon$ expansion.

On the other hand, the FRG can be introduced at any fixed $d$ in the large-$N$ limit \cite{LeDoussalWiese2004,LeDoussalWiese2003,LeDoussalWiese2002}. It would be interesting to see if at fixed $d=1$ an $N = \infty$ fixed-point describing the Derrida-Spohn solution with $\zeta=1/2$ can be found, and if $1/N$ corrections to it can be computed. This requires a better understanding of the interplay of replica-symmetry-breaking and functional RG \cite{LeDoussalMuellerWiese2008} at large $N$, and is an important challenge for the future.

\subsection{Moments of the partition sum\label{sec:KPZIntDimMoments}}
In the case $\dsf=1$ discussed in section \ref{sec:KPZD1}, an exact solution for the DP/KPZ problem could be obtained by analyzing the moments of the DP partition sum $Z$. In particular, the scaling exponents could be extracted from the ground state binding energies $E_n > 0$ of $n$ interacting bosons, the lowest eigenvalues of \eqref{eq:KPZMoments}
\bea
\label{eq:KPZMoments2}
E_n \Psi_n(u_1 \cdots u_n) = \sum_{j=1}^n \nabla_{u_j}^2 \Psi_n + 2 c \sum_{j<k}^n \delta^\dsf(u_j-u_k) \Psi_n,\quad\quad c = 2 \beta^3 D.
\eea
Let us now try and understand the energies $E_n$ for $\dsf > 2$. We shall see that they depend on a microscopic cutoff, but nevertheless some universality remains. 

\subsubsection{Second moment}
Introducing the relative coordinate $u := u_2-u_1$ in \eqref{eq:KPZMoments2}, we get the following equation for the two-boson ground state $\Psi_2$ with energy $E_2 > 0$:
\bea
E_2 \Psi_2(u) = 2 \nabla^2_u \Psi_2(u) + 2c \delta(u) \Psi_2(u).
\eea
Fourier transforming (cf.~appendix \ref{sec:AppendixNotations}) from $u$ to $p$, this gives
\bea
(E_2 + 2p^2)\tilde{\Psi}_2(p) = 2c \Psi_2(u=0),\quad\quad \Rightarrow  \quad\quad \tilde{\Psi}_2(p) = \frac{2c}{E_2 + 2p^2} \Psi_2(u=0).
\eea
By integrating over all $p$, one obtains a self-consistent equation for $\Psi_2(u=0)$:
\bea
\label{eq:KPZMomGS2SC1}
\Psi_2(u=0) = \int \frac{\rmd^\dsf p}{(2\pi)^\dsf} \frac{2c}{E_2 + 2p^2} \Psi_2(u=0).
\eea
Assuming $\Psi_2(u=0)$ is a finite, nonzero constant, it drops out from both sides: 
\bea
\label{eq:KPZMomGS2SC1b}
1 = \int \frac{\rmd^\dsf p}{(2\pi)^\dsf} \frac{2c}{E_2 + 2p^2} .
\eea
For $\dsf=1$, the integral is convergent, and one obtains the condition 
\bea
1 = \frac{c}{\sqrt{2E_2(\dsf=1)}}\quad\quad\quad \Rightarrow  \quad\quad\quad E_2(\dsf=1) = \frac{c^2}{2},
\eea
consistent with \eqref{eq:KPZGroundState1d}. For $\dsf \geq 2$, computing the $p$ integral in \eqref{eq:KPZMomGS2SC1} requires a regularization at large momenta. Let us perform this by introducing a hard cutoff $\Lambda$ in momentum space. One then obtains
\bea
\label{eq:KPZMomGS2SC2}
1 =& \int_0^\Lambda \frac{\rmd |p|}{(2\pi)^\dsf} \frac{2c\,S_\dsf\,|p|^{\dsf-1}}{E_2 + 2p^2} = \frac{2c \Lambda^\dsf}{(4\pi)^{\dsf/2} \Gamma(1+\dsf/2) E_2}\,_2F_1\left(1,\frac{\dsf}{2};1+\frac{\dsf}{2};-\frac{\Lambda^2}{E_2}\right),
\eea
where $S_\dsf = 2 \pi^{\dsf/2}/\Gamma(\dsf/2)$. This simplifies for integer dimensions. In $\dsf=2$ one finds that for any $c>0$, there is a solution given by
\bea
E_2(\dsf=2) = \frac{\Lambda^2}{e^{2\pi/c}-1}.
\eea 
In $\dsf \geq 3$, let us rewrite \eqref{eq:KPZMomGS2SC2} as an equation for the dimensionless binding energy $E_2' := E_2/\Lambda^2$:
\bea
\label{eq:KPZMomGS2SC3}
1 = \frac{2c \Lambda^{\dsf-2}}{(4\pi)^{\dsf/2}} g_\dsf(E_2'),\quad\quad \quad g_\dsf(x)=\frac{1}{x\,\Gamma(1+\dsf/2)}\,_2F_1\left(1,\frac{\dsf}{2};1+\frac{\dsf}{2};-\frac{1}{x}\right).
\eea
One finds that for $\dsf > 2$, $g_\dsf$ decays monotonously to zero for large $x$, and satisfies
\bea
\lim_{x\to 0 }g_\dsf(x) = \frac{2}{(\dsf-2)\Gamma(\dsf/2)}.
\eea
Hence, for $\dsf > 2$ \eqref{eq:KPZMomGS2SC3} only has a solution for $E_2'$ when
\bea
\frac{4 c \Lambda^{\dsf-2}}{(4\pi)^{\dsf/2} (\dsf-2) \Gamma(\dsf/2)} > 1 ,\quad\quad \Leftrightarrow c > c_\text{crit}(\dsf) := \frac{(4\pi)^{\dsf/2} (\dsf-2) \Gamma(\dsf/2)}{4 \Lambda^{\dsf-2}}.
\eea
For $c < c_\text{crit}(\dsf)$, there are no bound states and the system is in the high-temperature phase.  As expected, this phase only arises for $\dsf > 2$. The value $c_\text{crit}(\dsf)$ is non-universal, and depends on the cutoff $\Lambda$. However, the scaling of $E_2'$ for $c$ near $c_\text{crit}(\dsf)$ \textit{is} universal. To obtain it, let us expand $g_\dsf(x)$ near $x=0$:
\bea
g_\dsf(x) = \begin{cases}
\frac{2}{(\dsf-2)\Gamma(\dsf/2)}+\Gamma\left(1-\frac{\dsf}{2}\right) x^{\frac{\dsf}{2}-1}& \text{for}\quad 2 < \dsf < 4 \\
\frac{1}{\Gamma(\dsf/2)}\left(\frac{2}{\dsf-2}+ x \log x\right)& \text{for}\quad \dsf = 4 \\
\frac{1}{\Gamma(\dsf/2)}\left(\frac{2}{\dsf-2}-\frac{2}{\dsf-4} x\right)& \text{for}\quad \dsf > 4
\end{cases}.
\eea
Inserting this into \eqref{eq:KPZMomGS2SC3}, we obtain $E_2'$ for $c=c_\text{crit}+\delta c$, and $\delta c$ small
\bea
E_2' = \begin{cases}
\left[\frac{2 \sin \frac{(\dsf-2)\pi}{2}}{\pi(\dsf-2)} \frac{\delta c}{c_\text{crit}}\right]^{\frac{\dsf}{2}-1}& \text{for}\quad 2 < \dsf < 4 \\
-\frac{\delta c}{c_\text{crit}}\frac{1}{W\left(\frac{-\delta c}{c_\text{crit}}\right)}& \text{for}\quad \dsf = 4 \\
\frac{\dsf-4}{\dsf-2} \frac{\delta c}{c_\text{crit}}& \text{for}\quad \dsf > 4
\end{cases}.
\eea
Here $W$ is the Lambert $W$ function \cite{CorlessEtAl1996}. Observe that the rescaled energy $E_2'$ is entirely determined by $\delta c / c_\text{crit}$, and independent of the cutoff $\Lambda$. Hence, energy \textit{ratios} at different values of $c$ near the cutoff, $E_2(c_1)/E_2(c_2)$ are also independent of $\Lambda$.
Now let us discuss higher moments, and their universal properties.

\subsubsection{Third moment\label{sec:KPZMom3}}
The third moment $\Psi_3 (u_1, u_2,u_3)$ is, by translational invariance, a function of $x:=u_2-u_1$ and $y:=u_3-u_2$ only. \eqref{eq:KPZMoments2} then gives
\bea
E_3 \Psi_3(x,y) = \left[\nabla_x^2 + \nabla_y^2 + (\nabla_x-\nabla_y)^2\right]\Psi_3(x,y) + 2c \left[\delta(x)+\delta(y)+\delta(x+y)\right]\Psi_3(x,y).
\eea
Fourier transforming (cf.~appendix \ref{sec:AppendixNotations}) from $x$ to $p$ and from $y$ to $q$, we obtain
\bea
\label{eq:KPZMomGS3SC1}
E_3 \tPsi_3(p,q) = -\left[p^2+q^2+(p-q)^2\right] \tPsi_3(p,q) + 2c\left[ g(q) + g(p) + g(q-p) \right],
\eea
where the auxiliary function $g$ is defined as
\bea
g(q) := \int \rmd^\dsf y\, e^{-iqy} \Psi_3(0,y) = \int \rmd^\dsf x\, e^{-iqx} \Psi_3(x,0) = \int \rmd^\dsf x\, e^{-iqx} \Psi_3(x,-x).
\eea
The equality signs follow by symmetry of the ground state under the exchange $u_1 \leftrightarrow u_2 \leftrightarrow u_3$. \eqref{eq:KPZMomGS3SC1} can be solved for $\tPsi_3$, assuming $g$ is known:
\bea
\tPsi_3(p,q) = \frac{2c\left[ g(q) + g(p) + g(q-p) \right]}{E_3 + p^2+q^2+(p-q)^2}.
\eea
Integrating this over $p$, one finds a self-consistent equation for $g$:
\bea
\label{eq:KPZMomGS3SC2}
g(p) = \int \frac{\rmd^\dsf q}{(2\pi)^d} \tPsi_3(p,q) = \int \frac{\rmd^\dsf q}{(2\pi)^d} \frac{2c\left[ g(q) + g(p) + g(q-p) \right]}{E_3 + p^2+q^2+(p-q)^2}.
\eea
This is hard to solve in general. However, let us consider specific values of $\dsf$.

\paragraph{$\dsf=1$.}
Let us make the ansatz $g(p) = \frac{1}{p^2+\mu^2}$. Then \eqref{eq:KPZMomGS3SC2} reduces to
\bea
\frac{1}{p^2+\mu^2} = \frac{c \left\{2 E_3 \left(\mu ^2-\frac{4 \mu ^3}{\sqrt{3 p^2+2
   E_3}}+p^2\right)+\frac{E_3^2 \mu }{\sqrt{3 p^2+2 E_3}}+4
   \left[p^4+\mu ^4 \left(\frac{3 \mu }{\sqrt{3 p^2+2
   E_3}}-1\right)\right]\right\}}{\mu  \left[\mu ^2 \left(E_3-2 \mu
   ^2\right){}^2+4 p^6+4 E_3 p^4+E_3^2 p^2\right]}.
\eea
This is consistent if we choose
\bea
\mu = c,\quad\quad E_3 = 2c^2.
\eea
The value for $E_3$ agrees, as expected, with the result from \eqref{eq:KPZGroundState1d}.

\begin{figure}%
\centering
\includegraphics[width=0.45\columnwidth]{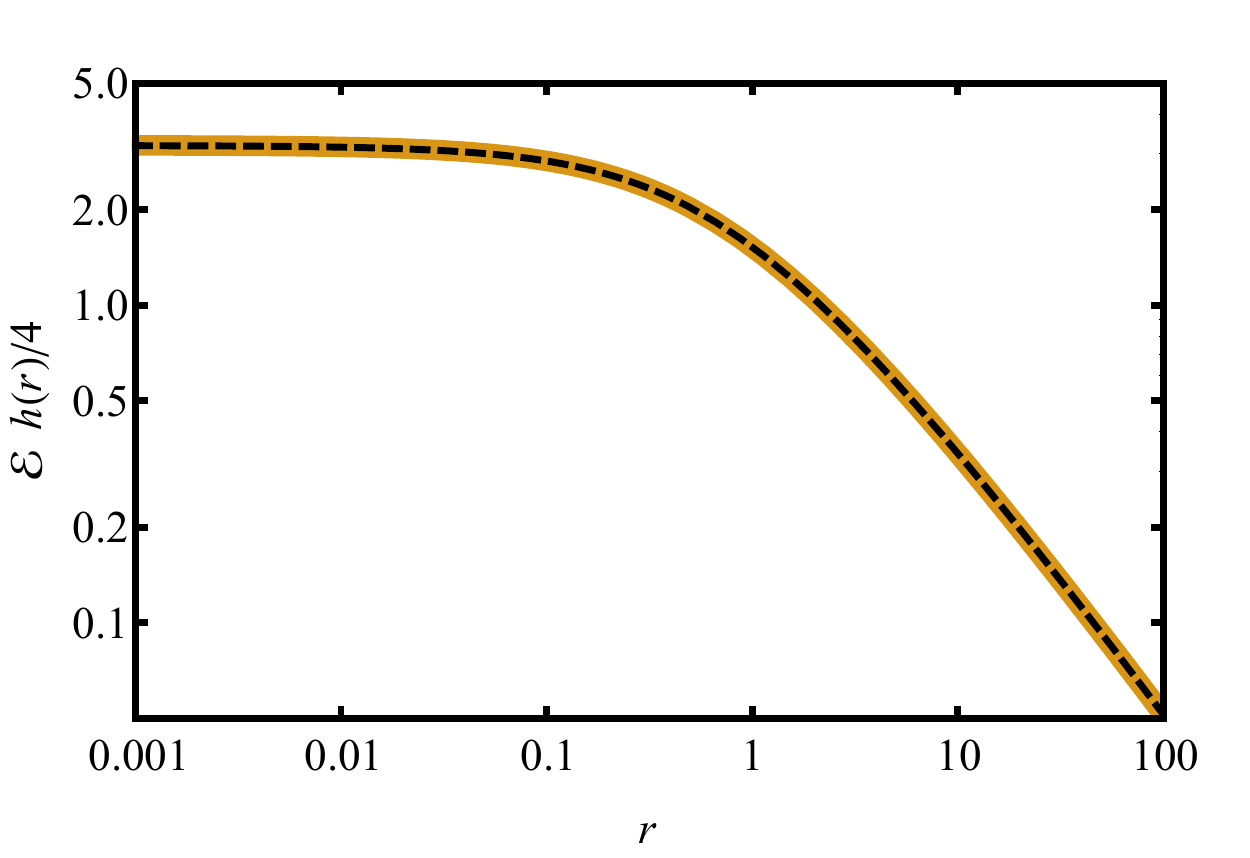}\quad
\includegraphics[width=0.45\columnwidth]{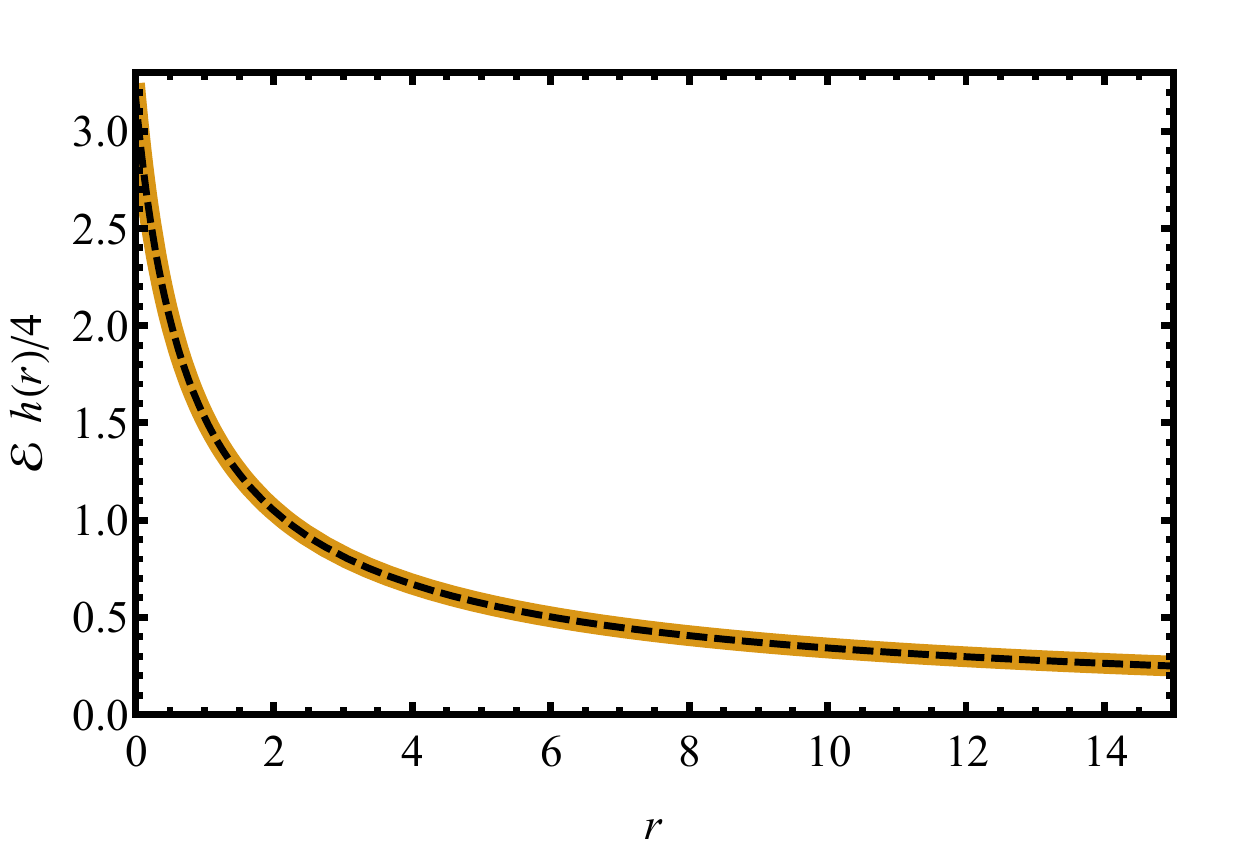}%
\caption{Resulting wave function from the variational ansatz \eqref{eq:KPZMomGS3d2Ans2} with parameters \eqref{eq:KPZMomGS3d2Ans2Vals}, normal and log-log plot. Thick yellow line: $\frac{1}{4}\mathcal{E}h(p)$. Dashed black line: $\int_0^\infty \frac{h(s) \rmd s}{\sqrt{4p^2+4p(1+s)+(1+2s)^2}}-\frac{1}{4}h(p)\ln(1+\frac{3}{2}p)$. Observe that both the small-$p$ and the large-$p$ behaviour is matched extremely well.}%
\label{fig:Z3VarBound}%
\end{figure}

\paragraph{$\dsf=2$.}
Let us collect the terms involving $g(p)$ in \eqref{eq:KPZMomGS3SC2}:
\bea
\label{eq:KPZMomGS3SC3}
g(p)\left[1-\int \frac{\rmd^\dsf q}{(2\pi)^\dsf} \frac{2c}{E_3 + p^2+q^2+(p-q)^2}\right] = \int \frac{\rmd^\dsf q}{(2\pi)^\dsf} \frac{4c \,g(q) }{E_3 + p^2+q^2+(p-q)^2}.
\eea
The integral on the left-hand side is a priori divergent in $\dsf=2$, unless one introduces a large-$q$ cutoff $\Lambda$. However, replacing the $1$ on the left-hand side by the self-consistent equation \eqref{eq:KPZMomGS2SC1b} for the second moment, one can write the coefficient of $g(p)$ as
\bea
\label{eq:KPZMomGS3SC3a}
1-\int \frac{\rmd^\dsf q}{(2\pi)^\dsf} \frac{2c}{E_3 + p^2+q^2+(p-q)^2} = 2c \int \frac{\rmd^\dsf q}{(2\pi)^d} \left[\frac{1}{E_2 + 2q^2}-\frac{1}{E_3 + p^2+q^2+(p-q)^2}\right].
\eea
The right-hand side of this equation has a finite limit even without an UV-cutoff.\footnote{This makes sense if we are considering $E_2, E_3 \ll \Lambda^2$, i.e. for sufficiently high temperatures only. When the temperature is so low that binding energies become comparable to the cutoff $\Lambda^2$, universality is lost and the moments $\overline{Z^n}$ are given essentially by the moments of the microscopic disorder distribution. This is basically the same observation as made in \cite{BustingorryLeDoussalRosso2010} for $\dsf=1$.} Computing it and inserting it into \eqref{eq:KPZMomGS3SC3} simplifies the latter to
\bea
g(p) \frac{c}{4\pi}\log\left( \frac{3 p^2 + 2E_3}{2E_2} \right) = 4c \int \frac{\rmd^\dsf q}{(2\pi)^d} \frac{g(q) }{E_3 + p^2+q^2+(p-q)^2}.
\eea
By rotational symmetry, $g$ can only depend on the absolute value of $p$. Performing the angular integrals in $\dsf=2$ and simplifying, one obtains
\bea
\label{eq:KPZMomGS3SC4b}
g(p) \frac{1}{8\pi}\ln\frac{3p^2+2E_3}{2E_2} = 2\int_0^\infty \frac{\rmd q\, q}{2\pi} \frac{g(q)}{\sqrt{4p^4+4p^2(E_3+q^2) +(E_3+2q^2)^2}}.
\eea
Observe that no dependence on the cutoff and on the value of the coupling $c$ remains; the energy scale is entirely determined by $E_2$ and the length scale is entirely determined by $E_3^{-1/2}$. Setting $p^2 = E_3\, r$, $q^2 = E_3\, s$, $g(p) = h(r)$ we obtain
\bea
\label{eq:KPZMomGS3SC4}
h(r) \frac{1}{4}\left[\log\frac{E_3}{E_2} +  \log\left(1+\frac{3}{2}r\right)\right] = \int_0^\infty  \frac{h(s) \, \rmd s}{\sqrt{1+4r+4s+4r^2+4s^2+4rs}}.
\eea
This is a linear eigenvalue equation 
\bea
\label{eq:KPZMomGS3SC5}
\mathcal{E}_3 h(r) = (H h)(r), \quad\quad\quad \mathcal{E}_3 := \log\frac{E_3}{E_2},
\eea
 where $H$ is a nonlocal Hamiltonian, acting on a function $h $ via
\bea
(Hh)(r) = \int_0^\infty \rmd s \frac{4 h(s)}{\sqrt{1+4r+4s+4r^2+4s^2+4rs}} - h(r)\log\left(1+\frac{3}{2}r\right)
\eea
The Hamiltonian $H$ is symmetric with respect to the ``natural'' scalar product defined by $\left<g|h\right>=\int_0^\infty g(r) h(r) \rmd r$, so its eigenfunctions are orthonormal. We can thus use the variational principle to estimate $\mathcal{E}_3$. For any ansatz $h$, we have
\begin{multline}
\ln \frac{E_{GS}^{(3)}}{E_{GS}^{(2)}} \int_0^\infty \rmd p\, h(p)^2 \\
\label{eq:KPZMomGS3d2Var1}
\geq \int_0^\infty \rmd p \int_0^\infty \rmd q \frac{4 h(p) h(q)}{\sqrt{4q^2+4p^2+4qp+4q+4p+1}}
-\int_0^\infty \rmd p \, h(p)^2\ln{\left(1+\frac{3}{2}p\right)}.
\end{multline}
Let us first try the ansatz $h(q) = \frac{1}{a+q}$, motivated by the result in $\dsf=1$. Evaluating the right-hand side of \eqref{eq:KPZMomGS3d2Var1} numerically as a function of $a$, one finds that the binding energy becomes maximal for $a \approx 1.37$, where it gives the bound
\beq
\ln \frac{E^{(3)}}{E^{(2)}} \geq 2.76 \Rightarrow E^{(3)} \geq E^{(2)} \cdot 15.8.
\eeq
Taking a more elaborate ansatz of the form
\beq
\label{eq:KPZMomGS3d2Ans2}
h(p) = \frac{c+\ln(p+b)}{a+p},
\eeq
one obtains the better estimate
\beq
\label{eq:KPZMomGS3d2Ans2Bound}
\mathcal{E} \geq 2.80474 \Rightarrow x = \frac{E^{(3)}}{E^{(2)}} \geq 16.52,
\eeq
obtained for the values
\beq
\label{eq:KPZMomGS3d2Ans2Vals}
a \approx 0.717 \quad b \approx 1.604 \quad c \approx 2.782.
\eeq
To check the quality of the approximation, observe in figure \ref{fig:Z3VarBound} the left-hand side and the right-hand side of \eqref{eq:KPZMomGS3SC5}. They agree very well, showing that we indeed constructed an approximate eigenfunction. It can also be checked analytically that the ansatz \eqref{eq:KPZMomGS3d2Ans2} satisfies the self-consistent equation \eqref{eq:KPZMomGS3SC4} in the tail.

\paragraph{$\dsf\geq 3$.}
Like in $\dsf=2$, the right-hand side of \eqref{eq:KPZMomGS3SC3a} has a finite limit in $\dsf=3$:
\bea
\int \frac{\rmd^3 q}{(2\pi)^3}\left[ \frac{1}{p^2+q^2+(q-p)^2+E_3} - \frac{1}{E_{2}+2q^2}\right] = \frac{\sqrt{2E_{2}}-\sqrt{3p^2+2E_3}}{16\pi}.
\eea
Performing the angular integrals, we find the generalization of \eqref{eq:KPZMomGS3SC4b} to $\dsf=3$:
\bea
g(p) \frac{\sqrt{3p^2+2E_3}-\sqrt{2E_{2}}}{16\pi} = 2\int_0^\infty \frac{\rmd q\, q}{8\pi^2\, p} g(q)\ln\frac{2p^2+2q^2+2qp+E_3}{2p^2+2q^2-2qp+E_3}
\eea
As for $d=2$, we can go to dimensionless variables $p^2=E_{3} r$, $q^2 = E_3 s$, $g(p) = h(r)$ and get the equation
\beq
\label{eq:KPZMomGS3d3SC}
\frac{g(r) \sqrt{r} }{16\pi}\left(\sqrt{3r+2}-\sqrt{2\frac{E_2}{E_3}}\right) = \int_0^\infty \frac{\rmd s}{8\pi^2} g(s)\ln\frac{2r+2s+2\sqrt{rs}+1}{2r+2s-2\sqrt{rs}+1}
\eeq
Like for the case $\dsf = 2$ above, this can be reformulated as an eigenvalue problem for a Hamiltonian which is symmetric with respect to the scalar product defined by $\left<g|h\right>=\int_0^\infty \frac{1}{\sqrt{r}} g(r) h(r) \rmd r$. One can then attempt a variational solution as above, but one finds that it fails. In particular, one finds numerically states with arbitrarily large $E_3$. This indicates the fact that in $\dsf=3$, the binding energy $E_3$ is cutoff-dependent, and that three-body bound states exist even when there are no two-body bound states.

The situation is even worse in $\dsf \geq 4$, where the right-hand side of \eqref{eq:KPZMomGS3SC3a} is divergent and requires UV regularization. I.e., for $\dsf \geq 4$, the three-body binding energy $E_3$ is completely nonuniversal and UV-cutoff-dependent.

\subsubsection{Fourth and higher moments}
For the general moment $\Psi_n(u_1...u_n)$, we have from \eqref{eq:KPZMoments2}
\beq
E_n \Psi_n =\sum_{j=1}^n\nabla_{u_j}^2 \Psi_n + 2 c \sum_{j<l} \delta(u_j-u_l)\Psi_n .
\eeq
Fourier-transforming (cf.~appendix \ref{sec:AppendixNotations}) from $u_1...u_n$ to $p_1...p_n$, we get
\beq
\label{eq:KPZMomGSnSC1}
(E_n+\sum_j p_j^2) \tPsi_n(p_1...p_n) = 2c\int \frac{\rmd^\dsf q}{(2\pi)^\dsf}\sum_{j<l}\tPsi_n(p_1...p_{j-1},q,p_{j+1},...p_{l-1},p_j+p_l-q,p_{l+1},...p_n).
\eeq
By symmetry, the terms in the sum have to be the same function (but with different arguments). We define 
\beq
g(p,p_3,p_4,...,p_n):=\int \frac{\rmd^\dsf q}{(2\pi)^\dsf} \tPsi_n(p,q-p,p_3,p_4,...p_n).
\eeq
Then, integrating \eqref{eq:KPZMomGSnSC1} we get a closed equation for $g$:
\beq
g(p,p_3,p_4,...,p_n) = 2c\int \frac{\rmd^\dsf q}{(2\pi)^\dsf}\frac{\sum_{j<l} g(k_j+k_l,k_1...\hat{k}_j...\hat{k}_l...k_n)}{E_n+\sum_j k_j^2},
\eeq
where we defined a new set of momenta
\beq
k_1=q,\quad k_2=p-q,\quad k_3=p_3,\quad ...\quad k_n=p_n.
\eeq
and $\hat{k}_j$ denotes an omitted entry.
Using the invariance of the ground state under center-of-mass translations, we have $p=-p_3-...-p_n$, and hence the first argument can be omitted everywhere. We then have:
\beq
g(p_3,p_4,...,p_n) = 2c\int \frac{\rmd^\dsf q}{(2\pi)^\dsf}\frac{\sum_{j<l}g (k_1...\hat{k}_j...\hat{k}_l...k_n)}{E_n+q^2+(p_3+...+p_n+q)^2+p_3^2+...+p_n^2}.
\eeq

Let us specialize this to the case $\overline{Z^4}$. For the fourth moments, we have a closed integral equation for
\beq
g(p_1+p_2,p_3,p_4) = \int_q \Psi(q,p_1+p_2-q,p_3,p_4) = \int_{xyz}e^{-i (p_1+p_2)x-i p_3 y-i p_4 z} \Psi_4(x,x,y,z).
\eeq
Since the ground state is only a function of position differences, i.e. the position of the center of mass is arbitrary, we have $p_1+p_2=-p_3-p_4$ and we can drop the first argument. 
The integral equation we then obtain reads:
\bea
\nn
 g(p_3,p_4) =& 2c\int_q \frac{1}{E_4+q^2+(q+p_3+p_4)^2+p_3^2+p_4^2} \times\\
\nn
& \quad \times \left[g(p_3,p_4)+g(-p_3-p_4-q,p_4)+g(-p_3-p_4-q,p_3)+ \right. \\
\nn
& \quad\quad \left. +g(q,p_4)+g(q,p_3)+g(q,-p_3-p_4-q)\right] \\
\label{eq:IntEqZ4}
=& 2 c\int_q \frac{\left[g(p_3,p_4)+2g(q,p_4)+2g(q,p_3)+g(q,-p_3-p_4-q)\right]}{E_4+q^2+(q+p_3+p_4)^2+p_3^2+p_4^2}.
\eea
The last equation follows by mapping $q\to -p_3-p_4-q$ in the integral.
It can now be checked that in $\dsf=1$, this integral equation is solved for $E_4 = 5c^2$ (in agreement with \eqref{eq:KPZGroundState1d}) with the wave function
\beq
g^{\dsf=1}(p_3,p_4)=\frac{45 c^2+2 \left(3 p_3^2+4 p_3 p_4+3
   p_4^2\right)}{\left(9 c^2+4 p_3^2\right) \left(9 c^2+4
   p_4^2\right) \left[4c^2+(p_3+p_4)^2\right]}
\eeq
In $\dsf=2$, we can transform \eqref{eq:IntEqZ4} in a standard eigenvalue equation by subtracting the 2-point part and rescaling by $E_4$ as in section \ref{sec:KPZMom3}. We obtain a dimensionless eigenvalue equation
\bea
\nn
\mathcal{E}\,g(p_3,p_4) = & \frac{2}{\pi}\int \rmd^2 q \frac{\left[2g(q,p_4)+2g(q,p_3)+g(q,-p_3-p_4-q)\right]}{1+q^2+(q+p_3+p_4)^2+p_3^2+p_4^2}  \\
\label{eq:KPZMomGS4SC}
& \quad\quad - g(p_3,p_4)\ln\left[1+\frac{3p_3^2+2p_3 p_4+3p_3^2}{2}\right],
\eea
where $\mathcal{E}=\ln\frac{E_4}{E_2}$. This is a much more complicated equation than for the third moment, since the function $g$ depends on 3 variables ($|p_3|$, $|p_4|$, and their relative angle). Although it should, in principle, fix $\mathcal{E}$ uniquely, it is not clear which ansatz one could choose for a variational approach. Higher moments become even more complicated.

\subsubsection{Summary}
After performing the present analysis of the moments of $Z$, I became aware of some recent studies on the same eigenstate equations, but with an entirely different motivation \cite{HammerPlatter2007,PlatterHammerMeissner2004,PlatterHammerMeissner2004b,HammerSon2004}. These studies consider \eqref{eq:KPZMoments2} as the Schr\"{o}dinger equation of $n$ \textit{physical} bosons with a short-ranged attractive interaction (realized e.g. in cold atoms \cite{GorlitzEtAl2001}), and derive very similar results\footnote{And also using very similar methods. The integral equations for the ground states of three \eqref{eq:KPZMomGS3SC4b} and four bosons \eqref{eq:KPZMomGS4SC} in that context are called Faddeev \cite{Faddeev1960,MalflietTjon1969} and Yakubovsky equations \cite{KamadaGloeckle1992,GloeckleKamada1993,PlatterHammerMeissner2004}, respectively.}. Here we can make the interesting connection to the moments $\overline{Z^n}$ of the KPZ equation, which was not recognized before. However, most of the other results (and some stronger versions) of the statements on the binding energies $E_n$ (and the transition temperatures $T_n$, i.e. the coupling strengths below which an $n$-body bound state appears) made above are already known:
\begin{itemize}
	\item In $\dsf=2$, the fact the ratios $E_n/E_2$ are universal for small $E_2$ is known from \cite{HammerSon2004}.\footnote{More precisely, as noted above, this holds as long as $E_n\ll \Lambda^2$, i.e. up to some critical $n_c$ above which the moments are non-universal. Since $E_2$ decreases with growing temperature, $n_c$ increases with growing temperature. There is a well-defined limit of infinitely small $E_2$ where all $E_n/E_2$ are universal.} Likewise, the bound \eqref{eq:KPZMomGS3d2Ans2Bound} on $E_3/E_2$ is very close to the extremely precise result $E_3/E_2 = 16.522688(1)$ given in \cite{HammerSon2004}. 	In \cite{PlatterHammerMeissner2004b,HammerSon2004}, a precise estimate for the universal ratio $E_4/E_2 = 197.3(1)$ was also obtained. This was done using a set of self-consistent integral equations similar to (but slightly more complicated than) \eqref{eq:KPZMomGS4SC}.
Furthermore, \cite{HammerSon2004} suggest the following large-$n$ asymptotic scaling for the energies
	\bea
	\label{eq:KPZMomD2Fit}
	\log \frac{E_n}{E_2} = 2.14792 n + \mO(\log n).
	\eea
	In particular, these results confirm $T_n=\infty$ for all $n$, as expected from the non-existence of a high-temperature phase established using perturbative RG methods (section \ref{sec:KPZRG}).
	\item In $\dsf = 3$, there are two non-universal parameters $E_2$, $E_3$ (consistent with what we observed applying the variational principle to \eqref{eq:KPZMomGS3d3SC}). However, there are claims that $E_4$ and presumbly all higher $E_n$ can be expressed in terms of these two energies \cite{PlatterHammerMeissner2004,HammerSon2004}. This would mean, in particular, that $T_2 \neq T_3$ but that $T_3=T_n$ for all $n\geq 3$. However, there are also claims that bound states of $n$ bosons exist for cases where all $n-1$-boson-systems are unbound \cite{HannaBlume2006}. This would mean that all transition temperatures $T_n$ are different, as in the mean-field case discussed in section \ref{sec:KPZMF}. It is not clear how to reconcile this and how to proceed further.
	\item The situation in higher dimensions is less clear. A natural guess would be that in each dimension, a new non-universal energy is introduced, i.e. that in $\dsf=4$ one would have three non-universal energies $E_2,E_3,E_4$ and all other energies have universal expressions in terms of these, independent of any UV cutoffs. This conjecture needs to be verified.
\end{itemize}
Interestingly, all these observations are similar to the behavior of directed polymers on hierarchical lattices. In \cite{CookDerrida1989b}, the transition temperatures $T_n$ for the moments $\overline{Z^n}$ were obtained as a function of the branching ratio $b$ of the hierarchical lattice (which plays a role similar to that of the space dimension $\dsf$). It was found that for $b \leq 2$, there are always bound states, and the polymer is always frozen (as in $\dsf \leq 2$). For a narrow range $2<b\leq 2.3$ , the three-body bound state is given uniquely by the two-body bound state, and the transition temperatures of the second and third moments coincide. Then, at a higher value of $b$ the transition temperature for the fourth moment splits off, etc. until for very high $b$, one recovers the mean-field situation where all transition temperatures are different (see the end of section \ref{sec:KPZMF}).

\begin{figure}%
\centering
\includegraphics[width=0.6\columnwidth]{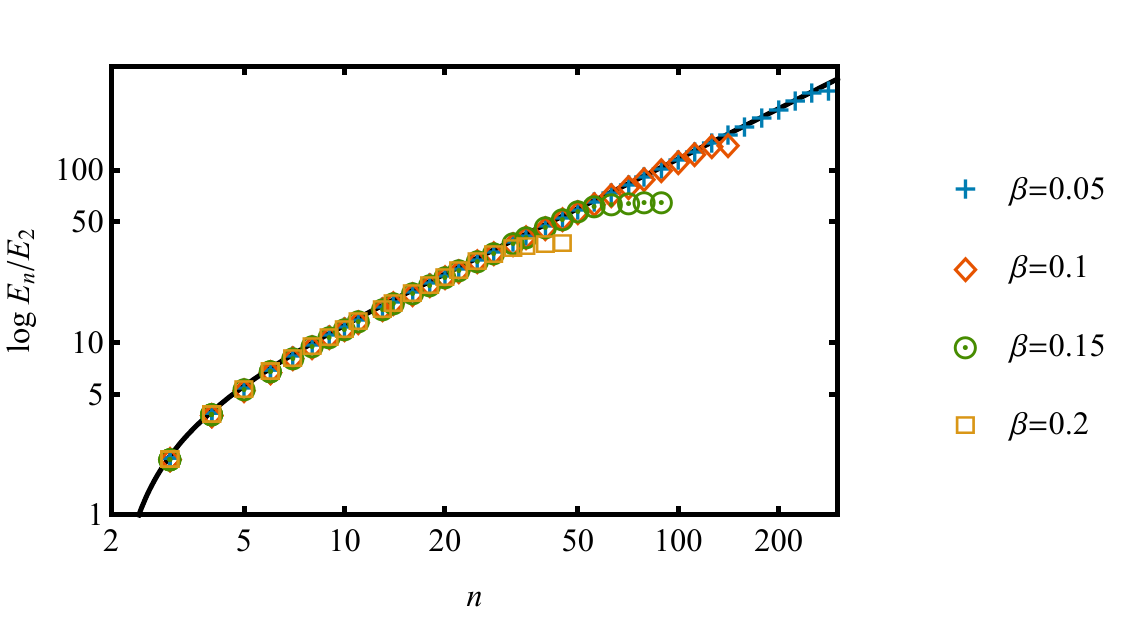}%
\caption{Universal binding energies $E_n/E_2$ on the hierarchical lattice with $b=2$, obtained using the recursion \eqref{eq:KPZMomB2HierarchRec1}, \eqref{eq:KPZMomB2HierarchRec2}. The solid curve is the fit \eqref{eq:KPZMomB2HierarchFit}. The critical moments $n_c$ above which the behavior of the moments becomes non-universal increase with increasing temperature, and are $n_c(\beta=0.2)\approx 30$, $n_c(\beta=0.15)\approx 50$, $n_c(\beta=0.1)\approx 110$, $n_c(\beta=0.05)\approx 200$.\label{fig:HierarchMomentsb2}}
\end{figure}

Even more surprisingly, the universal statistics of the bound-state energies on a hierarchical lattice with $b=2$ (which is the lower critical dimension on the hierarchical lattice) is very similar to the one observed in $\dsf=2$ (which is the lower critical cartesian space dimension). The moments $q_n := \overline{Z^n} / \overline{Z}^n$ can be computed for a hierarchical lattice with $b=2$ using the recursion \cite{CookDerrida1989b}
\bea
\nn
Z^{(2L)} =& Z^{(L),1}Z^{(L),2} + Z^{(L),3}Z^{(L),4}\\
\label{eq:KPZMomB2HierarchRec1}
q_n^{(2L)} = & \frac{1}{2^n}\sum_{k=0}^n \binom {n} {k} \left(q_k^{(L)}\right)^2 \left(q_{n-k}^{(L)}\right)^2,\quad\quad\quad q_0 = q_1 = 1, \quad\quad\quad 
q_n^{(1)} = e^{\frac{\beta^2}{2}n(n-1)}.
\eea
$Z^{(L),i}$, $i=1,...,4$ are four independent copies of the next-smaller sublattice, cf.~figure \ref{fig:HierarchLattice}. The ``binding energies'' $E_n$ are then defined as
\bea
\label{eq:KPZMomB2HierarchRec2}
E_n := \lim_{L\to\infty}\frac{1}{L} \log q_n^{(L)}.
\eea
Using the recursion formula \eqref{eq:KPZMomB2HierarchRec1}, they are easy to compute numerically. One observes that high moments are non-universal (and given by those of the initial distribution), but the ratios $E_n/E_2$ for $n < n_c$ are universal (see also \cite{MedinaKardar1992}). As $\beta \to 0$, $n_c \to \infty$. The universal moment ratios are shown in figure \ref{fig:HierarchMomentsb2}. Motivated by the discusion above, one can make the (new) observation that they are very well described by 
\bea
\label{eq:KPZMomB2HierarchFit}
\log \frac{E_n}{E_2} = 1.096(n-2) + \log \frac{n(n-1)}{2}.
\eea
This looks strikingly similar to the behaviour \eqref{eq:KPZMomD2Fit} observed in cartesian space for $\dsf=2$. It would be interesting to see if this behaviour can also be obtained analytically from \eqref{eq:KPZMomB2HierarchRec1}.

What do we learn from this on the KPZ equation? One can summarize the main new insights as follows.
\begin{itemize}
	\item Near the temperature where moments of $Z$ first start to deviate from the annealed behaviour (i.e. bound states in \eqref{eq:KPZMoments2} appear), there is some universality. The binding energies for high moments of $Z$ are given uniquely in terms of the (nonuniversal) binding energies for low moments. The number of nonuniversal energy scales increases with increasing $\dsf$; presumably, it is $\dsf-1$ (i.e. in $\dsf=2$, all $E_n$ can be expressed through $E_2$, in $\dsf=3$ one needs $E_2$ and $E_3$ etc.). 
	\item The most promising case is $\dsf=2$, where there is only one non-universal scale $E_2$. The universal growth of $E_n$ with large $n$ is very rapid (exponential as in \eqref{eq:KPZMomD2Fit}), which (by a generalization of the saddle-point argument discussed in section \ref{sec:KPZD1}) indicates a universal far tail of the free-energy distribution of the form $P(F)\sim e^{-F \log F}$. This large-$n$ behaviour does not permit any conclusion on the bulk of the distribution, and especially not on the scaling exponents. However, the fact that the far tail is universal gives reason to expect universality for the bulk, too.
	\item For $\dsf > 2$, one knows that the temperature where the three-body bound state appears is strictly above the temperature where the two-body bound state appears. The latter is larger than or equal to the true transition temperature to the frozen phase. Thus, in this case the bound states observed here are some universal precursors of the transition to the frozen phase, but still well inside the high-temperature phase.
	\item The behaviour of the moments of $Z$ observed in cartesian space and on a hierarchical lattice is very similar (in particular, the transition temperatures for the moments behave similarly when the dimension is increased, and the scaling of the binding energies at the lower critical dimension is very similar). The simple recursion \eqref{eq:KPZMomB2HierarchRec1} may allow further analytical progress for the case $b=2$, which is promising for understanding the behaviour in $\dsf=2$, too. Assuming these similarities between the hierarchical and the cartesian lattice persist in higher dimensions, this may be used to argue against the existence of an upper critical dimension of the KPZ equation.
\end{itemize}

\section{Directed polymers with complex random weights\label{sec:KPZComplex}}

\subsection{Physical motivation: Variable range hopping\label{sec:KPZCxVRH}}
In the previous sections, we considered the classical mechanics of an elastic interface in a disordered medium, modelled by a random potential. The partition sum of this model, at low temperature, is dominated by the local energy minima of the random energy landscape. The correlations of the disordered potential determine the statistics of these metastable states, and observables such as the roughness exponent, or correlation functions. 
The picture is much less clear for systems on smaller physical scales, e.g. electrons in a random medium, where quantum interference is
important. Their transition probability between two states is a sum over Feynman paths, whose weights are complex numbers, random due to the presence of disorder. In contrast to the partition sum of a classical system, where all contributions are positive and the low-energy states dominate, the dominant contributions to a Feynman sum with complex weights are less simple to identify.

To give a specific example, hopping conductivity in
disordered insulators in the strongly localized (variable-range-hopping) regime is described by the Nguyen-Spivak-Shklovskii
(NSS) model \cite{Nguyen1985,ShklovskiiPollak1991}. A \textit{locator expansion} \cite{Anderson1958,Nguyen1985,ShklovskiiPollak1991,GangopadhyayGalitskiMueller2013,Mueller2013} shows that the Green's function $J_{\omega}$, which describes the influence at site $b$ of an excitation localized at site $a$, at frequency $\omega$,
 is  the sum over interfering directed paths $\Gamma$ from  $a$ to $b$ \cite{MedinaKardar1991,MedinaKardar1992,MedinaKardarShapirWang1990,Mueller2013}
\begin{equation}
\label{eq:IntroNSS}
J_{\omega}(a,b) := \sum_{\Gamma:a \to b} e^{i\Phi_\Gamma}\prod_{j\in \Gamma} \frac{\left[\sgn \epsilon_j \right]^\mathsf{B}}{\epsilon_j - \omega} \ .
\end{equation}
Typically, we will consider $\omega=0$, i.e. the time-integrated Green's function. $\epsilon_j$ is the random energy of a defect at site $j$, and can be positive or negative. The Green's function for bosons is described by $\mathsf{B}=1$, and for fermions by $\mathsf{B}=0$ \cite{Mueller2013}. The additional phase factor $e^{i\Phi_\Gamma}$ arises in the presence of an external magnetic field. Then, $\Phi_\Gamma$ is the flux of the $B$-field through the area between path $\Gamma$ and an arbitrarily chosen reference path. 
The experimentally measurable conductivity is then given by $\sigma(a,b)\sim|J_{\omega=0}(a,b)|^2$. 

In absence of the magnetic field, for bosons $\mathsf{B}=1$, \eqref{eq:IntroNSS} is the sum over all directed paths with positive weights. Thus, $J$ is the partition sum of the standard directed polymer with endpoints fixed at $a$ and $b$, as discussed in section \ref{sec:KPZOverview}. On the other hand, for fermions, $\mathsf{B}=0$, and the weights of each site along each path can have positive or negative signs.\footnote{The relative amount of positive vs. negative signs can be varied by varying $\omega$ \cite{Mueller2013}.} $J$ is then a generalized ``partition sum'' of a directed polymer in $\dsf$ dimensions, except that positive Boltzmann weights $ e^{-\beta E_j}$ are replaced by weights with random signs (or more generally, in the case of an applied $B$-field, by complex random weights).

Let us first focus on the case of hopping conductivity of fermions in a two-dimensional insulator, e.g. a thin film. This corresponds to the directed polymer in $1+1$ dimensions, with random signs or complex random weights. The case with positive weights was discussed in section \ref{sec:KPZD1}. 
Numerical simulations \cite{MedinaKardarShapirWang1989,SomozaOrtunoPrior2007,PriorSomozaOrtuno2009} and an argument based on pairing of replicas \cite{MedinaKardarShapirWang1989} indicate that interference between paths does not modify the universality class of the model. In particular, in $\dsf=1$ the  numerically observed exponent for the free-energy fluctuations $\theta$ is still $\theta=1/3$ \cite{MedinaKardarShapirWang1989},
\bea
\overline{\left[\log J(L)\right]^2}-\overline{\log J(L)}^2 \propto L^{2\theta} \propto L^{2/3}.
\eea
Furthermore, the roughness exponent of the typical paths contributing to the conductance, $\zeta=2/3$, and the universal Tracy-Widom distribution of the free energy $\log |J|$ (see section \ref{sec:KPZD1}) is observed with great precision even in the presence of random signs \cite{SomozaOrtunoPrior2007,PriorSomozaOrtuno2009}. Let us thus take the conjecture that in $\dsf=1$, the directed polymer with positive, and with complex random weights are in the same universality class, for granted. This will be discussed further in section \ref{sec:KPZCxAnal}.

Adding a small, homogeneous, magnetic field modifies the conductance \eqref{eq:IntroNSS}. This \textit{magnetoconductance} can be measured experimentally,
and is different for the bosonic versus the fermionic case. 
In the bosonic case (only positive paths), the magnetic field adds a small amount of phase disorder\footnote{Even for a homogeneous magnetic field, since the amplitudes of different paths vary.}. Thus, the conductance decreases; one has \textit{negative magnetoconductance} or \textit{positive magnetoresistance} \cite{GangopadhyayGalitskiMueller2013,Mueller2013}. On the other hand, in the fermionic case (random signs), the phase disorder eliminates destructive interference between competing paths, and yields \textit{positive magnetoconductance} \cite{MedinaKardarShapirWang1990,MedinaKardar1992,IoffeSpivak2013}. Although the value and the sign of the magnetoconductance varies between these two cases, its scaling with the strength of the applied magnetic field $B$ is in both cases the same, and related to the directed-polymer roughness exponent $\zeta=2/3$ by \cite{GangopadhyayGalitskiMueller2013,IoffeSpivak2013}
\bea
\overline{\frac{\log |J^{B}(L)|}{\log |J^{B=0}(L)|}} \propto B^{\frac{2\zeta}{1+\zeta}} \propto B^{4/5}.
\eea

This example shows that understanding the statistics of directed polymers with random complex weights may help understanding conductance fluctuations in disordered insulators. In the following sections I will discuss some known and some new analytical results on the model of a DP with complex weights.

\subsection{Analytical approaches\label{sec:KPZCxAnal}}
Let us consider the multiplicative noise equation \eqref{eq:KPZMNZ}, with $\beta$ scaled away and with complex noise:
\bea
\label{eq:KPZMNZCx}
\partial_t Z_t(u) = \nabla^2_u Z_t(u)  - \left[\eta_1(u,t) + i \eta_2(u,t)\right] Z_t(u).
\eea
We assume $\eta_1$, $\eta_2$ to be independent white noises as in \eqref{eq:KPZNoise}, with possibly different amplitudes $c_1$ and $c_2$
\bea
\nn
\overline{\eta_1(u,t)} = 0,\quad\quad\quad \overline{\eta_1(u,t)\eta_1(u',t')} = 2 c_1 \delta^{\dsf}(u-u') \delta(t-t'),\\
\label{eq:KPZNoiseCx}
\overline{\eta_2(u,t)} = 0,\quad\quad\quad \overline{\eta_2(u,t)\eta_2(u',t')} = 2 c_2 \delta^{\dsf}(u-u') \delta(t-t').
\eea

\subsubsection{Non-integrability in $\dsf=1$}
The numerical results and heuristic arguments in \cite{MedinaKardarShapirWang1989,SomozaPriorOrtuno2006,SomozaOrtunoPrior2007,PriorSomozaOrtuno2009} indicate strongly that the paths dominating the partition sum of the DP with random signs, or complex random weights, have the same statistics\footnote{I.e. the same scaling of transversal fluctuations (roughness) and energy fluctuations.}, as those for a standard DP with positive random weights. 
It is tempting to try to formalize this conjecture using a Bethe-Ansatz calculation as discussed in section \ref{sec:KPZD1}. For the case of random phases, the equation \eqref{eq:KPZMoments} can be generalized for moments of the form $\overline{Z^n \left(Z^*\right)^m}$ as
\bea
\nn
\Psi_{n,m}(u_1 \cdots u_n,v_1\cdots v_m;t) := & \overline{Z(u_1,t)\cdots Z(u_n,t)Z^*(v_1,t)\cdots Z^*(v_m,t)}\\
\nn
\partial_t \Psi_{n,m}(u_1 \cdots u_n,v_1\cdots v_m;t) = & \sum_{j=1}^n \nabla_{u_j}^2 \Psi_{n,m} + \sum_{k=1}^m \nabla_{v_k}^2 \Psi_{n,m} +  \\
\nn
& + 2 c \sum_{j_1<j_2}^n \delta^\dsf(u_{j_1}-u_{j_2}) \Psi_{n,m} + 2 c \sum_{k_1<k_2}^m \delta^\dsf(v_{k_1}-v_{k_2}) \Psi_{n,m}  \\
\nn
& + 2 c' \sum_{j=1}^n\sum_{k=1}^m \delta^\dsf(u_{j}-v_{k}) \Psi_{n,m}.
\eea
Here, $c = \frac{c_1 - c_2}{2}$ and $c' = \frac{c_1 + c_2}{2}$.
Thus, the system with complex weights maps onto a gas of bosons with two ``flavours'', and pointwise $\delta$-function interactions. The interaction strengths for bosons of the same flavour $c$ and bosons of different flavours $c'$ are different, and only conincide when there is no phase disorder. However, the resulting two-flavour system is not integrable: One can write down the plane wave phase matching conditions whenever two points conincide, but one finds that the resulting scattering matrices do not satisfy the Yang-Baxter equation \cite{Sutherland2004}. Although some exact eigenstates for small $m,n$ and specific values of $c, c'$ can be found (e.g. for $m+n=3$, see \cite{GaudinDerrida1975}, and some special cases with $m=1$ or $n=1$ \cite{GurgenishviliKharadzeChobanyan1975}), these are far from being complete. Not even the ground state is known exactly: The naive generalization of \eqref{eq:KPZGroundState1d}, 
\bea
\nn
\Psi_{m,n}^{(0)} \propto \exp\left(-\frac{c}{2}\sum_{j_1<j_2}^n|u_{j_1}-u_{j_2}|-\frac{c}{2}\sum_{k_1<k_2}^m|v_{k_1}-v_{k_2}|-\frac{c'}{2}\sum_{j=1}^n\sum_{k=1}^m|u_j-v_k| \right),
\eea
is not an eigenstate. It satisfies the correct matching conditions whenever two points conincide, but the kinetic energy is different for different orderings of the $u,v$'s.
Thus, it does not seem feasible to make analytical progress using the Bethe ansatz methods discussed in section \ref{sec:KPZD1}; new approaches are needed to confirm this conjecture in a controlled way.

\subsection{Mean-field solution\label{sec:KPZCxMF}}
\begin{figure}%
\centering
\begin{minipage}[c]{0.4\textwidth}
\includegraphics[width=\columnwidth]{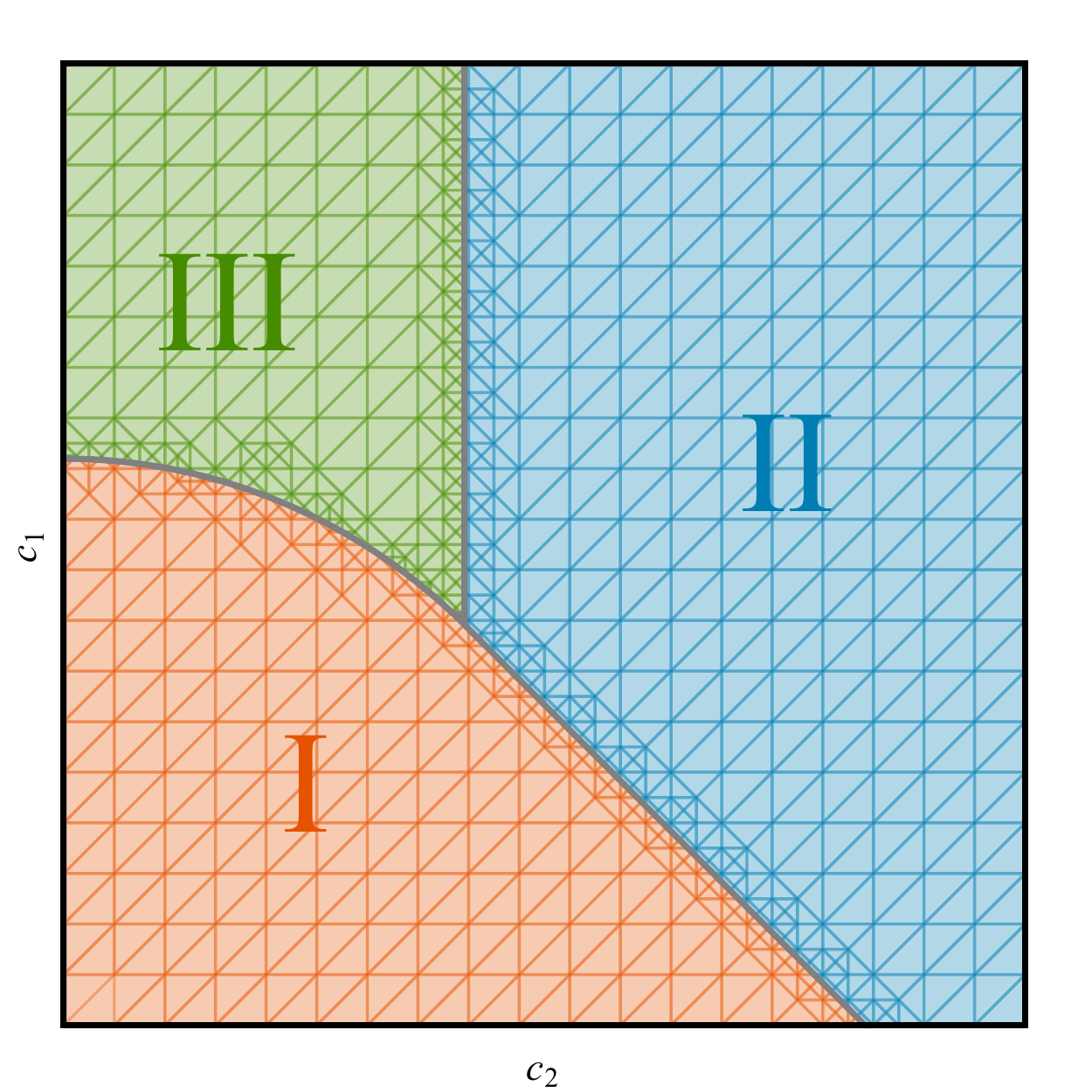}%
\end{minipage}\quad
\begin{minipage}[c]{0.4\textwidth}
\caption{Phase diagram of the directed polymer with complex random weights $e^{-c_1 V(x)-i c_2 V_2(x)}$ on the Cayley tree, following \cite{DerridaEvansSpeer1993}. The phase boundaries are shown for the case where $V(x)$, $V_2(x)$ are normally distributed; the horizontal axis shows $\beta$ (amplitude disorder) and the vertical axis shows $\gamma$ (phase disorder).\label{fig:PhaseDiagDPCx}}
\end{minipage}
\end{figure}

A case where analytical progress can be made is the mean-field limit. For a directed polymer with positive weights, we discussed in section \ref{sec:KPZMF} that an exact solution is possible on the Cayley tree. It yields a phase diagram with a high-temperature phase and a frozen, low-temperature phase. Using different methods, Derrida and coworkers also obtained an exact solution for the directed polymer on a tree with random complex weights \cite{DerridaEvansSpeer1993}. This exact solution shows that both the high-temperature phase (phase I), and the low-temperature phase (phase II) are stable to small amounts of random phase disorder. However, for sufficiently strong phase disorder, the high-temperature phase transitions to a new phase (denoted ``phase III'' by Derrida), which is dominated by interference effects (cf.~figure \ref{fig:PhaseDiagDPCx}). 

Using the expressions for the free energy obtained in \cite{DerridaEvansSpeer1993}, we can interpret the three phases as follows:
In the high-temperature phase, attraction between replicas is weak and quenched averages are equal to annealed averages:
\bea
\overline{\log |Z|} = \log |\overline{Z}|.
\eea
On the other hand, in phase III fluctuations of $Z$ are larger that its average. In this phase, replicas are bound in pairs, but attraction between replica pairs is weak. Quenched averages are equal to annealed averages over replica pairs, 
\bea
\overline{\log |Z|} = \frac{1}{2}\log \overline{|Z|^2}.
\eea
Finally, in the frozen phase II, attraction between replicas is strong and quenched averages are not related to annealed ones.
This phase diagram, and the phenomenology of the phases, is very similar to the random-energy-model at a complex temperature \cite{Derrida1991a}, and to a related toy model \cite{DobrinevskiLeDoussalWiese2011}, which I will review in section \ref{sec:KPZCxToy}.

\subsection{Perturbative RG\label{sec:KPZCxRG}}
\begin{figure}%
\centering
\begin{minipage}[c]{0.4\textwidth}
\includegraphics[width=\columnwidth]{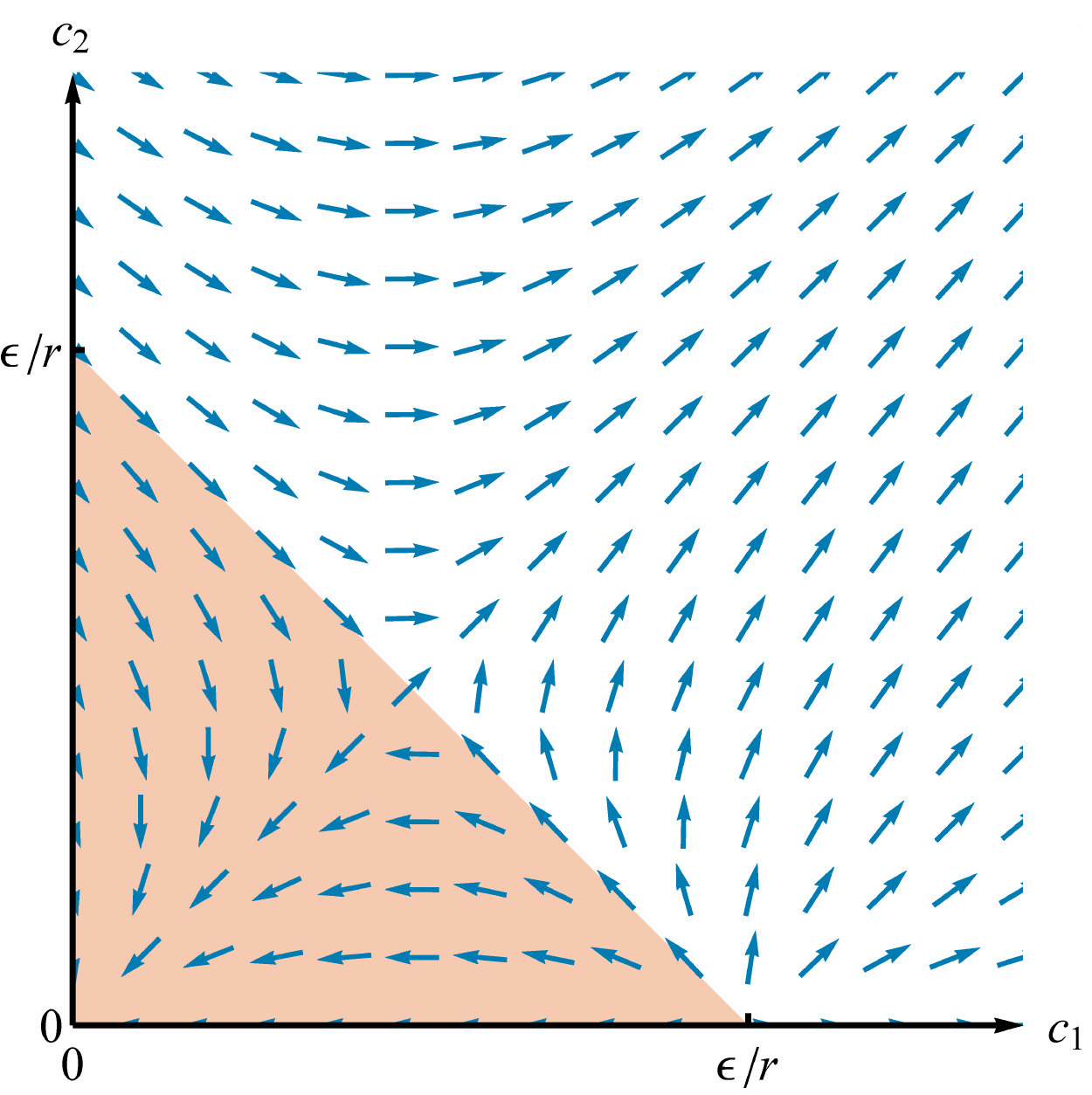}%
\end{minipage}\quad\quad
\begin{minipage}[c]{0.4\textwidth}
\caption{Perturbative RG flow of KPZ equation, for $\dsf > 2$, according to \eqref{eq:KPZCxRGBeta}. The shaded region is the high-temperature phase, where the flow goes to the Gaussian fixed point $c_1=c_2=0$ for large scales. The restriction to the real axis, $c_2=0$, corresponds to the flow of the standard KPZ equation \eqref{eq:KPZRGOneLoop}.}%
\label{fig:KPZRGFlow}%
\end{minipage}
\end{figure}

As mentioned in section \ref{sec:KPZRG}, the (perturbative) RG flow of the $Z^2\tZ^2$ vertex of the Martin-Siggia-Rose field theory of \eqref{eq:KPZMNZ} can be computed exactly in any dimension $\dsf$ \cite{Wiese1998}. This reflects the fact that the two-body problem problem with a $\delta$-interaction can be solved exactly in any $\dsf$. It is straightforward to generalize this calculation to the case where $Z$ and $\eta$ in \eqref{eq:KPZMNZ} are complex. Decomposing $Z= X+i Y$, $\eta = \eta_1 + i \eta_2$, we obtain from \eqref{eq:KPZMNZ}
\bea
\nn
\partial_t X =& \nabla^2 X - Y\eta_1 + X \eta_2, \\
\partial_t Y =& \nabla^2 Y + X\eta_1 + Y \eta_2.
\eea
Correlation functions of $Z$ can be expressed through the MSR field theory
\bea
\nn
\overline{\mO[X,Y,\tX,\tY]} =& \int \mD[X,Y,\tX,\tY]\,e^{-S[X,Y,\tX,\tY]} \mO[X,Y,\tX,\tY],\\
S[X,Y,\tX,\tY] := & \int_{u,t} \tX\left(\partial_t X + \nabla_u^2 X\right)+\tY\left(\partial_t Y + \nabla_u^2 Y\right) + c_1(X\tX + Y\tY)^2 + c_2(X\tY - Y\tX)^2
\eea
We see that $S$ contains two interaction vertices (in contrast to the single vertex $c Z^2 \tZ^2$ for the case of positive weights)
\beq
\label{eq:KPZCxRGIntVert}
c_1(X\tX + Y\tY)^2 + c_2(X\tY - Y\tX)^2.
\eeq
$c_1$ is the coupling characterizing the amplitude disorder, and $c_2$ is characterizing phase disorder. If we consider real noises only, $\eta_2=0$, then $c_2=0$. Generalizing the results of \cite{Wiese1998}, the $\beta$ function for $c_1$, $c_2$ can be computed exactly. The renormalized $c_1(\mu),c_2(\mu)$ are defined via four-point correlation functions $\overline{X^2 \tX^2}$, $\overline{X^2 \tY^2}$, etc.
Since the interaction verices \eqref{eq:KPZCxRGIntVert} preserve the number of external legs, the four-point correlation functions are given by a geometric series of bubble diagrams, which can be resummed exactly \cite{Wiese1998}.  One obtains \cite{Dobrinevski2012unpubl} the following $\beta$ function
\bea
\nn
-\mu \partial_\mu c_1 =& -\epsilon c_1 + r(\dsf) \left(c_1^2 + c_2^2\right),\\
\label{eq:KPZCxRGBeta}
-\mu \partial_\mu c_2 =& -\epsilon c_2 + 2 r(\dsf) c_1 c_2.
\eea
Here $\mu$ is the RG scale, $\epsilon := \dsf - 2$, and $r(\dsf)=\frac{4}{(8\pi)^{\dsf/2}}\Gamma(2-\dsf/2)$. Observe that for $c_2=0$, to order $\epsilon^2$ \eqref{eq:KPZCxRGBeta} reduces to \eqref{eq:KPZRGOneLoop}. The RG of \eqref{eq:KPZCxRGBeta} to large scales, i.e. to $\mu\to 0$, is shown in figure \ref{fig:KPZRGFlow}.

The new result \eqref{eq:KPZCxRGBeta} leads to the following conclusions:
\begin{itemize}
	\item For $\dsf > 2$ (i.e. $\epsilon > 0$), there is a high-temperature phase: When starting with sufficiently small $c_1, c_2$, the flow goes back to $c_1 = 0, c_2=0$. In particular, the high-temperature phase of the directed polymer with positive weights (obtained for $c_2=0, c_1 < c_{\text{crit}} = \epsilon/r$) is stable to the introduction of weak random phase disorder. In the mean-field limit, this high-temperature phase becomes the phase I which was observed in section \ref{sec:KPZCxMF}.
	\item On the other hand, the frozen phase II of the DP with positive weights seems unstable to the addition of small random phase disorder: As soon as one starts with $c_2 > 0, c_1 > \epsilon/r$, the flow leads to increasing $c_2$ and towards the axis $c_1 = c_2$.
\end{itemize}
While the result about the stability of the high-temperature phase is controlled perturbatively\footnote{In the sense that the series of bubble diagrams used to evaluate the renormalized coupling converges.}, the flow in the other regions leads to strong coupling. Thus, the perturbative result on the instability of phase II does not necessarily reflect the true behaviour of the system. The flow \eqref{eq:KPZCxRGBeta} does not show any hints of a possible phase III, which appeared in the mean-field result but not in $\dsf=1$. It could be that phase III appears only above another critical dimensions strictly larger than $2$, or only in the mean-field limit, or only at strong coupling which is out of reach of this perturbative treatment.

Since it is not known how to analyze the strong-disorder, frozen phase of the KPZ equation in any finite dimension, even in the absence of complex phases, I will now discuss a simpler toy model where more analytical progress can be made.

\subsection{A toy model: Particle in a complex landscape\label{sec:KPZCxToy}}

I will now discuss, following \cite{DobrinevskiLeDoussalWiese2011}, a toy model for a disordered system with interference: A particle in a complex random ``potential''. 
Let us consider a ``partition sum'' $Z$ defined as
\begin{equation}
\label{eq:KPZCxToyZ}
Z (w) =  \sqrt{\frac{\beta m^2}{2\pi}} \int_{-\infty}^{\infty}\rmd x\, e^{-\beta \big[V(x)+\frac{m^2}{2}(x-w)^{2}\big] -i \theta(x)}
\end{equation}
Here, $V(x)$ is a random potential and $\theta(x)$ a random
phase, both with translationally invariant correlations, 
and $\beta=1/T$ the inverse temperature. In the absence of random phases, this is the partition sum of a particle in a one-dimensional random potential $V(x)$ (in other words, a $0+1$-dimensional elastic interface). This is a drastic simplification compared to a $1+\dsf$ dimensional directed polymer, but it 
 reproduces some important physical features of realistic,
higher-dimensional models. For example, by moving $w$ one can observe shocks as for an elastic interface \cite{BalentsBouchaudMezard1996,LeDoussal2006,LeDoussalWiese2009}.

Due to the harmonic confinement, the partition sum \eqref{eq:KPZCxToyZ} decays quickly to zero for $|x - w| \gg 1/m$. Thus, $L := 1/m$ gives the typical length of the system sampled by the particle, and can be interpreted as a \textit{system size}. 
If one discretizes $x$, and draws $V(x)$ and $\theta(x)$ at each $x$ independently from a Gaussian distribution, \eqref{eq:KPZCxToyZ} becomes the partition sum of $L$ states with independent energies and phases. This is then essentially\footnote{I.e. up to correlations between the real and the imaginary part of the weight, which are present for the REM at complex temperature but not for \eqref{eq:KPZCxToyZ}.} the partition sum of the Random Energy Model, at a complex temperature, studied by Derrida in \cite{Derrida1991a}.  
There, the basic
distinction of three phases, high-temperature phase I, frozen phase
II, and strong-interference phase III, was established \cite{Derrida1991a}, 
exactly as for the directed polymer with complex random weights on a Cayley tree (see \cite{DerridaEvansSpeer1993}, section \ref{sec:KPZCxMF} and figure \ref{fig:PhaseDiagDPCx}). 

The complementary study in \cite{DobrinevskiLeDoussalWiese2011} focuses not on the phase diagram and the phase transitions, but rather on the behaviour of $Z$, when $w$ in  \eqref{eq:KPZCxToyZ} is varied, deep inside each phase. This allows a better understanding of the nature of these phases.
This analysis is based on the \textit{effective disorder correlator} $\Delta$, which proved useful for elastic interfaces (section \ref{sec:FRGReviewOneLoop}). Here we define it slightly differently, via the complex ``free energy'' or the complex ``effective potential'' $f$ for each realization of the random potential $V(x)$ and the random phases $\theta(x)$
\begin{equation}
\label{eq:KPZCxToyDefEffPot}
\beta \hat{V}(w) + i\hat{\theta}(w) := \beta f (w)\equiv  -\ln Z (w) \ .
\end{equation}
Note that $\hat{V}(w)$ is always unique, but $\hat{\theta}(w)$ is only
defined modulo $2\pi$. We will thus focus  on $\hat{\theta}'(w)$ which
is unambiguous. The effective disorder correlators for the potential and the phase are then defined by
\bea
\nonumber
\Delta_{V}(w_1-w_2) :=& \overline{\hat{V}'(w_1)\hat{V}'(w_2)}, \\
\label{eq:KPZCxToyDefDelta}
\Delta_{\theta}(w_1-w_2) :=&\overline{\hat{\theta}'(w_1)\hat{\theta}'(w_2)}\ .
\eea
We assume  $V(x)$ and $\theta(x)$ to be independent, and the distribution of $\theta$ to be symmetric under $\theta \to -\theta$. Then, the cross-correlator
$\overline{\hat{V}'(w_1)\hat{\theta}'(w_2)}$ vanishes.

Without going into the details of the calculations, I will review in the following sections the main new results of \cite{DobrinevskiLeDoussalWiese2011}.

\subsubsection{Interference-driven phase III}
Phases I and III are both high-temperature phases, in the sense that the centeral limit theorem holds. Thus, the entire length of the system $\Delta x \sim 1/m$ contributes to the partition sum, and the distribution of $Z$ in the limit $m\to 0$ is Gaussian. The distinction between these two phases is in the scaling of the mean of $Z$ versus the fluctuations of $Z$. In phase III, $\overline{Z}^2 \big/ \overline{\left(Z-\overline{Z}\right)^2} \to 0$ as $m\to 0$. Then, the distribution of $Z$ is dominated by its fluctuations (along the entire system length) and not by its mean.

It was shown in \cite{DobrinevskiLeDoussalWiese2011} that in this phase, $Z(w)$ tends (as $m\to 0$) to a Gaussian stochastic process in the complex plane, as a function of $w$. The mean of the limiting process is zero, and its two-point correlation functions (which determine the process uniquely, since it is Gaussian) are
\begin{multline}
\nn
\left( 
\begin{array}{cccc}
	\overline{Z(w)Z(w)} & \overline{Z(w)Z^*(w)} & \overline{Z(w)Z(w')} & \overline{Z(w)Z^*(w')} \\
	\overline{Z^*(w)Z(w)} & \overline{Z^*(w)Z^*(w)} & \overline{Z(w)Z(w')} & \overline{Z(w)Z^*(w')} \\
	\overline{Z(w')Z(w)} & \overline{Z(w')Z^*(w)} & \overline{Z(w')Z(w')} & \overline{Z(w')Z^*(w')} \\
	\overline{Z^*(w')Z(w)} & \overline{Z^*(w')Z^*(w)} & \overline{Z^*(w')Z(w')} & \overline{Z^*(w')Z^*(w')}
\end{array}
\right) = \\
\label{eq:KPZCxToyPh3SecondMom}
= m\sqrt{\frac{\beta}{2\pi}}\left( 
\begin{array}{cccc}
	q_+  & q_-  & q_+ f(\hat{w}) & q_- f(\hat{w}) \\
	q_-  & q_+  & q_- f(\hat{w}) & q_+ f(\hat{w}) \\
	q_+ f(\hat{w}) & q_- f(\hat{w}) & q_+ & q_-  \\
	q_- f(\hat{w}) & q_+ f(\hat{w}) & q_- & q_+
\end{array}
\right),
\end{multline}
where $\hat{w} := m \sqrt{\beta}(w-w')$, $f(\hat{w}) := e^{-\hat{w}^2/4}$, and $q_\pm$ are constants defined via
\bea
q_\pm := \int_{-\infty}^{\infty} \rmd x'\,\overline{e^{-\beta V(x)+ i\theta(x) - \beta V(x') \pm i\theta(x')}},
\eea
where $x$ is an arbitrary reference point. The finiteness of the $q_\pm$ is a necessary, but not a sufficient condition for the system to be in phase III (see \cite{DobrinevskiLeDoussalWiese2011} for details).

Using the Gaussianity of $Z$ and eq.~\eqref{eq:KPZCxToyPh3SecondMom} for its second moments, one can compute the effective disorder correlators $\Delta$ defined in \eqref{eq:KPZCxToyDefDelta} \cite{DobrinevskiLeDoussalWiese2011}. The expressions for the particularly interesting case $q_+ = 0$ (i.e. when the distribution of $Z$ is rotationally invariant in the complex plane), one obtains
\bea
\nn
\Delta_{V}(w) =& -\frac{m^2}{4\beta}\left[\frac{ \hat{w}^2}{e^{\frac{1}{2}\hat{w}^2}-1} + \log{\left(1-e^{-\frac{1}{2}\hat{w}^2}\right)}\right], \\
\label{eq:GaussianDeltaWW}
\Delta_{\theta}(w) =& -\beta\frac{m^2}{4} \log{\left(1-e^{-\frac{1}{2}\hat{w}^2}\right)},
\eea
or equivalently
\bea
\nn
\Delta_{ZZ}(w) :=& \Delta_V(w) - \beta^{-2} \Delta_\theta(w) =  -\frac{m^2}{4\beta}\frac{ \hat{w}^2}{e^{\frac{1}{2}\hat{w}^2}-1}, \\
\label{eq:GaussianDeltaZZs}
\Delta_{ZZ^*}(w) :=&\Delta_V(w) + \beta^{-2} \Delta_\theta(w) = -\frac{m^2}{4\beta}\left[\frac{ \hat{w}^2}{e^{\frac{1}{2}\hat{w}^2}-1} + 2 \log{\left(1-e^{-\frac{1}{2}\hat{w}^2}\right)}\right]\ .\qquad 
\eea
The correlator $\Delta_{ZZ^*}$ has a distinctive logarithmic singularity near $w=0$, which arises due to a finite density of zeroes of $Z$ on the $w$ axis \cite{Derrida1991a,DobrinevskiLeDoussalWiese2011}. 
The asymptotic Gaussianity of $Z$, and eq.~\eqref{eq:KPZCxToyPh3SecondMom} for its second moments, were checked explicitly in \cite{DobrinevskiLeDoussalWiese2011} for two examples:
\begin{itemize}
	\item A purely imaginary noise which is a Brownian, $V=0$ and $\overline{\left[\theta(x)-\theta(x')\right]^2}\propto |x-x'|$.
	\item A purely imaginary noise which is short-range correlated, $V=0$ and $\theta(x)$ independent for each $x$, and uniformly distributed on $[0;2\pi[$.
\end{itemize}
The calculation performed for these examples in \cite{DobrinevskiLeDoussalWiese2011} is based on the replica formalism. In this formalism, the dominance of fluctuations over the mean of $Z$ in this phase manifests itself as a binding of replicas in pairs.

\subsubsection{Low-temperature phase II}
For a sufficiently strong potential $V(x)$, the
modulus of the integrand in \eqref{eq:KPZCxToyZ} has a very broad
distribution. Then, the central-limit theorem does not hold. Instead, $Z$ is dominated by a few points, the minima of
$V(x)$. This so-called \textit{frozen} phase has been extensively
studied in the absence of random phases (see e.g. \cite{Derrida1981}). 
Distributions of $V$ where a frozen phase occurs in the model \eqref{eq:KPZCxToyZ} in the absence of complex phases include:
\begin{itemize}
	\item Long-range correlated random potentials $V(x)$, i.e.\
	$\overline{V(x)V(x')}=\sigma |x-x'|$. This is known as the  \textit{Sinai model}, which describes the diffusion of a random walker in a 1D random static force field \cite{Sinai1982,LeDoussalMonthus2003}.
	\item Short-range correlated random potentials $V(x)$, i.e.\
	$\overline{V(x)V(x')}=\sigma \delta(x-x')$, where the
	amplitude is rescaled logarithmically with the system size, or
	$m$: $\sigma \sim -\log m$. Freezing occurs  below some
	critical temperature, $\beta > \beta_c$, analogously to  the random energy model \cite{Derrida1981}.
\end{itemize}
Among the most interesting features of the frozen phase is the appearance of jumps between distant minima of $V(x)$ as the position $w$ of the harmonic well in \eqref{eq:KPZCxToyZ} is varied \cite{LeDoussal2006,LeDoussalWiese2006a,LeDoussalMiddletonWiese2009,LeDoussalWiese2008c}. 
In \cite{DobrinevskiLeDoussalWiese2011}, it was discussed how the structure of these shocks is changed upon introduction of random complex phases $\theta(x)$, following the standard treatment \cite{BernardGawedzki1998,LeDoussal2009} for the case without random phases.

The general picture is simple: In the frozen phase, for almost all values of $w$, there is a single minimum of the effective energy $V(x) + \frac{m^2}{2}(x-w)^2$, which dominates $Z$. If random phases are added, $Z$ will have precisely the phase at this point, i.e. $e^{i\theta(x)}$. At a few singular points $w^*$, contributions from two minima become equal. When moving $w$ through such a point, a shock occurs and the dominant position $x$ jumps from one minimum to the other one. 
Let us consider such a shock between two minima $x_1$, $x_2$ with energies $V_1:=V(x_1), V_2:= V(x_2)$ and phases $\theta_1:=\theta(x_1)$, $\theta_2:=\theta(x_2)$. One finds that in the limit $\beta \to \infty$ (i.e. strong amplitude disorder), a sharp shock occurs at the position $w^*$, where the parabolic effective potentials for a particle pinned at $x_1$ or $x_2$ meet:
\bea
\label{eq:KPZCxToyDefShockW}
V_1 + \frac{m^2}{2}(x_1-w^*)^2 = V_2 + \frac{m^2}{2}(x_2-w^*)^2.
\eea
One then finds that the effective potential \eqref{eq:KPZCxToyDefEffPot} can be written in terms of the jump
size $s:=\beta m^2 (x_2-x_1)$, the  phase difference, $\phi: =
\theta_2 - \theta_1$, and $w^{*}$, solution of \eqref{eq:KPZCxToyDefShockW}:
\begin{eqnarray}
\hat{\theta}'(w) &=& \frac{s}{2} \frac{\sin(\phi)}{\cos(\phi) + \cosh\left [ s (w-w^*)\right]}, \\
\label{eq:Ph2EffPot}
 -\hat{V}'(w) &= & \frac{s}{2\beta}\frac{\sinh\left [ s (w-w^*) \right]}{\cos(\phi) + \cosh\left[ s (w-w^*) \right]} + \frac{m^2}{2}(x_1+x_2-2w).
\end{eqnarray}
Observe that a pole arises in $\hat{V}'(w)$, $\hat{\theta}'(w)$ near $w=0$ when $\phi \to \pi$, i.e. when $Z$ jumps between opposite signs. This happens since $Z$ passes close to $0$ during such a jump.

An explicit calculation of the effective disorder correlator $\Delta$ is now more complicated, since it requires knowledge of the joint distribution $P(s,\phi)$ of shock sizes $s$ and phase jumps $\phi$ (some special cases can be found in \cite{DobrinevskiLeDoussalWiese2011}). However, one observes quite generally \cite{DobrinevskiLeDoussalWiese2011} that a logarithmic singularity arises in $\Delta_{ZZ^*}(w)$ near $w=0$, related to the probability density of jumps between opposite signs\footnote{$P(\phi)$ is the probability density for the phase jump angle $\phi$, i.e. it is supposed to be normalized as $\int_{-\pi}^{\pi}\rmd \phi\, P(\phi)=1$.},
\bea
\Delta_{ZZ^*}(w) = -2\pi P(\phi = \pm \pi) \log w + \mO(w)^0.
\eea
As in the case of phase III \eqref{eq:GaussianDeltaZZs}, $\Delta_{ZZ}$ is still regular. The logarithmic singularity in $\Delta_{ZZ^*}$ for $\phi \approx \pm \pi$ arises due to the poles in \eqref{eq:Ph2EffPot}, i.e. due to $Z$ being close to $0$, for these values of $\phi$.

\subsubsection{High-temperature phase I}
Finally, let us discuss the high-temperature phase I. As mentioned above, in this phase the modulus of the integrand in \eqref{eq:KPZCxToyZ} fluctuates little, and the central-limit theorem holds. In contrast to phase III, in the limit $m\to 0$, the fluctuations of $Z(w)$ around its mean $\overline{Z}$ are small. The disorder can be treated perturbatively, and one obtains regular disorder correlators \cite{DobrinevskiLeDoussalWiese2011}
\bea
\label{eq:HighTCorr}
\Delta_{V} (w) = & \sigma_V \frac{(\beta m^2)^{\frac{3}{2}}}{8\sqrt{\pi}}  \left(2-\hat{w}^{2}  \right)e^{-\frac{\hat{w}^2}{4}}, \\ 
\Delta_{\theta} (w) = & \sigma_\theta \frac{(\beta
m^2)^{\frac{3}{2}}}{8\sqrt{\pi}}  \left(2-\hat{w}^{2}  \right) e^{-\frac{\hat{w}^2}{4}}.
\eea
Here, $\hat{w}=m\sqrt{\beta}w$, and we assumed $\overline{V(x)V(x')} = \sigma_V \delta(x-x')$, $\overline{\theta(x)\theta(x')} = \sigma_\theta \delta(x-x')$. The correlators $\Delta$ in \eqref{eq:HighTCorr} are regular, and do not exhibit cusps or logarithmic singularities near $w=0$. Thus, no shocks or zeros of $Z$ appear in this phase.

\subsection{Summary}
In this section \ref{sec:KPZComplex} we saw that the model of directed polymers with complex random weights is of relevance for the study of quantum disordered systems. We saw that in some cases ($\dsf=1$, or the frozen phase II in mean-field limit), the addition of random phases does not modify the dominant contributions to the partition sum. On the other hand, in some cases (phase III in the mean-field limit or in the $\dsf=0$-dimensional toy model of section \ref{sec:KPZCxToy}) they lead to interesting new behaviour.

We also saw by explicit calculations in a toy model, that the renormalized disorder correlator $\Delta$ provides some insight into the fluctuations of $Z$, in particular the density of zeros of $Z$ which lead to a logarithmic singularity. As we saw in section \ref{sec:FRGReview}, $\Delta$ is the central object of the field theoretic treatment of disordered systems based on the functional renormalization group \cite{LeDoussal2006,WieseLeDoussal2006}. The present results may help to understand how to perform a functional RG treatment including complex phases, and thus to understand better disordered systems with quantum fluctuations.

Even for the simple model of a directed polymer with complex random weights, many important questions are still open. For example, it is not clear whether phase III can be observed in any finite dimension, and how it can be interpreted physically (e.g. for the application to variable-range-hopping discussed in section \ref{sec:KPZCxVRH}).
Furthermore, as we saw in section \ref{sec:KPZOverview} above, even for the directed polymer with positive weights the nature of the frozen phase, and the phase transition, are poorly understood for $1 < \dsf < \infty$. Understanding the influence of complex weights in this phase, either random or deterministic (important for understanding magnetoconductance, as discussed in section \ref{sec:KPZCxVRH}), is an interesting question. Lastly, for disordered electronic systems many-body effects can not be neglected in general, and yield interesting novel behavior (such as many-body localization, see \cite{BaskoEtAl2006,PalHuse2010} and many others). Understanding this interplay of disorder and strong interactions is an even more challenging avenue for further research.

%% file: Summary.tex
\chapter{Summary and Outlook\label{sec:Summary}}
In this thesis I discussed existing and new results on the theory of elastic interfaces in disordered media. As elaborated in the introduction (chapter \ref{sec:Introduction}), such models can be applied to describe and understand magnetic domain walls, crack fronts in fracture, fluid contact lines, and other experimentally relevant systems. We saw how quenched disorder leads to stochastic fluctuations on large scales, in particular a self-affine roughening of the interface. We also saw that the response to slow external loading proceeds not smoothly, but in avalanches, which have broad power-law distributions of sizes and durations. For a certain class of random media -- the mean-field case where the increments of the pinning force are independent -- we saw in section \ref{sec:BFM} how an exact solution for the avalanche statistics can be derived. In section \ref{sec:OneLoop}, we discussed realistic, short-range-correlated disorder. We identified the critical internal dimension of the interface $d=d_\mathrm{c}$, at and above which mean-field theory is valid. Below the critical dimension, we showed how the distributions of avalanche sizes, durations, and their average shape, can be computed in a controlled expansion in $\epsilon =d_\mathrm{c}-d$.

In both cases, we observed that different observables have different degrees of universality. Some observables -- such as the power-law exponents of the quasi-static avalanche size and duration distributions, and the scaling form of the average avalanche shape for short avalanches -- are extremely robust. They depend only on the internal dimension of the interface and the existence of long-ranged interactions or correlations, but not on the details of the large-scale cutoffs or the dynamics. For example, in the mean-field case, we observed in sections \ref{sec:BFMRetStat} and \ref{sec:BFMRetSize} that adding retardation effects does not modify the power-law exponent of the velocity distribution at quasi-static driving, and the average avalanche shape for short avalanches. 
Large avalanches also exhibit universality. The scaling forms computed in sections \ref{sec:OneLoopDurations} and \ref{sec:OneLoopShapeSize} for the avalanche duration and size distributions include not just the power-law regime for small avalanches, but also a nontrivial shape of the cutoff for large avalanches. It is universal in the limit $m\to 0$ for the case where the large-scale cutoff is a harmonic well with curvature $m^2$ (as in \eqref{eq:InterfaceEOM}), but will be different for a different type of cutoff (e.g. a finite system size). The conclusions for the average avalanche shape discussed in sections \ref{sec:OneLoopShapeTime} and \ref{sec:OneLoopShapeSize} are similar. The limiting forms of the average shape (at fixed duration, or at fixed size) for short avalanches are as universal as the power-law exponents for small avalanches of the size and duration distributions, and not cutoff-dependent. On the other hand, the average shape for long avalanches is cutoff-dependent. Presumably, modifying the dynamics of the interface (e.g. including retardation as discussed for the mean-field case in section \ref{sec:BFMRetardation}) would also modify the average shape for long avalanches, but not for short ones (though beyond the mean-field shape in section \ref{sec:BFMRetSize} this remains a conjecture to be verified).

There are many directions in which the present work can be extended. As observed previously for the roughness exponent $\zeta$ and for the universal effective disorder correlator $\Delta$, renormalization-group calculations to two-loop order can yield qualitatively new results compared to one-loop calculations \cite{ChauveLeDoussalWiese2000a,LeDoussalWieseChauve2002}. In particular, the difference between the roughness exponent for the static ground state and quasi-static depinning is only observed starting at two-loop order \cite{ChauveLeDoussalWiese2000a,LeDoussalWieseChauve2002}. Thus, it would be very interesting to extend the one-loop calculations for the avalanche size and duration distributions, as well as average avalanche shapes (presented in chapter \ref{sec:OneLoop}) to two loops. This is technically challenging, but will allow to understand the following questions: 
\begin{itemize}
	\item Do the scaling relations discussed in section \ref{sec:FRGScalingRelations}, which allow to express avalanche exponents in terms of interface roughness and dynamical exponents $\zeta$ and $z$, hold for quasi-static depinning, static ground states, or both?
	\item Is the form of the avalanche-size distribution different for static ``avalanches'', i.e. shocks between different ground states, and dynamical avalanches at quasi-static depinning?\footnote{Note that avalanche durations and shapes have no equivalent for the statics, hence cannot be compared.}
\end{itemize}
A further theoretical challenge suggested by the present work is the analysis of avalanches at a finite driving velocity. On the mean-field level, this does not pose problems and is discussed in chapter \ref{sec:BFM}. However, beyond mean-field, the problem is more difficult. Formally, one would need to repeat the calculations performed in sections \ref{sec:OneLoopDurations} - \ref{sec:OneLoopLongRange} including not only the diagrams starting with a single vertex at time $\ti$, but with two or more. This is cumbersome but certainly feasible. On the other hand, one would also need to understand the renormalization-group flow of the effective disorder correlator $\Delta$ at finite velocity. One attempt for doing that is considering the flow for $m\to 0$ at a fixed velocity \cite{ChauveGiamarchiLeDoussal2000}, which leads to complicated integro-differential equations. However, this may not be the best-adapted way of tackling the problem. It is known that a finite driving velocity introduces a finite correlation length $\xi$ of the interface \cite{DuemmerKrauth2005}, which for $m\to 0$ will be smaller than the correlation length $m^{-1}$ introduced by the mass. $\xi$ will remain finite even for $m=0$, as long as $v$ is strictly positive. It thus seems promising to look for an RG scheme with an RG scale $\mu$ containing both $v$ and $m$, and controlling the correlation length $\xi$ in such a way that $\mu \to 0 \Leftrightarrow \xi \to \infty$. In any case, a positive driving velocity $v$ will likely have some effect on the fixed point $\Delta_*$ of the renormalized disorder correlator, which needs to be understood in order to make predictions on the avalanche statistics. One simplifying aspect to keep in mind is that the Middleton theorem discussed in section \ref{sec:InterfaceMonot} still holds: Independently of the driving velocity, the interface traverses the same set of leftmost metastable states in the same order as $w$ is varied monotonously. This fact allowed an exact solution of the mean-field case at arbitrary monotonous driving, and carries over to short-range correlated disorder.

It is also important to extend the present discussion of elastic interfaces with a single-component height function, and directed polymers with a single internal dimension and $N$ components, to the general case of elastic manifolds with $d$ internal dimensions and $N>1$ components. This is particularly important for applying the theory to vortex lattices in superconductors, cf.~section \ref{sec:OtherExp}, which correspond to $d=2$ and $N=2$. There is some progress in understanding the FRG theory for the renormalized disorder correlator for the statics of an $N$-component interface \cite{LeDoussalWiese2005}. Still, a straightforward extension of the methods discussed here to study avalanches at depinning poses some difficulties, since the Middleton monotonicity property no longer holds, and avalanches have a more complicated structure (e.g.~one has to distinguish between avalanche sizes parallel to the displacement, and transverse to it).

Another important generalization, as mentioned in section \ref{sec:IntroInterfaceModel}, is the inclusion of thermal fluctuations. These lead to new, intriguing phenomena like thermally activated creep motion below the depinning threshold \cite{FeigelmanGeschkenbeinLarkinVinokur1989,BlatterGeshkenbeinVinokur1991,BlatterEtAl1994,LeDoussalVinokur1995,LemerleEtAl1998,ChauveGiamarchiLeDoussal2000,MuellerGorokhovBlatter2001,KoltonRossoGiamarchi2005,KoltonRossoGiamarchiKrauth2006,KoltonRossoGiamarchiKrauth2009}, which are also important for applications such as fracture \cite{Ponson2008,Ponson2009,PonsonBonamy2010}. They also leads to a universal thermal rounding of the depinning transition \cite{Middleton1992a,BustingorryKoltonGiamarchi2012b,BustingorryKoltonGiamarchi2008,BustingorryKoltonGiamarchi2012}.
Adding thermal noise in the equation of motion \eqref{eq:InterfaceEOM} breaks the monotonicity properties discussed in section \ref{sec:InterfaceMonot}, and makes analytical calculations much more difficult. It has been found that thermal fluctuations lead to a rounding of the cusp in the effective disorder correlator \cite{ChauveGiamarchiLeDoussal2000,LeDoussal2006}, and the temperature dependence of the pinning length has been obtained \cite{MullerGorokhovBlatter2001}, but the effects on avalanches remain to be investigated. In particular, it would be interesting to see if there are differences between thermally initiated avalanches during creep motion, and the avalanches at depinning considered here.
Similar questions can be posed for quantum fluctuations, which are important especially for vortex lattices in high-temperature superconductors (cf.~section \ref{sec:OtherExp}). They lead to novel phenomena such as quantum creep \cite{BlatterGeschkenbein1993} and quantum vortex lattice melting \cite{BlatterIvlev1993}. Some first steps towards the inclusion of thermal and quantum fluctuations in the present dynamical field theory formalism on the mean-field level have been taken in \cite{LeDoussalPetkovicWiese2012}. However, the validity of the approximations performed therein, and the extension beyond mean-field needs further study.

Another important goal for future research is establishing a closer connection to experiments. As we saw in section \ref{sec:Introduction}, some experimental observations show promising agreement with the results of the elastic interface model (e.g.~the power-law exponents for avalanche size and duration distributions in Barkhausen noise, and the roughness exponent of a brittle crack front). However, experimental data are still lacking, or need to be revisited, for comparison to many refined theoretical results (such as the form of the avalanche size and duration distributions, or the avalanche shapes at fixed avalanche size). Likewise, for verifying the analytical predictions made here, numerical simulations of the avalanche duration distribution, average avalanche shape, and related observables, in finite dimensions $1\leq d \leq 3$, are very important. Some work in this direction is in progress \cite{KoltonEtAl2013inpr}. On the other hand, from the theory side, some interesting observables need yet to be understood. One example is the statistics of the connected \textit{clusters}, in which an avalanche decomposes when elasticity is long-ranged. This is of relevance for comparison with fracture experiments (see section \ref{sec:Fracture}). It is also interesting to consider generalizations of the elastic interface model. For example, the phase field model with conserved dynamics is a better description for fluid invasion in porous media (see \cite{DubeEtAl1999,DubeEtAl2000,RostLaursonDubeAlava2007} and section \ref{sec:OtherExp}). Some of the methods developed here can possibly be generalized to elucidate its avalanche statistics.

Avalanches are also observed in another classical model of disordered systems, the \textit{random-field Ising model (RFIM)} (for a review, see \cite{Nattermann1998}). Its main difference to elastic interface models discussed in this thesis is the presence of \textit{nucleation}. This is important e.g. for describing Barkhausen noise in hard magnets, as mentioned in section \ref{sec:Barkhausen}. Some theoretical results on avalanches in the RFIM are available, indicating that (in contrast to the elastic interface models discussed here) large avalanches only occur when the disorder strength and the external magnetic field are tuned to be near a critical point \cite{DahmenSethna1996}. This critical point, known from the study of the equilibrium phase diagram of the RFIM, separates the ferromagnetic phase (where disorder is sufficiently weak, and spins align coherently in the same direction at zero temperature) from the paramagnetic phase (where disorder dominates over the ferromagnetic interaction, and at zero temperature spins typically align in the direction of the local random field) \cite{SchneiderPytte1977,Nattermann1998}. In the mean-field limit, the distribution of avalanche sizes $S$ in the RFIM is very similar to the one for mean-field elastic interfaces (i.e. the ABBM model): It has a $S^{-3/2}$ power law for small $S$, and decays exponentially for large $S$ \cite{DahmenSethna1996,SabhapanditEtAl1999}. Perturbative calculations for the the avalanche-size distribution, and for the avalanche shape beyond the mean-field limit are not yet available. Numerical simulations show an average avalanche shape similar to that of an elastic interface, and a non-trivial, fractal, spatial avalanche structure \cite{KuntzEtAl1999,MehtaMillsDahmenSethna2002,SethnaDahmenMyers2001}. Understanding better the common points and the differences between avalanches in the RFIM and in an elastic interface model is an important question for further work.

Lastly, a more fundamental problem is understanding the rough phase of the Kardar-Parisi-Zhang equation as discussed in chapter \ref{sec:KPZ}. Although this question does not have an immediate experimental motivation, it is a tantalizingly simple model whose solution has resisted all standard theoretical methods for over 25 years. New ideas and methods for studying systems with quenched disorder are required to systematically approximate its behaviour in the strong-coupling phase, in particular the roughness exponent, and to clarify whether an upper critical dimension exists. Likewise, the interplay between quantum fluctuations and disorder, which has only been briefly touched upon in this thesis, deserves further study.

%% file: AppendixNotations.tex
\chapter{Appendix}
\section{Notations\label{sec:AppendixNotations}}

Constants
\bea
\gamma = \gamma_E = 0.577216
\eea
Integrals
\bea
\int_x := \int\rmd^d x,\quad\quad\quad \int_p := \int\frac{\rmd^d p}{(2\pi)^d}.
\eea
Fourier Transforms
\bea
f(q) := \int \rmd^d x\, e^{-i q x}f(x),\quad\quad\quad f(x) = \int \frac{\rmd^d q}{(2\pi)^d}\,e^{i q x}f(q). 
\eea
Dynamical averages
\bea
\nn
\overline{O[\du,\tu]}^S :=& \int \mD[\du,\tu]\,e^{-S[\du,\tu]}O[\du,\tu],\\
\overline{O[\du,\tu]}^{\tree}:=& \int \mD[\du,\tu]\,e^{-S_{\tree}[\du,\tu]}O[\du,\tu], \quad\quad \text{with } S_{\tree} \text{ given in }\eqref{eq:IntVelActionCT1}
\eea
The one-loop correction $\delta_1 \overline{O}$ is defined in \eqref{eq:OneLoopDeltaG}. The one-loop correction including counterterms $\delta_c \overline{O}$ is given in \eqref{eq:OneLoopGenObs}:
\bea
\delta_c \overline{O} = \delta_1 \overline{O} + c_\eta (O) + c_\sigma (O) + c_m(O).
\eea
Other abbreviations:
\bea
b(\lambda) := \sqrt{1-4\lambda}.
\eea

\section{The BFM position theory\label{sec:BFMPositionTheory}}
One way to realize the BFM disorder \eqref{eq:BFM} is to take the force $F(u)$ to be a true, non-stationary Brownian starting at $u=0$ at each point in space. This amounts to setting
\bea
\Delta(u,u') = 2\sigma \min(u,u'),
\eea
and considering positive $u$ only. This automatically ensures that $u=0$ is a Middleton state. 
Another way is a ``stationary Brownian'', satisfying 
\bea
\Delta(u-u') \approx \Delta(0) - \sigma |u-u'|
\eea
for small $u$ and periodized on a cutoff scale $u \sim \Delta(0)/\sigma$. Then the BFM is valid on small scales $u \ll \Delta(0)/\sigma$. Here, a starting Middleton state needs to be prepared by pulling monotonously for a sufficiently long distance. For the velocity theory, both choices are equivalent and both lead to the equation of motion \eqref{eq:IntVelEOMBFM} for the interface velocity. However, when considering the interface position $u_{xt}$, the two choices lead to two different values for the average pinning force $\overline{u - w}$. This is discussed in detail in \cite{DobrinevskiLeDoussalWiese2012}, section V, where the distribution of pinning forces in both cases is obtained using functional determinants. In the following, we will focus on the simpler velocity theory, which is sufficient for the study of avalanches, which are unaffected by the mean pinning force.

\section{Average hysteresis loop of the ABBM model, perturbatively\label{sec:AppABBMHystPert}}
In this appendix I revisit the calculation of the backward branch of the quasi-static hysteresis loop of the ABBM model, which was done in section \ref{sec:HystABBMQS1Pt}. I compute the average $\overline{u(w)}$ perturbatively in the disorder, and show that this perturbative calculation only captures the asymptotics when $w$ is far away from the turning point; it does not capture the nontrivial shape near the turning point.

Let us return to the overdamped equation of motion \eqref{eq:BFMABBM},
\bea
\partial_t u(t) = -u(t) + w(t) +F(u(t)),
\eea
where $\overline{F(u)F(u')} = \Delta(u,u') = 2\sigma \min(u,u')$.
It can be solved iteratively:
\bea
\nn
u(t) =& \int_{-\infty}^t \rmd t_1\, e^{-(t-t_1)}\left[w(t_1)+F(u(t_1))\right]\\
= &  \int_{-\infty}^t \rmd t_1\, e^{-(t-t_1)}\left[w(t_1)+F\left(\int_{-\infty}^{t_1}\rmd t_2\,e^{-(t_1-t_2)}\left[w(t_2)+F(u(t_2))\right]\right)\right] = ...
\eea
Let us now expand in powers of the pinning force $F$ (or, equivalently, in powers of $\sqrt{\sigma}$). 
\bea
\nn
u(t) =& u^{(0)}(t) + u^{(1)}(t) + u^{(2)}(t) + \mO(\sigma)^{3/2}, \\
\nn
u^{(0)}(t) = & \int_{-\infty}^t \rmd t_1\, e^{-(t-t_1)}w(t_1), \\
\nn
u^{(1)}(t) = & \int_{-\infty}^t \rmd t_1\, e^{-(t-t_1)}F\left(u^{(0)}(t_1)\right), \\
\nn
u^{(2)}(t) = & \int_{-\infty}^t \rmd t_1\, e^{-(t-t_1)}F'\left(u^{(0)}(t_1)\right)\int_{-\infty}^{t_1}\rmd t_2\,e^{-(t_1-t_2)}F\left(u^{(0)}(t_2)\right).
\eea
Taking the disorder average, $u^{(1)}$ does not contribute (since $F$ has mean zero). Thus, the average particle position is
\bea
\label{eq:HystPertUAvg}
\overline{u(t)} = u^{(0)}(t) +\int_{-\infty}^t \rmd t_1\, e^{-(t-t_1)}\int_{-\infty}^{t_1} \rmd t_2\, e^{-(t_1-t_2)}\Delta'\left(u^{(0)}(t_1),u^{(0)}(t_2)\right) + \mO(\sigma)^{3/2}.
\eea
Now, let us choose a driving trajectory $w(t)$ such as in section \ref{sec:HystABBMQS1Pt}: A driving which slowly increases to $w=w_m$, and then slowly returns to $0<w<w_m$. We set
\bea
w(t) = \begin{cases}
v t & \text{for}\quad t<t_m \\
w_m - v(t-t_m) & \text{for} \quad t> t_m
\end{cases},
\eea
where $vt_m = w_m$. For small $v$, we have on both branches $u^{(0)}(t) = w(t) + \mO(v)$. Thus, for the ABBM choice of $\Delta$, we get from \eqref{eq:HystPertUAvg}
\bea
\overline{u(w)} = w + 2\sigma\int_{-\infty}^t \rmd t_1\, e^{-(t-t_1)}\int_{-\infty}^{t_1} \rmd t_2\, e^{-(t_1-t_2)}\theta\left(w(t_2)-w(t_1)\right) + \mO(\sigma)^{3/2}.
\eea
Now, for $t<t_m$ we have $w(t_1)> w(t_2)$ and hence $\theta\left(w(t_2)-w(t_1)\right)=0$. So, $\overline{u(w)}=w$ on the forward branch, in agreement with the result from \eqref{eq:Hyst1PtForward}. On backward branch, for $t > t_m$, we have $\theta\left(w(t_2)-w(t_1)\right)=1$ only for $t_2 \in [2t_m-t_1;t_1]$. Thus, the average shape of the backward branch is
\bea
\nn
\overline{u(w)} = & w + 2\sigma\int_{-\infty}^t \rmd t_1\, e^{-(t-t_1)}\int_{2t_m-t_1}^{t_1} \rmd t_2\, e^{-(t_1-t_2)} + \mO(\sigma)^{3/2} \\
= & w  +2\sigma\left(1-2e^{-(t-t_m)}+e^{-2(t-t_m)}\right).
\eea
On the backward branch, we have $t-t_m = (w_m-w)/v$. Hence, the exponential terms vanish quickly in the limit $v \to 0$, and the quasi-static result is
\bea
\overline{u(w)} = w + 2\sigma.
\eea
The exact result, obtained in \eqref{eq:Hyst1PtBackwardExact}, is
\bea
\overline{u(w)} = w  + 2\sigma + \frac{2(w-w_m)}{e^{(w_m-w)/\sigma}-1}.
\eea
We see that the perturbative calculation reproduces correctly the $\mO(\sigma)$ part of the exact result (which corresponds to the asymptotic value of the pinning force on the backward branch of the hysteresis loop). However, the perturbative calculation fails to capture the nontrivial, and non-perturbative in $\sigma$, shape of the hysteresis loop near the turning point.